\documentclass[letterpaper,12pt]{report}
\usepackage{tabularx} 
\usepackage{amsmath}  
\usepackage{graphicx} 
\usepackage{subcaption}
\usepackage[margin=1in,letterpaper]{geometry} 
\usepackage{cite} 
\usepackage{braket}
\usepackage[utf8]{inputenc}
\usepackage[greek,english]{babel}
\usepackage{alphabeta}
\usepackage{array} 
\usepackage{appendix}
\usepackage{float}
\usepackage{amssymb}
\usepackage{slashed}

\usepackage[final]{hyperref} 
\hypersetup{
	colorlinks=true,       
	linkcolor=blue,        
	citecolor=blue,        
	filecolor=magenta,     
	urlcolor=blue         
}
\title{\Huge{\textbf{Spontaneous Symmetry Breaking: From the Effective Action to Cosmological Phase Transitions in the Standard Model and Beyond\\}}}

\author{AUTHOR\\ \textbf{Apostolos Giovanakis}\thanks{agiovana@auth.gr}\\\\SUPERVISOR\\\textbf{Vasilis K. Oikonomou}\thanks{voikonomou@gapps.auth.gr}\\\\School of Physics\\\\Aristotle University of Thessaloniki}
\date{}

\begin{document}

\maketitle

\newpage

\pagenumbering{arabic}

\newpage

\begin{abstract}
The primary objective of this thesis is to investigate the cosmological phase transitions in the early Universe, with a focus on the electroweak phase transition in the Standard Model and its extensions. In the Standard Model, the spontaneously broken electroweak symmetry at zero temperature is restored in the early Universe due to finite-temperature effects. This phenomenon is studied using the effective potential at finite temperatures, which determines the true vacuum state of the theory, including quantum corrections. More specifically, the effective potential is initially defined by the effective action, which is the generating functional for the one-particle irreducible Feynman diagrams. Then, the background field method is introduced to provide a computational tool for the effective potential. Alternatively, the one-loop effective potential at zero temperatures is explicitly derived in non-Abelian gauge theories using the one-particle irreducible Feynman diagrams and can be later renormalized in different renormalization schemes. Moreover, symmetry restoration at high temperatures is studied by the finite-temperature field theory introduced to derive the Feynman rules at finite temperatures using the imaginary-time formalism. This formalism is applied to compute the one-loop effective potential at finite temperatures in various quantum field theories. As a result, the effective potential shows the symmetry restoration in the Standard Model and electroweak phase transition. However, perturbation theory breaks down around the critical temperature and thermal resummation is required to incorporate the dominant contribution of the ring diagrams to the one-loop effective potential. Furthermore, we present the theory of cosmological phase transitions, focusing on the concepts of thermal tunneling and bubble nucleation. We additionally discuss the observed baryon asymmetry of the Universe to formulate the conditions for baryogenesis, describe electroweak baryogenesis, and obtain the criterion for a strong first-order phase transition. Therefore, the one-loop effective potential in the Standard Model at finite temperatures is derived in detail including the ring corrections to study further the electroweak baryogenesis. Our results indicate that the electroweak phase transition is not strong enough to explain the observed baryon asymmetry of the Universe. On the other hand, the real singlet extensions of the Standard Model describe a strong enough electroweak phase transition and the observed baryon asymmetry of the Universe. These extensions are also discussed including a dimension-six operator, which originates from an effective field theory at a new physics scale. In particular, the parameter space of this singlet extension is examined extensively, while it is restricted by numerous phenomenological constraints, such as the invisible Higgs decay width. As a result, due to the presence of the higher-order operator, the real singlet extensions describe a strong electroweak phase transition in regions of the parameter space that were previously eliminated in the literature.

\end{abstract}

\tableofcontents

\newpage

\chapter*{Acknowledgments}
First and foremost, I would like to express my deepest gratitude to my supervisor, Prof. V.K. Oikonomou, who believed in me and introduced me to such a fascinating and multidimensional area of research. His suggestions and guidance sparked my interest and enthusiasm, leading me to delve deeper into the theory of cosmological phase transitions and finite-temperature field theory, which proved to be captivating and rewarding research areas. I am also grateful that he offered me the great opportunity to work on a research project, allowing me to enjoy the adventures and explore the challenges in research. Additionally, I extend my sincere appreciation to Fotis Fronimos for our insightful conversations on quantum field theory, cosmology, and physics in general. During the research project presented in the last chapter, sharing my ideas with Fotis was an invaluable assistance that enriched my understanding on numerous occasions. I am also immensely grateful to my family for their constant encouragement, understanding, and patience during this journey as they have been my source of strength. Last but not least, I am deeply thankful to my friends and close colleagues for their support and discussions throughout this difficult endeavor.

\chapter{Introduction}
The Standard Model is currently the most accepted and successful theory of physics which describes the elementary particles of our Universe and their interactions. More specifically, it describes three out of the four known fundamental interactions of nature: the electromagnetic, weak, and strong interactions, while it leaves gravity outside its scope. Each fundamental interaction in the Standard Model is related to a Lie group and the Standard Model is based on an \(SU(3) \times SU(2) \times U(1)\) gauge group. 

The electroweak theory in the Standard Model is a quantum field theory that possesses a \(SU(2) \times U(1)\) gauge symmetry between the weak and electromagnetic interactions. However, in a unified theory of weak and electromagnetic interactions, the gauge invariance of the theory is not reconciled with the addition of mass terms to the Standard Model Lagrangian. On the other hand, it is obvious that all fundamental particles in nature are not massless and it is crucial to introduce the essential mass terms through a mechanism that respects the Standard Model gauge symmetry. This mechanism is called (spontaneous) electroweak symmetry breaking and is associated with a scalar field in the Standard Model, which is called the Higgs field. The vacuum state of the theory is then described by the vacuum expectation value of the Higgs field. As a consequence, the \(W\) and \(Z\) gauge bosons and fermions in the Standard Model acquire a mass proportional to the Higgs vacuum expectation value after the electroweak symmetry breaking. In general, spontaneous symmetry breaking is a fundamental concept in particle physics and condensed matter physics and will be presented in detail in the next sections.

In the early Universe, matter can be described in terms of quantum fields as it was very compressed at very high temperatures. As a result, the Standard Model provides valuable insights into the history of the early Universe and the formation of cosmic structures by explaining the behavior of particles and their interactions under extreme conditions. From understanding the dynamics of the primordial plasma to predicting the abundances of elements produced during Big Bang nucleosynthesis, the Standard Model offers the framework to interpret various cosmological observations and phenomena. More specifically, the Standard Model formulated as a quantum field theory at non-zero temperatures predicts that a symmetry may not be spontaneously broken at high temperatures. For instance, the gauge symmetry of the electroweak theory is not spontaneously broken at high temperatures. In other words, this symmetry is restored at a high temperature which implies that a cosmological phase transition occurred in the early Universe at the end of the electroweak epoch. Additionally, the Standard Model predicts a second cosmological phase transition at a lower temperature called the quantum chromodynamics phase transition associated with the chiral symmetry in quantum chromodynamics.

In the Big Bang theory, the electroweak epoch plays an essential role in the physics of the Universe. First of all, one of the first stages in the history of the Universe is the Planck epoch, in which the current laws of physics may not be valid. The Planck epoch is then followed by the Grand Unification epoch at a temperature around \(10^{32}\) K. The electroweak epoch starts at a temperature around \(10^{28}\) K (\(\simeq 10^{15}\) GeV) and time \(10^{-36}\) seconds and ends at a temperature almost \(10^{15}\) K (\(\simeq 100\) GeV) and time \(10^{-10}\) seconds. Below the critical temperature around \(100\) GeV, the electroweak symmetry is spontaneously broken and the \(Z\) and \(W\) bosons acquire mass, whereas the symmetry is restored at temperatures above this critical temperature and the \(Z\) and \(W\) bosons are massless and the electroweak phase transition starts at the end of the electroweak epoch. In particular, the introduction of the Higgs field in the Standard Model plays an essential role in the cosmological history of the Universe. The vacuum expectation value of this scalar field determines the progress of this cosmological phase transition and accordingly describes the vacuum state of the Universe. Whether or not the vacuum expectation value of the Higgs field vanishes and the vacuum state is invariant under the gauge symmetry of the electroweak theory, the nature of the Universe changes significantly. 

Various field theories showcase that symmetries could be spontaneously broken in the early Universe through a first-order phase transition. This could be probed through its impact on the gravitational-wave spectrum at current and future gravitational-wave observatories as gravitational waves are generated during a first-order phase transition. Subsequently, due to the absence of new particle observations at the Large Hadron Collider (LHC) in the last decade, modern high-energy physics can rely on gravitational wave experiments and astrophysical observations to tackle fundamental issues such as the electroweak phase transition and baryogenesis. Much light on the cosmological history of the Universe is expected be shed by the current and future Cosmic Microwave Background (CMB) and gravitational wave experiments, such as the LISA, DECIGO, and the Einstein telescope. Therefore, the gravitational waves from a first-order phase transition could be a message from the physics of the early Universe.

Furthermore, if the Standard Model could describe a strong enough first-order phase transition in the early Universe, this could explain the abundance of matter rather than anti-matter in the Universe since it provides the non-equilibrium conditions for baryogenesis. This has been accomplished by numerous proposed extensions of the Standard Model. One of the simplest and most studied extensions to the Standard Model is the real singlet extension which solely includes a real singlet scalar field and is presented in detail in the last chapter.

The thermal history of the Universe clearly affects the evolution of the cosmological phase transitions as they depend on the temperature of the Universe. Thus, the phase transitions are studied in the framework of quantum field theories at non-zero temperatures, which is called finite-temperature field theory or thermal quantum field theory. Finite-temperature field theory combines the ordinary quantum field theory with statistical physics. Ordinary quantum field theory describes elementary fields and their interactions in a surrounding vacuum, at zero temperature, like in particle colliders, whereas statistical physics describes the properties of large ensemble systems, through parameters such as pressure, temperature, volume, entropy, etc. Therefore, in the early Universe, the dynamics of quantum fields are described by the finite-temperature field theory which is applied to the theory of cosmological phase transitions.

Finally, it was previously mentioned that the behavior of the vacuum expectation value of a scalar field determines the vacuum of the theory and spontaneous symmetry breaking. Nevertheless, the vacuum expectation value should be computed including radiative corrections. In fact, at the classical level, the state of the lowest energy is obtained by the minimization of the classical potential in the theory. In quantum field theory, however, higher-loop corrections can affect importantly the vacuum expectation value of the scalar field and generate spontaneous symmetry breaking. Consequently, it is essential to define a function that includes the quantum corrections to the classical potential in order to determine the true vacuum of the quantum theory. This function is called the effective potential and its minimum is identified as the expectation value of the scalar field in the true vacuum. It plays a major role in the implications of spontaneous symmetry breaking. Accordingly, the progress of a cosmological phase transition primarily depends on the effective potential at finite temperatures. Thus, the effective potential will be analyzed throughout this thesis in the case of zero temperature and finite temperature in the context of scalar, fermion, and gauge field theories, including the non-Abelian gauge theories.  

\section{Thesis Outline}
The subject of this thesis is the study of cosmological phase transitions focusing on the electroweak phase transition and baryogenesis in the Standard Model and beyond. As a result, this thesis is organized as follows: 
\begin{itemize}
    \item Chapter 1: the concept of spontaneous symmetry breaking and the Higgs mechanism is presented including a brief description of the Standard Model.
    \item Chapter 2: the effective action is introduced implying the mathematical definition of the effective potential. Then, the background field method is presented as an equivalent way to determine the effective potential. The one-loop effective potential is then computed in the case of scalar field theories and non-Abelian gauge theories using the Feynman diagrams. In this chapter, different regularization and renormalization schemes are also discussed in the context of the ultraviolet divergences in the on-loop effective potential.
    \item Chapter 3: the finite-temperature field theory is formulated to describe the finite-temperature effects in quantum field theories by computing the finite-temperature Green's functions and briefly derive the Feynman rules in the imaginary-time formalism. Thus, the effective potential at finite temperatures is derived explicitly and the symmetry restoration phenomenon is demonstrated using a simplified version of the Standard Model effective potential. Lastly, the contribution of the ring diagrams is discussed to formulate the thermal resummation in a scalar field theory.
    \item Chapter 4: the theory of cosmological phase transitions is shown to introduce the concept of bubble nucleation by thermal tunneling and then compute the transition probability per unit volume and per unit time from the false vacuum to the true one.
    \item Chapter 5: electroweak baryogenesis is presented to explain the generation of the observed baryon asymmetry in the Universe. In addition, the baryon and lepton number violation in the Standard Model is demonstrated to compute the rate of sphaleron transitions and derive approximately the condition for a strong first-order phase transition.
    \item Chapter 6: the finite-temperature effective potential at one loop in the Standard Model is derived explicitly to illustrate the electroweak phase transition incorporating the corrections by the ring diagrams. 
    \item Chapter 7: we introduce the singlet extensions to the Standard Model to describe a strong first-order electroweak phase transition and its consequences. This chapter presents the results of the paper written by the author and his supervisor Prof. V. K. Oikonomou. 
\end{itemize}    

\section{Spontaneous Symmetry Breaking}

The invariance of the action of a quantum system does not necessarily imply the invariance of the ground state of the system. In order to illustrate the concept of spontaneous symmetry breaking, we consider an action that is invariant under a particular symmetry transformation. In quantum theory, any symmetric system either has a unique and symmetric vacuum state, or a family of vacuum states which transform to each other under the symmetry group. Namely, if the vacuum state is degenerate, it can be represented by numerous eigenstates. When one of these states is chosen as the vacuum state of the system\footnote{Spontaneous symmetry breaking does not occur in quantum systems with a finite number of degrees of freedom because if there is a set of degenerate vacuum states, the chosen vacuum state can be expressed as a superposition of them which is invariant under the symmetry transformation \cite{Weinberg:1996kr}.}, the vacuum state will violate the symmetries of the action. As a result, the symmetry of the action is called spontaneously broken due to the asymmetric vacuum state. In general, spontaneous symmetry breaking is the phenomenon in which a stable state of a system, such as the vacuum state, is not invariant under the symmetry transformation of its action or Lagrangian and Hamiltonian \cite{Beekman:2019pmi}. As S. Coleman wrote "the laws of nature may possess symmetries which are not manifest to us because the vacuum state is not invariant under them" \cite{Coleman:1985rnk}.

A ferromagnetic material, such as iron, is a common example of a system that exhibits spontaneous symmetry breaking. The action of the system is invariant under spatial rotations. The magnetic moment of the system is described by the magnetization \(\vec{M}\) which is related to the direction of the spins in the ferromagnetic material. Below a critical temperature, a non-zero magnetization is acquired in the ground state, while domains with aligned spins are formed along a particular direction. Therefore, the rotational symmetry \(SO(3)\) is spontaneously broken due to the orientation of the magnetization and there is a set of degenerate vacua related to each other by spatial rotations. Above the critical temperature, the ferromagnet has a non-degenerate ground state with zero magnetization and spontaneous symmetry breaking does not occur.

Spontaneous symmetry breaking in field theory is always linked to the degeneracy of the vacuum. One distinguishes between the symmetric and asymmetric vacuum states introducing an operator that is not invariant under the symmetry transformations and has a non-vanishing expectation value in any degenerate vacuum state implying that the symmetry is broken. This operator is called the order parameter operator. In quantum field theory, this operator is usually assumed to be a scalar field operator \(\hat{\phi}\) as it will be clearly understood later.

\section{The Goldstone Model}
\begin{figure}[H]
    \centering
    \includegraphics[width=0.80\linewidth]{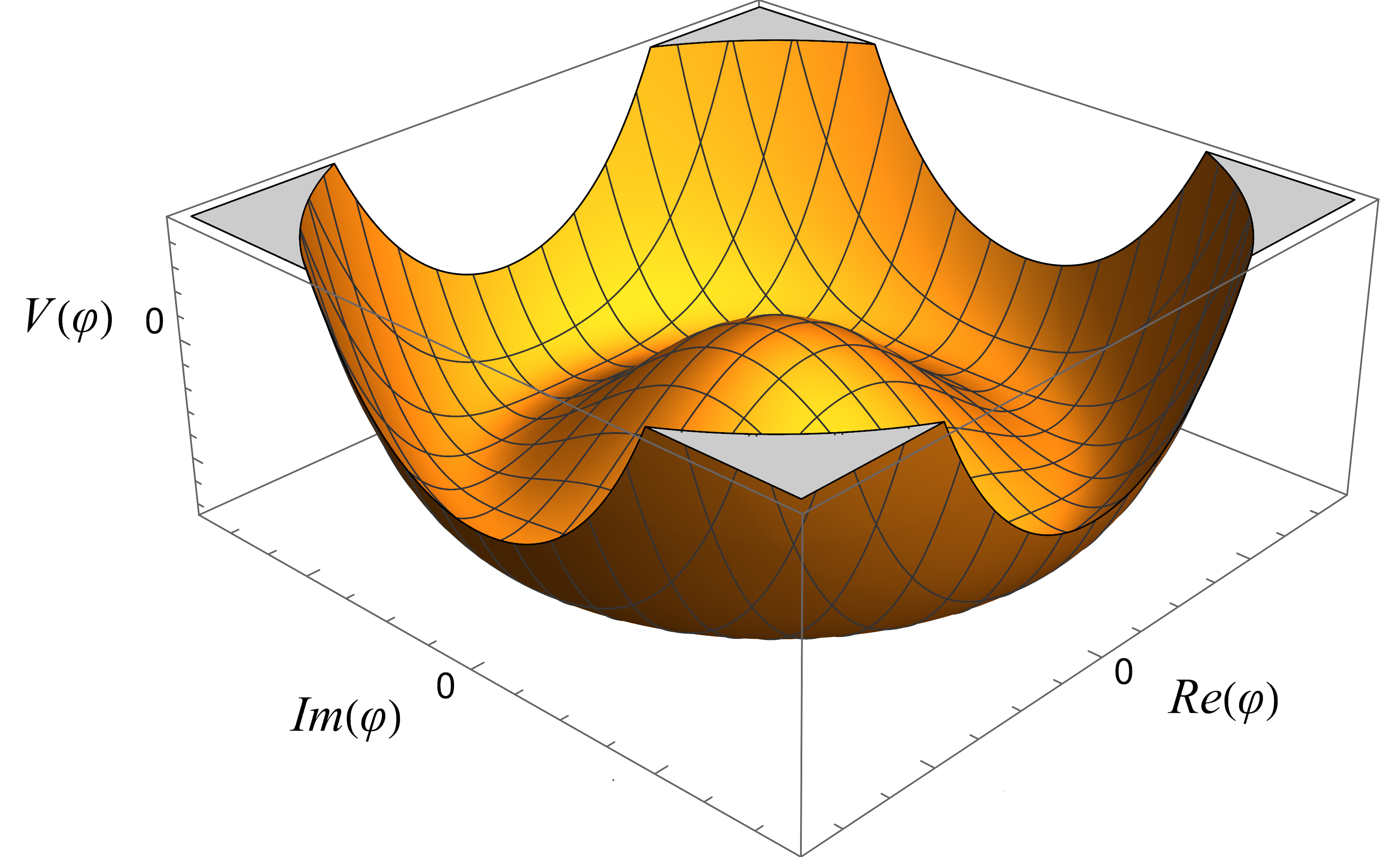}
    \caption{The potential energy density as a function of the imaginary and real part of the complex scalar field in the case of \(\mu^2 > 0\).}
    \label{higgs_graph}
\end{figure}

The classical theory of a self-interacting complex scalar field is an interesting example to present spontaneous symmetry breaking. The Lagrangian density of the so-called Goldstone model is written as \cite{Coleman:1985rnk}
\begin{equation}\label{lag_1}
    \mathcal{L} = \left( \partial^{\mu} \phi^{*} \right) \left( \partial_{\mu} \phi \right) - V(\phi),
\end{equation}
where the classical potential is
\begin{equation}\label{gol_pot}
    V (\phi) = - \mu^2 |\phi|^2 + \lambda |\phi|^4,
\end{equation}
where \(\mu\) and \(\lambda\) are free parameters. The theory is then invariant under the global \(U(1)\) phase transformations,
\begin{equation}
    \phi \to e^{i \theta} \phi
\end{equation}
with \(\theta\) an arbitrary constant. The Hamiltonian density of this theory can be written as
\begin{equation}
    \mathcal{H} = \left( \partial^{0} \phi^{*} \right) \left( \partial_{0} \phi \right) + \left( \nabla  \phi^{*} \right) \left( \nabla \phi \right) + V(\phi),
\end{equation}
where \(\lambda > 0\) to require that the energy density is bounded from below. Namely, the minimum energy density is determined by the constant field which minimizes the potential energy density. However, the location of the minima depends on the sign of \(\mu^2\). In the case of \(\mu^2 < 0\), spontaneous symmetry breaking does not occur as the potential is positive definite and there is only one minimum, the origin \(\phi = 0\). On the other hand, if \(\mu^2\) is positive, the potential is minimized for
\begin{equation}\label{gold_vac}
    \phi = \phi_0 = \frac{\mu}{\sqrt{2 \lambda}} e^{i \theta}
\end{equation}
with \(\theta \in [0, 2 \pi)\). In other words, the continuous set of vacua lies on the circle with radius \(|\phi_0|\) in the complex field plane as shown in Fig. \ref{higgs_graph}, and there is also a local maximum at the origin. Therefore, the \(U(1)\) symmetry is spontaneously broken, when one of these vacua is selected. For instance, we could take \(\theta = 0\). Now, after spontaneous symmetry breaking, the theory can be understood better by introducing two real scalar fields to study the deviations of the field from the vacuum,
\begin{equation}\label{param}
    \phi(x) = \phi_0 + \frac{1}{\sqrt{2}} \left( \chi_1 (x) + i \chi_2 (x) \right)
\end{equation}
and the Lagrangian density is expressed as
\begin{equation}\label{lag_after}
\begin{split}
    \mathcal{L} &= \frac{1}{2} \left( \partial^{\mu} \chi_1 \right) \left( \partial_{\mu} \chi_1 \right) - \frac{1}{2} \left( 4 \lambda \phi^2_0\right) \chi_1^2 \\
    & + \frac{1}{2} \left( \partial^{\mu} \chi_2 \right) \left( \partial_{\mu} \chi_2 \right) \\
    & - \sqrt{2} \lambda \phi_0  \left(\chi^2_1  + \chi^2_2 \right) \chi_1 - \frac{\lambda}{4} \left(\chi^2_1  + \chi^2_2 \right)^2,
\end{split} 
\end{equation}
where a field-independent term is neglected. In Fig. \ref{higgs_graph}, we could assume that the vacuum lies at \(\text{Re} \, \phi = \phi_0\) and \(\text{Im} \, \phi = 0\) for \(\theta = 0\). Then, \(\chi_2\) can be interpreted as a fluctuation along the plane of minimum potential energy, where a small displacement in this direction does not require energy due to the constant value of the potential because \(\phi_0 + i \chi_2\) is another vacuum for infinitesimal values of constant \(\chi_2\). In contrast, the \(\chi_1\) field can be visualized as a fluctuation in the direction orthogonal to the circle with radius \(|\phi_0|\), which quadratically increase the potential. Thus, after quantization, the fields \(\chi_1\) and \(\chi_2\) correspond to neutral spin-0 massive and massless particles, respectively. 

This is additionally shown in the Lagrangian (\ref{lag_after}) because the mass term of the field \(\chi_2\) is absent in the Lagrangian. It is important to mention that the Lagrangian densities (\ref{lag_1}) and (\ref{lag_after}) are equivalent and lead to the same physical predictions\footnote{They are solely equivalent for exact solutions of the theory.}.

Lastly, after spontaneous symmetry breaking, the massless spin-0 bosons, which are called Goldstone bosons, appear. This is clearly explained by the so-called Goldstone theorem that states that spontaneously broken continuous symmetries\footnote{This holds true for a field theory which is Lorentz invariant, local, and has a Hilbert space with positive definite scalar product.} imply the existence of a massless particle for each generator which breaks the symmetry. For instance, if a symmetry group \(G\) with \(n\) generators is spontaneously broken down to a group \(M\) with \(k\) generators, the number of massless particles is \(n - k\). However, massless scalar bosons have never been observed in nature. This problem can be addressed by coupling the scalar fields to gauge fields as it will be presented in the Higgs model. 

\section{The Higgs Model}
The Goldstone model can be generalized by introducing a gauge field to formulate the Higgs model \cite{Englert:1964et, Guralnik:1964eu, Kibble:1967sv, Higgs:1964pj} which is a \(U(1)\) gauge theory described by
\begin{equation}\label{lag_Hi}
    \mathcal{L} = - \frac{1}{4}  F_{\mu \nu} F^{\mu \nu} +  \left( D_{\mu} \phi \right)^* D^{\mu} \phi  + \mu^2 |\phi|^2 - \lambda |\phi|^4
\end{equation}
with
\[D_{\mu} = \partial_{\mu} - i e A_{\mu}\]
\[F_{\mu \nu} = \partial_{\mu} A_{\nu} - \partial_{\nu} A_{\mu}.\]
The Lagrangian is invariant under the \(U(1)\) gauge transformations:
\begin{equation}\label{gaug_tran}
    \phi \to e^{-i \theta (x)}\phi 
\end{equation}
\begin{equation}
    A_{\mu} \to A_{\mu} - \frac{1}{e} \partial_{\mu} \theta (x).
\end{equation}
Spontaneous symmetry breaking similarly occurs in this model in the case of \(\mu^2 > 0\), but the vector field \(A_{\mu}\) has zero vacuum expectation value to ensure Lorentz invariance. The vacuum expectation value of the scalar field is the same as in Eq. (\ref{gold_vac}) introducing the parameter
\begin{equation}
    \upsilon = \frac{\mu}{\sqrt{ \lambda}}.
\end{equation}
After spontaneous symmetry breaking, the scalar field is parameterized as in Eq. (\ref{param}) so that the Lagrangian is cast into the form
\begin{equation}\label{lag_Hib}
\begin{split}
    \mathcal{L} = & - \frac{1}{4}  F_{\mu \nu} F^{\mu \nu} + \frac{1}{2} e^2 \upsilon^2 A_{\mu} A^{\mu} \\
    & + \frac{1}{2} \left( \partial^{\mu} \chi_1 \right) \left( \partial_{\mu} \chi_1 \right) - \frac{1}{2} \left( 2 \lambda \upsilon^2\right) \chi_1^2 \\
    & + \frac{1}{2} \left( \partial^{\mu} \chi_2 \right) \left( \partial_{\mu} \chi_2 \right) \\
    & - e \upsilon A_{\mu} \partial^{\mu} \chi_2 + \mathcal{L}_{\text{int}},
\end{split} 
\end{equation}
where \(\mathcal{L}_{\text{int}}\) denotes the interaction Lagrangian density which includes the interaction terms apart from the mixed term between the vector field and the derivative of the massless scalar field in the last line. This mixed term is not included in the interaction Lagrangian because it is not of the same order as the terms in this interaction Lagrangian. As a consequence of the symmetry breaking, the vector field acquires a mass \(m_A = e \upsilon\) and the scalar field \(\chi_2\) is not physical since the number of degrees of freedom for the Lagrangian densities (\ref{lag_Hi}) and (\ref{lag_Hib}) are different from each other. In particular, we count four degrees of freedom for the former Lagrangian density which has one less degree of freedom than the latter. Therefore, the massless scalar field can be eliminated by the following parametrization in terms of new fields
\begin{equation}
    \phi (x) = \frac{1}{\sqrt{2}} \left( \upsilon + \chi_1 (x) \right) e^{i \chi_2 (x)}
\end{equation}
and by applying the gauge transformation (\ref{gaug_tran}) with \(\theta (x) = \chi_2 (x)\). As a result, the complex scalar \(\phi (x)\) is then given by
\begin{equation}
    \phi (x) = \frac{1}{\sqrt{2}} \left( \upsilon + \chi_1 (x) \right).
\end{equation}
When the field is transformed in this form, the gauge is called unitary or physical gauge since it immediately shows the particle spectrum, obtaining the following Lagrangian density
\begin{equation}\label{lag_Hic}
\begin{split}
    \mathcal{L} = & - \frac{1}{4}  F_{\mu \nu} F^{\mu \nu} + \frac{1}{2} e^2 \upsilon^2 A_{\mu} A^{\mu} \\
    & + \frac{1}{2} \left( \partial^{\mu} \chi_1 \right) \left( \partial_{\mu} \chi_1 \right) - \frac{1}{2} \left( 2 \lambda \upsilon^2\right) \chi_1^2 \\
    & - \lambda \upsilon \chi^3_1 - \frac{\lambda}{4} \chi^4_1 + \frac{1}{2} e^2 A_{\mu} A^{\mu} \left( 2 \upsilon \chi_1 + \chi_1^2 \right),
\end{split} 
\end{equation}
where the first two lines can be interpreted as the free-field Lagrangian density of a real scalar field \(\chi_1 (x)\) and a real massive vector field \(A_{\mu} (x)\). Namely, the massless scalar boson has been eliminated and its single degree of freedom has been transferred to the gauge boson as its longitudinal degree of freedom, while the two degrees of freedom of the complex scalar field \(\phi\) are finally shared by the real scalar field \(\chi_1 (x)\) and the vector field \(A_{\mu} (x)\). Hence, after quantization, the scalar field \(\chi_1 (x)\) corresponds to neutral spin-0 bosons of mass \(m_{\chi_1} = \sqrt{2 \lambda \upsilon^2}\), and the vector field to neutral gauge bosons of mass \(m_A = e \upsilon\). 

Overall, the Higgs mechanism explains the phenomenon by which the spontaneous symmetry breaking of a gauge symmetry leads to the absence of massless scalar fields and the presence of massive vector fields. The massive spin-0 boson, in this mechanism, is known as the Higgs boson or the Higgs scalar.

Lastly, we could briefly introduce a class of gauges which are called \(R_{\xi}\)-gauges to simplify our calculations \cite{tHooft:1972qbu, Fujikawa:1972fe, Yao:1973am}. The mixed term between the vector field and the derivative of the massless scalar field in the Lagrangian (\ref{lag_Hib}) can be removed by adding the following gauge-fixing Lagrangian.
\begin{equation}
    \mathcal{L}_{GF} = - \frac{1}{2 \xi} \left( \partial_{\mu} A^{\mu} + \xi e \upsilon \chi_2 \right)^2,
\end{equation}
which simply introduces a mixed interaction term and leads to an irrelevant total derivative \(\partial_{\mu} \left(A^{\mu} \chi_2 \right)\). As a result, this total derivative can be omitted and the Lagrangian (\ref{lag_Hib}) can be rewritten, including the gauge-fixing terms,
\begin{equation}\label{lag_Hib2}
\begin{split}
    \mathcal{L} = & - \frac{1}{4}  F_{\mu \nu} F^{\mu \nu} + \frac{1}{2} e^2 \upsilon^2 A_{\mu} A^{\mu} - \frac{1}{2 \xi} \left( \partial_{\mu} A^{\mu}\right)^2 \\
    & + \frac{1}{2} \left( \partial^{\mu} \chi_1 \right) \left( \partial_{\mu} \chi_1 \right) - \frac{1}{2} \left( 2 \lambda \upsilon^2\right) \chi_1^2 \\
    & + \frac{1}{2} \left( \partial^{\mu} \chi_2 \right) \left( \partial_{\mu} \chi_2 \right) - \frac{1}{2} \xi e^2 \upsilon^2 \chi_2^2 \\
    & + \mathcal{L}_{\text{int}},
\end{split} 
\end{equation}
where it is evident that the Goldstone boson is not a physical particle as it acquires a mass which is gauge-dependent. This class of gauges results in the following vector propagator written in terms of \(\xi\),
\begin{equation}\label{R-gauge}
    \Pi_{\mu \nu} (k) = \frac{1}{k^2 - e^2 \upsilon^2} \left( - g_{\mu \nu} + (1 - \xi) \frac{k_{\mu} k_{\nu}}{k^2 - \xi e^2 \upsilon^2} \right).
\end{equation}
We notice that setting \(\xi \to \infty\) yields to 
\begin{equation}
    \Pi_{\mu \nu} (k) = \frac{1}{k^2 - m_{A}^2} \left( - g_{\mu \nu} + \frac{k_{\mu} k_{\nu}}{m_{A}^2} \right),
\end{equation}
which is the known vector propagator in the unitary gauge with \(m_{A} = e \upsilon\). Moreover, this class includes the Feynman gauge with \(\xi = 1\), which leads to massive Goldstone bosons, and the Landau gauge with \(\xi = 0\), which has massless Goldstone bosons as shown above. The Landau gauge is mostly preferred in calculations with the effective potential of the Standard Model since in this gauge the scalar fields, such as the Higgs boson, are not coupled to the Goldstone bosons. In addition, in the Landau gauge, the physical scalar fields do not interact with the Fadeev-Popov ghost fields which have couplings and masses-squared proportional to the gauge parameter. In this study, the Landau gauge is adopted in the derivation of the effective potential in the Standard Model as the contributions of the Goldstone bosons and the massive bosons can be computed separately in this gauge.

\section{The Standard Model}\label{Standard Model}
In a unified theory of weak and electromagnetic interactions, the gauge invariance and renormalizability of the theory are not reconciled with the addition of mass terms to the Lagrangian. However, we could introduce the mass terms if the vacuum state of the theory is not invariant under the gauge transformations, but the Lagrangian of the renormalizable theory remains invariant. This is realized by the spontaneous symmetry breaking of the \(SU(2) \times U(1)\) symmetry in the Standard Model. In this section, we briefly present the Standard Model \cite{Glashow:1961tr, Weinberg:1967tq, Salam:1968rm}, a quantum field theory that describes the strong, the weak, and the electromagnetic interactions among elementary particles. A detailed and comprehensive review of the Standard Model can be found in Refs.\cite{Weinberg:1996kr, Langacker:1980js, Cheng:1984vwu, Schwartz:2014sze, Peskin:1995ev, Ryder:1985wq, Mandl:1985bg, Burgess:2006hbd}.

The Standard Model is based on the gauge symmetry group \(SU(3)_C \times SU(2)_L \times U(1)_Y\). The quantum field theory of strong interactions, quantum chromodynamics (QCD), is an \(SU(3)_C\) gauge theory that is associated with the color charge. The strong interactions are mediated by eight massless gauge bosons which are called gluons. However, the \(SU(3)_C\) gauge symmetry is not spontaneously broken due to the Higgs mechanism and this section is mainly focused on the gauge group \(SU(2)_L \times U(1)_Y\), which describes the electroweak interactions.  The \(SU(2)_L\) group has three generators \(\tau^a = \sigma^a/2\), where the \(\sigma^a\) are the Pauli matrices with \(a = 1, 2, 3\), while it introduces the gauge fields \(A^{a}_{\mu}\) with field strength,
\begin{equation}
    F^{a}_{\mu \nu} = \partial_{\mu} A^{a}_{\nu} - \partial_{\nu} A^{a}_{\mu} + g \epsilon^{abc} A^{b}_{\mu} A^{c}_{\nu},
\end{equation}
where \(g\) is the \(SU(2)_L\) gauge coupling. The \(U(1)_Y\), and \(SU(3)_C\) gauge couplings are denoted as \(g'\) and \(g_s\), respectively. In the \(U(1)_Y\) gauge group, the generator is called hypercharge, denoted as \(Y\), and the gauge field \(B_{\mu}\) is introduced with field strength,
\begin{equation}
    B_{\mu \nu} = \partial_{\mu} B_{\nu} - \partial_{\nu} B_{\mu}.
\end{equation}

The spontaneous symmetry breaking of \(SU(2)_L \times U(1)_Y \to U(1)_{EM}\) is essential in a unified electroweak theory in order to generate non-zero masses for the three weak bosons, and assist in introducing fermion masses, while photons remain massless. This mechanism requires the existence of a scalar field with a non-vanishing vacuum expectation value. Thus, a complex doublet \(H\) is introduced with hypercharge \(Y_H = 1/2\),
\begin{equation}\label{Hig_plus}
    H = \begin{pmatrix}
        \phi^{+} \\
        \phi^{0}
        \end{pmatrix},
\end{equation}
which is called the Higgs doublet. The Higgs doublet transforms under the \(SU(2)_L\) gauge transformations according to
\begin{equation}
    H \to e^{i \alpha^{a} (x) \tau^{a}} H
\end{equation}
and under the \(U(1)_Y\) transformations according to
\begin{equation}
    H \to e^{i \beta(x) Y_H} H,
\end{equation}
where the hypercharge is always multiplied by the identity which is omitted in the following. The \(\alpha^{a} (x)\) are the three \(SU(2)_L\) gauge parameters and \(\beta (x)\) is the 
the \(U(1)_Y\) gauge parameter. As a consequence, the covariant derivative on the Higgs doublet is
\begin{equation}
    D_{\mu} H = \partial_{\mu} H - i g A^{a}_{\mu} \tau^{a} H - i g' B_{\mu} Y_H H.
\end{equation}

\subsubsection{Gauge Bosons}
The Lagrangian density of the scalar and gauge sectors of the electroweak theory reads
\begin{equation}\label{L_SM_SG}
    \mathcal{L} = - \frac{1}{4}  F^{a}_{\mu \nu} F^{a \mu \nu} - \frac{1}{4}  B_{\mu \nu} B^{ \mu \nu} +  \left( D_{\mu} H \right)^{\dagger} D^{\mu} H - V(H^{\dagger} H),
\end{equation}
where the Higgs tree-level potential is
\begin{equation}\label{Higgs_potential}
    V(H^{\dagger} H) = - \mu^2 (H^{\dagger} H) + \lambda (H^{\dagger} H)^2,
\end{equation}
where the parameters \(\mu\) and \(\lambda\) are similar to those in Eq. (\ref{gol_pot}). Namely, imposing \(\mu^2 > 0\) and \(\lambda > 0\), the Higgs potential is minimized for
\begin{equation}
    H^{\dagger} H = \frac{\mu^2}{2 \lambda} =  \frac{\upsilon}{2}.
\end{equation}
After spontaneous symmetry breaking, without loss of generality, the Higgs vacuum expectation value in the quantum field theory could be written as
\begin{equation}\label{Higgs doublet VEV}
     \langle H \rangle = \frac{1}{\sqrt{2}}\begin{pmatrix}
       0\\
        \upsilon
        \end{pmatrix}.
\end{equation}
One observes that this vacuum is only invariant under the \(SU(2)_L \times U(1)_Y\) gauge transformations for
\begin{equation}
    \alpha^1 (x) = \alpha^2 (x) = 0, \quad \alpha^3 (x) = \beta (x)
\end{equation}
and the exponent in the \(SU(2)_L \times U(1)_Y\) gauge transformation is
\begin{equation}
    i \alpha^a (x) \tau^a + i \beta (x) Y_H =  \frac{i}{2} \beta (x) \begin{pmatrix}
        & 1 & 0 &\\
        & 0 & 0 & \\
    \end{pmatrix},
\end{equation}
which leads to
\begin{equation}
    \left(\tau^3 + Y_H \right)  \langle H \rangle = 0.
\end{equation}
This combination of generators is the unbroken generator associated with the massless photon and the conservation of the electric charge. Indeed, this vacuum state is only invariant under the \(U(1)\) electromagnetic gauge transformations as the electric charge will be later defined as \(Q = T^3 + Y\) resulting in
\begin{equation}
    Q \langle H \rangle = 0,
\end{equation}
which implies that the photon is massless. Furthermore, the Higgs doublet can be parameterized as
\begin{equation}\label{eq:1.2}
    H = \frac{1}{\sqrt{2}}\begin{pmatrix}
        \chi_1 + i \chi_2 \\
        \upsilon + h + i \chi_3
        \end{pmatrix},
\end{equation}
where \(h(x)\) is called the Higgs boson with mass \(m_H = \sqrt{2 \lambda \upsilon^2}\) as shown in the Higgs model and the scalar fields \(\chi_i\) are unphysical. The quantization of these fields is treated following the same procedure as in the Higgs model. Then, the Higgs potential (\ref{Higgs_potential}) is expressed as
\begin{equation}\label{Higgs_Goldstone}
\begin{split}
    V(H) & = \frac{m_H^2}{2}  h^2 + \lambda \upsilon h^3 + \frac{\lambda}{4} h^4 + \lambda \upsilon \left( \chi_1^2 + \chi_2^2 + \chi_3^2 \right) h \\ & + \frac{\lambda}{2}  \left(\chi_1^2 h^2 + \chi_2^2 h^2 + \chi_3^2 h^2 \right)
    + \frac{\lambda}{2}  \left(\chi_1^2 \chi_2^2 + \chi_1^2 \chi_3^2 + \chi_2^2 \chi_3^2 \right) + \frac{\lambda}{4}  \left(\chi_1^4 + \chi_2^4 + \chi_3^4 \right),
\end{split}
\end{equation}
where the constant terms are omitted. However, the Goldstone bosons can be eliminated with a gauge transformation which leads to the following form in the unitary gauge,
\begin{equation}
    H = \frac{1}{\sqrt{2}}\begin{pmatrix}
        0 \\ \upsilon + h
    \end{pmatrix}.
\end{equation}
Now the gauge boson mass terms are extracted by the scalar kinetic term in the Lagrangian density (\ref{L_SM_SG}),
\begin{equation}\label{DHDH}
    \left( D_{\mu}  H  \right)^{\dagger} D^{\mu}  H  =  \frac{1}{2} \begin{pmatrix}
       0 & \upsilon
        \end{pmatrix}
        \left( g A^{a}_{\mu} \tau^{a} + g' B_{\mu} Y_H \right) \left( g A^{b\mu} \tau^{b} + g' B^{\mu} Y_H \right) \begin{pmatrix}
       0 \\ \upsilon
        \end{pmatrix},
\end{equation}
where the interactions with the Higgs boson are neglected and (\ref{DHDH}) yields to
\begin{equation}\label{g_m_1}
     \left( D_{\mu}  H \right)^{\dagger} D^{\mu} H  = \frac{1}{2} \frac{\upsilon^2}{4} \left[ g^2 \left( A^{1}_{\mu} \right)^2 + g^2 \left( A^{2}_{\mu} \right)^2 + \left( - g A^{3}_{\mu} + g' B_{\mu} \right)^2\right],
\end{equation}
where the first two terms motivate us to introduce the following linear combinations for further convenience
\begin{equation}
    W^{\pm}_{\mu} = \frac{1}{\sqrt{2}} \left( A^{1}_{\mu} \mp i A^{2}_{\mu} \right),
\end{equation}
which immediately shows from Eq. (\ref{g_m_1}) that the \(W\)-boson mass term is
\begin{equation}
    \mathcal{L}^{W}_{m} = \frac{g^2 \upsilon^2}{4} W^{+}_{\mu} W^{- \mu}
\end{equation}
with
\begin{equation}\label{m_W}
    m_W^2 = \frac{g^2 }{4}\upsilon^2.
\end{equation}
However, the mixing term between \(A^{3}_{\mu}\) and \(B_{\mu}\) prevents us from reading off their masses. Therefore, these states can be rotated to a basis without mixing, writing the following term in the matrix form
\begin{equation}\label{p_Z_m}
   \frac{1}{2} \frac{\upsilon^2}{4} \left( - g A^{3}_{\mu} + g' B_{\mu} \right)^2 = \frac{1}{2} \begin{pmatrix}
       A^{3}_{\mu} & B_{\mu} 
        \end{pmatrix}
        \begin{pmatrix} 
        &g^2 \upsilon^2/4 & -g g' \upsilon^2/4 &\\ & -g g' \upsilon^2/4 & g'^2 \upsilon^2/4 &
         \end{pmatrix} 
        \begin{pmatrix}
         A^{3\mu}  \\ B^{\mu} 
        \end{pmatrix}.
\end{equation}
Namely, the above mass matrix\footnote{This is a submatrix of the well-known gauge boson mass matrix.} can be diagonalized to express the mass eigenstates in terms of the initial \(SU(2)_L\) and \(U(1)_Y\) gauge fields, 
\begin{equation}\label{rot}
    \begin{pmatrix}
       Z_{\mu} \\ A_{\mu}
        \end{pmatrix} = 
        \begin{pmatrix} 
        \cos \theta_W & -\sin \theta_W \\  \sin \theta_W & \cos \theta_W 
         \end{pmatrix} 
        \begin{pmatrix}
        A^{3}_{\mu} \\ B_{\mu}
        \end{pmatrix}
\end{equation}
or equivalently,
\begin{equation}
    B_{\mu} = \cos \theta_W A_{\mu} - \sin \theta_W Z_{\mu} 
\end{equation}
\begin{equation}
    A^{3}_{\mu} = \sin \theta_W A_{\mu} + \cos \theta_W Z_{\mu},
\end{equation}
where the weak mixing angle, \(\theta_W\), is defined as
\begin{equation}
    \cos \theta_W = \frac{g}{\sqrt{g^2 + g'^2}}, \quad \sin \theta_W = \frac{g'}{\sqrt{g^2 + g'^2}}
\end{equation}
with 
\begin{equation}
    g \sin \theta_W = g' \cos \theta_W.
\end{equation}
Therefore, Eqs. (\ref{p_Z_m}) and (\ref{rot}) give
\begin{equation}\label{Z_A_MASS}
    \frac{1}{2} \frac{\upsilon^2}{4}\left( - g A^{3}_{\mu} + g' B_{\mu} \right)^2 = 
        \frac{1}{2} \frac{g^2 + g'^2}{4} \upsilon^2 \begin{pmatrix}
       Z_{\mu} & A_{\mu}
        \end{pmatrix}
        \begin{pmatrix} 
        1 & 0 \\  0 & 0
         \end{pmatrix} 
        \begin{pmatrix}
       Z^{\mu} \\ A^{\mu}
        \end{pmatrix} = \frac{1}{2} m_Z^2 Z_{\mu} Z^{\mu},
\end{equation}
where the \(Z\)-boson and photon mass squared are
\begin{equation}\label{m_Z}
    m^2_Z = \frac{g^2 + g'^2}{4} \upsilon^2
\end{equation}
and
\begin{equation}\label{m_G}
    m^2_{\gamma} = 0,
\end{equation}
respectively. The gauge field \(A_{\mu}\) is identified with the electromagnetic field, if it is imposed that the electron charge\footnote{By convention, the electron has electric charge \( - 1\) in units of \(e\).} is
\begin{equation}
    e = g \sin \theta_W
\end{equation}
and the electric charge is
\begin{equation}
    Q = T^3 + Y.
\end{equation}
This is further concluded due to the last term in the covariant derivative for a fermionic field in an \(SU(2)\) representation with hypercharge \(Y\) which is cast into the form 
\begin{equation*}
    D_{\mu} = \partial_{\mu} - i\frac{ g}{\sqrt{2}} \left( W^{+}_{\mu} T^{+} + W^{-}_{\mu} T^{-} \right) - i Z_{\mu} \left( g \cos \theta_W T^{3} - g' \sin \theta_W Y \right) - i g \sin \theta_W A_{\mu} \left( T^{3} + Y \right),
\end{equation*}
where the generators are defined as
\begin{equation}
    T^{\pm} = T^1 \pm i T^2 = \frac{1}{2}(\sigma^1 \pm i \sigma^2) 
\end{equation}
and the gauge field \(A_{\mu}\) is coupled to the electric charges as an electromagnetic field so that  the covariant derivative is written as
\begin{equation}\label{cov_last}
    D_{\mu} = \partial_{\mu} - i\frac{ g}{\sqrt{2}} \left( W^{+}_{\mu} T^{+} + W^{-}_{\mu} T^{-} \right) - i Z_{\mu} \left(g \cos \theta_W T^{3} - g' \sin \theta_W Y \right) - i e A_{\mu} Q.
\end{equation}

One additionally observes that the photon coupling to the Higgs doublet justifies the labels in Eq. (\ref{Hig_plus}) since the upper component in the Higgs doublet has an electric charge equal to one, while the lower component has zero charge with
\begin{equation}
    Q \begin{pmatrix}
       \phi^{+} \\ \phi^{0}
        \end{pmatrix} = \begin{pmatrix}
       \phi^{+} \\ 0
        \end{pmatrix}.
\end{equation}

\subsubsection{Fermions}

In the Standard Model, the fermions are divided into at least three generations. Each generation of fermions consists of one charged lepton (\(e, \mu, \tau\)), one neutrino (\(\nu_{e}, \nu_{\mu}, \nu_{\tau}\)), one up-type quark (\(u, c, t\)), and one down-type quark (\(d, s, b\)). The right-handed fermions transform as singlets, whereas the left-handed fermions transform as doublets under the \(SU(2)_L\) gauge transformations. As a result, parity is violated by introducing the left-handed and right-handed fermions into these representations. The left-handed quarks are represented by \(SU(2)\) doublets,
\begin{equation}
    Q_{i} = \begin{pmatrix}
        u^{i}_{L} \\ d^{i}_{L}
    \end{pmatrix} = \begin{pmatrix}
        u_{L} \\ d_{L}
    \end{pmatrix}, \begin{pmatrix}
        c_{L} \\ s_{L}
    \end{pmatrix}, \begin{pmatrix}
        t_{L} \\ b_{L}
    \end{pmatrix},
\end{equation}
where \(i\) represents the generation with \(i = 1, 2, 3\). The left-handed leptons are also expressed as
\begin{equation}
    L_{i} = \begin{pmatrix}
        \nu^{i}_{e L} \\ e^{i}_{L}
    \end{pmatrix} = \begin{pmatrix}
        \nu_{e L} \\ e_{L}
    \end{pmatrix}, \begin{pmatrix}
        \nu_{\mu L} \\ \mu_{L}
    \end{pmatrix}, \begin{pmatrix}
        \nu_{\tau L} \\ \tau_{L}
    \end{pmatrix}.
\end{equation}
The hypercharge for each fermionic field is determined by the value of \(T^3\) and the electric charge as it was shown earlier. Therefore, the right-handed fermionic fields (\(T^3 = 0\)) have hypercharge equal to their electric charge, whereas the left-handed quarks and leptons have hypercharge \(Y = 1/6\) and \( Y = -1/2\), respectively, taking into account the value \(T^3 = \pm 1/2\). 

The coupling of the \(W\) and \(Z\) fields to fermions is immediately defined by the covariant derivative (\ref{cov_last}) and their charges above. Therefore, the gauge-invariant fermion kinetic terms are given by
\begin{equation}\label{kin_lag}
    \mathcal{L_\text{KF}} = i \sum_{i = 1}^{3} \left( \overline{Q}_{i} \slashed D Q_{i} + \overline{L}_{i} \slashed D L_{i} + \overline{u}^{i}_{R} \slashed D u^{i}_{R} + \overline{d}^{i}_{R} \slashed D d^{i}_{R} + \overline{e}^{ i}_{R} \slashed D e^{i}_{R} \right),
\end{equation}
where each covariant derivative is
\begin{equation}
    D_{\mu} Q_i = \left( \partial_{\mu} - i g A_{\mu}^{a} \tau^{a} - i\frac{g'}{6} B_{\mu} \right) Q_i, \quad D_{\mu} L_i = \left( \partial_{\mu} - i g A_{\mu}^{a} \tau^{a} + i\frac{g'}{2} B_{\mu} \right) L_i,
\end{equation}
\begin{equation}
    D_{\mu} u_R^{i} = \left( \partial_{\mu} - \frac{2} {3}g' B_{\mu} \right) u_R^{i}, \quad D_{\mu} d_R^{i} = \left( \partial_{\mu} + i\frac{g'}{3} B_{\mu} \right) e_R^{i}, \quad D_{\mu} e_R^{i} = \left( \partial_{\mu} +g' B_{\mu} \right) e_R^{i}.
\end{equation}
The right-handed neutrino is omitted as it is uncharged under the gauge group and its existence has not been experimentally confirmed. Using Eq. (\ref{cov_last}), the Lagrangian (\ref{kin_lag}) is expressed in terms of the gauge boson mass eigenstates as
\begin{equation}\label{LKF}
\begin{split}
    \mathcal{L_\text{KF}}  = &  i \sum_{i = 1}^{3} \left( \overline{Q}_{i} \slashed \partial Q_{i} + \overline{L}_{i} \slashed \partial L_{i} + \overline{u}^{i}_{R} \slashed \partial u^{i}_{R} + \overline{d}^{i}_{R} \slashed \partial d^{i}_{R} + \overline{e}^{i}_{R} \slashed \partial e^{i}_{R} \right) \\
    & + g \left( W^{+}_{\mu} J^{\mu+}_W + W^{-}_{\mu} J^{\mu-}_{W} + Z_{\mu} J^{\mu}_Z \right) + e A_{\mu} J^{\mu}_{EM},
\end{split}
\end{equation}
where the last two currents \(J_{EM}^{\mu}\) and \(J_{Z}^{\mu}\) are written in terms of all fermionic fields in contrast with the currents \(J^{\mu+}_W\) and \(J^{\mu-}_W\) which involve only the left-handed fermions, considering that only the left-handed fermions couple to the \(W\) bosons. It is remarkable that the current \(J_{EM}^{\mu}\) coupled to the photon field \(A_{\mu}\) is the common electromagnetic current. The explicit derivation and formulas for the currents can be found in Refs. \cite{Langacker:1980js, Burgess:2006hbd, Cheng:1984vwu, Schwartz:2014sze, Peskin:1995ev}. In addition, it is remarkable that the Higgs vacuum expectation value can be obtained owing to the charged weak current since the effective 4-Fermi interaction at low energy can be compared with the 4-Fermi theory and the Fermi coupling constant \(G_F\) is related to the Higgs vacuum expectation value,
\begin{equation}
    \frac{G_F}{\sqrt{2}} = \frac{g^2 }{8 m_W^2} \quad \Rightarrow \quad \upsilon = \left(\sqrt{2} G_F \right)^{-1/2},
\end{equation}
where the Fermi coupling constant is experimentally determined.

It was emphasized earlier that in the Standard Model Lagrangian, the fermion mass terms, such as 
\begin{equation*}
    m_{e} \left( \overline{e}_{L} e_{R} + \overline{e}_{R} e_{L} \right), 
\end{equation*}
are forbidden by the gauge symmetry since the left-handed and right-handed fermions are described by different \(SU(2)\) representations. However, the fermion masses can be generated by coupling the Higgs doublet to the fermion fields through Yukawa interactions. In the following, a single generation is considered to avoid the complications of the multiple generations, which will be mentioned later. For instance, the electron mass term comes from
\begin{equation}\label{ele_mss}
    y_e \overline{L} H e_{R} + \text{h.c.}\, ,
\end{equation}
which is gauge-invariant. 
As a result, in the unitary gauge, Eq. (\ref{ele_mss}) gives
\begin{equation}
    y \overline{L} H e_{R} + \text{h.c.} = \frac{y_e}{\sqrt{2}} \upsilon \overline{e}_L e_R + \frac{y_e}{\sqrt{2}} h \overline{e}_L e_R + \text{h.c.},
\end{equation}
where the electron mass can be defined as
\begin{equation}
    m_e = \frac{y_e}{\sqrt{2}}\upsilon.
\end{equation}
This method can be applied to obtain the masses for the charged leptons and the down-type quarks. The remaining fermion masses are determined by an alternative Yukawa interaction term which involves
\begin{equation}
    \tilde{H} \equiv i \sigma_2 H^*,
\end{equation}
where one can easily prove that \(\Tilde{H}\) is an \(SU(2)\) doublet with hypercharge \(Y = -1/2\). Then, the gauge-invariant term with the right-handed up-quark could be written as
\begin{equation}\label{quark_mass}
    y_{u} \overline{Q} \tilde{H} u_R + \text{h.c.} = m_{u} \overline{u}_L u_R + \frac{m_u}{\upsilon} h \overline{u}_L u_R + \text{h.c.}\,,
\end{equation}
where the up-quark mass is
\begin{equation}\label{m_u}
    m_u = \frac{y_u}{\sqrt{2}} \upsilon.
\end{equation}
Overall, the Yukawa interactions which introduce the fermion masses in the Standard Model are described by the Lagrangian density,
\begin{equation}
    \mathcal{L}_{Y} = - y_{d} \overline{Q} H d_R - y_{u} \overline{Q} \tilde{H} u_R -  y_{e} \overline{L} H e_R - y_{\nu} \overline{L} \tilde{H} \nu_R + \text{h.c.}\,,
\end{equation}
including the right-handed neutrino. Likewise, the other fermions masses can be obtained, but the three generations could introduce additional coupling terms and mixing between the generations. In particular, multiple generations introduce a level of complexity in the Yukawa couplings as the Yukawa interaction terms for the three generations can be written as
\begin{equation}
    \mathcal{L}_{Y} = - \sum_{i,j} \left( y^{ij}_{d} \overline{Q}_i H d_{j R} + y^{i j}_{u} \overline{Q}_i \tilde{H} u_{j R} +  y^{ij}_{e} \overline{L}_i H e_{jR} \right) + \text{h.c.}\,,
\end{equation}
where \(i, j = 1, 2, 3\) count the generations and the Yukawa couplings \(y_d\), \(y_u\), and \(y_e\) are replaced by \(3 \times 3\) matrices. In order to read off the fermion masses, we diagonalize the Yukawa coupling matrices. For instance, the quarks can be rotated among themselves as follows
\begin{equation}\label{change of basis}
    u^{i}_L \to \left(V_u \right)^{i}_j u^{j}_L, \quad d^{i}_L \to \left(V_d \right)^{i}_j d^{j}_L, \quad  u^{i}_R \to \left(U_u \right)^{i}_j u^{j}_R, \quad  d^{i}_R \to \left(U_d \right)^{i}_j d^{j}_R,
\end{equation}
with \(V_u, V_d, U_u, U_d \in U(3)\). Then, the Yukawa coupling matrices transform as
\begin{equation}
    y_u \to V_u^{\dagger}\, y_u \,U_u \quad \text{and} \quad y_d \to V_d^{\dagger}\, y_d \,U_d,
\end{equation}
which leads to diagonal Yukawa coupling matrices and the quark masses are cast into the form given by Eq. (\ref{m_u}). Similarly, we rotate the lepton fields\(L_i\) and \(e^i_R\) to obtain a diagonal Yukawa coupling matrix \(y_e\) and the mass of the electron, the muon, and the tau. In general, the fermion masses can be written as
\begin{equation}
    m_i = \frac{y_i}{\sqrt{2}} \upsilon,
\end{equation}
where \(i\) labels the fermions in the Standard Model with three generations. However, the currents coupled to the gauge bosons in Eq. (\ref{LKF}) are affected by the change of basis (\ref{change of basis}). More specifically, the currents \(J^{\mu}_{EM}\) and \(J^{\mu}_Z\) remain unaffected by this change of basis, whereas the charged weak currents are altered in the mass eigenbasis. Now if the mass eigenstates are transformed into the weak eigenstates with the matrices \(V_u, V_d, U_u\), and \(U_d\), only the combination \(V_d^{\dagger} U_d\) is observable as the charged weak current is flavor changing. This combination defines a unitary matrix known as the Cabibbo-Kobayashi-Maskawa (CKM) matrix, which is determined by three angles and one \(CP\)-violating phase. We did not analyze in detail the physics behind the multiple generations and the CKM matrix because the reader can explore these concepts in Refs. \cite{Langacker:1980js, Cheng:1984vwu, Schwartz:2014sze, Peskin:1995ev, Burgess:2006hbd, Ryder:1985wq} and many other books on the Standard Model.

Overall, the electroweak theory is an \(SU(2)_L \times U(1)_Y\) gauge theory described by a Lagrangian density which reads
\begin{equation}
\begin{split}
    \mathcal{L} = & - \frac{1}{4}  F^{a}_{\mu \nu} F^{a \mu \nu} - \frac{1}{4}  B_{\mu \nu} B^{ \mu \nu} \\
    & + \left( D_{\mu} H \right)^{\dagger} D^{\mu} H - V(H^{\dagger} H) \\
    & + i \sum_{i = 1}^{3} \left( \overline{Q}_{i} \slashed D Q_{i} + \overline{L}_{i} \slashed D L_{i} + \overline{u}^{i}_{R} \slashed D u^{i}_{R} + \overline{d}^{i}_{R} \slashed D d^{i}_{R} + \overline{e}^{ i}_{R} \slashed D e^{i}_{R} \right)\\
    & - \sum_{i,j}^{3} \left( y^{ij}_{d} \overline{Q}_i H d_{j R} + y^{i j}_{u} \overline{Q}_i \tilde{H} u_{j R} +  y^{ij}_{e} \overline{L}_i H e_{jR}  \right) + \text{h.c.}
\end{split}   
\end{equation}
and it is determined by the \(U(1)_Y\) gauge coupling \(g'\), the \(SU(2)_L\) gauge coupling \(g\), the Higgs vacuum expectation value \(\upsilon\), the Higgs mass \(m_H\), the three CKM mixing angles, the CKM \(CP\)-violating phase and the nine Yukawa couplings in the fermion sector\footnote{The neutrino masses are omitted here since they introduce additional parameters, such as at least three Yukawa couplings. Namely, there are more than 19 (or 25) parameters in the Standard Model, if the neutrino masses are included. See Ref. \cite{Burgess:2006hbd} to understand how to introduce neutrino masses and the additional free parameters.}. The Standard Model additionally includes the free parameters from QCD, the \(SU(3)_C\) gauge coupling \(g_s\) and the QCD vacuum angle. For the sake of completeness, we finally present the full Lagrangian of the Standard Model, 
\begin{equation}
\begin{split}
    \mathcal{L}_{\text{SM}} = & - \frac{1}{4}  G^{a}_{\mu \nu} G^{a \mu \nu} - \frac{1}{4}  F^{a}_{\mu \nu} F^{a \mu \nu} - \frac{1}{4}  B_{\mu \nu} B^{ \mu \nu} \\
    & + \left( D_{\mu} H \right)^{\dagger} D^{\mu} H - V(H^{\dagger} H) \\
    & + i \sum_{i = 1}^{3} \left( \overline{Q}_{i} \slashed D Q_{i} + \overline{L}_{i} \slashed D L_{i} + \overline{u^{i}}_{R} \slashed D u^{i}_{R} + \overline{d^{i}}_{R} \slashed D d^{i}_{R} + \overline{e^{ i}}_{R} \slashed D e^{i}_{R} \right)\\
    & - \sum_{i,j}^{3} \left( y^{ij}_{d} \overline{Q}_i H d_{j R} + y^{i j}_{u} \overline{Q}_i \tilde{H} u_{j R} +  y^{ij}_{e} \overline{L}_i H e_{jR} \right) + \text{h.c.}\,,
\end{split}   
\end{equation}
including the strong interactions, where \(G^{\mu \nu}\) is the \(SU(3)\) gluon field strength and the form of each covariant derivative depends on the representation of the fermion field\footnote{For instance, the covariant derivative of the doublets \(Q_i\) reads \begin{equation}
     D_{\mu} Q_i = \left( \partial_{\mu} -i g_s G_{\mu} - i g A_{\mu} - i\frac{g'}{6} B_{\mu} \right) Q_i.
\end{equation}} \cite{Burgess:2006hbd}.


\chapter{Effective Action}\label{Effective Action}
Our previous approach to spontaneous symmetry breaking was presented at a classical level since we determined the vacuum state of the scalar fields by the minimization of the classical potential energy density. However, quantum corrections could generate spontaneous symmetry breaking changing the vacuum expectation value which is computed at the classical approach \cite{Coleman:1973jx, Goldstone:1962es,Jona-Lasinio:1964zvf}. Instead of the classical potential, a new function could be introduced to obtain the vacuum expectation value of a quantum field operator in the full quantum theory. At the tree-level approximation, this function should coincide with the classical potential and it would include radiative corrections in higher orders. This function can be defined as the effective potential which is minimized at the vacuum expectation value of the quantum field operator, including quantum corrections. 

The effective potential is defined in terms of the so-called effective action which encodes the symmetries of the full quantum theory. In general, an effective action is defined as a functional of fields to result in the same Green's functions and \(S\)-matrix elements as the action that describes the full theory. Using effective actions instead of full theory actions offers the benefit of simplifying calculations by concentrating solely on the relevant degrees of freedom for a given problem. The effective action frequently has fewer degrees of freedom than the action of the full theory, is non-renormalizable and its range of validity is restricted. This concept was first introduced by Euler and Heisenberg (1936)  to describe light-by-light scattering and other phenomena in Quantum Electrodynamics \cite{Heisenberg:1936nmg}. They derived the known Euler-Heisenberg Lagrangian as an approximation neglecting the dynamics of the electromagnetic field, but could we formulate an exact approach to the effective action? A 1PI effective action is loosely defined as the action that leads to all of the physical predictions of a full quantum theory when it is used at tree level. In the 1PI effective action, all the fields have been integrated out in contrast with the general effective action. 

The 1PI effective action is identified as the Legendre transform of the generating functional of all connected Feynman diagrams which is essential to evaluate the \(S\)-matrix elements of a theory. As a result, the generating functionals are introduced since they demonstrate the connection between the full, the connected, and the one-particle irreducible (1PI) Green's functions (or \(n\)-point functions) that are obtained by simple mathematical operations. More specifically, the logarithm of the generating functional for the full Green's functions is the generating functional for all connected Green's functions whose Legendre transform is the generating functional for 1PI Green's functions. The latter implies that the vacuum expectation value of a quantum field operator minimizes the effective action and its non-zero value signals spontaneous symmetry breaking. In addition, the effective potential is derived for a constant background field and it can be interpreted as the generating functional of all 1PI diagrams with zero external momenta.

Finally, the background field method is presented which clearly provides an interpretation and a useful computational tool for the effective potential. Then the one-loop contribution to the effective potential is computed for scalar field theories and is extended to non-Abelian gauge theories evaluating the 1PI diagrams with a single loop and zero external momenta. Thus, we proceed to the regularization and renormalization of the aforementioned theories to deal with the ultraviolet divergences in the one-loop effective potential.


\section{Generating Functionals}
Consider the action \(S[\phi]\) of a single scalar field in the presence of an external classical source \(J(x)\) and the following definitions and results can be similarly generalized to various field theories of fermion and gauge fields. The vacuum-to-vacuum transition amplitude in the presence of this source is defined as the so-called generating functional
\begin{equation}\label{def_func}
    Z[J] = \bra{0_{\text{out}}} 0_{\text{in}}\rangle_J = \int  \mathcal{D} \phi \, \exp \left[i S[\phi] + i \int d^{4}x \, J(x) \phi(x) \right].
\end{equation}
This functional can be expanded in powers of the source and rewritten as,
\begin{equation*}
    Z[J] = \int \mathcal{D} \phi \, e^{i S[\phi]} \left( 1 + i \int d^{4}x \, J(x) \phi(x) + \frac{i^2}{2!} \int d^{4}x_1 \, d^4 x_2 \, \phi (x_1) \phi (x_2) J(x_1) J(x_2) + \, ... \, \right),
\end{equation*}
which subsequently leads to
\begin{equation}\label{Z_G}
\begin{split}
    \frac{Z[J]}{Z[0]} & = \sum_{n = 0}^{\infty} \frac{i^n}{n!}   \int d^{4}x_{1}... \, \int d^{4}x_{n} \, J(x_1)...J(x_n) \bra{\Omega} T \{\hat{\phi}(x_1)...\hat{\phi}(x_n)\} \ket{\Omega} \\
    & = \sum_{n = 0}^{\infty} \frac{i^n}{n!}   \int d^{4}x_{1}... \, \int d^{4}x_{n} \, J(x_1)...J(x_n) G^{(n)} (x_1,...,x_n) ,
\end{split}
\end{equation}
where it is normalized to \(Z[0] = 1\). This generating functional is the quantum field theory equivalent of the partition function in statistical physics. It provides comprehensive information about the system. Having an exact closed-form expression for the generating functional for a given field theory implies the complete solution of that theory. According to the functional calculus, Eq. (\ref{Z_G}) can lead to solve for the full Green's functions,
\begin{equation}
    G^{(n)} (x_1,...,x_n) = (-i)^n \left. \left(\frac{\delta}{\delta J(x_1)} \right)...\left(\frac{\delta}{\delta J(x_n)}\right) Z[J] \right|_{J=0},
\end{equation}
where the Green's function is equal to
\begin{equation}
    G^{(n)} (x_1,...,x_n) = \frac{\int  \mathcal{D} \phi \, e^{i S[\phi]} \phi(x_1)...\phi(x_n)}{\int  \mathcal{D} \phi \, e^{i S[\phi]}}.
\end{equation}

As established by the Lehmann-Symanzik-Zimmermann (LSZ) formalism, the computation of scattering amplitudes involves only the connected Green’s functions \(G^{(n)}_c (x_1, ..., x_n)\) which correspond to connected Feynman diagrams\footnote{A connected diagram is defined as the diagram, where every line is connected to at least one external line.}. Therefore, the generating functional for connected Green's functions is defined as,
\begin{equation}\label{def_W}
    i W[J] = \sum_{n = 0}^{\infty} \frac{i^n}{n!}   \int d^{4}x_{1}... \, \int d^{4}x_{n} \, J(x_1)...J(x_n) G^{(n)}_c (x_1,...,x_n) ,
\end{equation}
where the connected \(n\)-point function, which is the sum of all connected diagrams with \(n\) external lines, is given by
\begin{equation}\label{connected_Green}
    G^{(n)}_c (x_1,...,x_n) = (-i)^n \left. \left(\frac{\delta}{\delta J(x_1)} \right)...\left(\frac{\delta}{\delta J(x_n)}\right) i W[J] \right|_{J=0}
\end{equation}
and can be also written in the presence of the external source (\(J \neq 0 \)). For instance, the one-point function in the presence of the source is written as
\begin{equation}
     \bra{\Omega} \hat{\phi} (x) \ket{\Omega}_J = \frac{\delta W[J]}{\delta J(x)}.
\end{equation}
The generating functional (\ref{def_func}) is related to the generating functional of all connected diagrams (\ref{def_W}) as
\begin{equation}
    Z[J] = 1 + iW[J] - \frac{1}{2} W[J]^2 - \frac{1}{6}iW[J]^3 + \, ... \,= \sum_{n = 0}^{\infty} \frac{i^n}{n!}  \left(W[J]\right)^n ,
\end{equation}
which is a widely known relation and is well-proven in Ref. \cite{Itzykson:1980rh}. This leads to the following expression,
\begin{equation}\label{Z_W}
    Z[J] = e^{i W[J]} .
\end{equation}

\begin{figure}[H]
    \centering
    \includegraphics[width=0.65\linewidth]{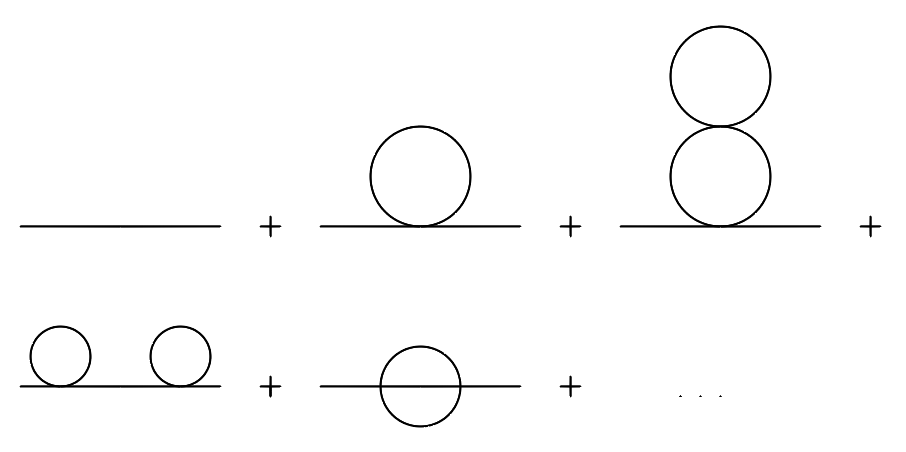}
    \caption{The Feynman diagrams that contribute to the perturbative corrections to the propagator in the \(\phi^4\)-theory. The fourth diagram is not 1PI because one can produce two separate non-trivial diagrams by cutting the line between the two loops.}
    \label{fig:1.1}
\end{figure}

Now, the connected Feynman diagrams can be decomposed into one-particle irreducible (1PI) subdiagrams which are the building blocks of correlation functions. The 1PI diagrams are a significant subclass of connected Feynman diagrams, which cannot be disconnected into two non-trivial diagrams by cutting a single internal line (see Fig. \ref{fig:1.1}). The 1PI diagrams are conventionally evaluated with: i) no propagators on the external lines and ii) no energy-momentum-conserving delta function. We will show that a sum of connected diagrams can be obtained by constructing tree diagrams with 1PI diagrams as vertices connected with full propagators.
We initially define the Legendre transform of \(W[J]\),
\begin{equation}\label{legendre}
    \Gamma [\phi_c] = W[J_{\phi_c}] - \int d^4 x \, \phi_c (x) J_{\phi_c} (x) ,
\end{equation}
where \(J_{\phi_c}\) is defined as an implicit functional of \(\phi_c\),
\begin{equation}\label{J_impl}
      \left. \frac{\delta W[J]}{\delta J(x)} \right|_{J=J_{\phi_c}} = \phi_c (x),
\end{equation}
where the so-called classical field or background field \(\phi_c (x)\) is the expectation value of the quantum field operator \(\hat{\phi}\) in the presence of the source,
\begin{equation}
    \phi_c (x) =  \bra{\Omega} \hat{\phi} (x) \ket{\Omega}_J
\end{equation}
with the source \(J = J_{\phi_c} (x)\). In the absence of the source, the background field is simply the vacuum expectation value
\begin{equation}
    \left. \phi_c (x) \right|_{J = 0} =  \bra{\Omega} \hat{\phi} (x) \ket{\Omega}.
\end{equation}
We could assume that \(\bra{\Omega} \hat{\phi} (x) \ket{\Omega} = 0\) without loss of generality since Lorentz invariance requires that \(\bra{\Omega} \hat{\phi} (x) \ket{\Omega} = \bra{\Omega} \hat{\phi} (0) \ket{\Omega} = c\) and if \(c \neq 0\), we redefine \( \hat{\phi} \to  \hat{\phi} - c\) to obtain \(\bra{\Omega} \hat{\phi} (x) \ket{\Omega} = 0\). From Eqs. (\ref{legendre}) and (\ref{J_impl}) it follows that
\begin{equation}\label{derivative_1}
    \frac{\delta \Gamma [\phi_c]}{\delta \phi_c (x)} = \int d^4 y \, \left[ \frac{\delta W[J_{\phi_c}]}{\delta J_{\phi_c}(y)} \frac{\delta J_{\phi_c}(y)}{\delta \phi_c(x)} - \phi_c (y) \frac{\delta J_{\phi_c}(y)}{\delta \phi_c(x)} \right] - J_{\phi_c}(x) = - J_{\phi_c}(x).
\end{equation}
This motivates us to define \(\phi_c\) as an implicit functional of \(J\) and the solution to 
\begin{equation}\label{ran}
    \left. \frac{\delta \Gamma[\phi_c]}{\delta \phi_c (x)} \right|_{\phi_c = \phi_J} = -J(x),
\end{equation}
which results in the inverse Legendre transform
\begin{equation}\label{inv_legendre}
    W[J] = \Gamma [\phi_J] + \int d^4 x \, \phi_J (x) J(x).
\end{equation}
Differentiating Eq. (\ref{inv_legendre}) with respect to the external current
\begin{equation}
    \frac{\delta W[J]}{\delta J (x)} = \int d^4 y \, \left[ \frac{\delta \Gamma [\phi_J]}{\delta \phi_J (y)} \frac{\delta \phi_J(y)}{\delta J(x)} + J (y) \frac{\delta \phi_J (y)}{\delta J (x)} \right] + \phi_J (x) = \phi_J (x),
\end{equation}
which agrees with Eq. (\ref{J_impl}). Namely, this relation between \(\phi_c (x)\) and \(J(x)\) is invertible. 

\begin{figure}[H]
    \centering
    \includegraphics[width=0.85\linewidth]{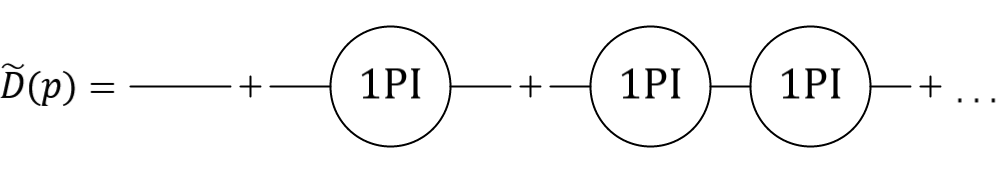}
    \caption{Graphical representation of the geometric series in Eq. (\ref{geomseries}), where the 1PI-circle represents the \(i \Sigma (p)\) as the sum of all 1PI two-point diagrams.}
    \label{selfenergy}
\end{figure}

\begin{figure}[H]
    \centering
    \includegraphics[width=0.45\linewidth]{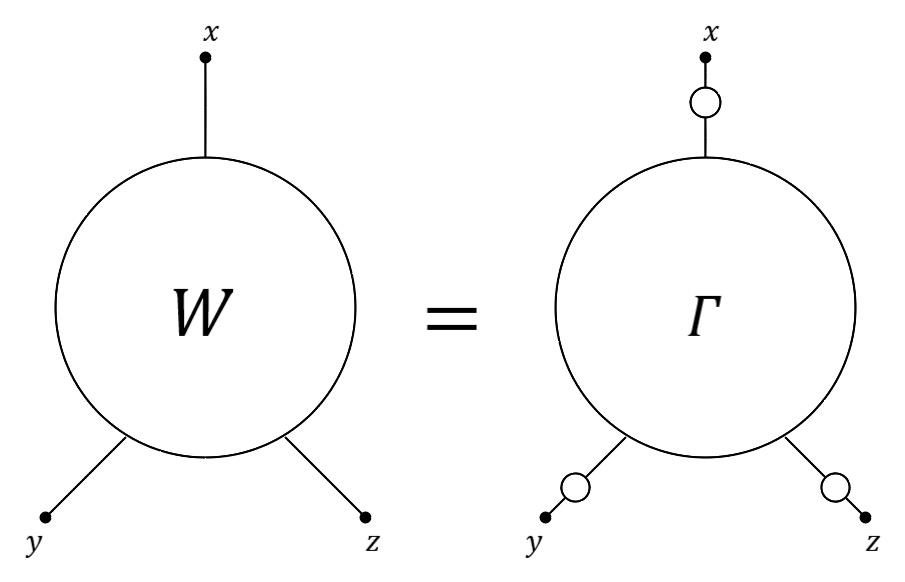}
    \caption{On the left-hand side, the \(W\)-circle illustrates the sum of three-point connected diagrams, while the \(\Gamma\)-circle on the right-hand side corresponds to the 1PI three-point function and the three small circles represent the sum of the two-point connected diagrams.}
    \label{W_G}
\end{figure}

\begin{figure}[H]
    \centering
    \includegraphics[width=0.75\linewidth]{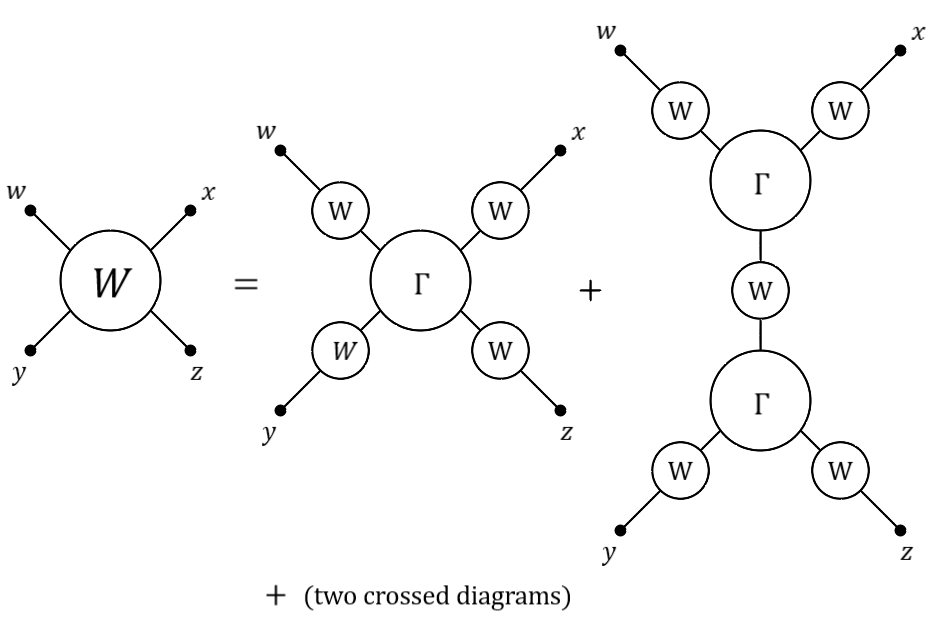}
    \caption{The four-point connected function which is the W-circle on the left leads to the four-point 1PI function which is the fourth derivative of \(i \Gamma[\phi_c]\) represented by the \(\Gamma\)-circle in the first term on the right-hand side which is connected to the full propagators. Similarly, in the second term the \(\Gamma\)-circle indicates the 1PI three-point function.}
    \label{4W_G}
\end{figure}

Lastly, we will prove the relation between the Legendre transform of the connected generating functional and the 1PI Green's functions which correspond to the 1PI diagrams. This is also diagrammatically demonstrated in Refs. \cite{Itzykson:1980rh, Abbott:1981ke}. We firstly vary Eq. (\ref{derivative_1}) with respect to \(J(y)\) and we omit the redundant subscripts related to the Legendre transform in Eq. (\ref{legendre}) and (\ref{inv_legendre}) for simplicity,
\begin{equation}\label{before_dev_2}
    \frac{\delta^2 \Gamma [\phi_c]}{\delta J(y) \delta \phi_c (x)} = - \delta^{(4)} (x-y) \Rightarrow \int d^4 z \, \frac{\delta \phi_c (z)}{\delta J(y)} \frac{\delta^2 \Gamma [\phi_c]}{\delta \phi_c (z) \delta \phi_c (x)} = -\delta^{(4)} (x-y),
\end{equation}
where from Eq. (\ref{J_impl}) we have
\begin{equation}\label{dev_W2}
    \frac{\delta^2 W[J]}{\delta J(y) \delta J(x)} = \frac{\delta \phi_c (x)}{\delta J(y)}.
\end{equation}
Thus, Eqs. (\ref{before_dev_2}) and(\ref{dev_W2}) lead to
\begin{equation}\label{inverse}
    \delta^{(4)} (x-y) = - \int d^4 z \, \frac{\delta^2 W[J]}{\delta J(y) \delta J(z)}  \frac{\delta^2 \Gamma [\phi_c]}{\delta \phi_c (z) \delta \phi_c (x)},
\end{equation}
which means that the second derivative of \(\Gamma [\phi_c]\) is the inverse of the second derivative of the generating functional for connected Green's functions. If we set \(J = 0\), then Eq. (\ref{dev_W2}) is written as
\begin{equation}
    i G^{(2)}_c (x, y) = \left. \frac{\delta^2 W[J]}{\delta J(x) \delta J(y)} \right|_{J = 0}
\end{equation}
and the second derivative of \(\Gamma [\phi_c]\) is denoted as
\begin{equation}
     \left. \frac{\delta^2\Gamma [\phi_c]}{\delta \phi_c (x) \delta \phi_c (y)} \right|_{\phi_c=0} = - i \Gamma^{(2)} (x,y)
\end{equation}
in order to express Eq. (\ref{inverse}) as
\begin{equation}
    \int d^4 z \, G^{(2)}_c (y, z) \Gamma^{(2)} (z,x) = - \delta^{(4)} (x-y).
\end{equation}
In momentum space, this relation is expressed as
\begin{equation}\label{inv_relation_p}
    G^{(2)}_c (p, -p) \Tilde{\Gamma}^{(2)} (p,-p) = -1,
\end{equation}
where \( G^{(2)}_c (p, -p) = \tilde{D} (p)\) is the full propagator of the field \(\phi\),
\begin{equation}\label{def_prop}
    \tilde{D} (p) = \frac{i}{p^2 - m^2 + \Sigma (p)},
\end{equation}
where \(\Sigma (p)\) is the self-energy. If Eq.(\ref{def_prop}) is expanded as a geometric series,
\begin{equation}\label{geomseries}
\begin{split}
    \Tilde{D}(p) = & \frac{i}{p^2 - m^2 } + \frac{i}{p^2 - m^2} i \Sigma (p) \frac{i}{p^2 - m^2}\\
    & + \frac{i}{p^2 - m^2} i \Sigma (p) \frac{i}{p^2 - m^2}  i \Sigma (p) \frac{i}{p^2 - m^2} + ...,
\end{split}
\end{equation}
where \(i \Sigma (p)\) is the sum of all 1PI Feynman two-point diagrams, as demonstrated in Fig. \ref{selfenergy}.
Therefore, in view of Eqs. (\ref{inv_relation_p}) and (\ref{def_prop}), we deduce that
\begin{equation}
    \tilde{\Gamma}^{(2)} (p,-p) = i \left( p^2 - m^2 + \Sigma(p) \right).
\end{equation}
In addition, higher derivatives of the Legendre transform (\ref{legendre}) can be evaluated using the following chain rule
\begin{equation}
    \frac{\delta }{\delta J(x)} = \int d^4 w \, \frac{\delta \phi_c (w)}{\delta J(x)} \frac{\delta }{\delta \phi_c (w)} = i\int d^4 w \, D (x,w) \frac{\delta }{\delta \phi_c (w)}
\end{equation}
where \(D(x,y)\) is the full propagator in position space. From Eq. (\ref{inverse}) we see that
\begin{equation}
\begin{split}
    \frac{\delta^3 W[J]}{\delta J(x) \delta J(y) \delta J(z)} &= i\int d^4 w \, D(x,w) \frac{\delta }{\delta \phi_c (w)} \frac{\delta^2 W[J] }{\delta J (y) \delta J(z)} \\
    & = - i\int d^4 w \, D(x,w) \frac{\delta }{\delta \phi_c (w)} \left(\frac{\delta^2 \Gamma[\phi_c] }{\delta \phi_c (y) \delta \phi_c(z)}\right)^{-1}
\end{split}
\end{equation}
Using the following differentiation rule
\begin{equation}
    \frac{\delta}{\delta x}  F^{-1}(x)= - F^{-1}(x) \frac{\delta F(x)}{\delta x} F^{-1} (x)
\end{equation}
and
\begin{equation}
    \frac{\delta^3 W[J]}{\delta J(x) \delta J(y) \delta J(z)} = - G^{(3)} (x,y,z),
\end{equation}
we arrive at
\begin{equation*}
\begin{split}
    G^{(3)}_c (x,y,z) & = - i\int d^4 w \, D(x,w) \int d^4 u \, \int d^4 s \, (-i D(y,u)) \frac{\delta^3 \Gamma[\phi_c] }{\delta \phi_c(w) \delta \phi_c (u) \delta \phi_c (s)} (-i D(s,z)) \\
    & = i\int d^4 w \, d^4 u \,  d^4 s\, D(x,w) D(y,u) D(z,s) \frac{\delta^3 \Gamma[\phi_c] }{\delta \phi_c(w) \delta \phi_c (u) \delta \phi_c (s) }.
\end{split}
\end{equation*}
As a result, the connected three-point function is
\begin{equation}
    G^{(3)}_c (x,y,z) = \int d^4 w \, d^4 u \,  d^4 s\, D(x,w) D(y,u) D(z,s) \frac{\delta^3 i \Gamma[\phi_c] }{\delta \phi_c(w) \delta \phi_c (u) \delta \phi_c (s) }.
\end{equation}
This relation expresses that the third derivative of \(i \Gamma[\phi_c]\) is the connected Green's function with all three full propagators removed as shown in Fig \ref{W_G}. Namely, this amputated Green's function is the 1PI three-point function, denoted as 
\begin{equation}
    \Gamma^{(3)} (x,y,z) = \frac{\delta^3 i \Gamma[\phi_c]}{\delta \phi_c (x) \delta \phi_c (y) \delta\phi_c (z)}.
\end{equation}
Likewise, the fourth derivative of \(W [J]\) is related to the fourth derivative of \(i\Gamma [\phi_c]\) which identifies as the 1PI four-point function. In Fig. \ref{4W_G} the connected four-point function (left-hand side) is constructed from a 1PI diagram (\(\Gamma\)-circle in the first term) connected with the full propagators in the external legs (W-circle on the right) and from three one-particle reducible parts which can be generated by each other through "crossing".
Therefore, this can be generalized to higher derivatives and the claim is proven inductively so that
\begin{equation}
     \Gamma^{(n)} (x_1,..,x_n) = \left. \left(\frac{\delta}{\delta \phi_c (x_1)} \right)...\left(\frac{\delta}{\delta \phi_c (x_n)}\right) i \Gamma[\phi_c]  \right|_{\phi_c =0} 
\end{equation}
for \(n \geq 2\), where the \(\Gamma^{(n)} (x_1,..,x_n)\) are called the 1PI Green's functions or 1PI \(n\)-point functions. The induction proof is discussed in detail in Refs. \cite{Peskin:1995ev, Brandenberger:1984cz}. Equivalently, the generating functional for 1PI Green's functions is equal to
\begin{equation}\label{1PI_funct}
    i\Gamma (\phi_c) = \sum_{n=2}^{\infty} \frac{1}{n!} \int d^{4}x_{1} \, ...  \int d^{4}x_{n} \, \phi_c(x_1)...\phi_c(x_n) \Gamma^{(n)} (x_1,..., x_n),
\end{equation}
where the 1PI \(n\)-point function is the sum of the tree-level vertex plus the higher-loop 1PI diagrams with \(n\) external lines (with \(n \geq 3\)). One observes that the sum (\ref{1PI_funct}) starts from \(n = 2\) due to the equation (\ref{ran})
\begin{equation}
    \Gamma^{(1)} (x) = i \left. \frac{\delta \Gamma[\phi_c]}{\delta \phi_c (x)} \right|_{\phi_c = 0} = 0.
\end{equation}

In conclusion, the connected diagrams in the full theory are constructed from trees of 1PI diagrams connected with the full propagators, implying that the 1PI Green's functions are the building blocks of correlation functions. Hence, in the computation of a connected \(n\)-point function in the full theory, we use the tree-level Feynman rules but replace each \(n\)-vertex (with \(n \geq 3\)) of the action for the full theory with the 1PI diagrams represented by the 1PI \(n\)-point function and replace the free propagator with the full propagator. This theorem will be clearly understood in the next section, connecting the 1PI generating functional with the classical action for the full theory.

\section{1PI Effective Action}
\begin{figure}[h!]
    \centering
    \includegraphics[width=0.85\linewidth]{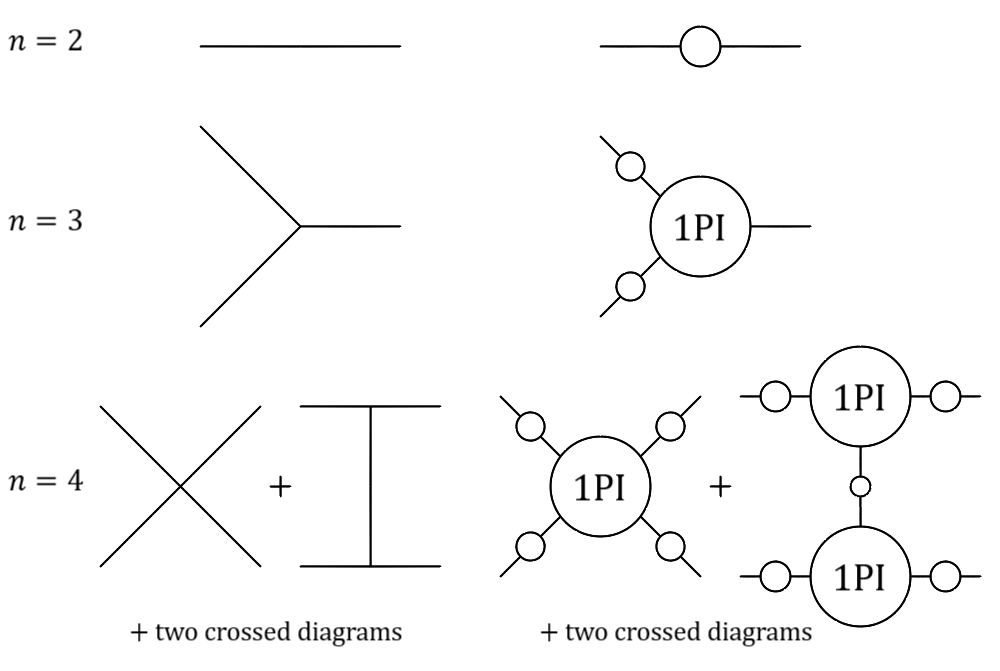}
    \caption{The left-hand side illustrates the Feynman diagrams in the theory described by the effective action, while the right-hand side shows the connected diagrams of the full theory expressed in terms of the 1PI diagrams connected with the full propagators.}
    \label{table_G}
\end{figure}

The previous section motivates us to interpret the generating functional of 1PI diagrams as an action that replaces the classical action \(S[\phi]\) and reproduces the physical predictions of the full quantum theory when it is used at tree level. In other words, this so-called effective action used at tree level should produce the same correlation functions as the classical action would do, including higher loop corrections. This is shown, if the sum of all connected diagrams is obtained by all tree diagrams of the theory described by the effective action, expressing their vertices and propagators in terms of the Green's functions of the full theory. This is illustrated in Fig. \ref{table_G} for the \(n\)-point Green's functions with \(n = 2, 3, 4\) which were also presented explicitly in the previous section.

The above interpretation can be further explained using the saddle-point approximation \cite{Weinberg:1996kr, Fukuda:1974ey, Schwartz:2014sze, Peskin:1995ev, Itzykson:1980rh, Ryder:1985wq ,Jackiw:1974cv}. The generating functional \(Z[J]\), restoring the Planck's constant in the relation (\ref{Z_W}), reads
\begin{equation}
    \exp{\left[\frac{i}{\hbar}W[J]\right]} = \int  \mathcal{D} \phi \, \exp \left[\frac{i}{\hbar} \left( S[\phi] +  \int d^{4}x \, J(x) \phi(x) \right)\right],
\end{equation}
where \(Z[0] = 1\). Analogously, we define the generating functional \(\tilde{W}[J]\), so that
\begin{equation}\label{limith_1}
    \exp{\left[i\Tilde{W}[J]\right]} = \int  \mathcal{D} \phi \, \exp \left[\frac{i}{\hbar} \left( \Gamma[\phi_c] +  \int d^{4}x \, J(x) \phi_c (x) \right)\right]
\end{equation}
and we expand this generating functional in powers\footnote{This expansion and the following analysis could be written in terms of an arbitrary constant \(g\) instead of the Planck's constant \cite{Weinberg:1996kr}.} of \(\hbar\) \cite{Weinberg:1996kr}.
\begin{equation}
    \Tilde{W}[J] = \sum_{l = 0} \hbar^{l-1} W^{(l)}[J],
\end{equation}
where \(l\) is the number of loops in the connected diagrams, while the connected tree-level diagrams scale as \(\hbar^{-1}\) and the loops as \(\hbar^{l-1}\). Considering the \(\hbar \to 0\) limit, according to the stationary phase approximation, the right-hand side of (\ref{limith_1}) is given by the value of the integrand at the saddle-point,
\begin{equation}\label{W(0)}
    \exp{\left[\frac{i}{\hbar}W^{(0)}[J]\right]} = \exp \left[\frac{i}{\hbar} \left( \Gamma[\phi_J] +  \int d^{4}x \, J(x) \phi_J (x) \right)\right],
\end{equation}
where the limit \(\hbar \to 0\) requires that the field configurations contributing to the path integral are exactly those that extremize the exponent, so that
\begin{equation}\label{extremize}
    \frac{\delta }{\delta \phi_c} \left. \left( \Gamma [\phi_c] + \int d^4 x \, \phi_c (x) J(x) \right) \right|_{\phi_c = \phi_J} = 0 \Rightarrow \left. \frac{\delta \Gamma[\phi_c]}{\delta \phi_c} \right|_{\phi_c = \phi_J} = -J(x).
\end{equation}
Then, Eq. (\ref{W(0)}) leads to the inverse Legendre transform 
(\ref{inv_legendre})
\begin{equation}
    W^{(0)} [J] = \Gamma [\phi_J] + \int d^4 x \, \phi_J (x) J(x) 
\end{equation}
and we deduce that \(W^{(0)}[J] = W[J]\). Therefore, all connected diagrams can be computed with the generating functional of connected Green's functions using \(\Gamma [\phi_c]\) in the \(\hbar \to 0\) limit, while Eq. (\ref{W(0)}) is rewritten as
\begin{equation}\label{limith}
     \lim_{\hbar \to 0} \exp \left[\frac{i}{\hbar} W[J] \right]= \lim_{\hbar \to 0}  \int  \mathcal{D} \phi \, \exp \left[ \frac{i}{\hbar} \left( \Gamma[\phi_c] +  \int d^{4}x \, J(x) \phi_c (x) \right)\right],
\end{equation}
where the left-hand side of this equation is the generating functional in a theory described by the action \(\Gamma[\phi_c]\) on the right. Thus, replacing the action of the full theory \(S[\phi]\) by the action \(\Gamma [\phi_c]\) satisfying Eq. (\ref{extremize}) and working at the tree-level leads to the full quantum theory. Namely, the generating functional for 1PI Green's function can be expressed as
\begin{equation}\label{g_s_h}
    \Gamma [\phi_c] = S[\phi_c] + \mathcal{O} (\hbar),
\end{equation}
which is called the 1PI effective action whose vertices are sums of 1PI diagrams in the full quantum theory and the effective equation of motion is
\begin{equation}\label{EOM}
    \frac{\delta \Gamma[\phi_c]}{\delta \phi_c(x)} = -J(x)
\end{equation}
and in the absence of the external source,
\begin{equation}\label{EOM_vacuum}
    \frac{\delta \Gamma[\phi_c]}{\delta \phi_c(x)} = 0,
\end{equation}
which means that the vacuum expectation value of the quantum field operator, \(\bra{\Omega} \hat{\phi} (x) \ket{\Omega}\), in the true vacuum of the theory minimizes the effective action. Namely, the value of the effective action away from its minimum represents the action for the quantum field theory in the presence of an external source. This can be also understood in terms of expectation values. When an external current \(J\) is present, the minimum of the action shifts to the expectation value \(\bra{\Omega} \hat{\phi} (x) \ket{\Omega}_J \). This means that the effective action shows how the minimum changes as the theory is modified \cite{Schwartz:2014sze}. In other words, only the true vacuum state, with vanishing external current, corresponds to the solution to the quantum theory with a given classical action.

Additionally, let us consider a Lagrangian which is invariant under a symmetry transformation. Spontaneous symmetry breaking occurs when a non-zero vacuum expectation value is developed which is not invariant under this symmetry transformation, even if the source is zero\footnote{As it was stated earlier, the zero external current implies that the vacuum expectation value is zero.}. As a consequence, spontaneous symmetry breaking is signaled by
\begin{equation}\label{EOM_vacuum_1}
   \frac{\delta \Gamma[\phi_c]}{\delta \phi_c(x)}  = 0
\end{equation}
for a non-zero vacuum expectation value.

\section{Effective Potential}
In momentum space, the generating functional for 1PI Green's functions is
\begin{equation}\label{b_trans}
    i\Gamma [\phi_c] = \sum_{n=2}^{\infty} \frac{1}{n!} \int \, \prod_{i=1}^n \left[\frac{d^{4}p_{i}}{(2 \pi)^4} \Tilde{\phi}_c(p_i)\right] (2 \pi)^4 \delta^{(4)} \left( \sum_{i = 1}^n p_i \right) \Tilde{\Gamma}^{(n)} (p_1, ..., p_n),
\end{equation}
where, in position space, the 1PI Green's functions are written as,
\begin{equation}
    \Gamma^{(n)} (x_1,..., x_n) = \int \prod_{i=1}^n \left[ \frac{d^{4}p_{i}}{(2 \pi)^4} e^{i p_i x_i} \right](2 \pi)^4 \delta^{(4)} \left( \sum_{i = 1}^n p_i \right) \Tilde{\Gamma}^{(n)} (p_1, ..., p_n), 
\end{equation}
where \(\Tilde{\Gamma}^{(n)} (p_1, ..., p_n) \) is the sum of all 1PI diagrams with \(n\) external lines with momenta \(p_1,..., p_n\). 

Now, we consider field theories in which the vacuum expectation values are invariant under translations as we are not typically interested in the spontaneous breakdown of momentum conservation \cite{Coleman:1973jx}. Hence, in momentum space, the background field \(\phi_c (x) = \phi_c\) is
\begin{equation}
    \Tilde{\phi}_c (p_i) = (2 \pi)^4 \delta^{(4)} (p_i) \phi_c
\end{equation}
and Eq. (\ref{b_trans}) can be computed as
\begin{equation}\label{form_G}
\begin{split}
    i\Gamma [\phi_c] &= \sum_{n=2}^{\infty} \frac{1}{n!} \int \,  \prod_{i=1}^n \left[ \frac{d^{4}p_{i}}{(2 \pi)^4} (2 \pi)^4 \, \delta^{(4)} (p_i) \phi_c \right] (2 \pi)^4 \delta^{(4)} \left( \sum_{i = 1}^n p_i \right) \Tilde{\Gamma}^{(n)} (p_1, ..., p_n) \\
    &= \sum_{n=2}^{\infty} \frac{1}{n!} \phi^{n}_c \int \,  \prod_{i=1}^n \left[ d^{4}p_{i} \, \delta^{(4)} (p_i)  \right] (2 \pi)^4 \delta^{(4)} \left( \sum_{i = 1}^n p_i \right) \Tilde{\Gamma}^{(n)} (p_1, ..., p_n) \\
    &=  \sum_{n=2}^{\infty} \frac{1}{n!} \phi^{n}_c   (2 \pi)^4 \delta^{(4)} \left( 0 \right) \Tilde{\Gamma}^{(n)} (0, ..., 0) \\
    &= (VT) \sum_{n=2}^{\infty} \frac{1}{n!} \phi^{n}_c  \Tilde{\Gamma}^{(n)} (p_i = 0),
\end{split}
\end{equation}
where in the last line a finite space-time region of volume \(VT = (2 \pi)^4 \delta^{(4)} \left( 0 \right)\) is considered. Nevertheless, an alternative expansion of the effective action is in powers of momentum, about the point where all external momenta vanish,
\begin{equation}\label{expand_p}
    \Gamma [\phi_c] = \int d^4 x \, \left[- V_{\text{eff}} (\phi_c) + \frac{1}{2} Z (\phi_c) \left(\partial_{\mu} \phi_c \right)^2 + ... \right],
\end{equation}
where the kinetic term includes the field renormalization \(Z (\phi_c)\) and the function \(V_{\text{eff}} (\phi_c)\) is called the effective potential since one observes the similarities between the expression of the effective action and the classical action in terms of the negative non-derivative part of the Lagrangian density which is defined as the potential energy density in the classical field theory. Therefore, in a translationally invariant theory, the effective action is
\begin{equation}\label{G_V}
    \Gamma [\phi_c] = - \int d^4 x \, V_{\text{eff}} (\phi_c)
\end{equation}
and the effective potential can be cast into the form
\begin{equation}\label{gen_eff}
    V_{\text{eff}} (\phi_c) = i \sum_{n=2}^{\infty} \frac{1}{n!} \, \phi^{n}_c  \, \Tilde{\Gamma}^{(n)} (p_i = 0).
\end{equation}
As a result, the effective potential can be understood as the generating functional of 1PI diagrams with zero external momenta. In addition, Eq. (\ref{EOM_vacuum}) implies that the vacuum expectation value of the quantum field operator is the solution to the equation of motion in the absence of the external current,
\begin{equation}
    \frac{\partial V_{\text{eff}} (\phi_c)}{\partial \phi_c} = 0.
\end{equation}
Let's consider a field theory described by the Lagrangian,
\begin{equation}
    \mathcal{L} = \frac{1}{2} \partial^{\mu} \phi \partial_{\mu} \phi - V (\phi),
\end{equation}
where the classical potential is
\begin{equation}
    V (\phi) = \frac{\lambda_k}{k!} \phi^k.
\end{equation}
To the lowest order in \(\hbar\) only the tree diagrams contribute to the effective potential. Each term in the classical potential corresponds to a single tree-level 1PI diagram since all tree diagrams with more than one vertex are not 1PI. Thus, the 1PI tree-level diagram with \(k\) external lines contributes a factor of \(- i \lambda_k \phi^k_c\) which leads to the tree-level contribution,
\begin{equation}
    V_0 (\phi_c) = \frac{i}{k!} \left(- i \lambda_k \phi^k_c \right) =  \frac{\lambda_k}{k!} \phi^k_c.
\end{equation}
This result will be further illustrated in the next sections in Fig. \ref{fig:1.2}. Hence, at tree level, the effective potential is identified as the classical potential,
\begin{equation}\label{classical_effective}
    V_{\text{eff}} (\phi_c) = V_0 (\phi_c) + \mathcal{O}(\hbar)
\end{equation}

Lastly, the classical potential can be interpreted as the energy density for that state in which the field is \(\phi\). On the other hand, the effective potential is proved to be the expectation value of the energy density in the class of all (normalized) states \(\ket{\psi}\) which minimize the expectation value of the Hamiltonian density, i.e. $\delta \bra{\psi} \hat{\mathcal{H}}\ket{\psi} = 0$, satisfying \(\bra{\psi} \hat{\phi} (x) \ket{\psi} = \phi_c\). A proof can be found in Ref. \cite{Brandenberger:1984cz}. This physical interpretation of the effective potential is also closely related to two fundamental issues that triggered a great deal of debate in the past decades: the presence of imaginary parts in the effective potential and the convexity problem in the formulation of the effective action as the Legendre transform of the generating functional for connected Green's functions \cite{Sher:1988mj}. The convexity problem states that the effective action is identified with the generating functional for 1PI Green's functions only if the effective potential is convex \cite{Sher:1988mj, ORaifeartaigh:1976bhs,Brandenberger:1984cz, Dannenberg:1987fw}. This problem is resolved as presented in Ref. \cite{Dannenberg:1987fw}. In the following, both problems are not discussed further, and the effective action is directly identified with the 1PI generating functional.

\section{Background Field Method}

The background field method is an alternative way to compute the effective action which can be derived from the action of the full theory via a Legendre transform \cite{Abbott:1981ke, Schwartz:2014sze}. Initially, we shift the field \(\phi \to \phi + \phi_b \) by an arbitrary background field configuration \(\phi_b\). In the background field method, the generating functional for all connected Green's functions follows
\begin{equation}
    e^{i W_b [\phi_b,J]} = \int  \mathcal{D} \phi \, \exp{\left(i S[\phi + \phi_b] + i \int d^{4}x J(x) \phi(x) \right)}.
\end{equation}
Then, one easily proves the following
\begin{equation}
    \frac{\delta W_b[\phi_b, J]}{\delta J} = \phi_{J;b}.
\end{equation}
Now the shift \(\phi \to \phi - \phi_b\) in the path integral leads to the relation between the background field and conventional generating functional for all connected Green's functions
\begin{equation}\label{def_Wb}
    W_b [\phi_b, J] = W[J] - \int d^4x \, J (x) \phi_b (x)
\end{equation}
and the differentiation with respect to the source implies that
\begin{equation}\label{34.35}
    \frac{\delta W_b[\phi_b]}{\delta J} = \frac{\delta W[J]}{\delta J} - \phi_b \Rightarrow \phi_{J;b} (x) = \phi_{J} (x) - \phi_b (x).
\end{equation}
This relation demonstrates that if we shift \(\phi \to \phi + \phi_b \) in the path integral, the expectation value of a field in the presence of an external source shifts by \(- \phi_b\). Finally, the 1PI effective action can be expressed through the Legendre transform with the additional dependence on the background field,
\begin{equation}
    \Gamma_b [\phi_b, \phi] = W_b [\phi_b, J_{\phi;b}] - \int d^4 x \, J_{\phi;b} (x) \phi(x).
\end{equation}
Using Eq. (\ref{def_Wb}) one gets
\begin{equation}\label{34.38}
    \Gamma_b [\phi_b, \phi] = W[J_{\phi;b}] - \int d^4 x \, J_{\phi;b} (x) (\phi (x) + \phi_b (x)) .
\end{equation}
If we consider the result of Eq. (\ref{34.35}) with \(\phi = \phi_{J;b}\) and \(J_{\phi_J} = J\), Eq. (\ref{34.38}) is written as
\begin{equation}
    \Gamma_b [\phi_b, \phi_{J;b}] = W[J] - \int d^4 x \, J(x) \phi_J (x) = \Gamma[\phi_J] = \Gamma [\phi_{J;b} + \phi_b].
\end{equation}
Thus, this relation can be rewritten as it holds for any source:
\begin{equation}
    \Gamma_b [\phi_b, \phi] = \Gamma [\phi + \phi_b]
\end{equation}
and it is concluded that \(\Gamma[\phi_b] = \Gamma_b [\phi_b, 0]\), meaning that the 1PI effective action is determined by the background field effective action \(\Gamma_b [\phi_b, 0]\) which is computed by summing all 1PI vacuum graphs in the presence of the background field \cite{Abbott:1981ke}. In this approach, the field \(\phi\) could be understood as the quantum fluctuations around a classical background field \(\phi_b\).

Now we present a functional approach to define the effective action as the Legendre transform of the generating functional of all connected diagrams and prove the relation between the classical and effective action in (\ref{g_s_h}). This approach was developed by R. Jackiw \cite{Jackiw:1974cv}. Let's define a modified action as
\begin{equation}
    \Tilde{S}[\phi, J] = S[\phi] + \int d^4 x \,  \phi(x) J(x),
\end{equation}
where \(S[\phi]\) is the classical action. The generating functional is then written as
\begin{equation}\label{pathint}
     \exp{\left[\frac{i}{\hbar}W[J]\right]} = \int  \mathcal{D} \phi \, \exp \left[\frac{i}{\hbar} \left( S[\phi] + \int d^4 x \,  \phi(x) J(x) \right) \right],
\end{equation}
where we restored the Planck's constant and in the limit \(\hbar \to 0\) the path integral is dominated by the classical solution \(\phi_{s}\) which extremizes the exponent,
\begin{equation}\label{classicaleq}
    \frac{\delta S[\phi]}{\delta \phi(x)} = - J(x),
\end{equation}
which defines \(\phi_s\) as an implicit functional of \(J\). Expanding the modified action around the classical solution \(\phi_s = \phi  - \Tilde{\phi}\):
\begin{equation}
\begin{split}
     \Tilde{S}[\phi,J] &= S[\phi_s] + \int d^4 x \, \phi_s (x) J(x) + \frac{1}{2} \int d^4 x \, d^4 y \, \Tilde{\phi} (x)  \left.\frac{\delta^2 S[\phi]}{\delta \phi(x) \phi(y)} \right|_{\phi_s} \Tilde{\phi}(y) + ...\\
    & = S[\phi_s] + \int d^4 x \, \phi_s (x) J(x) - \frac{1}{2} \int d^4 x \, \Tilde{\phi} (x)  \left[ \partial^2 + V'' (\phi_s) \right] \Tilde{\phi}(x) + ...,
\end{split}
\end{equation}
where the functional differentiation of the action for a scalar field theory leads to
\begin{equation}
    \left.\frac{\delta^2 S[\phi]}{\delta \phi(x) \phi(y)} \right|_{\phi = \phi_s} = - \left[ \partial^2 + V'' (\phi_s) \right] \delta^{(4)} (x-y)
\end{equation}
and in the saddle-point approximation, the path integral gives
\begin{equation}\label{sad}
\begin{split}
    e^{\frac{i}{\hbar}W[J]} & \simeq e^ {\frac{i}{\hbar} \left[S[\phi_s] + \int d^4 x \,  \phi_s (x) J(x) \right]} \int  \mathcal{D} \Tilde{\phi} \, e^{- \frac{i}{2\hbar} \int d^4 x \, \Tilde{\phi} (x) \left[ \partial^2 + V'' (\phi_s) \right] \Tilde{\phi} (x) } \\
    & = e^ {\frac{i}{\hbar} \left[S[\phi_s] + \int d^4 x \,  \phi_s (x) J(x) \right]} \left( \det \left[ \partial^2 + V'' (\phi_s) \right] \right)^{-1/2},
\end{split}
\end{equation}
where in the first line the integral in the second exponent is evaluated after a Wick rotation to the Euclidean space 
and the field \(\tilde{\phi}\) was additionally rescaled to \(\tilde{\phi} \to \sqrt{\hbar} \Tilde{\phi}\) to form the loop expansion. Next, the following formula is used
\begin{equation}
     \det A  = e^{Tr \ln A}
\end{equation}
and Eq. (\ref{sad}) yields to
\begin{equation}\label{back}
    W[J] = S[\phi_s] + \int d^4 x \,  \phi_s (x) J(x)  + \frac{i \hbar}{2} Tr \ln \left[ \partial^2 + V'' (\phi_s) \right] + \mathcal{O} \left( \hbar^2\right).
\end{equation}
In order to obtain the Legendre transform of Eq. (\ref{back}) we compute initially at the classical limit,
\begin{equation}
    \frac{\delta W[J]}{\delta J(x)} = \int d^4 y \, \left(\frac{\delta S[\phi_s]}{\delta \phi_s (y)} \frac{\delta \phi_s (y)}{\delta J(x)} + J(y) \frac{\delta \phi_s (y)}{\delta J(x)}\right)  + \phi_s (x) = \phi_s (x),
\end{equation}
according to the classical equation (\ref{classicaleq}). Therefore, at the one-loop approximation this is written as
\begin{equation}\label{c = s + h}
    \frac{\delta W[J]}{\delta J(x)} = \phi_c (x) = \phi_s(x) + \mathcal{O} (\hbar) 
\end{equation}
and the Legendre transform of the generating functional for all connected Green's functions can be written as 
\begin{equation}
    \Gamma[\phi_c] = S[\phi_s] + \int d^4 x \,  \phi_s (x) J(x) - \int d^4 x \,  \phi_c (x) J(x) + \frac{i \hbar}{2} Tr \ln \left[ \partial^2 + V'' (\phi_s) \right] + \mathcal{O} \left( \hbar^2\right)
\end{equation}
and expanding the functionals around the field \(\phi_c\) leads to 
\begin{equation}\label{last_gam}
    \Gamma[\phi_c] = S[\phi_c] + \frac{i \hbar}{2} Tr \ln \left[ \partial^2 + V'' (\phi_c) \right] + \mathcal{O} \left( \hbar^2\right).
\end{equation}
This relation yields to (\ref{g_s_h}),
\begin{equation}
    \Gamma[\phi_c] = S[\phi_c]  + \mathcal{O} \left( \hbar\right),
\end{equation}
where the one-loop correction has already been computed in (\ref{last_gam}). It is very intriguing that at the limit \(\hbar \to 0\) the expectation value \(\phi_c\) of the quantum field operator is equal to the classical solution \(\phi_s\) on account of (\ref{c = s + h}).

The one-loop correction in (\ref{last_gam}) can be computed by finding all the eigenvalues of the operator in the trace. In a translationally invariant theory, the second derivative of the classical potential is constant which leads to  
\begin{equation}\label{trace}
\begin{split}
    Tr \ln \left( \partial^2 + V'' (\phi_c) \right) & = \int d^4 x \, \bra{x} \ln \left( \partial^2 + V'' (\phi_c) \right) \ket{x} \\
    & = \int d^4 x \int d^4 p \int d^4 k \, \langle x | p \rangle \bra{p}  \ln \left( \partial^2 + V'' (\phi_c) \right) \ket{k} \langle k | x \rangle.
\end{split}
\end{equation}
The action of the logarithm operator on a momentum state is 
\begin{equation}
    \ln \left( \partial^2 + V '' (\phi_c) \right) \ket{k} = \ln \left( - k^2 + V '' (\phi_c) \right) \ket{k}
\end{equation}
and Eq. (\ref{trace}) is 
\begin{equation}
    Tr \ln \left( \partial^2 + V'' (\phi_c) \right) = \int d^4 x \int \frac{d^4 k}{(2 \pi)^4} \,  \ln \left( -k^2 + V'' (\phi_c) \right),
\end{equation}
where we have used \(\langle p | k \rangle = \delta^{(4)} (p - k) \) and the normalization
\begin{equation*}
    |\langle x | k \rangle|^2 = \frac{1}{(2 \pi)^4} .
\end{equation*}
Thus, after a Wick rotation to the Euclidean space, \(p^0 = i p^0_E\), the effective potential at the one-loop approximation is\footnote{A field-independent term could be inserted to have a dimensionless argument for the logarithm, but this term is omitted to compare and discuss this result in the next section.}
\begin{equation}
    V_{\text{eff}} (\phi_c) = V_0 (\phi_c) + \frac{1}{2} \int \frac{d^4 p_E}{(2 \pi)^4} \,  \ln \left( p_E^2 + V'' (\phi_c) \right).
\end{equation}
The effective mass-squared is then defined as \(m^2_{\text{eff}} (\phi_c) = V ''(\phi_c)\) to rewrite the effective potential as
\begin{equation}
    V_{\text{eff}} (\phi_c) = V_0 (\phi_c) + \frac{1}{2} \int \frac{d^4 p_E}{(2 \pi)^4} \,  \ln \left( p_E^2 + m^2_{\text{eff}} (\phi_c) \right).
\end{equation}
This motivates us to interpret the field \(\phi_c\) as a background field, while the effective mass introduces a correction to the mass of the scalar particle due to the interaction of the particle with this background field \cite{Zee:2003mt}. The higher-loop effective potential can be computed following this functional method \cite{Jackiw:1974cv} which is highly advantageous in higher loop corrections compared to evaluating the 1PI Feynman diagrams as shown in the next section.

\section{One-loop Effective Potential}
In this section, we demonstrate a straightforward method \cite{Coleman:1973jx} to derive the one-loop effective potential for scalar field theories and non-Abelian gauge theories using the 1PI Feynman diagrams based on the formula (\ref{gen_eff}).
\subsubsection{Scalar Field Theory}
\begin{figure}[h!]
    \centering
    \includegraphics[width=0.55\linewidth]{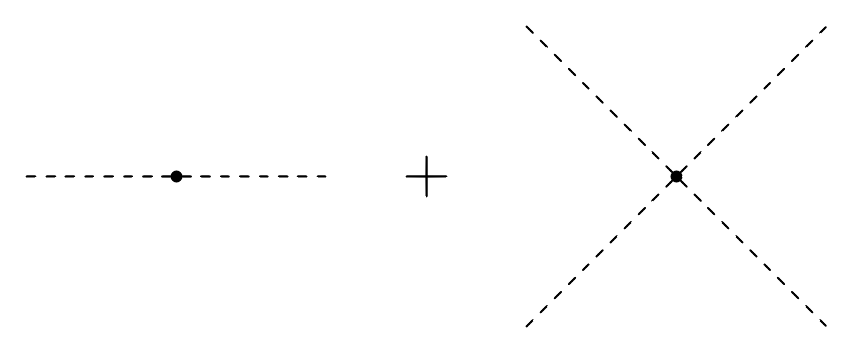}
    \caption{The tree-level diagrams which correspond to the zero-loop contribution to the effective potential.}
    \label{fig:1.2}
\end{figure}

\begin{figure}[H]
    \centering
    \includegraphics[width=0.85\linewidth]{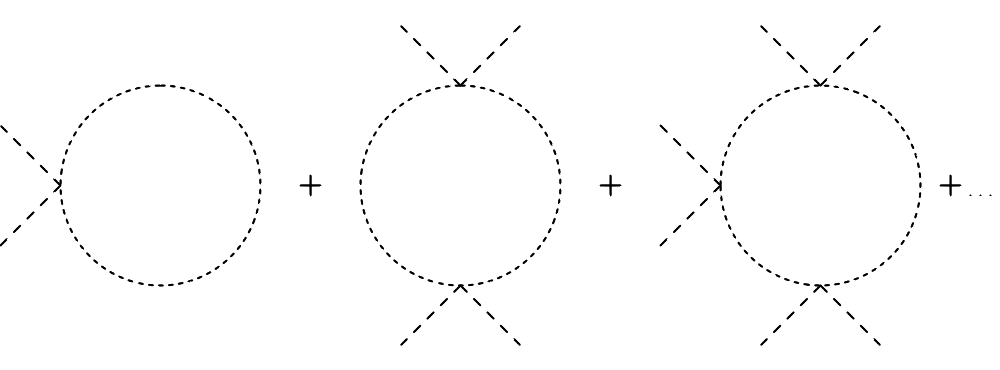}
    \caption{The 1PI diagrams with a single loop and zero external momenta which are summed to obtain the one-loop correction to the effective potential.}
    \label{fig:1.3}
\end{figure}

Consider a self-interacting real scalar field, described by the Lagrangian density
\begin{equation}\label{lag_scalar}
    \mathcal{L} = \frac{1}{2} \partial^{\mu} \phi \partial_{\mu} \phi - \frac{1}{2} m^2 \phi^2 - \frac{\lambda}{4 !} \phi^4.
\end{equation}
We previously discussed that the effective potential consists of the zero-loop contribution, which is the classical potential, and higher loop corrections. Eq. (\ref{classical_effective}) shows that the zero-loop contribution corresponds to the classical potential, represented by Fig. \ref{fig:1.2}, and it is given by
\begin{equation}
    V_0 (\phi_c) = \frac{1}{2} m^2 \phi_c^2 + \frac{\lambda}{4 !} \phi_c^4.
\end{equation}
Then, the one-loop correction is the sum of all 1PI Feynman diagrams with a single loop and zero external momenta, as illustrated in Fig. \ref{fig:1.3}, where the \(n^{th}\) diagram is comprised of \(n\) propagators, \(n\) vertices, and \(2n\) external lines. 
Therefore, Eq. (\ref{gen_eff}) is written as 
\begin{equation}\label{gen_eff_2}
    V_{\text{eff}} (\phi_c) = V_0(\phi_c) + V_1 (\phi_c) =\frac{1}{2} m^2 \phi_c^2 + \frac{\lambda}{4 !} \phi_c^4 + i \sum_{n=1}^{\infty} \frac{1}{(2n)!} \, \phi^{2n}_c  \, \Tilde{\Gamma}^{(2n)} (p_i = 0) ,
\end{equation}
where the 1PI \(2n\)-point function is the sum of all 1PI diagrams with \(2n\) external lines with zero momenta and is computed as
\begin{equation}
    \Gamma^{(2n)} (p_i = 0) = \frac{(2n)!}{2n} \int \frac{d^4 p}{(2 \pi)^4} \left(\frac{ - i \lambda}{2}\right)^n \left(\frac{i}{p^2 - m^2 + i\epsilon} \right)^n \, ,
\end{equation}
according to the following Feynman rules stated here: 
\begin{itemize}
    \item The \(n\) propagators contribute a factor of \(i \left(p^2 - m^2 + i \epsilon\right)^{-n}\).
    \item Each vertex contributes a factor of \(- i \lambda/2\), where \(1/2\) is a Bose statistics factor since interchanging the two external lines of the vertex does not change the diagram.
    \item There is a global symmetry factor \((2n)!/2n\), where the factor of \((2n)!\) corresponds to the number of ways to distribute \(2n\) particles in \(2n\) external lines. Then, the factor \(1/n\) comes from the symmetry of the diagram under the discrete group of rotations \(\mathbb{Z}_n\) and the factor \(1/2\) from the symmetry of the diagram under reflections. 
    \item There is a momentum integration over the loop, which is the same for each diagram as all external momenta are zero.
\end{itemize}
Consequently, the one-loop contribution to the effective potential is given by
\begin{equation}
\begin{split}
    V_1 (\phi_c) & = i \sum_{n=1}^{\infty} \int \frac{d^4 p}{(2 \pi)^4} \frac{1}{2 n} \left(\frac{ - i \lambda}{2}\right)^n \left(\frac{i}{p^2 - m^2 + i\epsilon} \right)^n \phi^{2n}_c \\
    & = i \sum_{n=1}^{\infty} \int \frac{d^4 p}{(2 \pi)^4} \frac{1}{2 n} \left(\frac{ \lambda \phi^2_c/2}{p^2 - m^2 + i\epsilon} \right)^n \, .
\end{split}
\end{equation}
According to the Taylor series of logarithms, this expression is cast into the form
\begin{equation}\label{beforeWick}
    V_1 (\phi_c) = - \frac{i}{2}  \int \frac{d^4 p}{(2 \pi)^4} \ln \left[1 - \frac{\lambda \phi^2_c/2}{p^2 - m^2 + i\epsilon} \right] ,
\end{equation}
where after a Wick rotation to the Euclidean space, \( p_{E} = \left(-i p^0, \Vec{p} \right) \), this can be rewritten as
\begin{equation}\label{afterWick}
    V_1 (\phi_c) = \frac{1}{2}  \int \frac{d^4 p_{E}}{(2 \pi)^4} \ln \left[1 + \frac{\lambda \phi^2_c/2}{p_{E}^2 + m^2} \right]
\end{equation}
The background field method in the previous section also ends up in Eq. (\ref{afterWick}) introducing the effective mass-squared,
\begin{equation}\label{meff_1}
    m^2_{\text{eff}} (\phi_c) = m^2 + \frac{\lambda}{2}  \phi^2_c.
\end{equation}
Omitting a field-independent term can lead to the final formula
\begin{equation}\label{finalV1}
    V_1 (\phi_c) = \frac{1}{2}  \int \frac{d^4 p_{E}}{(2 \pi)^4} \ln \left[p_{E}^2 + m^2_{\text{eff}} (\phi_c) \right]
\end{equation}
and the effective potential can be cast as
\begin{equation}\label{final_Veff_scalar}
    V_{\text{eff}}(\phi_c) = \frac{1}{2} m^2 \phi_c ^2 + \frac{\lambda}{4 !} \phi_c^4 + \frac{1}{2}  \int \frac{d^4 p_{E}}{(2 \pi)^4} \ln \left[p_{E}^2 + m^2_{\text{eff}} (\phi_c) \right].
\end{equation}

If the Lagrangian density describes a theory with \(N\) complex scalar fields
\begin{equation}
    \mathcal{L} = \partial^{\mu} \phi^a \partial_{\mu} \phi^{\dagger}_{a} - V (\phi_a, \phi^{\dagger}_a),
\end{equation}
the effective potential is a trivial generalization of the previous formula (\ref{final_Veff_scalar}),
\begin{equation}\label{final_Veff_complex_scalar}
    V_{\text{eff}}(\phi_c) = V_0 (\phi_a, \phi^{\dagger}_a) + \frac{1}{2} Tr \int \frac{d^4 p_{E}}{(2 \pi)^4} \ln \left[p_{E}^2 + M^2 (\phi_a, \phi^{\dagger}_b) \right],
\end{equation}
where the effective mass-squared is defined as
\begin{equation}\label{meff_2}
    (M^2)^{b}_a \equiv V^a_b = \frac{\partial^2 V_0 }{\partial \phi^{\dagger}_a \partial \phi^b}
\end{equation}
and each complex field has two degrees of freedom with \(Tr (\mathbf{1}) = 2 N\) so that \(Tr (M^2) = 2 V^a_a\).

\subsubsection{Non-Abelian Gauge Theories}
Similarly, we compute the one-loop effective potential in the case of non-Abelian gauge theories \cite{Coleman:1985rnk, Sher:1988mj, Coleman:1973jx, Brandenberger:1984cz}. The Lagrangian density of the theory is written as
\begin{equation}\label{Lag_non_Abelian}
    \mathcal{L} = - \frac{1}{4} Tr \left( F_{\mu \nu} F^{\mu \nu} \right) +  Tr \left(i \overline{\psi} \slashed D \psi \right) - Tr \left(\overline{\psi} \Gamma \phi \psi \right) + \frac{1}{2} Tr \left( D_{\mu} \phi \right)^{\dagger} D^{\mu} \phi - V (\phi).
\end{equation}
Given a non-Abelian gauge group \(G\) with generators \(\tau_{\alpha}\), a set of spinor fields \(\psi_{a}\), and a set of scalar fields \(\phi_i\), each of which forms a representation of the group, we then define
\begin{equation*}
    A_{\mu} = A^{\alpha}_{\mu} \tau_{\alpha},
\end{equation*}
\begin{equation}
    F_{\mu \nu} = \partial_{\mu} A_{\nu} - \partial_{\nu} A_{\mu} + i g [A_{\mu}, A_{\nu}],
\end{equation}
\begin{equation*}
    D_{\mu} = \partial_{\mu} + i g A_{\mu}.
\end{equation*}
In an appropriate gauge, the one-loop effective potential as a function of the expectation values of the scalar fields, \(\phi_i\), consists of
\begin{equation}
    V_{\text{eff}} (\phi_c) = V_0 (\phi_c) + V_{s} (\phi_c) + V_{f} (\phi_c) + V_{g} (\phi_c) + V_{c} (\phi_c) ,
\end{equation}
where \( V_0 (\phi_c)\) is the classical potential (tree-level approximation), \(V_{c} (\phi_c)\) is a quartic polynomial
including all counterterms, \(V_{s} (\phi_c)\), \(V_{f} (\phi_c)\) and \(V_{g} (\phi_c)\) are the contributions from the scalar-field, the fermion, and the gauge-field loops, respectively. The first two terms have been already computed as the effective potential (\ref{final_Veff_complex_scalar}) in the case of the complex scalar fields.

\begin{figure}[h!]
    \centering
    \includegraphics[width=0.95\linewidth]{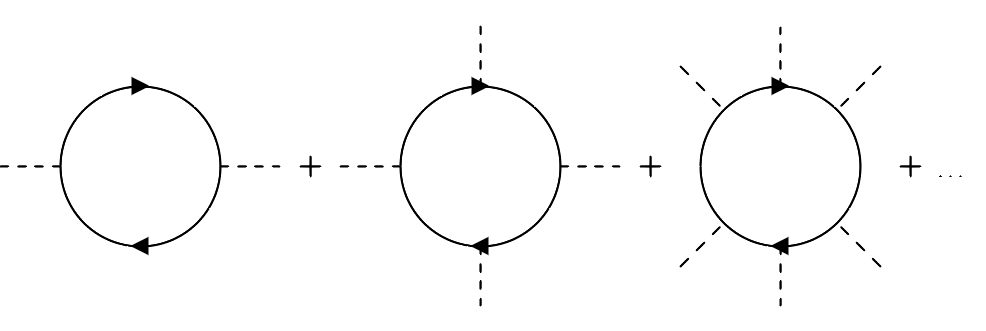}
    \caption{The 1PI fermion diagrams which contribute to the one-loop correction (\ref{fermion_eff}).}
    \label{fig:1.4}
\end{figure}

The contribution of the fermion loops comes from the following part of the Lagrangian:
\begin{equation}\label{Lag_ferm}
    \mathcal{L}_f = i \overline{\psi}_a \slashed \partial \psi^a - \overline{\psi}_a m^a_b \psi^b,
\end{equation}
where the mass matrix \(m^a_b (\phi^i_c) = \Gamma^a_{bi} \phi^i_c\) is a function of the scalar fields\footnote{The \(\Gamma^a_{bi}\) could also denote a matrix of Yukawa coupling constants.}, such as
\begin{equation}
    m^a_b (\phi^i_c) = A^a_{b} (\phi^i_c)  + i \gamma_5 B^a_b (\phi^i_c),
\end{equation}
where \(A\) and \(B\) are Hermitian matrices and linear functions of the scalar fields and \(\gamma_5 = i \gamma_0 \gamma_1 \gamma_2 \gamma_3\). The same method for the calculation of the effective potential can be adopted here as in the scalar field theory, but the Feynman diagrams with an odd number of vertices do not contribute to the one-loop effective potential because the trace of an odd number of \(\gamma\) matrices is zero. Hence, the one-loop correction is given by the sum of the 1PI diagrams which are presented in Fig. \ref{fig:1.4}.
For instance, according to the Feynman rules, the first diagram with two external lines in Fig. \ref{fig:1.4} corresponds to
\begin{equation}
    -  \frac{1}{2} Tr_s \left( m^{a}_b \frac{\slashed p^2}{p^4} m^{b}_a \right) = - \frac{1}{2} Tr_s \left( \phi^i_{c} \Gamma^a_{bi} \frac{\slashed p^2}{p^4} \Gamma^{b}_{aj} \phi^j_c \right) = - \frac{1}{2} \frac{Tr\left[(\Gamma \phi_c )^2\right]}{p^2} Tr_s(\mathbf{1}),
\end{equation}
where the minus sign comes from the fermion loop, the trace runs over the spinor indices, and \(\slashed p^2 = p^2 \mathbf{1}\). Analogously, the next terms in the one-loop correction are obtained by the following contributions: 
\begin{itemize}
    \item The \(2n\) propagators contribute a factor of \(Tr_s \left[i^{2n} \slashed p^{2n} \left(p^2 + i \epsilon \right)^{-2n} \right]\). 
    \item The \(2n\) vertices yield a factor \(Tr \left( - i^{2n} m^{2n} \right)\), where the trace refers to the different fermionic fields.
    \item There is a combinatorial factor \(1/2n\) from the cyclic and anticyclic symmetry of the diagrams in Fig. \ref{fig:1.4}.
    \item There is a momentum integration over the loop, which is the same for each diagram as all external momenta are zero.
\end{itemize}
Thus, the one-loop correction is computed as
\begin{equation}\label{fermion_eff}
\begin{split}
     V_f (\phi_c) & = i \sum_{n=1}^{\infty} \int \frac{d^4 p}{(2 \pi)^4} \frac{1}{2 n} Tr \left[ \left( - i m \right)^{2n} \right] Tr_s \left[\left( \frac{i \slashed p}{p^2 + i \epsilon} \right)^{2n} \right] \\
     & = - i \sum_{n=1}^{\infty} \int \frac{d^4 p}{(2 \pi)^4} \frac{1}{2 n} \frac{Tr \left( m^{2n} \right)}{p^{2n}} Tr_s \left(\mathbf{1}\right) \\
     & = - i Tr \sum_{n=1}^{\infty}  \int \frac{d^4 p}{(2 \pi)^4} \frac{n_f}{2 n} \left(\frac{  m^2}{p^{2}}\right)^n,
\end{split}
\end{equation}
where the trace \(Tr_s(\mathbf{1}) = n_f\) counts the number of degrees of freedom of the fermions. The Dirac (Weyl) fermions have \(n_f = 4\) \((n_f = 2)\) degrees of freedom. Using the logarithmic Taylor series, this expression yields to
\begin{equation}\label{beforeWick_ferm}
    V_f (\phi_c) = \frac{in_f}{2} Tr  \int \frac{d^4 p}{(2 \pi)^4} \ln \left[ 1 - \frac{  m^2}{p^{2}}\right],
\end{equation}
After a Wick rotation, the final expression for the contribution of the fermion loops can be written as
\begin{equation}\label{fermi_eff}
      V_f (\phi_c) = - \frac{n_f}{2} Tr  \int \frac{d^4 p_E}{(2 \pi)^4} \ln \left[ p^2_E + m^2(\phi_c)\right].
\end{equation}

\begin{figure}[h!]
    \centering
    \includegraphics[width=0.95\linewidth]{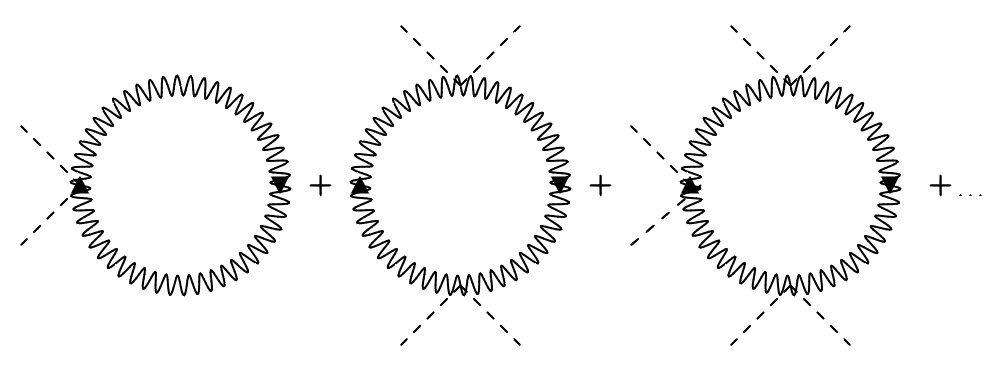}
    \caption{The 1PI gauge-field diagrams which contribute to the one-loop correction (\ref{gauge_eff}).}
    \label{fig:1.5}
\end{figure}

Finally, we compute the one-loop correction due to the gauge-field loops. The 1PI diagrams that contribute to the one-loop correction are depicted in Fig. \ref{fig:1.5}. 
In the Landau gauge, the gauge-boson propagator is expressed as
\begin{equation}
    \Pi_{\mu\nu} = - \frac{i}{p^2 + i \epsilon} \left( g_{\mu\nu} - \frac{p_{\mu}p_{\nu}}{p^2} \right).
\end{equation}
In the effective potential, the \(A^2 \phi^2\) vertex solely contributes to the one-loop order, coming from the fourth term in the Lagrangian which is expressed as
\begin{equation}
    \frac{1}{2} Tr \left[\left( i g A_{\mu} \phi \right)^{\dagger} i g A^{\mu} \phi \right] = \frac{1}{2} g_{\alpha} g_{\beta} Tr \left[ (\tau_{\alpha} A^{\alpha}_{\mu} \phi)^{\dagger} \tau_{\beta} A^{\mu \beta} \phi \right] = \frac{1}{2} M^2_{\alpha \beta} A^{\alpha}_{\mu} A^{\mu \beta} ,
\end{equation}
where \(M^2\) is the gauge boson effective-coupling constant matrix\footnote{This means that after the spontaneous symmetry breaking with the Higgs vacuum expectation value (\ref{Higgs doublet VEV}), the effective mass squared \(M^2\) is the gauge boson mass matrix.}, such as the squared effective masses (\ref{meff_1}) and (\ref{meff_2}), and it is given by 
\begin{equation}
    M^2_{\alpha \beta} = g_{\alpha} g_{\beta} Tr \left[ (\tau_{\alpha} \phi)^{\dagger} \tau_{\beta}  \phi \right] ,
\end{equation}
where \(g_{\alpha}\) is the coupling constant\footnote{If the non-Abelian group is simple, such as \(SO(10)\) and \(SU(5)\), all the gauge couplings are equal.} associated with the gauge field \(A^{\alpha}_{\mu}\) and the trace runs over the indices in the representation of the scalar fields. Following similar calculations to the previous cases, we conclude that the combinatorial factor for the \(n^{th}\) diagram in Fig. \ref{fig:1.5} with \(n\) propagators and \(n\) vertices is written as
\begin{equation}
    \frac{1}{2 n} \frac{Tr \left( M^{2n} \right)}{p^{2n}} Tr \left(\Delta^n\right),
\end{equation}
where \(\Delta_{\mu \nu}\) is defined as
\begin{equation}\label{delta_de}
    \Delta_{\mu \nu} = g_{\mu\nu} - \frac{p_{\mu}p_{\nu}}{p^2}.
\end{equation}
It satisfies
\begin{equation}
    p^{\mu} \Delta_{\mu \nu} = 0
\end{equation}
and
\begin{equation}
    \Delta^{\mu \nu} \Delta_{\nu\lambda} = \Delta^{\mu}_{\,\, \lambda} \quad \Rightarrow \quad \Delta^n = \Delta
\end{equation}
for $n\in \mathbb{N}^{+}$. This leads to
\begin{equation}\label{delta_trace}
    Tr\left(\Delta^{n} \right) =  Tr\left(\Delta\right) = \Delta^{\mu}_{\,\, \mu} = \delta^{\mu}_{\, \mu} - \frac{p^{\mu}p_{\mu}}{p^2} = \delta^{\mu}_{\, \mu} - 1 = 3.
\end{equation}
Notice that this trace coincides with the number of degrees of freedom of a massive gauge boson\footnote{In the Feynman gauge, this factor does not coincide with the number of degrees of freedom and reduces from 3 to 2 due to the contributions of the unphysical fields to the one-loop effective potential \cite{Delaunay:2007wb}.}. Therefore, in the Landau gauge, the one-loop contribution to the effective potential from the gauge-field loops is written as
\begin{equation}\label{gauge_eff}
      V_g (\phi_c) = \frac{3}{2} Tr  \int \frac{d^4 p_E}{(2 \pi)^4} \ln \left[ p^2_E + M^2(\phi_c)\right],
\end{equation}
which is the final result after a Wick rotation to the Euclidean space.

Now it is important to notice that the effective potential is gauge-dependent \cite{Jackiw:1974cv}. This is clearly demonstrated, if the effective potential is computed in the \(R_{\xi}\) gauge, which introduces a gauge fixing term in the Lagrangian, and the effective potential depends explicitly on the gauge parameter \(\xi\) \cite{Fujikawa:1972fe, Yao:1973am, DiLuzio:2014bua}. Although the effective potential is not a physical quantity, all observable quantities should be gauge invariant. Nevertheless, the gauge dependence of the finite-temperature effective potential, which will be presented later, may propagate to the calculations of the phase transitions, while the critical temperature and the true vacuum of the theory are obtained by the effective potential. As a result, a number of gauge-independent methods have been developed to compute the effective potential and determine the gauge-invariant observables of the phase transitions in the Standard Model and other new physics theories \cite{Yao:1973am, DiLuzio:2014bua, Dolan:1974gu, Nielsen:1975fs, Aitchison:1983ns, Espinosa:2016uaw, Ekstedt:2018ftj, Nielsen:2014spa, Alexander:2008hd, DelCima:1999gg, Kobes:1990dc}. In this work, the gauge dependence of the effective potential (as an active topic in research) is not discussed further to continue our introduction to the one-loop effective potential.

\section{Renormalization}

The calculation of quantum corrections to the classical potential involves adding up all of the 1PI diagrams and integrating over the internal momenta. For instance, each individual 1PI diagram in Fig. \ref{fig:1.3} has a polynomial infrared singularity, whereas the one-loop contribution to the effective potential (\ref{finalV1}) has only a logarithmic ultraviolet divergence\footnote{In fact, the divergence in the effective potential is infrared since we neglected a field-independent term which depends on the momentum in Eq. (\ref{afterWick}).}. In the case of a renormalizable theory, the parameters of the theory absorb these divergences. More specifically, we proceed with regularizing the theory and absorbing all infinities by the appropriate counterterms, that we have not considered in our analysis yet. The definition of the renormalized parameters then determine how these infinities are absorbed by the counterterms. As a result, the theory is finite and it is expressed in terms of the renormalized parameters. Next, we will examine the cut-off and dimensional regularization schemes and discuss various renormalization schemes. A general and concise review of renormalization can be found in a number of quantum field theory books, such as in \cite{Mandl:1985bg, Schwartz:2014sze, Peskin:1995ev, Itzykson:1980rh, Ryder:1985wq}. A more detailed review of the renormalized one-loop effective potential is discussed in Refs. \cite{Sher:1988mj, Coleman:1973jx, Brandenberger:1984cz}.

\subsection{Cut-off Regularization}
We present the cut-off regularization scheme in the case of a massless real scalar field to renormalize the effective potential as well as showcase the impact of the quantum corrections on the vacuum state which at a classical level resides at the origin (\(\phi_c = 0\)) \cite{Sher:1988mj, Coleman:1973jx, Brandenberger:1984cz}. The Lagrangian of the theory is written in terms of the renormalization counterterms as 
\begin{equation}
    \mathcal{L} = \frac{1}{2} \left( 1 + \delta Z \right) \partial^{\mu} \phi \partial_{\mu} \phi - \frac{1}{2} \delta m^2 \phi^2 - \frac{\lambda + \delta \lambda}{4 !} \phi^4,
\end{equation}
where \(\delta Z\), \(\delta m^2\), and \(\delta \lambda\) are the wave-function, mass, and coupling constant renormalization counterterms, respectively. The renormalized mass of the scalar field is conventionally defined in terms of the 1PI two-point function\footnote{This definition may differ from other approaches 
in the literature up to a factor of \(i\), depending on the definition of the generating functional for the 1PI Green's functions.} at zero momentum and according to Eq. (\ref{gen_eff}), 
\begin{equation}\label{m_R}
    m^2_{R} = i \Gamma^{(2)} (p = 0) = \left. \frac{d^2 V_{\text{eff}}}{d \phi_c^2} \right|_{\phi_c = 0}.
\end{equation}
In addition, the definition of the renormalized self-coupling is the 1PI four-point at zero momentum multiplied by \(i\),
\begin{equation}\label{l_R}
    \lambda_R = i \Gamma^{(4)} (p = 0) = \left. \frac{d^4 V_{\text{eff}}}{d \phi_c^4} \right|_{\phi_c = 0}
\end{equation}
and the conventional definition of the wave-function renormalization is
\begin{equation}
    Z (0) = 1,
\end{equation}
where the wave-function renormalization is not necessary because we compute the one-loop effective potential for constant field configuration. It is important to notice that the physics should not depend on the subtraction point \(\phi_c = 0\) and we could choose another point as will be shown later \cite{Ramond:1981pw}.

Initially, the divergent integral in the one-loop contribution to the effective potential (\ref{finalV1}) is computed imposing a momentum-space cut-off at \(p^2_E = \Lambda^2\). The integration over the angular variables is then performed using a very useful formula,
\begin{equation}\label{useful_formula}
    \int d^n p \, f(\rho) = \frac{\pi^{n/2}}{\Gamma \left( \frac{n}{2} \right)} \int d \rho \, f(\rho) \rho^{n/2 - 1},
\end{equation}
where \(\rho = p_E^2 \) so that Eq. (\ref{finalV1}) can be written as
\begin{equation}\label{omit}
\begin{split}
    V_1 (\phi_c)  = & \frac{1}{32 \pi^2} \int_{0}^{\Lambda^2} d \rho \, \rho \ln \left[ \rho + m^2_{\text{eff}} (\phi_c) \right] \\
   = &   \frac{1}{64 \pi^2}  \left( \Lambda^4 - m^4_{\text{eff}} (\phi_c) \right) \ln \left[m^2_{\text{eff}} (\phi_c) + \Lambda^2 \right] + \frac{m^4_{\text{eff}}(\phi_c)}{64 \pi^2}  \ln \left[m^2_{\text{eff}}  (\phi_c) \right]  \\
    & + \frac{1}{64 \pi^2} \left(  m^2_{\text{eff}} (\phi_c) \Lambda^2 - \frac{\Lambda^4}{2} \right) \\
    =&   \frac{\Lambda^4}{64 \pi^2} \left( 1 - \frac{m^4_{\text{eff}} (\phi_c) }{\Lambda^4}\right) \ln \left[\frac{m^2_{\text{eff}}(\phi_c)}{\Lambda^2} + 1 \right] +  \frac{m^4_{\text{eff}} (\phi_c)}{64 \pi^2} \ln \left[\frac{m^2_{\text{eff}} (\phi_c)}{\Lambda^2}  \right] \\
    & +  \frac{m^2_{\text{eff}} (\phi_c)\Lambda^2}{64 \pi^2} +  \frac{\Lambda^4}{64 \pi^2} \ln \Lambda^2 - \frac{\Lambda^4}{128 \pi^2},
\end{split}
\end{equation}
where in the second line the following formula for the indefinite integral was applied \cite{Gradshteyn:1943cpj},
\begin{equation}
    \int d x\, x \ln (x + a)  = \frac{1}{2} \left( x^2 - a^2 \right) \ln ( x + a ) - \frac{1}{2} \left( \frac{x^2}{2} - a x\right).
\end{equation}
Now the first logarithm in the last equality of Eq. (\ref{omit}) is expanded with \(m^2_{\text{eff}} \ll \Lambda^2\) to group the terms which vanish in the limit \(\Lambda \to \infty\), leading to
\begin{equation}\label{omit2}
\begin{split}
    V_1 (\phi_c)  =& \frac{\Lambda^4}{64 \pi^2} \left( 1 - \frac{m^4_{\text{eff}} (\phi_c) }{\Lambda^4}\right) \left( \frac{m^2_{\text{eff}} (\phi_c) }{\Lambda^2} - \frac{m^4_{\text{eff}} (\phi_c) }{2\Lambda^4} + \mathcal{O} \left(\frac{m^6_{\text{eff}} (\phi_c) }{\Lambda^6}\right)\right) \\
    & +  \frac{m^4_{\text{eff}} (\phi_c)}{64 \pi^2} \ln \left[\frac{m^2_{\text{eff}} (\phi_c)}{\Lambda^2}  \right] +  \frac{m^2_{\text{eff}} (\phi_c)\Lambda^2}{64 \pi^2} +  \frac{\Lambda^4}{64 \pi^2} \ln \Lambda^2 - \frac{\Lambda^4}{128 \pi^2} \\
    =&  \frac{\Lambda^4}{64 \pi^2} \left( \frac{m^2_{\text{eff}} (\phi_c) }{\Lambda^2} - \frac{m^4_{\text{eff}} (\phi_c) }{2\Lambda^4} + \mathcal{O} \left(\frac{m^6_{\text{eff}} (\phi_c) }{\Lambda^6}\right)\right) \\
    & +  \frac{m^4_{\text{eff}} (\phi_c)}{64 \pi^2} \ln \left[\frac{m^2_{\text{eff}} (\phi_c)}{\Lambda^2}  \right] +  \frac{m^2_{\text{eff}} (\phi_c)\Lambda^2}{64 \pi^2} +  \frac{\Lambda^4}{64 \pi^2} \ln \Lambda^2 - \frac{\Lambda^4}{128 \pi^2} \\
    =&  \frac{m^4_{\text{eff}} (\phi_c)}{64 \pi^2} \left( \ln \left[\frac{m^2_{\text{eff}} (\phi_c)}{\Lambda^2}  \right] - \frac{1}{2} \right)+  \frac{m^2_{\text{eff}} (\phi_c)\Lambda^2}{32 \pi^2} \\
    & + \frac{\Lambda^4}{64 \pi^2} \mathcal{O} \left(\frac{m^6_{\text{eff}} (\phi_c) }{\Lambda^6}\right) + \frac{\Lambda^4}{64 \pi^2} \ln \Lambda^2 - \frac{\Lambda^4}{128 \pi^2} 
\end{split}
\end{equation}
and in the last line of Eq. (\ref{omit2}) the field-independent terms and the vanishing terms in the limit \(\Lambda \to \infty\) are omitted to obtain
\begin{equation}\label{one_reg}
    V_1 (\phi_c) =  \frac{m^4_{\text{eff}} (\phi_c)}{64 \pi^2} \left[ \ln \left(\frac{m^2_{\text{eff}} (\phi_c)}{\Lambda^2}  \right) - \frac{1}{2} \right]+  \frac{m^2_{\text{eff}} (\phi_c)\Lambda^2}{32 \pi^2}.
\end{equation}
Hence, the effective potential including the renormalization counterterms is cast as
\begin{equation}\label{eff_count}
    V_{\text{eff}} (\phi_c) = \frac{1}{2} \delta m^2 \phi^2 + \frac{\lambda + \delta \lambda}{4 !} \phi^4  +  \frac{\lambda \Lambda^2 \phi^2_c}{64 \pi^2} + \frac{\lambda^2 \phi_c^4}{256 \pi^2} \left[ \ln \left(\frac{\lambda \phi_c^2}{2\Lambda^2}  \right) - \frac{1}{2} \right].
\end{equation}

Now the mass renormalization condition (\ref{m_R}) requires that the renormalized mass vanishes,
\begin{equation}
    m^2 =\left. \frac{d^2  V_{\text{eff}} }{d \phi^2_c} \right|_{\phi_c = 0} = 0.
\end{equation}
If we impose the renormalization condition (\ref{l_R}), a logarithmic divergence appears in the fourth derivative of the effective potential at \(\phi_c = 0\). As a consequence, the renormalized coupling constant is redefined at an arbitrary non-symmetric point \(\phi_c = \mu_r\) with
\begin{equation}
    \lambda = \left. \frac{d^4 V_{\text{eff}}}{d \phi_c^4} \right|_{\phi_c = \mu_r},
\end{equation}
where \(\mu_r\) is an arbitrary non-zero mass scale. As a result, the counterterms are computed as
\begin{equation}\label{count_m}
    \delta m^2 = - \frac{\lambda \Lambda^2}{32 \pi^2}
\end{equation}
and
\begin{equation}\label{count_la}
    \delta \lambda = -\frac{11 \lambda^2}{32 \pi^2} - \frac{3 \lambda^2}{32 \pi^2} \ln \frac{\lambda \mu_r^2}{2 \Lambda^2}.
\end{equation}
In the view of Eqs. (\ref{count_m}) and (\ref{count_la}), the renormalized one-loop effective potential (\ref{eff_count}) reads
\begin{equation}\label{eff_cut}
    V_{\text{eff}} (\phi_c) = \frac{\lambda }{4 !} \phi^4 + \frac{\lambda ^2 \phi_c^4}{256 \pi^2} \left[ \ln \left(\frac{ \phi_c^2}{\mu_r^2}  \right) - \frac{25}{6} \right],
\end{equation}
which is the so-called Coleman-Weinberg (CW) potential \cite{Coleman:1973jx}.

It should be also ensured that the renormalization scale \(\mu_r\) should not affect the physics. If a different scale \(\mu'_r\) is chosen, the renormalized coupling constant should be redefined as
\begin{equation}
    \lambda' = \left. \frac{d^4 V_{\text{eff}}}{d \phi_c^4} \right|_{\phi_c = \mu'_r} = \lambda + \frac{3 \lambda^2}{32 \pi^2} \ln 
    \frac{\mu_r^{\prime 2}}{\mu_r^2},
\end{equation}
which is simply a redefinition of the renormalized self-coupling and the renormalized effective potential (\ref{eff_cut}) can be expressed as
\begin{equation}\label{re_eff_cut}
    V_{\text{eff}} (\phi_c) = \frac{\lambda' }{4 !} \phi^4 + \frac{\lambda'^{2} \phi_c^4}{256 \pi^2} \left[ \ln \left(\frac{ 
   \phi_c^2}{\mu'^{2}_r}  \right) - \frac{25}{6} \right] + \mathcal{O} \left( \lambda^3 \right).
\end{equation}

This particular field theory was examined to showcase that the classical minimum at \(\phi_c = 0\) can be altered due to quantum corrections. Namely, two questions arise to be answered: Is still the vacuum expectation value at the origin \(\langle \hat{\phi} \rangle = 0 \)? Is spontaneous symmetry breaking generated by quantum corrections in this theory? Indeed, the second term in the effective potential (\ref{eff_cut}) can be negative for small field values, and a non-zero minimum is formed, whereas the classical minimum converts to a maximum. This happens when
\begin{equation*}
    \lambda \ln \frac{\phi_c^2}{\mu_r^2} = \frac{11}{3} \lambda - \frac{32 \pi^2}{3}.
\end{equation*}
where the logarithm on the left-hand side is unexpectedly very large for all values of the coupling constant due to \(\frac{32 \pi^2}{3} \approx 105\), and higher-order terms in our expansion are required since the validity of the one-loop approximation is broken. Therefore, it is required that the loop-expansion parameter \(|\lambda \ln (\phi_c^2/\mu_r^2)|\) is small in order to remain valid the one-loop expansion \cite{Coleman:1973jx}. However, spontaneous symmetry breaking can be generated by one-loop corrections following this approach in massless scalar electrodynamics and numerous non-Abelian gauge theories \cite{Coleman:1973jx}. In this work, we will apply this formalism to the Standard Model and extensions to the Standard Model.

Finally, this renormalization scheme can be similarly extended to theories including fermions and gauge bosons. Nevertheless, the cut-off regularization explicitly breaks gauge invariance in the case of gauge bosons \cite{Schwartz:2014sze}, while the dimensional regularization preserves the local symmetries in the Lagrangian such as gauge invariance, and we will prefer to work on this scheme as well as on the modified minimal subtraction renormalization scheme.

\subsection{Dimensional Regularization}
Dimensional regularization was first formulated by G. t’Hooft and M. Veltman \cite{tHooft:1972tcz}. The main motivation behind this regularization scheme is based on the fact that an integral with the following form
\begin{equation}
    \int \frac{d^{n} p}{(2 \pi)^n} \, \frac{1}{( p^2 - X + i \epsilon )^2}
\end{equation}
is divergent only if \(n \geq 4 \), but it is convergent for \(n < 4\). In the case of convergence, after a Wick rotation, the result can be obtained by analytically continuing all the formulas to \(n\) dimensions \cite{Leibbrandt:1975dj, Bollini:1972bi,Ashmore:1972uj}.

In the context of the dimensional regularization, the one-loop contribution to the effective potential (\ref{finalV1}) is written as
\begin{equation}
    V_1 (\phi_c) = \frac{1}{2} (\mu^2)^{2-\frac{n}{2}} \int \frac{d^n p_{E}}{(2 \pi)^n} \ln \left[p_{E}^2 + m^2_{\text{eff}} (\phi_c) \right],
\end{equation}
where \(\mu\) is a scale with unit mass dimension balancing the mass dimensions of the integration measure. In this case, the tadpole is computed as
\begin{equation}\label{tadpolereg}
    \frac{dV_1 (\phi_c)}{d m^2_{\text{eff}}} = \frac{1}{2} (\mu^2)^{2-\frac{n}{2}} \int \frac{d^n p_{E}}{(2 \pi)^n} \frac{1}{ p_{E}^2 + m^2_{\text{eff}}}.
\end{equation}
Applying the known formula of dimensional regularization,
\begin{equation}
    \int \, d^n p \frac{(p^2)^{\alpha}}{\left(p^2 + M^2\right)^{\beta}} = \pi^{\frac{n}{2}} (M^2)^{\frac{n}{2}+\alpha-\beta} \frac{\Gamma \left(\alpha + \frac{n}{2} \right) \Gamma \left(\beta - \alpha - \frac{n}{2} \right)}{\Gamma\left(\frac{n}{2}\right) \Gamma (\beta)},
\end{equation}
Eq. (\ref{tadpolereg}) is written as\footnote{The following property is used:
\begin{equation*}
    x \Gamma (x) = \Gamma (x+1).
\end{equation*}}
\begin{equation}
\begin{split}
    \frac{dV_1 (\phi_c)}{d m^2_{\text{eff}}} & = \frac{(\mu^2)^{2-\frac{n}{2}}}{2(2 \pi)^n} \pi^{\frac{n}{2}}(m^2_{\text{eff}})^{\frac{n}{2}-1} \Gamma\left(1-\frac{n}{2}\right) \\
    & = -\frac{1}{32\pi^2} \frac{1}{\frac{n}{2}-1} \left(\frac{1}{4 \pi \mu^2}\right)^{\frac{n}{2}-2} 
    (m^2_{\text{eff}})^{\frac{n}{2}-1} \Gamma\left(2-\frac{n}{2}\right).
\end{split}
\end{equation}
After the integration, the regularized potential is cast into the form
\begin{equation}\label{before_expand_V_1}
\begin{split}
    V_1 (\phi_c) & = -\frac{1}{32\pi^2} \frac{1}{\frac{n}{2}\left(\frac{n}{2}-1\right)} \left(\frac{1}{4 \pi \mu^2}\right)^{\frac{n}{2}-2} 
    (m^2_{\text{eff}})^{\frac{n}{2}} \Gamma\left(2-\frac{n}{2}\right) \\
    & = -\frac{1}{32\pi^2} \frac{m^4_{\text{eff}}}{\frac{n}{2}\left(\frac{n}{2}-1\right)} \left(\frac{m^2_{\text{eff}}}{4 \pi \mu^2}\right)^{\frac{n}{2}-2}  
    \Gamma\left(2-\frac{n}{2}\right),
\end{split}
\end{equation}
and the gamma function can be expanded as follows
\begin{equation}\label{gamma_expansion}
    \Gamma (x) = \frac{1}{x} - \gamma_E + \mathcal{O} (x),
\end{equation}
where \(\gamma_E = 0.5772...\) is called the Euler-Masccheroni constant \cite{Gradshteyn:1943cpj}. A convenient choice for four-dimensional theories is to set \(n = 4 - 2 \epsilon\) such that\footnote{A factor of \(2\) for \(\epsilon\) conveniently cancels a factor of \(1/2\), which appears in the expressions.} \(\epsilon \to 0\) is equivalent to \(n \to 4\). Then, we expand the expressions in powers of \(\epsilon\) such that
\begin{equation}\label{expandn22}
    \left(\frac{m^2_{\text{eff}}}{4 \pi \mu^2}\right)^{-\epsilon}=e^{-\epsilon \log \left(\frac{m^2_{\text{eff}}}{4 \pi \mu^2}\right)} = 1 - \epsilon \log \left(\frac{m^2_{\text{eff}}}{4 \pi \mu^2}\right) + \mathcal{O} \left( \epsilon^2 \right).
\end{equation}
and
\begin{equation}\label{thirdexp}
    \frac{1}{\frac{n}{2}\left(\frac{n}{2}-1\right)}= \frac{1}{2-3\epsilon+\epsilon^2} = \frac{1}{2} \frac{1}{1-\frac{3}{2}\epsilon+ \mathcal{O}\left(\epsilon^2\right)} = \frac{1}{2} \left(1+\frac{3}{2} \epsilon - \mathcal{O}\left(\epsilon^2\right)\right)
\end{equation}
Using (\ref{gamma_expansion}), (\ref{expandn22}) and (\ref{thirdexp}), Eq. (\ref{before_expand_V_1}) becomes
\begin{equation}
\begin{split}
    V_1 (\phi_c) & =- \frac{m^4_{\text{eff}}}{64 \pi^2} \left(1+\frac{3}{2} \epsilon - \mathcal{O}\left(\epsilon^2\right)\right) \left( 1 - \epsilon \ln \left( \frac{m^2_{\text{eff}}}{4\pi\mu^2}\right) + \mathcal{O} \left(\epsilon^2\right) \right) \left(\frac{1}{\epsilon} - \gamma_E+\mathcal{O}(\epsilon) \right)\\
    &=- \frac{m^4_{\text{eff}}}{64 \pi^2} \left(1+\frac{3}{2} \epsilon - \epsilon \ln \left( \frac{m^2_{\text{eff}}}{4\pi\mu^2}\right) + \mathcal{O} \left(\epsilon^2\right) \right) \left(\frac{1}{\epsilon} - \gamma_E+\mathcal{O}(\epsilon) \right)\\
    &= - \frac{m^4_{\text{eff}}}{64 \pi^2} \left(\frac{1}{\epsilon} - \gamma_E - \ln \left( \frac{m^2_{\text{eff}}}{4\pi\mu^2}\right) + \frac{3}{2} + \mathcal{O} \left(\epsilon \right) \right). 
\end{split}
\end{equation}
Thus, the regularized one-loop effective potential is expressed as
\begin{equation}\label{bef_CUV}
    V_1 (\phi_c) = \frac{m^4_{\text{eff}}}{64 \pi^2} \left[- \left(\frac{1}{2-\frac{n}{2}} - \gamma_E + \ln 4 \pi \right) + \ln \frac{m^2_{\text{eff}}}{\mu^2} - \frac{3}{2} + \mathcal{O} \left(\frac{n}{2} - 2 \right) \right].
\end{equation}

\subsection{Modified Minimal Subtraction Scheme}
The previous integrals have singularities arose as poles in \(1/(n-4)\) and it is necessary to subtract them out. A renormalization scheme is a prescription for subtraction. In the framework of the effective potential and cosmological phase transitions, the modified minimal subtraction (\(\overline{\rm MS}\)) scheme is widely adopted as a renormalization scheme in which the counterterms should not have finite parts \cite{tHooft:1973mfk, Bardeen:1978yd}. In this renormalization scheme, we subtract the divergent term in the potential (\ref{bef_CUV}),
\begin{equation}\label{Ccc}
    C = - \frac{m^4_{\text{eff}}}{64 \pi^2}\left(\frac{1}{2-\frac{n}{2}} - \gamma_E + \ln 4 \pi \right),
\end{equation}
in order to be absorbed by the counterterms. Thus, in the \(\overline{\rm MS}\) renormalization scheme, the one-loop effective potential is written as
\begin{equation}
    V_1 (\phi_c) = \frac{m^4_{\text{eff}}}{64 \pi^2} \left( \ln \frac{m^2_{\text{eff}} (\phi_c)}{\mu^2} - \frac{3}{2} \right).
\end{equation}
Namely, in the case of a massless scalar field, such as in the Lagrangian (\ref{count_la}), the counterterms are
\begin{equation}
    \delta m^2 = 0
\end{equation}
and
\begin{equation}
    \delta \lambda = \frac{3 \lambda^2}{32 \pi^2} \left(\frac{1}{2-\frac{n}{2}} - \gamma_E + \ln 4 \pi \right)
\end{equation}
and the one-loop effective potential yields to
\begin{equation}
    V_{\text{eff}} (\phi_c) = \frac{\lambda }{4 !} \phi^4 + \frac{\lambda ^2 \phi_c^4}{256 \pi^2} \left[ \ln \left(\frac{ \phi_c^2}{2 \mu}  \right) - \frac{3}{2} \right].
\end{equation}

In the case of fermionic fields, the dimensional regularization requires a trace operation as \(Tr (\mathbf{1}) = f(n)\), so that the matrix representation of the Clifford algebra behaves as if its dimensions are equal to \(f(n)\) \cite{Collins:1984xc}. A detailed discussion on the spinor algebra in higher dimensions is contained in Ref. \cite{
Polchinski:1998rr}. In particular, the Dirac and the Weyl fermions correspond to \(f(n) = 2^{n/2}\) and \(f(n) = 2^{n/2-1}\), respectively, for an even number \(n\). Then any change in \(f(n)\) amounts to a renormalization-group transformation since the difference \(f(n) - f(4)\) is solely relevant for divergent diagrams \cite{Collins:1984xc}. It is common to set \(f(n) = f(4)\) for all \(n\) values. Now the renormalized one-loop effective potential is similarly computed as in the scalar theory and Eq. (\ref{fermi_eff}) yields to
\begin{equation}\label{before_CUV_2}
    V_1 (\phi_c) = - f(n) \frac{m^4}{64 \pi^2} \left[- \left(\frac{1}{2-\frac{n}{2}} - \gamma_E + \ln 4 \pi \right) + \ln \frac{m^2}{\mu^2} - \frac{3}{2} + \mathcal{O} \left(\frac{n}{2} - 2 \right) \right].
\end{equation}
In the framework of the \(\overline{\rm MS}\) renormalization scheme, we subtract the divergent part, 
\begin{equation}
    C = f(n) \frac{m^4}{64 \pi^2}\left(\frac{1}{2-\frac{n}{2}} - \gamma_E + \ln 4 \pi \right),
\end{equation}
leading to the final expression of the CW potential for fermions
\begin{equation}\label{before_CUV_3}
    V_1 (\phi_c) = - f(4) \frac{m^4 (\phi_c)}{64 \pi^2} \left( \ln \frac{m^2 (\phi_c)}{\mu^2} - \frac{3}{2} \right).
\end{equation}
Likewise, in the gauge theories, the potential (\ref{gauge_eff}) results in
\begin{equation}
    V_1 (\phi_c) = Tr (\Delta) \frac{M^4}{64 \pi^2} \left[- \left(\frac{1}{2-\frac{n}{2}} - \gamma_E + \ln 4 \pi \right) + \ln \frac{M^2}{\mu^2} - \frac{5}{6} + \mathcal{O} \left(\frac{n}{2} - 2 \right) \right],
\end{equation}
where \(Tr(\Delta) = n - 1\) according to the derivation (\ref{delta_trace}). Then, in the \(\overline{\rm MS}\) scheme, after subtracting the divergent term
\begin{equation}
    C = - Tr (\Delta) \frac{M^4}{64 \pi^2}\left(\frac{1}{2-\frac{n}{2}} - \gamma_E + \ln 4 \pi \right),
\end{equation}
the CW potential for gauge bosons can be expressed as
\begin{equation}
    V_1 (\phi_c) = \frac{3 M^4(\phi_c)}{64 \pi^2} \left( \ln \frac{M^2(\phi_c)}{\mu^2} - \frac{5}{6} \right).
\end{equation}

\subsection{On-Shell Scheme}
We will briefly discuss the on-shell (OS) renormalization scheme in order to develop it further in the context of the Standard Model effective potential in Chapter \ref{Electroweak Phase Transition}. In the OS renormalization scheme, the relations among the bare parameters in the Lagrangian and the physical parameters do not alter at the one-loop approximation \cite{Anderson:1991zb, Iliopoulos:1974ur}. In fact, this scheme offers the benefit of having a predetermined prescription for the renormalization scales for each field and it also improves the computational efficiency because the boundary conditions for the scalar mass and the vacuum expectation value fixed at tree level remain valid at the one-loop correction for the same set of bare parameters.

Initially, the theory could be regularized by a momentum space cut-off as shown earlier. For instance, let us consider the scalar field theory (\ref{lag_scalar}) and we impose the renormalization conditions so that the tree-level relations among the vacuum expectation value and the mass of the scalar field are not altered by higher loop corrections. These renormalization conditions are expressed as
\begin{equation}\label{ren_con1}
    \left. \frac{d \left( V_1 + V_1^{c}\right)}{d \phi_c} \right|_{\phi_c = \upsilon} = 0
\end{equation}
\begin{equation}\label{ren_con_2}
    \left. \frac{d^2 \left( V_1 + V_1^{c}\right)}{d \phi_c^2} \right|_{\phi_c = \upsilon} = 0,
\end{equation}
where \(\phi_c = \upsilon\) is the global minimum of the tree-level potential and the potential \(V_1^{c}\) includes all counterterms and it can be written as
\begin{equation}
    V_1^{c} (\phi_c) = \delta \Omega + \frac{\delta m^2}{2} \phi_c^2 + \frac{\delta \lambda}{4} \phi_c^4.
\end{equation}
where the counterterm \(\delta \Omega \) is related to the vacuum energy or the cosmological constant. Next, the regularized one-loop correction in the effective potential (\ref{one_reg}) reads as
\begin{equation}\label{scal_reg}
    V_1 (\phi_c) =  \frac{m^4_{\text{eff}} (\phi_c)}{64 \pi^2} \left[ \ln \left(\frac{m^2_{\text{eff}} (\phi_c)}{\Lambda^2}  \right) - \frac{1}{2} \right]+  \frac{m^2_{\text{eff}} (\phi_c)\Lambda^2}{32 \pi^2}.
\end{equation}
Now the renormalization conditions (\ref{ren_con1}) and (\ref{ren_con_2}) are imposed to cancel the infinities in the one-loop correction (\ref{scal_reg}) against those in the potential \(V_1^c\). Consequently, the renormalized effective potential in a scalar field theory is
\begin{equation}
    V_{\text{eff} }(\phi_c) = V_0 (\phi_c) + \frac{1}{64 \pi^2} \left[ m^4_{\text{eff}}(\phi_c)\left( \ln  \frac{m^2_{\text{eff}} (\phi_c)}{m^2_{\text{eff}}(\upsilon)} - \frac{3}{2} \right) + 2 m^2_{\text{eff}} (\phi_c) m^2_{\text{eff}}(\upsilon) \right].
\end{equation}
The OS scheme can be also modified slightly to define the OS-like renormalization scheme. However, both OS schemes suffer from an infrared divergence that originates from the Goldstone bosons as they acquire a zero mass at zero temperature. This subtlety can be rectified, by the methods in Refs. \cite{Martin:2014bca, Elias-Miro:2014pca}.

\chapter{Finite-Temperature Field Theory}\label{Finite-Temperature Field Theory}

Ordinary quantum field theory describes interactions and fields in a surrounding vacuum, at zero temperature. Namely, scattering matrix elements are expressed in terms of Green's functions at zero temperature, predicting the results of particle physics experiments, such as in LHC. Nevertheless, the assumption of zero temperature was violated in the early Universe, where the matter and radiation density were very high. The finite-temperature field theory \cite{Kapusta:2006pm, Bellac:2011kqa, Laine:2016hma, Landsman:1986uw, Kubo:1957mj, Matsubara:1955ws, Dolan:1973qd, Martin:1959jp} can be then applied to explain quantum processes at non-zero temperatures which are often called finite temperatures.

A finite-temperature quantum field theory requires replacing the common time-ordering by the contour ordering which orders the field operators along a contour in the complex time plane. In this chapter, we present the general formulation of finite-temperature field theory and then the imaginary time formalism is discussed extensively to derive the Feynman rules at finite temperatures.

The effects of finite temperature in a system could generate spontaneous symmetry breaking. This has been mentioned in the very first pages of this thesis presenting a well-known system that exhibits spontaneous symmetry breaking due to finite temperature, a ferromagnet. The action of a ferromagnet is invariant under spatial rotations. Below a critical temperature, the magnetization does not vanish and the rotational symmetry is spontaneously broken. In other words, in a ferromagnetic material, the rotational symmetry is restored at high temperatures. This is also observed in superconductivity. Similarly, a gauge symmetry could be restored at high temperatures in the early Universe. This was first suggested by Kirzhnits (1972) \cite{Kirzhnits:1972iw} and Kirzhnits and Linde (1972) \cite{Kirzhnits:1972ut, Kirzhnits:1974as, Kirzhnits:1976ts}. These finite-temperature effects in quantum field theory are studied using the effective potential at finite temperature which is derived explicitly in this chapter for scalar field theories and non-Abelian gauge theories. This formalism was developed further by Weinberg \cite{Weinberg:1974hy}, Bernard \cite{Bernard:1974bq} and Dolan and Jackiw \cite{Dolan:1973qd}.

\section{Introduction}

Normally, there are three types of ensembles in equilibrium statistical physics:
\begin{itemize}
    \item The microcanonical ensemble describes an isolated system with fixed energy \(E\), particle number \(N\), and volume \(V\).
    \item The canonical ensemble describes a system in contact with a heat reservoir at temperature \(T\): the system can freely exchange energy with the reservoir, but the particle number and volume are fixed quantities.
    \item In the grand canonical ensemble the system can exchange energy and particles with the reservoir, while the temperature, volume, and chemical potential are fixed. In quantum field theories, observables are often computed using the grand canonical ensemble as particles can be created and destroyed.
\end{itemize}
Consider a system described by a Hamiltonian \(\hat{H}\) and a number of conserved and mutually commuting charges \(\hat{Q}_i\), which respect \([\hat{Q}_i, \hat{H}] = 0\). The equilibrium state of the system in a large volume \(V\) is described by the density matrix,
\begin{equation}
    \hat{\rho} = \frac{1}{Z} \exp \left(-\beta \hat{H} + \beta\sum_i  \mu_i \hat{Q}_i \right),
\end{equation}
where \(\mu_i\) are the chemical potentials and the partition function of the system \(Z\) equals to
\begin{equation}\label{def_ZZ}
    Z = Tr\left[\exp \left(-\beta \hat{H} + \beta\sum_i  \mu_i \hat{Q}_i \right)\right],
\end{equation}
where \(\beta\) is the inverse temperature as usual (\(k_B = 1\)). 

The density matrix formalism can be highly useful in quantum statistical mechanics \cite{Kapusta:2006pm, Bellac:2011kqa, Laine:2016hma}. For instance, in the grand canonical ensemble, the expectation value of an operator \(\hat{A}\) can be computed as,
\begin{equation}\label{vevA}
    \langle \hat{A} \rangle = Tr (\hat{A} \hat{\rho}).
\end{equation}
Furthermore, the energy density is defined, on account of (\ref{vevA}) as
\begin{equation}
    E = \frac{\langle \hat{H} \rangle}{V} = - \frac{1}{V} \frac{\partial \ln Z}{\partial \beta}.
\end{equation}
In general, the grand potential can be additionally introduced as 
\begin{equation}
    \Omega = - \frac{1}{\beta} \ln Z
\end{equation}
and
\begin{equation}
    \Omega = E - TS - \sum_i Q_i \mu_i .
\end{equation}
This expression leads to
\begin{equation}
    d \Omega = - S dT - P dV - \sum_i Q_i d\mu_i 
\end{equation}
and the following thermodynamic definitions are derived,
\begin{equation}
    P = - \left(\frac{\partial \Omega }{\partial V}\right)_{T,\, \mu_i}\, , \quad  S = - \left(\frac{\partial \Omega }{\partial T}\right)_{V,\, \mu_i} \, , \quad  Q_i = - \left(\frac{\partial \Omega }{\partial \mu_i}\right)_{T,\, V}.
\end{equation}

\section{Partition Function}
Let \(\hat{\phi} (\Vec{x}, 0) \) be a scalar field operator in the Schrödinger picture at time \(t = 0\) and let \(\hat{\pi} (\Vec{x},0)\) be its conjugate momentum operator. The state \(\ket{\phi}\) is defined such that
\begin{equation}\label{d_phi}
    \hat{\phi} (\Vec{x}, 0) \ket{\phi} = \phi(\Vec{x}) \ket{\phi},
\end{equation}
which means that \(\ket{\phi}\) is the eigenstate of the field operator with eigenvalue \(\phi (\Vec{x})\). It is assumed that the above states form a complete set at any given time writing the completeness and orthogonality conditions as
\begin{equation}
    \int d \phi (\Vec{x}) \, \ket{\phi} \bra{\phi} = \mathbf{1}
\end{equation}
and 
\begin{equation}
     \langle \phi_a | \phi_b \rangle  = \prod_{\Vec{x}} \delta \left( \phi_a (\Vec{x}) - \phi_b (\Vec{x}) \right).
\end{equation}
Likewise, the state \(\ket{\pi}\) is defined as the eigenstate of the conjugate momentum operator with
\begin{equation}
    \hat{\pi} (\Vec{x}, 0) \ket{\pi} = \pi(\Vec{x}) \ket{\pi},
\end{equation}
where \(\pi(\Vec{x})\) is the eigenvalue. The completeness and orthogonality conditions are expressed as
\begin{equation}
    \int d \pi (\Vec{x}) \, \ket{\pi} \bra{\pi} = \mathbf{1}
\end{equation}
and 
\begin{equation}
     \langle \pi_a | \pi_b \rangle  = \prod_{\Vec{x}} \delta \left( \pi_a (\Vec{x}) - \pi_b (\Vec{x}) \right).
\end{equation}
As a result, the partition function (\ref{def_ZZ}) can be rewritten as
\begin{equation}\label{trace_Z}
    Z = \sum_{a} \int d \phi_a \, \bra{\phi_a} \exp \left(-\beta \hat{H} + \beta\sum_i  \mu_i \hat{Q}_i \right) \ket{\phi_a},
\end{equation}
where the sum runs over all the states defined in (\ref{d_phi}) and the trace in (\ref{def_ZZ}) indicates the integration over all fields \(\phi_a (\Vec{x})\). One observes that the partition function resembles a transition amplitude, which can be expressed in quantum field theory as a path integral. Then, one derives, according to the definition of the path integral, the partition function \cite{Kapusta:2006pm} which can be written as
\begin{equation}\label{gen_Z}
    Z = \int \mathcal{D} \pi \int_{P} \mathcal{D} \phi \, \exp \left[ \int_{0}^{\beta} d \tau \, \int d^3 x \, \left(i \pi \frac{\partial \phi}{\partial \tau} - \mathcal{H} (\pi, \phi) + \sum_i  \mu_i \mathcal{Q}_i (\pi, \phi) \right) \right],
\end{equation}
where the imaginary time \(\tau = i t\) is considered above and \(P\) denotes the periodicity in the integration over the fields \(\phi\) which is expressed as \(\phi(\Vec{x}, 0) = \phi (\Vec{x}, \beta)\) for all \(\Vec{x}\). The Hamiltonian density and the conserved charge density are defined by
\begin{equation}
    H = \int d^3 x \, \mathcal{H} (\pi, \phi)
\end{equation}
and
\begin{equation}
    Q_i = \int d^3 x \, \mathcal{Q}_i (\pi, \phi),
\end{equation}
respectively. Our approach to finite-temperature field theory will be mostly based on imaginary time formalism. This formalism exploits the connection between the inverse temperature and the imaginary time. This is clearly illustrated in the next section including an external classical source.

Now if the Hamiltonian in (\ref{gen_Z}) describes a real scalar field theory (\ref{lag_scalar}), the Hamiltonian density is then expressed as
\begin{equation}
    \mathcal{H} = \frac{1}{2} \pi^2 + \frac{1}{2} \left( \nabla \phi \right)^2 + \frac{1}{2} m^2 \phi^2 + V_0 (\phi),
\end{equation}
where the momentum conjugate is
\begin{equation}
    \pi = \frac{\partial \mathcal{L}}{\partial \left(\partial_0 \phi \right)} = \frac{\partial \phi}{\partial t} = i \frac{\partial \phi}{\partial \tau}.
\end{equation}
The Lagrangian does not possess any continuous symmetries and there are no conserved charges in the partition function (\ref{gen_Z}). Then, the expression (\ref{gen_Z}) can be written as a Lagrangian path integral,
\begin{equation}\label{part_SE}
    Z = \int_{P} \mathcal{D} \phi \, \exp \left(- S_E \right).
\end{equation}
where the Euclidean action \(S_E\) is defined on an imaginary time interval \(0 < \tau < \beta\) and is given by
\begin{equation}
    S_E = \int_{0}^{\beta} d \tau \int d^3 x \, \mathcal{L}_E,
\end{equation}
where \(\mathcal{L}_E\) is the Euclidean Lagrangian density,
\begin{equation}
    \mathcal{L}_E = - \mathcal{L} \left( t \to - i \tau \right).
\end{equation}
If we assume that the interaction Lagrangian is zero, the Euclidean action reads
\begin{equation}\label{euclid}
    S_E =  \int_{0}^{\beta} d \tau  \int d^3 x \, \left( \frac{1}{2}\left(\frac{\partial \phi}{\partial \tau} \right)^2 + \frac{1}{2} \left( \nabla \phi \right)^2 + \frac{1}{2} m^2 \phi^2 \right).
\end{equation}
The Fourier expansion of the field is
\begin{equation}\label{four_ex}
    \phi (\Vec{x}, \tau) = \frac{1}{\beta V} \sum_{n = - \infty}^{\infty} \sum_{\Vec{p}} e^{ - i \left( \Vec{p} \cdot \Vec{x} - \omega_n \tau \right) } \phi_n (\Vec{p}) \, ,
\end{equation}
where the normalization is conventionally chosen and the periodicity condition in the path integral (\ref{gen_Z}) yields
\begin{equation}\label{periodic}
  \phi(\Vec{x}, 0) = \phi (\Vec{x}, \beta) \quad \Rightarrow \quad e^{i \beta \omega_n } = 1 \quad \Rightarrow \quad\omega_n = 2\pi n T.
\end{equation}
Moreover, the reality of the field implies that
\begin{equation}\label{real}
    \phi_{-n} (-\Vec{p}) = \phi^{*}_{n} (\Vec{p}).
\end{equation}
We then substitute Eq. (\ref{four_ex}) into Eq. (\ref{euclid}) using the relation (\ref{real})
\begin{equation}
    S_E =  \frac{1}{2 \beta V} \sum_n \sum_{\Vec{p}} \left( \omega_n^2 +  \Vec{p}^{\, 2} + m^2 \right) \phi_n (\Vec{p})  \phi_n^{*} (\Vec{p}).
\end{equation}
Thus, the partition function (\ref{part_SE}) is expressed as
\begin{equation}
    Z = \int_{P} \mathcal{D} \phi \, \prod_{\Vec{p}} \exp \left[- \frac{1}{2 \beta V} \sum_n  \left( \omega_n^2 +  \Vec{p}^{\, 2} + m^2 \right) \phi_n (\Vec{p})  \phi_n^{*} (\Vec{p}) \right].
\end{equation}
which is a path integral similar to the generating functional (\ref{sad}). This partition function is also a common path integral in quantum statistical mechanics and finite-temperature field theory \cite{Kapusta:2006pm, Bellac:2011kqa, Laine:2016hma} and we do not present its derivation here. In fact, the finite temperature effective potential for a scalar field theory is almost identical to the logarithm of this partition function which will be derived in a later section. Therefore, the logarithm of the partition function is cast into the form
\begin{equation}\label{lnz}
    \ln Z = - V \int \frac{d^3 p}{(2 \pi)^3} \, \left[ \frac{\beta}{2 }\omega + \ln \left(1 - e^{-\beta \omega} \right) \right].
\end{equation}
with \(\omega = \sqrt{\Vec{p}^{\, 2} + m^2}\). It is worth computing the free energy in this case as it will be very useful later. In statistical mechanics, the free energy is defined as
\begin{equation}
    \Omega = - T \ln Z
\end{equation}
and from Eq. (\ref{lnz}), the free energy is
\begin{equation}\label{free_energy}
    \Omega = V \int \frac{d^3 p}{(2 \pi)^3} \, \left[ \frac{\omega}{2 } + \frac{1}{\beta}\ln \left(1 - e^{-\beta \omega} \right) \right].
\end{equation}
This integral will be computed explicitly later as it plays a crucial role in the effective potential. In addition, one easily proves the known thermodynamic relations for bosons
\begin{equation}
    E_0 = - \frac{\partial}{\partial \beta} \ln Z_0 =  V \int \frac{d^3 p}{(2 \pi)^3} \, \frac{\omega}{2 },
\end{equation}
which is the vacuum energy and the pressure is
\begin{equation}
    P_0 = T \frac{\partial}{\partial V} \ln Z_0 =  - \int \frac{d^3 p}{(2 \pi)^3} \, \frac{\omega}{2}.
\end{equation}

\section{Green's Functions}

\subsection{Generating Functional}
In this section, we will repeat our discussion about the partition function (\ref{gen_Z}), but it will be formulated differently to express the generating functional for the full Green's functions at finite temperature. 
Let \(\hat{\phi} (x)\) be a single scalar field operator in the Heisenberg picture such that
\begin{equation}
    \hat{\phi} (x) = e^{i \hat{H} t} \hat{\phi} (\Vec{x}, 0) e^{-i \hat{H} t}.
\end{equation}
The eigenstate of the field operator at time \(t\), denoted as \(\ket{\phi(\Vec{x}); t}\), satisfy
\begin{equation}\label{eigenstate}
    \hat{\phi} (x) \ket{\phi (\Vec{x}); t} = \phi(\Vec{x}) \ket{\phi (\Vec{x}); t}
\end{equation}
with eigenvalue \(\phi (\Vec{x})\). The time evolution of this state is derived from the time evolution of the Heisenberg field operator,
\begin{equation}
    \ket{\phi (\Vec{x}); t} = e^{i \hat{H}t} \ket{\phi (\Vec{x}); t = 0}.
\end{equation}
This eigenstate is written differently under the action of the operator \(e^{- \beta \hat{H}}\) so that
\begin{equation}
    e^{-\beta \hat{H}} \ket{\phi (\Vec{x}); t}  =  e^{ i \hat{H}\left(i \beta \right)} \ket{\phi (\Vec{x}); t} = \ket{\phi (\Vec{x}); t + i \beta},
\end{equation}
which translates into an imaginary time shift. 

Now it is essential to comment on the implications of the complex time in the path integral formalism. In particular, this can be understood better considering the Feynman-Matthews-Salam (FMS) formula \cite{Matthews:1955zi, QMP},
\begin{equation*}
    \bra{\phi_1 (\Vec{x}); t_1}T\{F [\hat{\phi}]\}\ket{\phi_2 (\Vec{x}); t_2} = N \int \mathcal{D} \pi \int \mathcal{D} \phi \, F [\phi]  \exp \left[ \int_{t_1}^{t_2} d t \, \int d^3 x \, \left( \pi \Dot{\phi} - \mathcal{H} (\pi, \phi)  \right) \right],
\end{equation*}
where the normalization depends on the time interval. The volume and the boundary conditions in the path integral are
\begin{equation}
    \phi(\Vec{x}, t_1) = \phi_1(\Vec{x}) \,, \quad \quad \phi(\Vec{x}, t_2) = \phi_2(\Vec{x})
\end{equation}
and the momentum integration in the path integral is not constrained by any boundary condition. If the times \(t_1\) and \(t_2\) are complex, we consider that the time integration above goes over a contour with a complex time interval from \(z = t_1\) to \(z = t_2\). This complex contour should be restricted to going monotonically downward or parallel to the real axis. More specifically, this complex contour alters the time-ordering prescription that is frequently adopted in quantum field theory. This time-ordering prescription, denoted as \(T_c\), orders the operators along a given path \(C\) in the complex time plane \cite{Schwinger:1960qe}. If the contour \(C\) is parameterized as a function \(t = z (\tau)\), where \(\tau\) is a real parameter, the time-ordering \(T_c\) coincides with the standard time-ordering along \(\tau\). Accordingly, we define the contour step function and the contour delta function \cite{Niemi:1983nf},
\begin{equation}
    \theta_c ( t_1 - t_2) = \theta (\tau_1 - \tau_2)
\end{equation}
and 
\begin{equation}
    \delta_ c (t_1 - t_2) = \left( \frac{\partial z}{\partial \tau} \right)^{-1} \delta (\tau_1 - \tau_2).
\end{equation}
For instance, the time-ordered product of two field operators is defined as
\begin{equation}\label{time_or_1}
    T_c\{\hat{\phi (x)} \hat{\phi}(y)\} = \theta_c (x^0 - y^0) \hat{\phi} (x) \hat{\phi} (y) + \theta_c (y^0 - x^0) \hat{\phi} (y) \hat{\phi} (x)
\end{equation}
and differentiation leads to
\begin{equation}\label{time_or_2}
    \partial_t T_c\{\hat{\phi (x)} \hat{\phi}(y)\} = \delta_c (x^0 - y^0) [\hat{\phi} (x), \hat{\phi} (y)] + T_c\{\partial_t \hat{\phi} (y) \hat{\phi} (x)\}.
\end{equation}
If a c-number function \(J(x)\) is defined on the contour \(C\), then the functional differentiation is written as
\begin{equation}
     \frac{\delta J(y)}{\delta J(x)} = \delta_c ( x^0 - y^0) \delta^{(3)} (\Vec{x} - \Vec{y})
\end{equation}
with
\begin{equation*}
    \delta^{(4)}_c (x - y ) \equiv \delta_c ( x^0 - y^0) \delta^{(3)} (\Vec{x} - \Vec{y}).
\end{equation*}
An additional requirement for the FMS formula in this formalism is to ensure that the time arguments in \(F[\phi]\) must lie on the contour since they did also lie in the real interval \([t_1, t_2]\) in the ordinary quantum field theory. Consequently, the partition function can be expressed as a path integral, such as (\ref{gen_Z}), defining the contour in the complex time plane as it was described previously.

Finally, the above discussion implies the definition of the generating functional in the presence of an external classical source which is given by
\begin{equation}
    Z[J;\beta] = N \int \mathcal{D} \pi \int \mathcal{D} \phi \, \exp  \left[ i \int_C d^4 x \, \left( \pi(x) \Dot{\phi}(x) - \mathcal{H} (x) + \phi (x) J(x) \right) \right], 
\end{equation}
where the contour is defined as an arbitrary contour\footnote{The contour is arbitrary, but it is required that it goes monotonically downward or parallel to
the real axis.} starting at \(z = t_i\) and finishing at \(z = t_i - i \beta\). The integration over the field in the path integral is constrained by
\begin{equation}
    \phi (\Vec{x}, t_i - i\beta) = \phi (\Vec{x}, t_i),
\end{equation}
which is identical to the periodicity condition in the path integral (\ref{gen_Z}) substituting \(t \to - i \tau\) (\(t_i = 0\)). Moreover, this generating functional can be written as a Lagrangian path integral similar to (\ref{part_SE}),
\begin{equation}\label{Z_LA}
    Z[J; \beta] = N' \int  \mathcal{D} \phi \, \exp \left[- S_E + i \int_C d^{4}x \, J(x) \phi(x) \right],
\end{equation}
where the normalization \(N\) cancels the unwanted contributions of the path integral which are usually infinite \cite{Landsman:1986uw}. On account of Eq. (\ref{trace_Z}), Eq. (\ref{Z_LA}) could be interpreted as a statistical average, such as the expectation value (\ref{vevA}), while the trace has been replaced by a functional over the field \(\phi\). This allows us to write the generating functional as the statistical average
\begin{equation}
    Z[J;\beta] = Z[0;\beta] \biggl \langle T_c \exp \left( i \int_C d^4 x\, J(x) \phi (x) \right) \biggr \rangle .
\end{equation}
The generating functional (\ref{Z_LA}) can be finally written as an expansion in powers of the external source following the proof in (\ref{Z_G}),
\begin{equation}\label{Z_J_B}
    \frac{Z[J;\beta]}{Z[0;\beta]} = \sum_{n = 0}^{\infty} \frac{i^n}{n!}   \int_C d^{4}x_{1}... \, \int d^{4}x_{n} \, J(x_1)...J(x_n) G^{(c)} (x_1,...,x_n),
\end{equation}
where the so-called thermal (full) Green's function is defined as the statistical average of the ordered product of the \(n\) field operators
\begin{equation}
    G^{(c)} (x_1, ..., x_n) \equiv \langle T_c \{\hat{\phi} (x_1) ... \hat{\phi} (x_n) \} \rangle,
\end{equation}
where \(x^0 = t\) may be complex and the integration over time should follow the path \(C\) in the complex time plane. From Eq. (\ref{Z_J_B}) the thermal Green's functions can be also expressed as
\begin{equation}\label{G_c_dZ}
    G^{(c)} (x_1, ..., x_n) = \frac{1}{Z[0;\beta]}(-i)^n  \left. \left(\frac{\delta}{\delta J(x_1)} \right)...\left(\frac{\delta}{\delta J(x_n)}\right) Z[J;\beta] \right|_{J=0},
\end{equation}
where the normalization factor is \(Z[0; \beta] = Z = Tr \left(e^{-\beta \hat{H}}\right)\) which is later set to \(Z = 1\).

\subsection{Scalar Fields}
In this subsection, we present the generating functional and the thermal Green's functions in scalar field theories. It is particularly interesting that not all the contours are allowed if it is required for the Green's functions to be analytic with respect to their time arguments \cite{MillsR.G.J.}. For instance, the two-point Green function is decomposed using Eq. (\ref{time_or_1}) as
\begin{equation}\label{def_G_c}
    G^{(c)} (x, y) = \theta_c (x^0 - y^0) G_{+} (x,y) + \theta_c (y^0 - x^0) G_{-} (x, y),
\end{equation}
where
\begin{equation}\label{cor_G}
    G_{+} (x, y) = \langle \hat{\phi} (x) \hat{\phi} (y) \rangle = G_{-} (y, x).
\end{equation}
If we consider the complete set of states \(\ket{n}\) with eigenvalues \(E_n\):
\begin{equation*}
    \hat{H} \ket{n} = E_n \ket{n},
\end{equation*}
the correlation function (\ref{cor_G}) can be expressed in the spectral form 
\begin{equation}
    G_{+} (x^0,y^0) =  Z^{-1} \sum_{m} \sum_{n} |\bra{m} \hat{\phi}(0) \ket{n}|^2 e^{- i E_n (x^0 - y^0)} e^{i E_m (x^0 - y^0 + i \beta)}
\end{equation}
at the point where all spatial coordinates vanish. The convergence of the sum is ensured by the condition \(- \beta < \text{Im} \left(x^0 - y^0\right) < 0\) which implies the existence of \(G_{+} (x^0, y^0) \equiv G_{+} (x^0 - y^0)\) as an analytic function. This condition also requires \(\theta_c (x^0 - y^0) = 0\) for \(\text{Im} \left(x^0 - y^0 \right) > 0\). Similarly, the existence of \(G_{-} (x^0 - y^0)\) is ensured by the condition \(0 < \text{Im} \left(x^0 - y^0\right) < \beta\), which requires \(\theta_c (y^0 - x^0 ) = 0\) for \(\text{Im} \left(x^0 - y^0\right) < 0\). Thus, the final condition for the convergence of the Green's function (\ref{def_G_c}) defined on the strip
\begin{equation}
    - \beta \leq \text{Im} \left( x^0 - y^0 \right) \leq \beta
\end{equation}
is that the function \(\theta_c (x^0 - y^0)\) is defined such that \(\theta_c (x^0 - y^0) = 0\) for \(\text{Im} (x^0 - y^0) > 0\). This requires that a point that moves along \(C\) has a monotonically decreasing or constant imaginary time. This condition also ensures the existence of all higher-order thermal Green's functions \cite{MillsR.G.J., Martin:1959jp}.

The definition of the correlation functions \(G_{+}\) and \(G_{-}\) leads to a significant periodicity relation. This relation is derived by the definitions (\ref{vevA}) and (\ref{cor_G}),
\begin{equation}\label{KMS1}
    \begin{split}
        G_{-} (\Vec{x}, \Vec{x}\,'; t, t') & = \langle \hat{\phi} (\Vec{x}\,',t') \hat{\phi} (\Vec{x}, t)  \rangle \\
        & = Z^{-1} Tr \left[ e^{- \beta \hat{H}}  \hat{\phi} (\Vec{x}\,',t')  \hat{\phi} (\Vec{x}, t) \right] \\
        & = Z^{-1} Tr \left[\hat{\phi} (\Vec{x}, t) e^{- \beta \hat{H}}  \hat{\phi} (\Vec{x}\,',t') \right] \\
        & = Z^{-1} Tr \left[ e^{- \beta \hat{H}} e^{\beta \hat{H}} \hat{\phi} (\Vec{x},t) e^{- \beta \hat{H}} \hat{\phi} (\Vec{x}\,', t') \right] \\
        & = Z^{-1} Tr \left[ e^{- \beta \hat{H}}  \hat{\phi} (\Vec{x},t- i \beta) \hat{\phi} (\Vec{x}\,', t') \right],
    \end{split}
\end{equation}
where the correlation function \(G_{+} (\Vec{x}, \Vec{x}\,'; t - i \beta, t')\) is defined as
\begin{equation}\label{KMS2}
        G_{+} (\Vec{x}, \Vec{x}\,'; t - i \beta, t') = \langle \hat{\phi} (\Vec{x}, t - i \beta)  \hat{\phi} (\Vec{x}\,',t')  \rangle = Z^{-1} Tr \left[ e^{- \beta \hat{H}}  \hat{\phi} (\Vec{x},t- i \beta) \hat{\phi} (\Vec{x}\,', t') \right].
\end{equation}
From Eqs. (\ref{KMS1}) and (\ref{KMS2}) the following periodicity relation\footnote{This relation holds true in the case of conserved charges because the result remains the same using the general definition of the grand canonical average (\ref{vevA}).} is proved
\begin{equation}\label{KMS_relation}
    G_{+} (\Vec{x}, t - i \beta) =  G_{-} (\Vec{x}, t),
\end{equation}
which is the well-known Kubo-Martin-Schwinger (KMS) relation \cite{Martin:1959jp,Kubo:1957mj, Umezawa:1982nv}. 

Now we can compute the generating functional and the two-point function for a free real scalar field described by the Lagrangian (\ref{lag_scalar}) which satisfies the equal-time commutation relation
\begin{equation}
    \left[\hat{\phi} ( \Vec{x}, t), \Dot{\hat{\phi}} (\Vec{x}\,', t ) \right] = i \delta^{(3)} (\Vec{x} - \Vec{x}\,').
\end{equation}
The equation of motion of a free scalar field is
\begin{equation}
    \left( \partial^2 + m^2 \right) \hat{\phi} (x) = 0,
\end{equation}
where the derivative with respect to time is defined as the directional derivative on a given path \(C\). From Eqs. (\ref{time_or_1}) and (\ref{time_or_2}), one deduces the Dyson-Schwinger equation for the generating functional,
\begin{equation}
    i \left( \partial^2 + m^2 \right) \frac{\delta Z_0 [J; \beta]}{\delta J(x)} + J(x) Z_0 [J;\beta] = 0,
\end{equation}
where \(Z_0 [J;\beta]\) is the generating functional for the free field. This equation coincides with the Dyson-Schwinger equation at zero temperature \cite{Schwartz:2014sze} and its solution reads
\begin{equation}
    Z_0 [J; \beta] = Z_0 [0, \beta] \exp \left[ -\frac{i}{2} \int_C d^4 x \int_C d^4 x' \, J(x) D^{(c)} (x - x') J(x')\right],
\end{equation}
where \(D^{(c)} (x - x')\) is the propagator that satisfies the equation
\begin{equation}\label{eq_D}
     \left( \partial^2 + m^2 \right) D^{(c)} (x - x') = - \delta_c^{(4)} (x - x').
\end{equation}
As a result, Eqs. (\ref{G_c_dZ}) and (\ref{eq_D}) lead to the two-point Green's function for the free theory
\begin{equation}\label{two_G_D}
    G^{(c)} (x, x') = i D^{(c)} (x - x')\, ,
\end{equation}
which reproduces the same differential equation for the Feynman propagator as the one at zero temperature,
\begin{equation}
     \left( \partial^2 + m^2 \right) G^{(c)} (x, x')  = - i \delta_c^{(4)} (x - x') \, .
\end{equation}
Next, the free scalar field is expanded in terms of the annihilation and creation operators
\begin{equation}\label{expand_phi}
    \hat{\phi} (x) = \int \frac{d^3 p}{(2 \pi)^{3} 2 \omega_p}\left( \hat{a} (p) e^{-i p x} + \hat{a}^{\dagger} (p) e^{i p x} \right),
\end{equation}
where \(\omega_p = \sqrt{p^{\, 2} + m^2}\). Then, the equal-time commutation relation translates into
\begin{equation}
    \left[\hat{a}(p), \hat{a}^{\dagger} (k) \right] = 2 \omega_p (2 \pi)^3  \delta^{(3)} (\Vec{p} - \Vec{k}).
\end{equation}
Then, one deduces the expectation values in the case of an ideal gas \cite{Matsubara:1955ws}, which is very common in quantum statistical mechanics,
\begin{equation}\label{a_a_1}
    \langle \hat{a}^{\dagger} (p) \hat{a}(k) \rangle = 2 \omega_p (2 \pi)^3 n_B ( \omega_p) \delta^{(3)} ( \Vec{p} - \Vec{k})
\end{equation}
and
\begin{equation}\label{a_a_2}
    \langle \hat{a}(p)\hat{a}^{\dagger} (k)  \rangle = 2 \omega_p (2 \pi)^3 \left(1 + n_B ( \omega_p) \right) \delta^{(3)} ( \Vec{p} - \Vec{k}),
\end{equation}
where the Bose-Einstein distribution function, \(n_B(\omega)\), is
\begin{equation}\label{B-E-fun}
n_B (\omega) = \frac{1}{e^{\beta \omega} - 1 }.
\end{equation}
The correlation functions (\ref{cor_G}) in the free theory can be reformulated using Eqs. (\ref{expand_phi}), (\ref{a_a_1}), and (\ref{a_a_2}),
\begin{equation}\label{G_tilde}
    \Tilde{G}_{+} (p) = \rho (p) \left( 1+ n_B ( \omega_p) \right) =  \tilde{G}_{-} (p) e^{\beta \omega_p},
\end{equation}
where the Fourier transforms read
\begin{equation}
    G_{\pm} (x,y) = \int \frac{d^4 p}{(2 \pi)^4} e^{ - i p (x - y)} \Tilde{G}_{\pm} (p),
\end{equation}
and \(\rho (p)\) is the spectral density given by \cite{Bellac:2011kqa}
\begin{equation}\label{spectral_density}
    \rho (p) = \Tilde{G}_{+} (p) - \Tilde{G}_{-} (p) = 2 \pi \left( \theta (\omega_p) - \theta (- \omega_p) \right) \delta (p^2-m^2).
\end{equation}
Therefore, the two-point function, according to (\ref{def_G_c}) and (\ref{two_G_D}), can be expressed as
\begin{equation}\label{final_G}
    G^{(c)} (x, y) = \int \frac{d^4 p}{(2 \pi)^4} \, \rho (p) e^{-i p(x-y)} \left( \theta_c (x^0 - y^0 ) + n_B (\omega_p) \right),
\end{equation}
which depends on the chosen contour \(C\). The two-point Green's function for interacting fields has a similar spectral form since the relation (\ref{spectral_density}) retains its validity in the case of interacting fields due to the KMS relation \cite{Landsman:1986uw}.

\subsection{Fermion Fields}
The thermal Green's functions can be further developed for fermion and gauge fields, although it is mostly a generalization of the scalar field theory \cite{Kapusta:2006pm, Bellac:2011kqa, Landsman:1986uw, Dolan:1973qd}. First, we compute the Green's function for fermion fields. We define the Green's function for fermion fields replacing Eqs. (\ref{def_G_c}) and (\ref{cor_G}) by,
\begin{equation}
    S_{\alpha \beta}^{(c)} (x - y) \equiv \langle T_c \{\psi_{\alpha} (x) \overline{\psi}_{\beta} (y)\} \rangle = \theta_c (x^0 - y^0) S^{+}_{\alpha \beta} - \theta_c (y^0 - x^0) S^{-}_{\alpha \beta}\, ,
\end{equation}
where \(\alpha\) and \(\beta\) denote the spinor indices and the so-called reduced Green's functions\footnote{We previously referred to the reduced Green's functions for scalar fields as correlation functions.} are
\begin{equation}
    S^{+}_{\alpha \beta } (x - y)= \langle \psi_{\alpha} (x) \overline{\psi}_{\beta} (y) \rangle\, ,
\end{equation}
and obey the following KMS relation,
\begin{equation}
    S_{\alpha \beta}^{+} (\Vec{x}, t - i \beta) = - S^{-}_{\alpha  \beta} (\Vec{x}, t) \,.
\end{equation}

Now the two-point function for the free fermion fields is computed following the same method as in the case of scalar fields (\ref{final_G}). The two-point function for a fermion field satisfies the differential equation
\begin{equation}
    \left(i \slashed \partial - m \right)_{\alpha \sigma} S_{\sigma \beta }^{(c)} (x - y) = i \delta^{(4)}_c (x - y) \delta_{\alpha \beta} \, ,
\end{equation}
which has the same form as the zero-temperature case and we could define the thermal Green's function \(S^{(c)} (x-y)\) such that
\begin{equation}
    S^{(c)}_{\alpha \beta} (x - y) \equiv \left( i \slashed \partial + m \right)_{\alpha \beta} S^{(c)} (x - y) \, ,
\end{equation}
where one easily proves that \(S^{(c)} (x - y)\) satisfies the equation (\ref{eq_D}) as the two-point function for a free scalar field does. As a result, the expression for this Green's functions is computed as
\begin{equation}
    S^{(c)} (x - y) = \int \frac{d^4 p}{(2 \pi)^4} \, \rho(p) e^{- ip (x - y)} \left( \theta_c (x^0 - y^0) - n_F (\omega_p) \right) \, ,
\end{equation}
where the Fermi-Dirac distribution function, \(n_F (\omega)\), is
\begin{equation}\label{F-D-fun}
    n_F (\omega) = \frac{1}{e^{\beta \omega} + 1} \, .
\end{equation}

\section{Imaginary Time Formalism}

In this section, the Feynman rules are discussed in the finite-temperature field theory. However, at finite temperatures, the propagators represented by lines in the Feynman diagrams depend on the chosen contour in the complex time plane. Hence, the choice of the contour results in a particular formulation of quantum field theories in which the Feynman rules are derived \cite{Matsumoto:1984au}. 

At the beginning of this chapter, the partition function (\ref{gen_Z}) was expressed as a path integral selecting the contour on the imaginary axis \(t = - i \tau\). This contour is called Matsubara contour, named after Matsubara \cite{Matsubara:1955ws}, and it has been widely applied in quantum statistical mechanics. Namely, on this contour, the field operators are time-ordered in terms of the real variable \(\tau = i t\). 

On the Matsubara contour, the Green's functions have imaginary-time arguments. If we set \(t - t' = - i \tau\) with \(-\beta \leq \tau \leq \beta\), the two-point functions for scalar and fermion fields can be cast into the form
\begin{equation}\label{combo_G}
    G ( \Vec{x}, \tau) = \int \frac{d^4 p}{(2 \pi)^4} \, \rho (p) e^{i \Vec{p} \cdot \Vec{x}} e^{- \tau p^0} \left[ \theta (\tau) + \eta n (p^0) \right], 
\end{equation}
where \(\eta\) is defined as \(\eta_B = 1\) \((\eta_F = -1)\) for bosons (for fermions). Additionally, \(n (p^0)\) stands either for the Bose-Einstein distribution function (\ref{B-E-fun}) or the Fermi-Dirac distribution (\ref{F-D-fun}), and it can be written as\footnote{The contribution of the chemical potential to the Green's functions is considered to vanish in the previous cases, but the chemical potential can be included in (\ref{combo_G}) by replacing \(p^0 \to p^0 -\mu\). See Appendix \ref{Appendix A}.}
\begin{equation}
    n (\omega) = \frac{1}{e^{\beta \omega} - \eta} .
\end{equation}
Then, the Green's function (\ref{combo_G}) can be expressed as
\begin{equation}
    G ( \Vec{x}, \tau) = G_{+} (\Vec{x}, \tau ) \theta (\tau) + G_{-} (\Vec{x}, \tau ) \theta (- \tau).
\end{equation}
One easily proves using the KMS relations (\ref{KMS1}) and (\ref{KMS2}) that 
\begin{equation*}
    G (\tau + \beta) = \eta G(\tau) \quad \text{for} \, - \beta \leq \tau \leq 0
\end{equation*}
and
\begin{equation*}
    G (\tau - \beta) = \eta G(\tau) \quad \text{for} \, 0 \leq \tau \leq \beta \, .
\end{equation*}
Namely, this propagator for bosons (fermions) is periodic (anti-periodic) in the time variable \(\tau\) with period \(\beta\). Now the Fourier transform of (\ref{combo_G}) is written as
\begin{equation}\label{G_143}
    \Tilde{G} (\Vec{p},\omega_n) = \int_{\alpha - \beta}^{\alpha} d \tau \int d^3 x \, e^{i \omega_n \tau - i \Vec{p} \cdot \Vec{x}} G (\Vec{x}, \tau) \, ,
\end{equation}
which does not depend on \(\alpha\) for \(0 \leq \alpha \leq \beta \). The periodicity and anti-periodicity conditions for bosons and fermions are cast into the form
\begin{equation}
    \psi(\vec{x}, 0) = \eta \psi (\vec{x}, -i \beta),
\end{equation}
and the discrete frequencies satisfy the condition
\begin{equation}
    e^{i \omega_n \beta} = \eta^{-1},
\end{equation}
which results in
\begin{equation}
    \omega_n = 2  n \pi\beta^{-1} \quad \text{for bosons},
\end{equation}
\begin{equation}\label{matsubara_2}
    \omega_n = (2  n + 1) \pi\beta^{-1} \quad \text{for fermions},
\end{equation}
where the discrete frequencies \(\omega_n\) are called Matsubara frequencies. The condition for the bosonic frequencies is the same as the condition (\ref{periodic}) as they originate from the same periodicity condition. If we insert Eq. (\ref{combo_G}) into Eq. (\ref{G_143}), the two-point function in momentum space can be written as
\begin{equation}\label{momentum_G}
    \Tilde{G} (\Vec{p}, \omega_n) = \int_{- \infty}^{\infty} \frac{d p_0}{2 \pi}\, \frac{\rho (p)}{ p_0 - i \omega_n} 
     = \frac{1}{\Vec{p}^{\, 2} + m^2 + \omega_n^2}
\end{equation}
and the Euclidean propagator can be defined as
\begin{equation}
    G (\Vec{x}, \tau) = i \Delta (\Vec{x}, -i \tau ) \, .
\end{equation}
The inverse Fourier transform of (\ref{G_143}) is cast as
\begin{equation}
     G (\Vec{x}, \tau) = \frac{1}{\beta} \sum_{n = - \infty}^{\infty} \int \frac{d^3 p}{(2 \pi)^3} \, e^{- i \omega_n \tau + i \Vec{p} \cdot \Vec{x}}  \Tilde{G} (\Vec{p}, \omega_n) \, ,
\end{equation}
and using Eq. (\ref{momentum_G}), one deduces the two-point function,
\begin{equation}\label{iD_1}
     G (\Vec{x}, \tau) = \frac{1}{\beta} \sum_{n = - \infty}^{\infty} \int \frac{d^3 p}{(2 \pi)^3} \, e^{- i \omega_n \tau + i \Vec{p} \cdot \Vec{x}} \frac{1}{\Vec{p}^{\, 2} + m^2 + \omega_n^2} \, ,
\end{equation}
which is defined for any real parameter \(\tau\) and is periodic (anti-periodic) for bosons (fermions) with period \(\beta\). The Matsubara frequencies in this expression are either for bosons or for fermions. Finally, the Euclidean propagator is
\begin{equation}\label{iD_2}
     i\Delta (x) = \frac{i}{\beta} \sum_{n = - \infty}^{\infty} \int \frac{d^3 p}{(2 \pi)^3} \, e^{- i p x} \frac{i}{p^2 - m^2} \, ,
\end{equation}
where in the integral the four-momentum is \(p^{\mu} = \left( i \omega_n, \Vec{p} \right)\). Similarly, the fermion Euclidean propagator can be written as
\begin{equation}\label{iD_3}
    i S (x) = \frac{i}{\beta} \sum_{n = - \infty}^{\infty} \int \frac{d^3 p}{(2 \pi)^3} \, e^{- i p x} \frac{i}{\gamma^{\mu} p_{\mu} - m} \, ,
\end{equation}
where \(\gamma^{\mu}\) is the usual gamma matrix. This formalism can be extended analogously to gauge fields \cite{Landsman:1986uw}. In fact, the gauge boson propagator in the imaginary time formalism can be obtained by a substitution rule which can be derived by observing the similarities of the above expressions to the corresponding ones in the zero-temperature quantum field theory. In particular, the two-point function (\ref{iD_1}) can be understood as the ordinary two-point function in the quantum field theory \cite{Schwartz:2014sze} substituting \cite{Landsman:1986uw}
\begin{equation}\label{substi}
    \int \frac{d p_0}{2 \pi} f(p_0) \to \frac{i}{\beta} \sum_{n = - \infty}^{\infty} f \left(p_0 = i \omega_n \right) \, ,
\end{equation}
where \(\omega_n\) are the Matsubara frequencies. As a consequence, the imaginary time formalism at the limit \(\beta \to \infty\) reduces to the ordinary quantum field theory, applying the reverse substitution rule \cite{Landsman:1986uw}.

In conclusion, the Feynman rules in the imaginary formalism are simply stated in momentum space and they can be read off by the Euclidean propagators (\ref{iD_2}) and (\ref{iD_3}). These Feynman rules are briefly formulated as \cite{Kapusta:2006pm, Laine:2016hma, Landsman:1986uw}:
\begin{itemize}
    \item Draw the Feynman diagrams and determine the combinatoric factor for each diagram as in the quantum field theory at zero temperature.
    \item Assign a propagator to each line with:
    \begin{equation*}
    \begin{split}
        & \text{-- the boson propagator} \quad : \quad \frac{i}{p^2 - m^2}; \, \, p^{\mu} = \left( 2 n \pi i \beta^{-1}, \Vec{p} \right). \\
        & \text{-- the fermion propagator} \quad :  \quad \frac{i}{\slashed p - m}; \, \, p^{\mu} = \left( (2 n +1) \pi i \beta^{-1}, \Vec{p} \right).
    \end{split}
    \end{equation*}
    \item  Assign to each loop a factor:
    \begin{equation*}
        \frac{i}{\beta} \sum_{n = - \infty}^{\infty} \int \frac{d^3 p}{(2 \pi)^3}.
    \end{equation*}
    \item Include the usual vertex factors from the coefficients of the interaction terms in the action and conserve energy and momentum at each vertex with:
    \begin{equation*}
        - i \beta (2 \pi)^3 \delta_{n, 0} \delta^{(3)} \left( \sum_{i} \Vec{p}_i \right) .
    \end{equation*} 
\end{itemize}

In the imaginary time formalism, the infinite summations in the Green's functions can be performed by a standard method \cite{Landsman:1986uw,Morley:1978aq, Quiros:1999jp} which replaces the frequency sums by contour integrals. In the case of bosons, the frequency sums (\ref{substi}) can be computed using the function
\begin{equation*}
    \frac{\beta}{2} \coth{ \left( \frac{\beta z}{2} \right)}  \, ,
\end{equation*}
which has poles at \(z = i \omega_n\) and is analytic and bounded. Subsequently, the frequency sums (\ref{substi}) are written as
\begin{equation}\label{E11}
    \frac{1}{\beta} \sum_{n = - \infty}^{\infty} f(p^0 = i \omega_n) = \frac{1}{2 \pi i \beta } \int_{\gamma} d z \, f(z) \frac{\beta}{2} \coth{ \left( \frac{\beta z}{2} \right)}  \, ,
\end{equation}
where the contour \(\gamma\) encircles anticlockwise the poles on the imaginary axis. It is additionally assumed that the function \(f(z)\) is analytic in the neighborhood of the imaginary axis to ensure the validity of the formula (\ref{E11}). Now the contour can be deformed to a new contour which consists of a first line starting at \(- i \infty + \epsilon\) and ending at \(i \infty + \epsilon\) and a second line starting at \(i \infty - \epsilon\) and going to \(- i \infty - \epsilon \). Expressing the hyperbolic cotangent in terms of the exponential, the formula (\ref{E11}) can be rewritten as
\begin{equation}\label{E22}
    \frac{1}{\beta} \sum_{n = - \infty}^{\infty} f(p^0) = \int_{- i\infty}^{i \infty}  \frac{dz}{4 \pi i} \left[ f(z) + f(-z) \right] + \int_{-i\infty + \epsilon}^{i \infty + \epsilon}  \frac{d z}{2\pi i} \left[ f(z) + f(-z) \right] n_B (z).
\end{equation}
All the singularities of the functions \(f(z)\) and \(f(-z)\) in the right half plane can be encircled clockwise, if, in the second integral, the contour is deformed to a contour, labeled as \(C\). As a result, Eq. (\ref{E22}) reads
\begin{equation}
    \frac{1}{\beta} \sum_{n = - \infty}^{\infty} f(p^0) = \int_{- i\infty}^{i \infty}  \frac{dz}{4 \pi i} \left[ f(z) + f(-z) \right] + \int_C \frac{d z}{2\pi i} \left[ f(z) + f(-z) \right] n_B (z)
\end{equation}
and it is generalized by the following expression for both bosons and fermions,
\begin{equation}\label{expr_hen}
    \frac{1}{\beta} \sum_{n = - \infty}^{\infty} f(p^0) = \int_{- i\infty}^{i \infty}  \frac{dz}{4 \pi i} \left[ f(z) + f(-z) \right] + \eta \int_C \frac{d z}{2\pi i} n (z) \left[ f(z) + f(-z) \right] ,
\end{equation}
where the distribution function \(n(z)\) stands for either the Bose-Einstein or the Fermi-Dirac distribution function. The expression (\ref{expr_hen}) showcases an intriguing property of these frequency sums which will be observed in detail later. This property indicates that the frequency sums are decomposed into a temperature-dependent term which coincides with the term computed in the quantum field theory at zero temperature, and a temperature-dependent term which vanishes at the limit \(\beta \to \infty\).

The notable benefit of the imaginary time formalism is that it leads to a perturbation expansion that is represented by Feynman diagrams identical to those in the theory at zero temperature. On the other hand, the computation of the Green's functions with imaginary-time arguments is an inherent drawback of this formalism, and real-time correlation functions can be obtained by an analytic continuation to the real axis. Namely, a contour may be chosen to evaluate directly the correlation functions in the real time and establish the real time formalism \cite{Kapusta:2006pm, Laine:2016hma, Landsman:1986uw,Morley:1978aq, Quiros:1999jp}. This formalism will not be developed here as this work focuses on the imaginary time formalism for the computational simplicity in the finite-temperature effective potential. Nonetheless, the imaginary time and real time formalism should lead to the same physical predictions \cite{Kobes:1990kr, Aurenche:1991hi, vanEijck:1992mq, Evans:1990hy, Evans:1990qh}.

\section{Finite-Temperature Effective Action}
In this section, the formalism of the effective action is extended to finite-temperature field theories \cite{Brandenberger:1984cz, Dolan:1973qd, Weinberg:1974hy, Bernard:1974bq}. This is formulated as a direct extension of the zero-temperature formalism, which was presented in Chapter \ref{Effective Action} since the major difference in the finite temperature formalism is the contour ordering and the discretization of the Matsubara frequencies. In particular, the generating functional (\ref{Z_J_B}) in the presence of an external classical current can be directly expressed as
\begin{equation}
    Z[J;\beta] = e^{i W[J; \beta]} \, ,
\end{equation}
where \(W[J; \beta]\) is the generating functional for thermal connected Green's functions. Likewise, the generating functional of the 1PI diagrams at finite temperature is defined by the Legendre transform
\begin{equation}\label{legendre_b}
    \Gamma [\phi_c; \beta] = W[J;\beta] - \int_C d^4 x \, \phi_c(x) J(x) \, ,
\end{equation}
where the integral along the time variable follows a given path \(C\) in the complex time plane and the background field, \(\phi_c (x)\), is obtained by
\begin{equation}\label{dWdJ}
    \frac{\delta W[J; \beta]}{\delta J(x)} = \phi_c (x)
\end{equation}
where the thermal connected one-point function is computed using the grand canonical average in the presence of an external source. In the absence of this source, the background field at a finite temperature reads
\begin{equation}
    \left. \phi_c (x) \right|_{J = 0}  = Z^{-1} Tr \left( e^{-\beta \hat{H}} \hat{\phi} (x) \right) \,.
\end{equation}
One obtains, from Eqs. (\ref{legendre_b}) and (\ref{dWdJ}), the functional differential equation
\begin{equation}\label{EOM_T}
    \frac{\delta \Gamma[\phi_c; \beta]}{\delta \phi_c(x)} = - J (x) \,.
\end{equation}
The 1PI generating functional at finite temperature is easily identified as the effective action at finite temperature following the same procedure as in the previous chapter. Therefore, the equation of motion (\ref{EOM_vacuum}) at zero temperature is translated into the finite-temperature case (\ref{EOM_T}).

Now the translational invariance of the vacuum state is required and the finite-temperature effective potential is defined similarly to the zero-temperature case (\ref{expand_p}) and (\ref{G_V}), resulting in
\begin{equation}
    \Gamma [\phi_c; T] = - \int_{C} d^4 x \,V_{\text{eff}}  (\phi_c, T) \, .
\end{equation}
Subsequently, the finite-temperature effective potential is derived naturally by the zero-temperature quantum field theory and is given by the sum of all 1PI diagrams with zero external momenta as it is demonstrated by the zero-temperature effective potential (\ref{gen_eff}). 

Let's suppose that the Lagrangian density possesses an internal symmetry. Lorentz invariance can impose that the background field  \(\phi_c (x)\) vanishes at \(J (x) = 0\), but symmetry violation is signaled by a non-zero background field \(\phi_c (x)\), which satisfies the condition
\begin{equation}\label{T_EOM_vacuum}
    \frac{\delta \Gamma[\phi_c; T]}{\delta \phi_c(x)} = 0 \, .
\end{equation}
Hence, the differential equation (\ref{T_EOM_vacuum}) can be rewritten in terms of the finite-temperature effective potential as
\begin{equation}\label{V_EOM}
    \frac{\partial V_{\text{eff}}  (\phi_c, T)}{\partial \phi_c} = 0
\end{equation}
and spontaneous symmetry breaking occurs when the solution of (\ref{V_EOM}) is a non-zero background field \(\phi_c = \upsilon (T)\). 

Lastly, it is important to emphasize that the above definitions of the thermal Green's functions should be formulated by choosing a contour in the complex time plane and the Feynman rules at finite temperature should be modified as it was discussed in the previous section.

\section{Finite-Temperature Effective Potential}
The previous section illustrates that the finite-temperature effective potential is calculated following a method very similar to that used in the quantum field theory at zero temperature. In fact, at finite temperatures, the thermal 1PI diagrams are represented by the same 1PI diagrams as the ones at zero temperatures, but they are computed using the Feynman rules in the previous section. In this section, we derive the finite-temperature one-loop effective potential for scalar field theories and non-Abelian gauge theories in the imaginary time formalism \cite{Brandenberger:1984cz, Dolan:1973qd, Weinberg:1974hy, Bernard:1974bq, Quiros:1999jp}. 

In general, the finite-temperature one-loop effective potential consists of the zero-loop contribution which is the classical potential, and the one-loop contribution which is temperature dependent such that,
\begin{equation}\label{decom_Veff}
    V_{\text{eff}} (\phi_c, T) = V_0 (\phi_c) + V_1 (\phi_c, T).
\end{equation}
Furthermore, the effective potential at finite temperature includes the effective potential at zero temperature which was presented in the previous chapter, and as it will be shown soon the full finite-temperature one-loop effective potential is decomposed into
\begin{equation}
    V_{\text{eff}} (\phi_c, T) = V_0 (\phi_c) + V_1 (\phi_c) + V^T_1 (\phi_c, T) \, ,
\end{equation}
which agrees completely with the expression (\ref{expr_hen}) which was discussed in an earlier section expressing the frequency sums in terms of contour integrals.

\subsection{Scalar Field Theories}
First, the one-loop effective potential is computed for a real scalar field theory which is described by the Lagrangian density (\ref{lag_scalar}). In this finite-temperature field theory, the sum of all 1PI Feynman diagrams with a single loop and zero external momenta is still represented by the sum of the diagrams in Fig. \ref{fig:1.3}. As a result, the one-loop effective potential at finite temperature is given by (\ref{gen_eff}) following the Feynman rules in the imaginary time formalism,
\begin{equation}\label{ini_V1}
\begin{split}
    V_1{(\phi_c, T)} & = i \sum_{n=1}^{\infty} \frac{i}{\beta} \sum_{n = - \infty}^{\infty} \int \frac{d^3 p}{(2\pi)^3} \, \frac{1}{2 n} \left(-\frac{ \lambda \phi^2_c/2}{\omega_n^2 + \Vec{p}^{\, 2} + m^2 } \right)^n \\
    & = - \frac{i}{2} \frac{i}{\beta} \sum_{n = - \infty}^{\infty} \int \frac{d^3 p}{(2\pi)^3} \, \ln \left(1 + \frac{\lambda \phi^2/2}{\omega_n^2 + \Vec{p}^{\, 2} + m^2} \right) \\
    & = \frac{1}{2 \beta} \sum_{n = - \infty}^{\infty} \int \frac{d^3 p}{(2\pi)^3} \, \ln \left(\frac{\omega_n^2 + \Vec{p}^{\, 2} + m_{\text{eff}}^2 (\phi_c)}{\omega_n^2 + \Vec{p}^{\, 2} + m^2} \right) \\
    & = \frac{1}{2 \beta} \sum_{n = - \infty}^{\infty} \int \frac{d^3 p}{(2\pi)^3} \, \ln \left(\omega_n^2 + \Vec{p}^{\, 2} + m_{\text{eff}}^2 (\phi_c) \right),
\end{split}
\end{equation}
where the effective mass (\ref{meff_1}) was introduced in the second line and a field-independent term was omitted in the last line. It is remarkable that the final expression in (\ref{ini_V1}) can be obtained directly by simply applying the substitution rule (\ref{substi}) which clearly indicates the transition from the finite-temperature to the zero-temperature theory. Eq. (\ref{ini_V1}) can be rewritten in a compact form as
\begin{equation}\label{initial_V1}
    V_1{(\phi_c, T)} = \frac{1}{2 \beta} \sum_{n = - \infty}^{\infty} \int \frac{d^3 p}{(2\pi)^3} \, \ln \left(\omega_{n}^2 + \omega^2 \right),
\end{equation}
where \(\omega\) follows the relation
\begin{equation}
    \omega^2 = \Vec{p}^{\, 2} + m_{\text{eff}}^2 (\phi_c).
\end{equation}

The frequency sum (\ref{initial_V1}) is divergent, but the infinite part is field-independent. The finite part is computed by the method presented in Ref. \cite{Dolan:1973qd}. First, we define the function
\begin{equation}\label{fun_u}
    \upsilon (\omega) = \sum_{n = - \infty}^{\infty} \ln \left(\omega_{n}^2 + \omega^2 \right) 
\end{equation}
and we differentiate with respect to \(\omega\),
\begin{equation}\label{du}
    \frac{\partial \upsilon}{\partial \omega} = \sum_{n = - \infty}^{\infty} \frac{2 \omega}{\omega_{n}^2 + \omega^2} \, .
\end{equation}
Using the identity,
\begin{equation}\label{fun_5}
\begin{split}
    f(y) = \sum_{n = 1}^{\infty} \frac{y}{y^2 + n^2} & = - \frac{1}{2y} + \frac{\pi}{2} \coth{\pi y} \\
    & = - \frac{1}{2y} + \frac{\pi}{2} + \frac{\pi e^{-2 \pi y}}{1 - e^{-2 \pi y}}
\end{split}   
\end{equation}
with \(y = \beta \omega / 2 \pi\), the expression (\ref{du}) can be written as
\begin{equation}
    \frac{\partial \upsilon }{\partial \omega} = 2 \beta \left[\frac{1}{2} + \frac{e^{- \beta \omega}}{1 - e^{-\beta \omega}} \right] \, .
\end{equation}
After integration, the initial function (\ref{fun_u}) is cast into the form
\begin{equation}\label{final_u}
    \upsilon (\omega) = \beta \omega + 2 \ln \left(1 - e^{-\beta \omega}\right) + c \, ,
\end{equation}
where the constant \(c\) includes all the \(\omega\)-independent terms. Therefore, we substitute the result (\ref{final_u}) into the one-loop correction in the effective potential (\ref{initial_V1}), obtaining 
\begin{equation}\label{last_V1}
    V_1 (\phi_c, T) = \int \frac{d^3 p}{(2 \pi)^3} \, \left[\frac{\omega}{2} + \frac{1}{\beta} \ln \left(1 - e^{-\beta \omega}\right) \right].
\end{equation}
The temperature-independent part of (\ref{last_V1}) is identified as the zero-temperature one-loop effective potential (\ref{final_Veff_scalar}) as it was suggested in an earlier section. This can be easily shown by proving the identity
\begin{equation}\label{int_1}
    \omega \int_{- \infty}^{\infty} \frac{d x}{2 \pi i} \frac{1}{- x^2 + \omega^2 - i \epsilon} = \frac{1}{2} \, .
\end{equation}
If we close the integration interval \((-\infty, + \infty)\) in the complex \(x\)-plane along an anticlockwise contour, the integral (\ref{int_1}) can be evaluated by choosing the pole of the integrand at \(x = - \sqrt{\omega^2 - i \epsilon}\) with a residue \((2 \omega)^{-1}\), and according to the Residue theorem, the identity (\ref{int_1}) holds true. Then, the identity (\ref{int_1}) is integrated with respect to \(\omega\) to give
\begin{equation}\label{int_2}
    -\frac{i}{2} \int_{- \infty}^{\infty} \frac{d x}{ 2 \pi} \ln \left( -x^2 + \omega^2 - i \epsilon \right) = \frac{\omega}{2} + c,
\end{equation}
where \(c\) is the constant of integration. Thus, the first integral in (\ref{last_V1}) is given by using the identity (\ref{int_2}),
\begin{equation}
     \int \frac{d^3 p}{(2 \pi)^3} \, \frac{\omega}{2} = - \frac{i}{2} \int \frac{d^4 p}{(2 \pi)^4} \, \ln \left( - p_0^2 + \omega^2 - i \epsilon \right) 
\end{equation}
and after a Wick rotation, dropping the \(i \epsilon\) in the logarithm, the zero-temperature one-loop effective potential is reproduced,
\begin{equation}
    \int \frac{d^3 p}{(2 \pi)^3} \, \frac{\omega}{2} = \frac{1}{2} \int \frac{d^4 p_E}{(2 \pi)^4} \, \ln \left[ p_E^2 + m_{\text{eff}}^2 (\phi_c) \right]\, .
\end{equation}

Now we focus on the temperature-dependent term in (\ref{last_V1}) which can be written using the formula (\ref{useful_formula}),
\begin{equation}\label{pre_therm}
\begin{split}
    \frac{1}{\beta} \int \frac{d^3 p}{(2 \pi)^3} \, \ln \left(1 - e^{-\beta \omega}\right) & = \frac{\pi^{3/2}}{\Gamma\left(\frac{3}{2}\right)} \frac{1}{\beta}\int_{0}^{\infty} \frac{d \rho}{(2 \pi)^3} \,   \sqrt{\rho} \ln \left(1 - e^{- \beta \sqrt{\rho + m^2_{\text{eff}}(\phi_c)}} \right) \\
    & = \frac{1}{4 \pi^2 \beta } \int_{0}^{\infty} d \rho \, \sqrt{\rho} \ln \left(1 - e^{-\sqrt{ \beta^2 \rho + \beta^2 m^2_{\text{eff}}(\phi_c)}} \right)\, ,
\end{split}
\end{equation}
where \(\rho = \vec{p}^{\,2}\). Then, Eq. (\ref{pre_therm}) yields, setting \(x =  \beta\sqrt{\rho} \),
\begin{equation}
    \frac{1}{\beta} \int \frac{d^3 p}{(2 \pi)^3} \, \ln \left(1 - e^{-\beta \omega}\right) = \int_{0}^{\infty} d x \, x^2 \ln \left[ 1 - \exp\left(- \sqrt{x^2 + y^2} \right)\right] \, ,
\end{equation}
and the final result is expressed in terms of the so-called bosonic thermal function,
\begin{equation}
    \frac{1}{\beta} \int \frac{d^3 p}{(2 \pi)^3} \, \ln \left(1 - e^{-\beta \omega}\right) = \frac{1}{2 \pi^2 \beta^4 } J_B \left(m_{\text{eff}}^2 (\phi_c) \beta^2\right) \, ,
\end{equation}
where the bosonic thermal function is defined as
\begin{equation}\label{thermal_B}
    J_B (y^2) = \int_{0}^{\infty} d x \, x^2 \ln \left[ 1 - \exp\left(- \sqrt{x^2 + y^2} \right)\right] \, ,
\end{equation}
This thermal function admits a low-temperature and a high-temperature expansion which will be discussed in detail at the end of this section. Therefore, the temperature-dependent one-loop contribution to the effective potential reads
\begin{equation}\label{V_scalar_1_T}
    V_{1}^{T} (\phi_c, T) = \frac{T^4}{2 \pi^2} \int_{0}^{\infty} dx \, x^2 \ln \left[ 1 - \exp \left( - \sqrt{x^2 + \frac{m_{\text{eff}}^2  (\phi_c)}{T^2} }\right) \right] \, .
\end{equation}

Finally, one notices the surprising agreement of the finite temperature one-loop contribution (\ref{last_V1}) with the free energy\footnote{In the free energy (\ref{free_energy}), a temperature-independent term was neglected which is identical to the tree-level contribution in the effective potential \cite{Kapusta:1989bd}} (\ref{free_energy}). In fact, the finite-temperature effective potential can be interpreted as the free energy density of a quantum system in most of the cases \cite{Sher:1988mj, Brandenberger:1984cz, Kapusta:2006pm, Laine:2016hma, Landsman:1986uw, Kapusta:1989bd}. In particular, the relation between the free-energy density and the effective potential is derived explicitly for a scalar field theory in Refs. \cite{Laine:2016hma, Kapusta:1989bd}.

\subsection{Non-Abelian Gauge Theories}
The previous calculation can be similarly generalized to non-Abelian gauge theories \cite{Dolan:1973qd, Weinberg:1974hy, Bernard:1974bq}. Initially, we consider the Lagrangian density (\ref{Lag_ferm}) for the fermion fields to calculate the contribution of the fermion loops. The sum of all 1PI Feynman diagrams with a single loop and zero external momenta in Fig. \ref{fig:1.4} is evaluated following the Feynman rules in the imaginary time formalism leading to the one-loop effective potential at finite temperature given by (\ref{gen_eff}),
\begin{equation}\label{ini_V1_F}
\begin{split}
    V_1{(\phi_c, T)} & = n_f \frac{i}{2} \frac{i}{ \beta} \sum_{n = - \infty}^{\infty} \int \frac{d^3 p}{(2\pi)^3} \, \ln \left(\frac{\omega_n^2 + \Vec{p}^{\, 2} + m^2 (\phi_c)}{\omega_n^2 + \Vec{p}^{\, 2} + m^2} \right) \\
    & = -\frac{n_f}{2 \beta} \sum_{n = - \infty}^{\infty} \int \frac{d^3 p}{(2\pi)^3} \, \ln \left(\omega_n^2 + \Vec{p}^{\, 2} + m^2 (\phi_c) \right) \, ,
\end{split}
\end{equation}
where \(\omega_n\) are the Matsubara frequencies for fermions and a field-independent term was omitted in the last line. As a result, Eq. (\ref{ini_V1_F}) is rewritten as
\begin{equation}\label{V1_T_F}
    V_1 (\phi_c, T) = - \frac{n_f}{2 \beta} \sum_{n = - \infty}^{\infty} \int \frac{d^3 p}{(2\pi)^3} \, \ln \left(\omega_{n}^2 + \omega^2 \right)
\end{equation}
with
\begin{equation}
    \omega^2 = \Vec{p}^{\, 2} + m^2 (\phi_c) \, .
\end{equation}
The sum over \(n\) in (\ref{V1_T_F}) is performed following a similar method to the previous one with the function (\ref{fun_u}). However, the sum in (\ref{V1_T_F}) is different from the one in (\ref{initial_V1}) due to the odd integers \(k = 2n+1\) in the fermionic Matsubara frequencies (\ref{matsubara_2}). Subsequently, the function \(\upsilon (\omega)\) for fermions is defined as
\begin{equation}\label{u_11}
    \upsilon (\omega) = \sum_{n = - \infty}^{\infty} \ln \left(\omega_{n}^2 + \omega^2 \right) = 2 \sum_{n = 1}^{\infty} \ln\left[ \frac{\pi^2 n^2}{\beta^2} + \omega^2\right],
\end{equation}
where the integer \(n = 1, 3, ...\) was redefined. This subtlety can be tackled by decomposing the function (\ref{fun_5}) into two pieces for even and odd integers,
\begin{equation*}
    \sum_{m = 2, 4, ...} \frac{y}{y^2 + m^2} = \sum_{n = 1}^{\infty} \frac{y}{y^2 + 4 n^2} = \frac{1}{2} f \left(\frac{y}{2} \right)
\end{equation*}
\begin{equation*}
    \sum_{m = 1, 3, ...} \frac{y}{y^2 + m^2} = f(y) - \frac{1}{2} f \left(\frac{y}{2} \right)\, ,
\end{equation*}
and using (\ref{fun_5}), the correct sum is obtained
\begin{equation}\label{inview}
    \sum_{m = 1, 3, ...} \frac{y}{y^2 + m^2} = \frac{\pi}{4} - \frac{\pi}{2}\frac{1}{e^{\pi y}+ 1} \, ,
\end{equation}
As a result, the derivative of (\ref{u_11}) with respect to the frequency is
\begin{equation}
    \frac{\partial \upsilon}{\partial \omega} = \frac{4 \beta}{\pi} \sum_{n = 1} \frac{y}{y^2 + n^2} \, ,
\end{equation}
where \(y = \beta \omega / \pi\). In view of (\ref{inview}) the derivative yields to
\begin{equation}
     \frac{\partial \upsilon}{\partial \omega} = 2 \beta \left[ \frac{1}{2} - \frac{1}{e^{\beta \omega} + 1} \right] \, ,
\end{equation}
and integrating this result with respect to the frequency, one obtains
\begin{equation}\label{final_u2}
    \upsilon (\omega) = \beta \omega +  2 \ln \left(1 + e^{-\beta \omega}\right) + c \, ,
\end{equation}
where the constant \(c\) includes all the frequency-independent terms. Therefore, we substitute (\ref{final_u2}) into the one-loop contribution to the effective potential (\ref{V1_T_F}), leading to
\begin{equation}\label{last_V1_2}
    V_1 (\phi_c, T) = - n_f \int \frac{d^3 p}{(2 \pi)^3} \, \left[\frac{\omega}{2} + \frac{1}{\beta} \ln \left(1 + e^{-\beta \omega}\right) \right].
\end{equation}
The temperature-independent term in (\ref{last_V1_2}) coincides with the zero-temperature one-loop contribution to the effective potential (\ref{fermi_eff}) as it was also proved in the scalar field theory.

The second integral in (\ref{last_V1_2}) represents the temperature-dependent term in the effective potential which is expressed in terms of the so-called fermionic thermal function,
\begin{equation}
    - \frac{n_f}{\beta} \int \frac{d^3 p}{(2 \pi)^3} \, \ln \left(1 + e^{-\beta \omega}\right) = - \frac{n_f}{2 \pi^2 \beta^4 } J_F \left(m^2 (\phi_c) \beta^2\right) \, ,
\end{equation}
where the fermionic thermal function is defined as
\begin{equation}
    J_F (y^2) = \int_{0}^{\infty} d x \, x^2 \ln \left[ 1 + \exp\left(- \sqrt{x^2 + y^2} \right)\right] \, ,
\end{equation}
Hence, the temperature-dependent contribution of the fermion loops reads
\begin{equation}\label{eff-T-fermion}
    V_{1}^{T} (\phi_c, T) = \frac{T^4}{2 \pi^2} \int_{0}^{\infty} d x \, x^2 \ln \left[ 1 + \exp \left( - \sqrt{x^2 + \frac{m^2  (\phi_c)}{T^2} }\right) \right].
\end{equation}

Finally, the finite-temperature one-loop effective potential for the scalars and fermions can be easily generalized to the case of gauge bosons in the non-Abelian gauge theory (\ref{Lag_non_Abelian}). More specifically, the contribution of the gauge boson loops can be computed by the tadpole diagram using the effective mass for the gauge bosons \cite{Dolan:1973qd, Bernard:1974bq, Quiros:1999jp} and alternatively by using the substitution rule (\ref{substi}). In this case, the gauge boson propagator in the Landau gauge is given by
\begin{equation}
    \Pi_{\mu \nu} (p) ^{(\alpha \beta)} = \Delta_{\mu \nu} G^{(\alpha \beta)} (p),
\end{equation}
where \(\Delta_{\mu \nu}\) has been defined in (\ref{delta_de}). As a result, the temperature-dependent term in the one-loop effective potential is written in terms of the bosonic thermal function as
\begin{equation}\label{eff-T-gauge}
    V^{T}_1 (\phi_c, T) = Tr \left(\Delta\right) \frac{1}{\beta} \int \frac{d^3 p}{(2 \pi)^3} \, \ln \left(1 - e^{-\beta \omega}\right) = \frac{Tr \left(\Delta\right) }{2 \pi^2 \beta^4 } J_B \left(M^2 (\phi_c) \beta^2\right) \, ,
\end{equation}
where the thermal function for bosons is given by (\ref{thermal_B}). Hence, the finite temperature one-loop contribution to the effective potential reads
\begin{equation}
     V_{1} (\phi_c, T) = Tr \left(\Delta\right) \left[\frac{1}{2} \int \frac{d^4 p_E}{(2 \pi)^4} \, \ln \left[p^2 + M^2 (\phi_c) \right] +  \frac{T^4}{2 \pi^2 } J_B \left(\frac{M^2 (\phi_c)}{T^2}\right) \right] \, .
\end{equation}
The first term does not depend on temperature and coincides with the zero-temperature effective potential (\ref{gauge_eff}), while the second term is the same as in the scalar-field case multiplied by the number of degrees of freedom of the gauge field.

\subsection{Thermal Functions}
This section is focused on the temperature-dependent component of the effective potential which is strongly determined by the thermal functions. The thermal functions for both bosons and fermions can be written as
\begin{equation}\label{thermal_functi}
J_{B/F} \left(y^2\right) = \int_{0}^{\infty} dx \, x^2 \ln \left[ 1 \mp \exp \left( - \sqrt{x^2 + y^2 }\right) \right] \, ,
\end{equation}
where the subscript \(B\) \((F)\) stands for bosons (fermions). The thermal functions can be computed numerically, but they also admit a high-temperature expansion \cite{Dolan:1973qd, Quiros:1999jp},
\begin{equation}\label{bosonthermalfunction_1}
\begin{split}
    J_B(y^2) =& - \frac{\pi^4}{45} + \frac{\pi^2}{12} y^2 - \frac{\pi}{6} y^3 - \frac{1}{32} y^4 \log \left(\frac{y^2}{a_b} \right) \\
    & - 2 \pi^{7/2} \sum_{l = 1}^{\infty} (-1)^l \frac{\zeta(2 l + 1)}{(l + 1)!} \Gamma \left(l + \frac{1}{2} \right) \left(\frac{y^2}{4 \pi^2} \right)^{l+2}
\end{split} 
\end{equation}
and
\begin{equation}\label{fermionthermalfunction_1}
\begin{split}
    J_F(y^2)  = & \, \frac{7\pi^4}{360} - \frac{\pi^2}{24} y^2 - \frac{1}{32} y^4 \log \left(\frac{y^2}{a_f} \right) \\ 
    & - \frac{\pi^{7/2}}{4} \sum_{l = 1}^{\infty} (-1)^l \frac{\zeta(2 l + 1)}{(l + 1)!} \left( 1 - 2^{-2l -1} \right) \Gamma \left(l + \frac{1}{2} \right) \left(\frac{y^2}{\pi^2} \right)^{l+2},
\end{split}  
\end{equation}
where 
\begin{equation*}
    \begin{split}
        a_b & = 16\pi^2 \exp{\left(3/2 - 2 \gamma_E \right)},\\
        a_f &= \pi^2 \exp{\left(3/2 - 2 \gamma_E\right)},
    \end{split}
\end{equation*}
and \(\zeta\) denotes the Riemann \(\zeta\)-function. One notices that both thermal functions (\ref{thermal_functi}) vanish at zero temperature which implies that the temperature-dependent part of the effective potential always vanishes at zero temperature as expected. Nevertheless, the thermal functions at the high-temperature expansion (\ref{bosonthermalfunction_1}) and (\ref{fermionthermalfunction_1}), including the first few \(y\)-dependent terms, have an indefinite or divergent value at zero temperature as they are polynomials with respect to \(y\), which go to infinite as the temperature decreases. This fact indicates the limits of the validity of the high-temperature approximation and is obviously caused by the violation of the condition, \(|y^2| \ll 1\), for the high-temperature expansion. Numerical analysis shows that the high-temperature expansion up to the logarithmic term is accurate to better than \(5 \% \) for \(y \leq 1.6\) for fermions and \(y \leq 2.2\) for bosons \cite{Laine:2016hma, Anderson:1991zb}.
This approximation for both the potential and its derivatives also agrees with the exact form to better than approximately \( 10 \% \) for the values \(y \lesssim (1 - 3)\), depending on the function \cite{Curtin:2016urg}. As a result, in our applications, the high-temperature expansion is adopted extensively including terms up to the logarithm, while its validity can be constantly checked. 

\begin{figure}[H]
    \centering
    \includegraphics[width=0.85\linewidth]{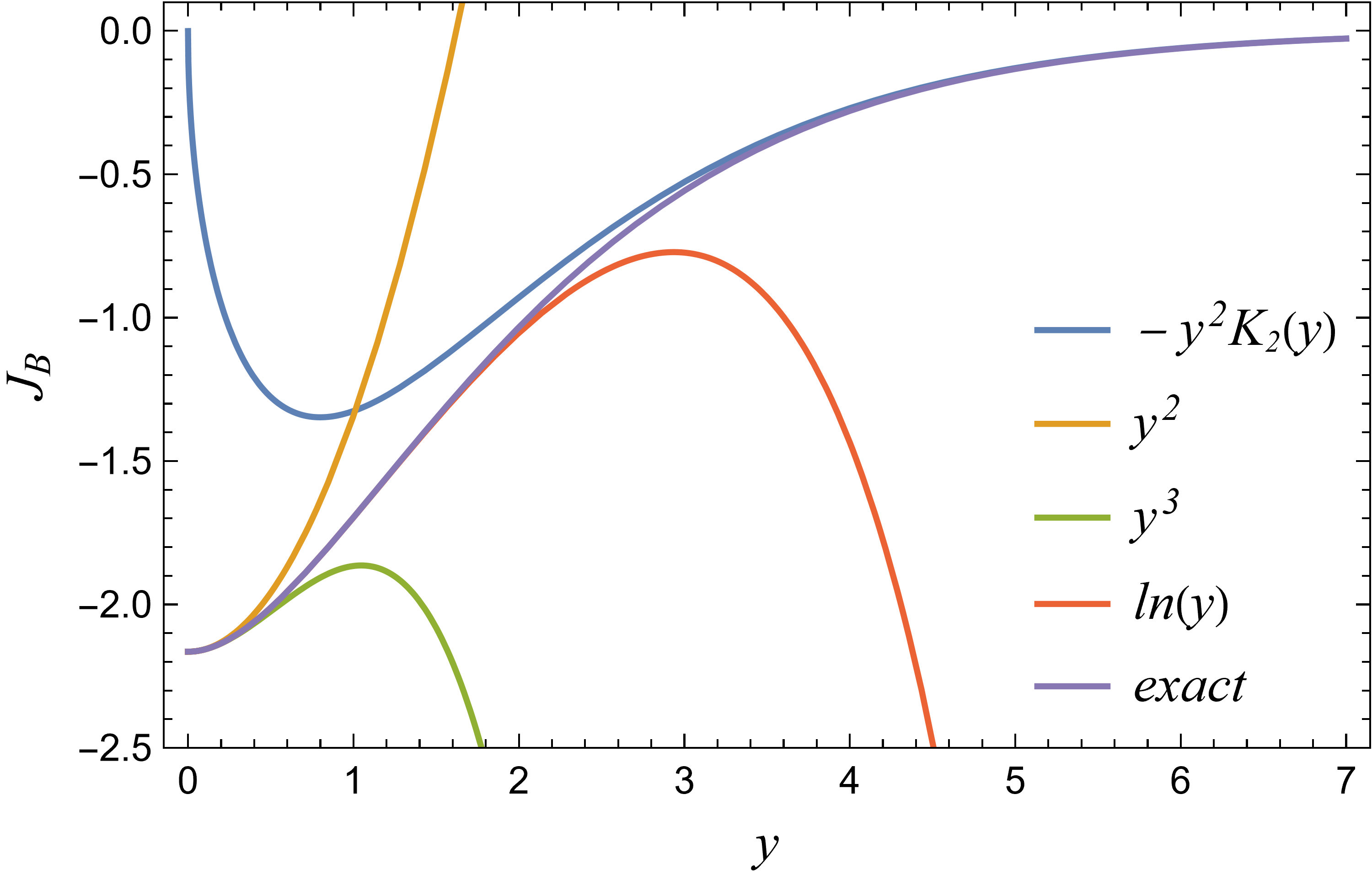}
    \caption{The exact numerical result (purple) of the thermal function for bosons and a number of approximations for this function: the low-temperature approximation in (\ref{low_tem}) (purple), the high-temperature approximation expanded up to the second term \(\mathcal{O}(y^2)\) (orange), up to the third term \(\mathcal{O}(y^3)\) (green) and up to the logarithmic term \(\mathcal{O}(y^4)\) (red).}
    \label{therm_fun}
\end{figure}

On the other hand, at the low-temperature limit, \(|y^2| \gg 1\), the thermal functions can be expanded in terms of the modified Bessel function of the second kind \cite{Laine:2016hma, Curtin:2016urg, Anderson:1991zb },
\begin{equation}
    J_{B} (y^2) = \tilde{J}_B^{(m)} (y^2)  = - \sum_{n = 1}^{m} \frac{1}{n^2} y^2 K_2 (y n)
\end{equation} 
and
\begin{equation}
    J_{F} (y^2) = \tilde{J}_F^{(m)}  (y^2)  = - \sum_{n = 1}^{m} \frac{(-1)^n}{n^2} y^2 K_2 (y n) \, ,
\end{equation}
which can be truncated at \(m = 2\) or \(3\), remaining a very good approximation \cite{Curtin:2016urg}. The modified Bessel function of the second kind is given by
\begin{equation}
    K_2 (y) = e^{- y}\sqrt{\frac{\pi}{2 y} } \left( 1 + \frac{15 }{8 y} + \,... \, \right) \,,
\end{equation}
For instance, the low-temperature expression for bosons can be written as 
\begin{equation}
    J_B (y^2) = -y^2 K_2 (y) \left[ 1 + \mathcal{O} (e^{-y}) \right] \,,
\end{equation}
where the higher-order terms are suppressed due to the exponential. Hence, the temperature effects are generally suppressed by high values \(|y^2| \gg 1\), as a result, in the low-temperature approximation the following function is used
\begin{equation}\label{low_tem}
    J_B (y^2) \approx -y^2 e^{- y}\sqrt{\frac{\pi}{2 y} } \left( 1 + \frac{15 }{8 y}\right) \,.
\end{equation}
The thermal bosonic function at the low-temperature and high-temperature expansions are compared in the case of the different approximations in Fig. \ref{therm_fun} which also shows the exact numerical evaluation.

\section{Symmetry Restoration}
In the 1970s, it was proposed that a spontaneously broken gauge symmetry could be restored at a high temperature \cite{Kirzhnits:1972iw, Kirzhnits:1972ut, Kirzhnits:1974as}. Namely, temperature corrections could generate spontaneous symmetry breaking in a gauge theory. This phenomenon is especially crucial in the Standard Model to describe the electroweak spontaneous symmetry breaking related to a cosmological phase transition, which is the main focus of this thesis. As a result, a gauge theory will be investigated here to demonstrate the mechanism of symmetry restoration.

To begin with, we present a model that simplifies the Standard Model considering only the contribution of the top quark and the gauge bosons to the one-loop corrections in the effective potential \cite{Anderson:1991zb}. This is a sufficient approximation in the case of a Higgs boson mass lower than the W-boson mass and we are not interested in the accuracy of its results. The derivation of the effective potential of this model is not shown here as the complete Standard Model effective potential will be discussed extensively in Chapter \ref{Electroweak Phase Transition}.

Firstly, the Higgs tree-level potential (\ref{Higgs_potential}) is written in terms of the Higgs boson as
\begin{equation}
    V_0 (h) = - \frac{1}{2} \mu^2 h^2 + \frac{\lambda}{4} h^4 \, ,
\end{equation}
where \(h\) is the Higgs boson and the Higgs doublet is
\begin{equation}
     H = \frac{1}{\sqrt{2}}\begin{pmatrix}
       \chi_1 + i \chi_2\\
        \phi_c + h + i \chi_3
        \end{pmatrix}\, ,
\end{equation}
where \(\chi_1\), \(\chi_2\), and \(\chi_3\) are the Goldstone bosons and \(\phi_c\) is a constant background field. Subsequently, the zero loop contribution to the effective potential in terms of the background field is
\begin{equation}
    V_0 (\phi_c) = - \frac{1}{2} \mu^2 \phi_c^2 + \frac{\lambda}{4} \phi_c^4,
\end{equation}
where the Higgs mass is \(m_H = \sqrt{2 \mu^2}\) and the Higgs minimum is given by
\begin{equation}
    \upsilon = \frac{\mu}{\sqrt{\lambda}} \, .
\end{equation}
The temperature-dependent term in the effective potential is written applying the high--temperature expansions (\ref{bosonthermalfunction_1}) and (\ref{fermionthermalfunction_1}) including the logarithmic term,
\begin{equation}\label{bosonthermalfunction_2}
\begin{split}
    V^{T}_1 (\phi_c, T) = & \sum_{i = Z, W} n_i \left[\frac{m_{i}^2 T^2}{24} - \frac{ \left(m_{i}^2 \right)^{3/2}}{12 \pi}T - \frac{m_{i}^4}{64 \pi^2} \log \left(\frac{m_{i}^2}{a_bT^2}\right)\right] \\
    & - 12 \left[ - \frac{m_{t}^2 T^2}{48} - \frac{m_{t}^4}{64 \pi^2} \log \left(\frac{m_{t}^2}{a_f T^2}\right) \right],
\end{split}
\end{equation}
where \(m_i^2 = m_i^2 (\phi_c)\) is the effective mass-squared for each particle \(i\) and \(n_Z = 3\), \(n_W = 6\), and \(n_t = 12\) are the corresponding numbers of degrees of freedom. The field-independent terms were also neglected in the above expression. As a result, the one-loop effective potential at finite temperature in this model is cast as \cite{Anderson:1991zb}
\begin{equation}\label{Veff_first}
    V_{\text{eff}} (\phi_c, T) = D (T^2 - T^2_0) \phi_c^2 - ET \phi_c^3 + \frac{\lambda (T)}{4} \phi_c^4,
\end{equation}
where the coefficients are expressed as
\begin{equation}\label{DD}
    D = \frac{2 m^2_W + m^2_Z + 2 m_t^2}{8 \upsilon^2},
\end{equation}
\begin{equation}
    E = \frac{2 m_W^3 + m^3_Z}{4 \pi \upsilon^3},
\end{equation}
\begin{equation}
    B = \frac{3}{64 \pi^2 \upsilon^4} \left(2 m_W^4 + m_Z^4 - 4 m_t^4 \right),
\end{equation}
\begin{equation}\label{TT}
    T^2_0 = \frac{m^2_H - 8 B \upsilon^2 }{4D},
\end{equation}
\begin{equation}\label{LL}
    \lambda (T) = \lambda - \frac{3}{16 \pi^2 \upsilon^4} \left( 2 m_W^4 \ln \frac{m_W^2}{A_B T^2} + m^4_Z \ln \frac{m^2_Z}{A_B T^2} - 4 m_t^4 \ln \frac{m^2_t}{A_F T^2} \right),
\end{equation}
where \(\ln A_{B/F} = \ln a_{b/f} -3/2\) and in the above coefficients, all masses are the masses at the Higgs minimum at zero temperature. The expression (\ref{Veff_first}) may seem too abstract and complex, but it will be understood in great detail in Chapter \ref{Electroweak Phase Transition}.

\begin{figure}[H]
    \centering
    \includegraphics[width=0.85\linewidth]{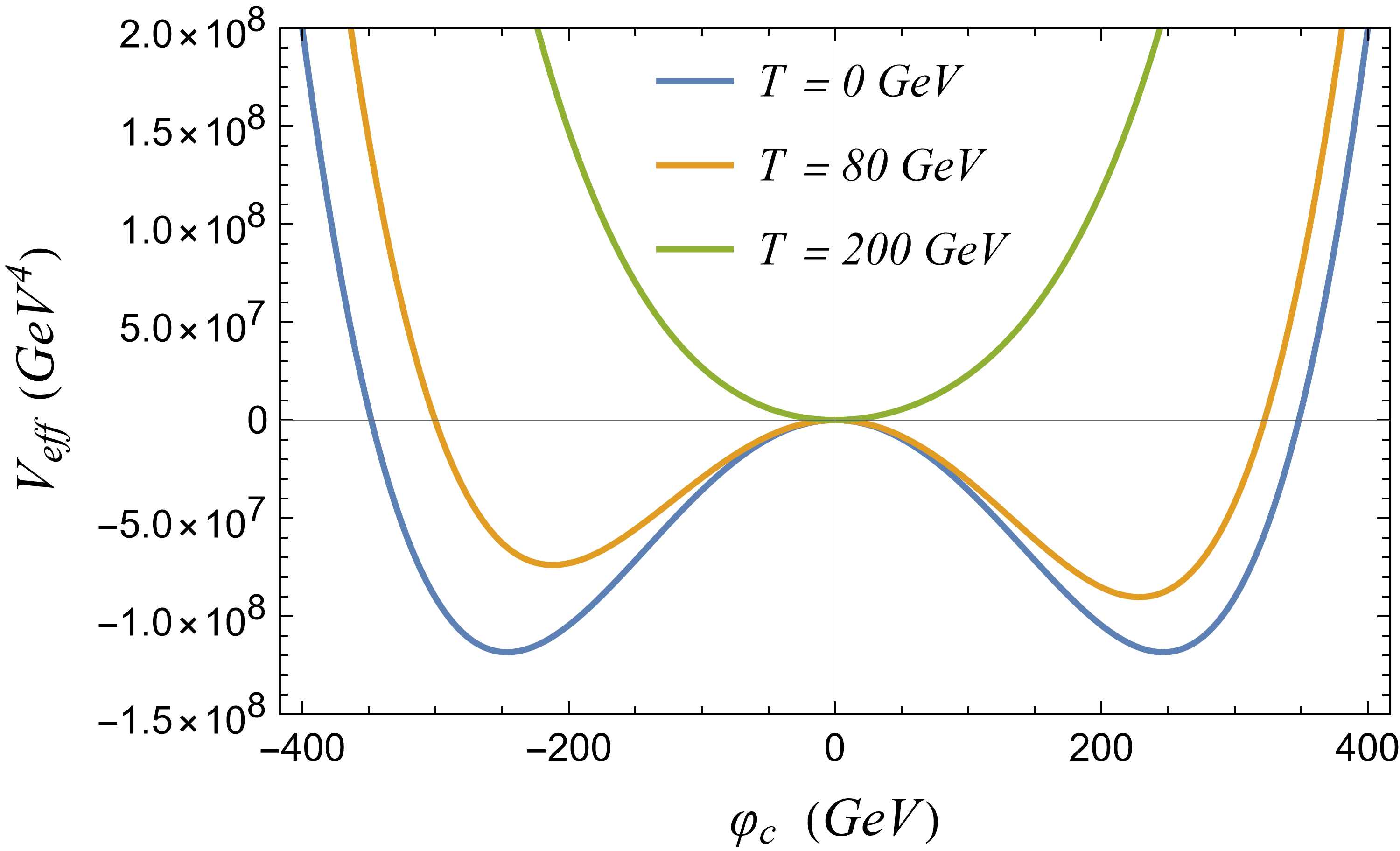}
    \caption{The finite-temperature one-loop effective potential (\ref{Veff_first}) at temperature \(T = 80\) GeV and \(T = 200\) GeV and the tree-level potential which significantly contributes to the one-loop effective potential at zero temperature.}
    \label{Symmetry_Restoration_1}
\end{figure}

\begin{figure}[H]
    \centering
    \includegraphics[width=0.85\linewidth]{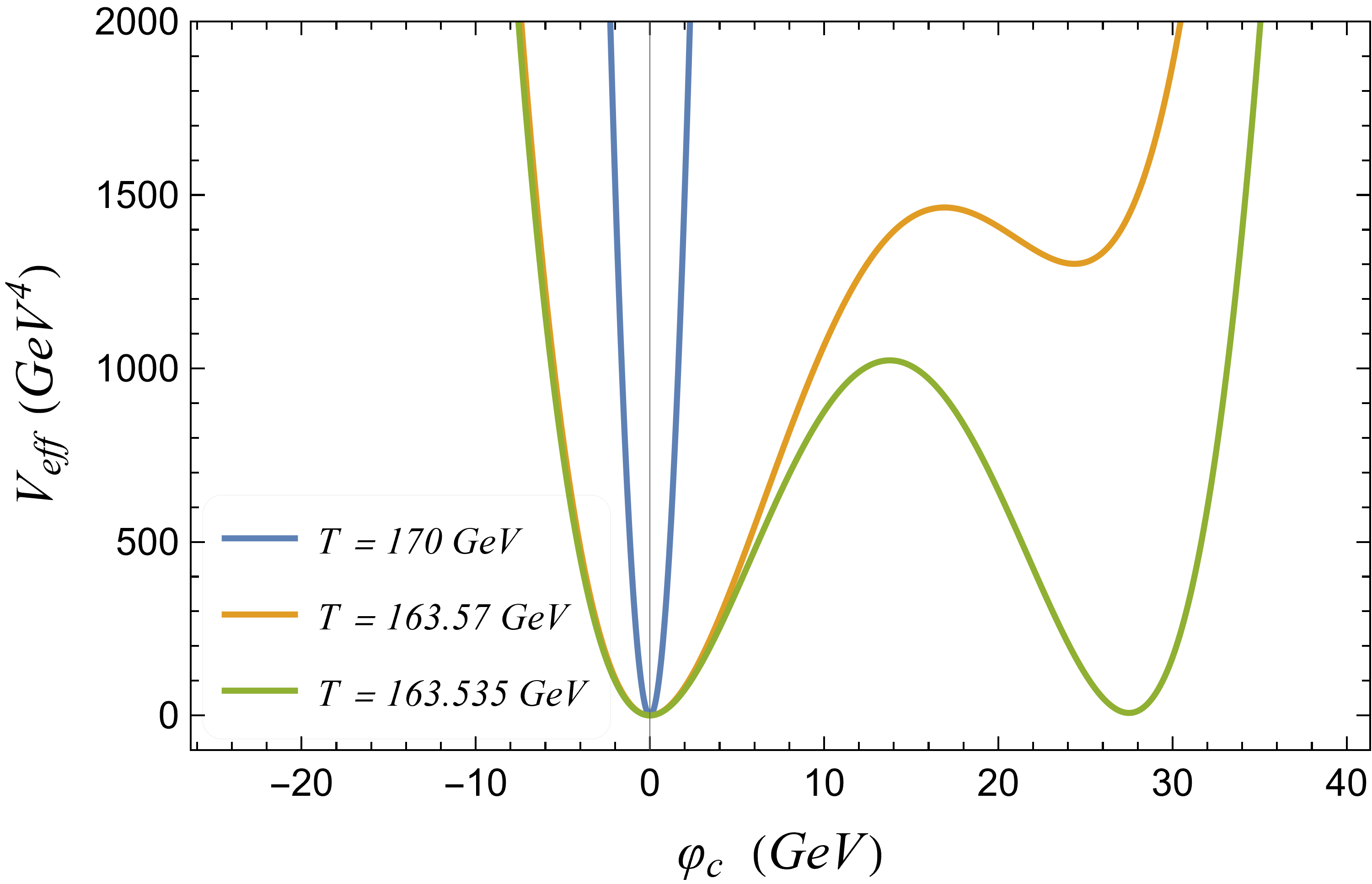}
    \caption{The finite-temperature one-loop effective potential (\ref{Veff_first}) at temperature \(T = 170\) GeV and \(T = 163.57\) GeV and \(T = 163.535\) GeV.}
    \label{Symmetry_Restoration_2}
\end{figure}

The effective potential (\ref{Veff_first}) clearly illustrates the mechanism of symmetry restoration in Fig. \ref{Symmetry_Restoration_1}, where it was computed using \(m_H = 125\) GeV, \(m_W = 80\) GeV, \(m_Z = 91\) GeV and \(m_t = 173\) GeV. Namely, at a high temperature, the effective potential is minimized for \(\phi_c = 0\), whereas at zero temperature its minimum is at a non-zero background field. As a consequence, the \(SU(2)_L \times U(1)_Y\) gauge symmetry is spontaneously broken at zero temperature, while at high temperatures the vacuum expectation value of the Higgs field vanishes, and the gauge symmetry is restored. Furthermore, one notices that the broken and unbroken phases can be distinguished by defining a critical temperature \(T_c\). Around the critical temperature, \(T>T_c\), a second local minimum \(\phi_m\) in the effective potential appears forming a barrier between the two local minima, \(\phi_c = 0\) and \(\phi_m \ne 0\), which implies the existence of a local maximum \(\phi_M\) as illustrated in Fig. \ref{Symmetry_Restoration_2}. At the critical temperature, the two local minima are degenerate, and spontaneous symmetry breaking is generated. In this case, the critical temperature is \(T_c = 163.535\) GeV as shown in Fig. \ref{Symmetry_Restoration_2} and the barrier is generated by the negative \(\phi_c^3\) term in the effective potential coming from the high-temperature expansion.
Below the critical temperature, the second local minimum with non-zero Higgs vacuum expectation value becomes the absolute minimum, while the minimum at the origin is metastable. At the temperature \(T = T_0\), the effective potential is written as
\begin{equation}\label{def_T0}
     V_{\text{eff}} (\phi_c, T_0) = - ET_0 \phi_c^3 + \frac{\lambda (T_0)}{4} \phi_c^4.
\end{equation}
This implies that the barrier vanishes and the origin becomes a local maximum. Moreover, one observes that in the true vacuum of the theory, the vacuum expectation value of the Higgs field as a function of temperature is zero at \(T > T_c\) and suddenly at \(T < T_c\) acquires discontinuously a non-zero value. This transition from the unbroken gauge symmetry to the spontaneously broken gauge symmetry, \(SU(2)_L \times U(1)_Y \to U(1)_{EM}\), can be considered a cosmological phase transition from the unbroken phase with \(\langle \hat{h} \rangle = 0\) to the broken phase with \(\langle \hat{h} \rangle \ne 0\). This phase transition is characterized by the barrier between the two degenerate local minima and it is called first-order phase transition. First-order phase transitions play a crucial role in the Standard Model and gravitational-wave physics.

\begin{figure}[H]
    \centering
    \includegraphics[width=0.85\linewidth]{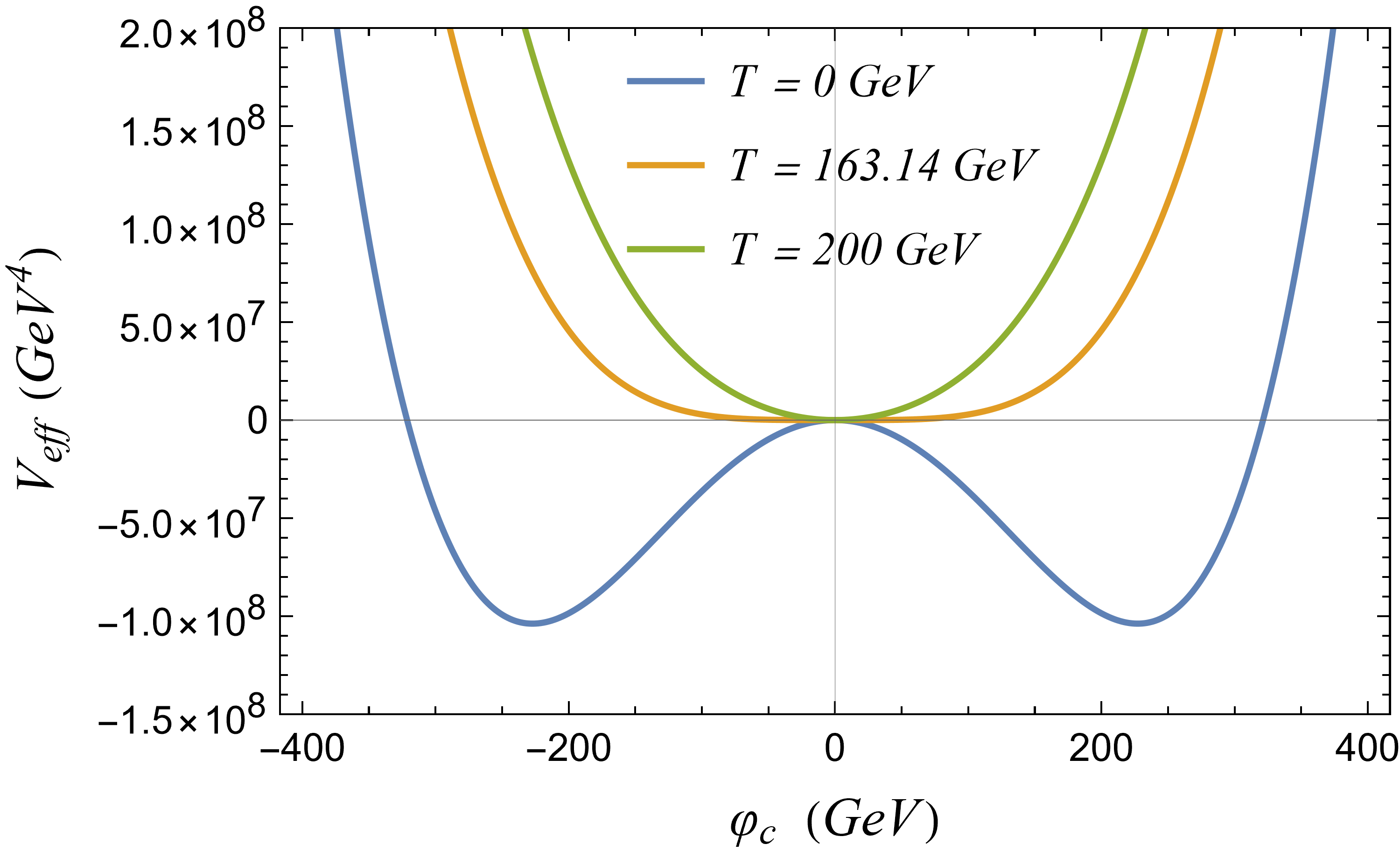}
    \caption{The one-loop effective potential (\ref{Veff_aaa}) is simply evaluated using the previous values for the coefficients (\ref{DD}), (\ref{TT}) and (\ref{LL}).}
    \label{Symmetry_Restoration_3}
\end{figure}

An alternative scenario that exhibits symmetry restoration is demonstrated using the following finite-temperature one-loop effective potential which does not contain a cubic term:
\begin{equation}\label{Veff_aaa}
    V_{\text{eff}} (\phi_c, T) = D (T^2 - T^2_0) \phi_c^2 + \frac{\lambda (T)}{4} \phi_c^4
\end{equation}
This effective potential is studied in the most general case without specifying the underlying theory. More specifically, at zero temperature, the effective potential is
\begin{equation}
    V_{\text{eff}} (\phi_c, 0) = - D T^2_0 \phi_c^2 + \frac{\lambda}{4} \phi_c^4
\end{equation}
and the symmetry of the theory is spontaneously broken since the effective potential is minimized at 
\begin{equation}
    \phi_0 = \pm \sqrt{\frac{2 D T_0^2}{\lambda}} 
\end{equation}
and the origin, \(\phi_c = 0\), corresponds to a local maximum. While the temperature increases, the stationary points of the potential are
\begin{equation}\label{sol1}
    \phi_0 (T) = 0
\end{equation}
and
\begin{equation}\label{sol2}
    \phi_0 (T)  = \pm \sqrt{\frac{2 D (T_0^2 - T^2)}{\lambda}} \, ,
\end{equation}
The origin is a local maximum and the second stationary point remains the minimum for \(T < T_0\) as it does not exist for \(T > T_0\). Namely, at high temperatures, \(T> T_0\), the origin is the single global minimum of the effective potential and the minimum (\ref{sol2}) disappears. This is illustrated clearly in Fig. \ref{Symmetry_Restoration_3}. 
However, both points (\ref{sol1}) and (\ref{sol2}) are identical to \(\phi_c = 0\) at \(T = T_0\), and the potential is given by
\begin{equation}
    V_{\text{eff}} (\phi_c, T_0) = \frac{\lambda(T_0)}{4} \phi_c^4 \, .
\end{equation} 
As a result, the symmetry is restored at temperatures \(T > T_0\) and one could define the critical temperature \(T_0\). In other words, as the temperature declines, the system transitions from a symmetric phase to a phase with a spontaneously broken symmetry. During this transition, the minimum expectation value \(\phi_0 (T)\) is a continuous function with respect to temperature and this process is considered as a second-order phase transition. Finally, a barrier is not formed in this phase transition as it is expected due to the absence of the cubic field term in the effective potential.

In summary, finite-temperature effects can generate spontaneous symmetry breaking and restore a broken gauge symmetry at high temperatures. This process occurs through a phase transition from the unbroken phase to the broken phase as temperature decreases.

\section{Thermal Resummation}

\begin{figure}[H]
    \centering
    \includegraphics[width=0.65\linewidth]{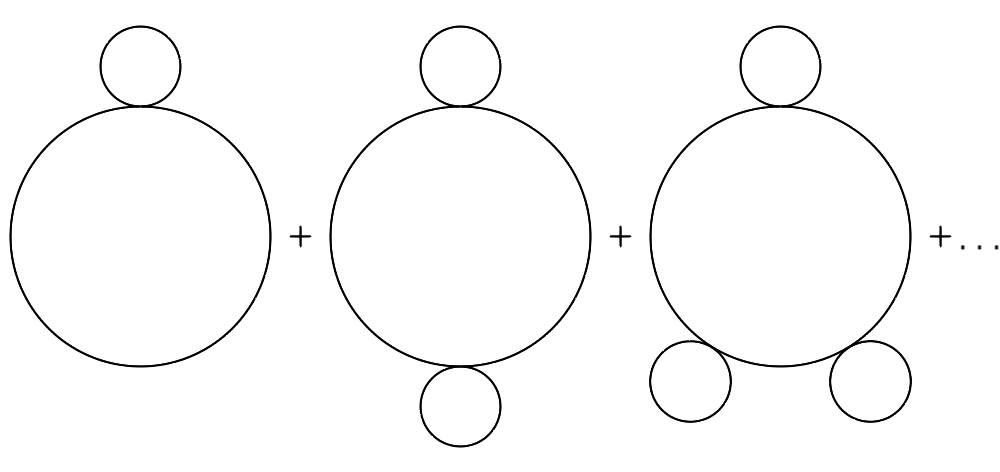}
    \caption{Ring diagrams to leading order.}
    \label{ring_1}
\end{figure}

Symmetry restoration at high temperatures implies that the ordinary perturbation theory breaks down near the critical temperature \cite{Weinberg:1974hy}. If perturbation theory were to remain valid, the temperature-dependent radiative corrections should be incapable of restoring the symmetry with the presence of the temperature-independent potential. In particular, the one-loop approximation in terms of small coupling constants, breaks down at a high temperature, due to the appearance of higher-loop infrared divergent diagrams which are associated with the zero bosonic Matsubara frequencies. Therefore, the dominant infrared contributions to the effective potential from higher-order diagrams should be included at all orders in the perturbative expansion \cite{Dolan:1973qd,Parwani:1991gq, Arnold:1992rz, Espinosa:1992gq, Fendley:1987ef, Gross:1980br, Altherr:1989rk, Kapusta:1984dx, Carrington:1991hz}. These higher-order diagrams are illustrated in Fig. \ref{ring_1} and are called ring diagrams, which are \(N\)-loop diagrams, where \(N-1\) of them are attached to the central one. These diagrams are resummed in the infrared limit with zero momenta in the small loops by using the full propagators in this limit \cite{Dolan:1973qd}. In practice, the so-called thermal resummation is a method that includes higher-order corrections to the effective mass which is replaced by the thermal mass
\begin{equation}\label{def_thermal_mass}
    M^2 (\phi_c, T) = m^2 (\phi_c) + \Pi (\phi_c, T) \,,
\end{equation}
where \(m^2 ( \phi_c)\) is the effective mass squared and \(\Pi(\phi_c, T)\) is the temperature-dependent self-energy in the infrared limit as described below.

In addition, one observes that the next-to-leading order term in the one-loop effective potential (\ref{V_scalar_1_T}) is proportional to \(\lambda^{3/2}\) as it was also shown in Eq. (\ref{Veff_aaa}). On the other hand, it is shown that the so-called ring diagram contributions are also of the same order and not \(\lambda^2\) \cite{Bellac:2011kqa}. Namely, the next-higher order correction is not the two-loop correction as was expected, but the ring diagram which is of order \(\lambda^{3/2}\). 

\begin{figure}[H]
    \centering
    \includegraphics[width=0.45\linewidth]{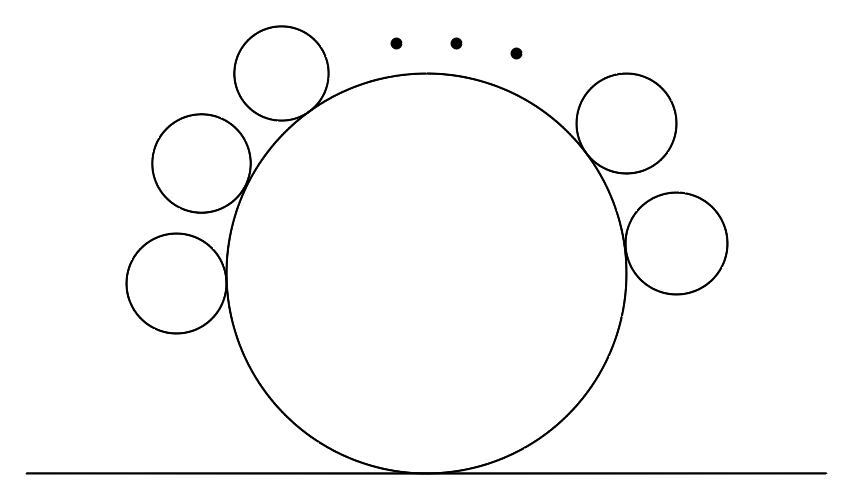}
    \caption{The \(N\)-loop ring diagram contribution to the self-energy of order \(\lambda^N\).}
    \label{ring_2}
\end{figure}

\begin{figure}[H]
    \centering
    \includegraphics[width=0.75\linewidth]{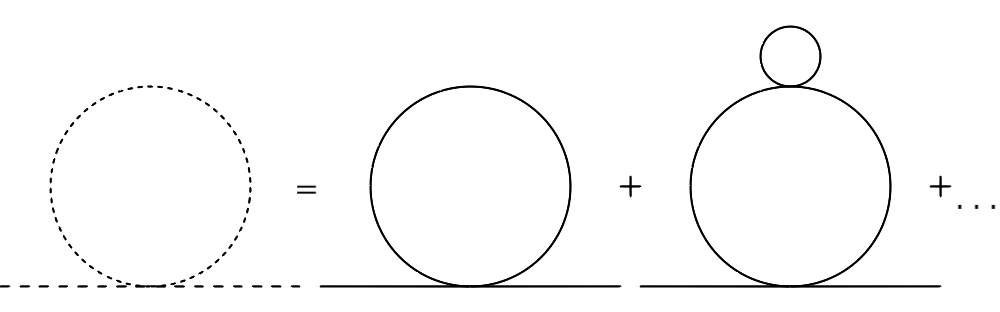}
    \caption{Resummed self-energy. The dotted line represents the full propagator (\ref{full_D_1}) and the solid line represents the free propagator.}
    \label{ring_3}
\end{figure}

The discussion below is based on Ref. \cite{Carrington:1991hz} and a scalar field theory is initially considered with Lagrangian density,
\begin{equation}\label{usual_scalar}
    \mathcal{L} = \frac{1}{2} \partial_{\mu} \phi \partial^{\mu} \phi  + \frac{1}{2} \mu^2 \phi^2 - \frac{\lambda}{4} \phi^4 \,
\end{equation}
The temperature-dependent self-energy, denoted as \(\pi (\omega_n, \vec{p})\), is introduced through the inverse of the full propagator at finite temperature,
\begin{equation}\label{full_D_1}
    D^{-1} (\omega_n, \vec{p}) = \omega_n^2 + \vec{p}^{\, 2} + m^2 + \pi (\omega_n, \vec{p}) 
\end{equation}
where the self-energy satisfies a Dyson equation \cite{Carrington:1991hz}. First, the one-loop self-energy is computed at finite temperature in the infrared limit: \(\omega_n = 0\) and \(\vec{p} \to 0\), where \(p = (i \omega_n, \Vec{p}\, )\) is the external momentum. If we consider a massless scalar field, at order \(\lambda^N\) the infrared-divergent diagram in Fig. \ref{ring_2} is the most important contribution to the self-energy in the infrared limit and the one-loop self-energy is given by
\begin{equation}\label{int_11}
    \pi^{(1)} (\omega_n, \vec{k}) = \pi^{(1)} (0) = 3 \lambda T \sum_{n = - \infty}^{\infty} \int \frac{d^3 p}{(2\pi)^3} \, \frac{1}{\omega_n^2 + \vec{p}^{\, 2}} = \frac{\lambda T^2}{4}\,  ,
\end{equation}
where \(\pi^{(1)} (\omega_n = 0, \vec{p} \to 0) \equiv \pi^{(1)} (0) \) and \(\omega_n\) are the bosonic Matsubara frequencies.
A detailed calculation of the integral (\ref{int_11}) can be found in Refs. \cite{Kapusta:2006pm, Bellac:2011kqa, Laine:2016hma, Dolan:1973qd,
Fendley:1987ef}, where this integral is often calculated using
\begin{equation}
    \sum_{n = - \infty}^{\infty} \frac{1}{n^2 + y^2} = \frac{\pi}{y} \coth{\pi y} = \frac{\pi}{y} \left( 1 + \frac{2}{e^{2\pi y} - 1} \right)
\end{equation}
and the integral formula
\begin{equation}
    \int_{0}^{\infty} d x \, \frac{x}{e^x -1} = \frac{\pi^2}{6}\, .
\end{equation}
The integral (\ref{int_11}) will be alternatively computed later as the derivative of the thermal function (\ref{bosonthermalfunction_1}). The final expression does not depend on momentum and one sums over \(N\) explicitly. Thus, the self-energy is expressed as a function of the propagator with an effective mass squared \(\pi^{(1)} (0)\), and the most important contribution to the ring diagrams in Fig. \ref{ring_1} is obtained self-consistently in the infrared limit by,
\begin{equation}
    \pi (\omega_n, \vec{p}) = \pi (0) = 3 \lambda T \sum_{n = - \infty}^{\infty} \int \frac{d^3 p}{(2\pi)^3} \, \frac{1}{\omega_n^2 + \vec{p}^{\, 2} + \pi^{(1)} (0)} \, .
\end{equation}
In the case of a massive scalar field, the temperature-dependent self-energy is expressed similarly as a function of the propagator with an effective mass squared \(m^2 (\phi_c) + \pi^{(1)} (0)\). This is depicted in Fig. \ref{ring_3} and the one-loop self-energy can be written as
\begin{equation}
    \pi^{(1)} (\omega_n, \vec{k}) = \pi^{(1)} (0) = 3 \lambda T \sum_{n = - \infty}^{\infty} \int \frac{d^3 p}{(2\pi)^3} \, \frac{1}{\omega_n^2 + \vec{p}^{\, 2} + m^2(\phi_c)} \, ,
\end{equation}
which is a very similar integral to (\ref{int_11}) and yields to
\begin{equation}\label{p10}
     \pi^{(1)} (0) = \frac{\lambda T^2}{4} \left[1 + \mathcal{O} \left( \frac{m (\phi_c)}{T} \right) \right] \, .
\end{equation}
After the summation over \(N\), the temperature self-energy is computed as
\begin{equation}\label{p0}
    \pi (\omega_n, \vec{k}) = \pi (0) = 3 \lambda T \sum_{n = - \infty}^{\infty} \int \frac{d^3 p}{(2\pi)^3} \, \frac{1}{\omega_n^2 + \vec{p}^{\, 2} + m^2(\phi_c) + \pi^{(1)} (0)} \, ,
\end{equation}
This expression can be expanded in powers of \(m/T\) substituting Eq. (\ref{p10}) and omitting the higher order terms that leads to
\begin{equation}\label{final_p0}
    \pi (0) = \frac{\lambda T^2}{4} \left[1 + \mathcal{O} \left( \frac{m (\phi_c)}{T} \right) \right] \, . \, .
\end{equation}
The next-to-leading order terms in (\ref{p10}) contribute significantly if \(m (\phi_c) \sim T\), but then the one-loop self-energy is \(\pi^{(1)} (0) \sim \lambda m^2 (\phi_c) < m^2 (\phi_c)\), given the fact that \(\lambda < 1\). As a result, it is sufficient to omit the higher order corrections in (\ref{p10}) as they do not contribute to the self-energy (\ref{p0}). 

The inverse  of the full propagator (\ref{full_D_1}) can be finally written using the temperature-dependent self-energy (\ref{final_p0}) as
\begin{equation}\label{DD11}
    D^{-1} (\omega_n , \vec{p}) = \omega_n^2 + \vec{p}^{\, 2} + m^2(\phi_c) + \frac{\lambda T^2}{4} \left[1 + \mathcal{O} \left( \frac{m (\phi_c)}{T} \right) \right]
\end{equation}
The higher-order terms are important in (\ref{DD11}) when \(m(\phi_c) \sim T\) leading to \(\pi(0) \sim \lambda m^2 (\phi_c)\) which is small compared to the effective mass squared. Subsequently, at \(m(\phi_c) \sim T\), the self-energy does not contribute to the full propagator and the finite-temperature one-loop effective potential is derived following the same method as (\ref{initial_V1}) and the ring corrections do not contribute to the effective potential. Namely, the zeroth order in the self-energy (\ref{final_p0}) can be applied extensively to compute the ring diagram contributions because this approximation is sufficient in the regime in which the ring diagram contributions do not vanish. Therefore, the corrections in (\ref{DD11}) are omitted and the inverse of the full propagator at finite temperature is expressed as
\begin{equation}\label{D_before}
     D^{-1} (\omega_n , \vec{p}) = \omega_n^2 + \vec{p}^{\, 2} + m^2(\phi_c) + \frac{\lambda T^2}{4} \, ,
\end{equation}
and this is rewritten as
\begin{equation} \label{D_11AA}
     D^{-1} (\omega_n , \vec{p}) =  \left( \omega_n^2 + \omega^2 \right) \left( 1 + \frac{\pi(0)}{\omega_n^2 + \omega^2} \right) \, ,
\end{equation}
where \(\omega^2 = \vec{p}^{\, 2} + m^2(\phi_c)\). Then, we insert inverse of the full propagator in the one-loop effective potential (\ref{initial_V1}) and the full finite-temperature effective potential (\ref{decom_Veff}) generally reads
\begin{equation}\label{VE}
    V_{\text{eff}} (\phi_c, T) = V_0 (\phi_c) + \frac{1}{2 \beta} \sum_{n = - \infty}^{\infty} \int \frac{d^3 p}{(2\pi)^3} \, \ln \left(\omega_n^2 + \vec{p}^{\, 2} + m^2(\phi_c) + \pi(0)\right)   \, ,
\end{equation}
where the first term is the tree-level contribution. The temperature-dependent part of the effective potential is included in the second term which improves the previous result (\ref{V_scalar_1_T}) using the thermal mass (\ref{def_thermal_mass}),
\begin{equation}\label{T-dependent_V}
    V_{1}^{T} (\phi_c, T) = \frac{T^4}{2 \pi^2} \int_{0}^{\infty} dx \, x^2 \ln \left( 1 - e^{  - \sqrt{x^2 + \frac{M^2  (\phi_c)}{T^2}}} \right) = \frac{T^4}{2 \pi^2} J_B \left(\frac{M^2 (\phi_c)}{T^2}\right) \, ,
\end{equation}
where \(\Pi (\phi_c, T) \equiv \pi (0)\) is the temperature-dependent self-energy, and the thermal mass is written as
\begin{equation}\label{thermal mass scalar}
    M^2 (\phi_c, T) = m^2 (\phi_c) + \frac{\lambda T^2}{4}  \,.
\end{equation}
Alternatively, we insert Eq. (\ref{D_11AA}) in (\ref{initial_V1}) and the second term in (\ref{VE}) can be decomposed into the logarithm in (\ref{initial_V1}) and a second logarithm which involves the ring corrections in the effective potential. Subsequently, the effective potential (\ref{VE}) can be cast into the form
\begin{equation}\label{fin_V}
    V_{\text{eff}} (\phi_c, T) = V_0 (\phi_c) + V_1 (\phi_c, T) - \frac{1}{2 \beta} \sum_{n = -\infty}^{\infty} \int \frac{d^3 p}{(2 \pi)^3} \, \sum_{N = 1}^{\infty} \frac{1}{N} \left( -\frac{\pi(0)}{\omega_n^2 + \omega^2}\right)^N \, ,
\end{equation}
where the second logarithm was expanded in the Taylor series to obtain the third term in (\ref{fin_V}) which includes the ring corrections shown in Fig. \ref{ring_1} and is given by
\begin{equation}
    V_{r} (\phi_c, T) = - \frac{1}{2 \beta} \sum_{n = -\infty}^{\infty} \int \frac{d^3 p}{(2 \pi)^3} \, \sum_{N = 1}^{\infty} \frac{1}{N} \left( -\frac{\pi(0)}{\omega_n^2 + \omega^2}\right)^N \, .
\end{equation}
It is obvious that in the low-temperature limit, the ring corrections are small, which implies that it is sufficient to consider the high-temperature (or massless) limit in the calculation of the temperature-dependent self-energy.

Lastly, it is important to mention that the temperature-dependent self-energy can be also computed by the second derivatives of the high-temperature expansion of the temperature-dependent term in the effective potential as it is demonstrated in the next paragraph \cite{Dolan:1973qd, Altherr:1989rk, Bellac:2011kqa}.

\subsection{What about the higher-loop diagrams?}
\begin{figure}[H]
    \centering
    \includegraphics[width=0.25\linewidth]{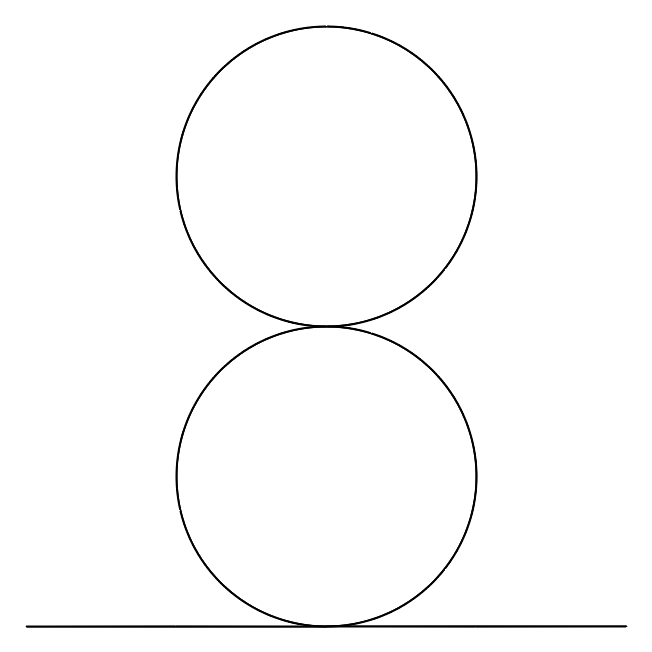}
    \caption{The two-loop diagram which consists of one 1-VB and one 2-VB.}
    \label{figure_8}
\end{figure}

\begin{figure}[H]
    \centering
    \includegraphics[width=0.35\linewidth]{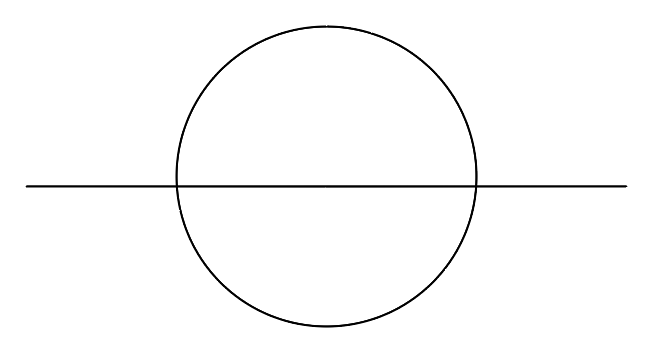}
    \caption{The sunset diagram is a two-loop contribution that is subleading in temperature.}
    \label{sunset}
\end{figure}

What about the higher-loop diagrams that were not computed in this thermal resummation? We will address this question by investigating the high-temperature behavior of some higher-loop diagrams in the scalar field theory \cite{Curtin:2016urg, Parwani:1991gq, Arnold:1992rz, Espinosa:1992gq, Fendley:1987ef, Dolan:1973qd}. 
\begin{figure}[H]
    \centering
    \includegraphics[width=0.35\linewidth]{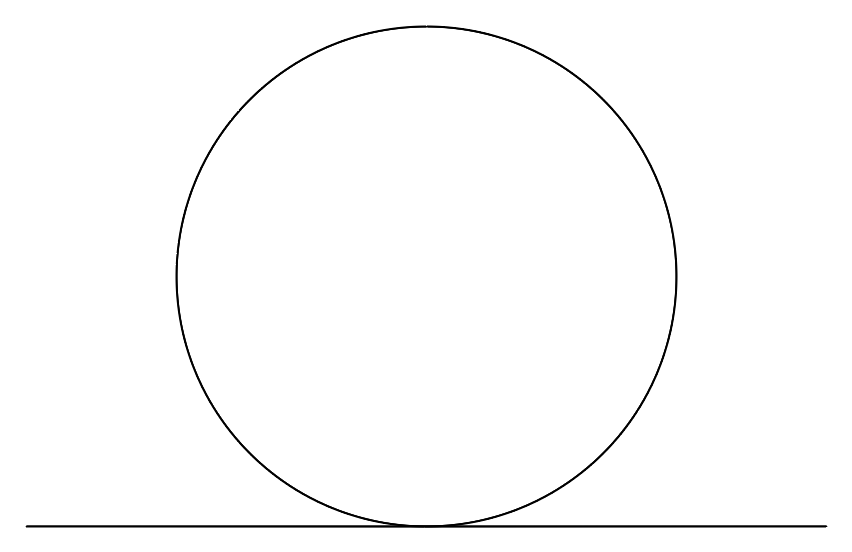}
    \caption{One-loop self-energy diagram.}
    \label{one-loop self}
\end{figure}
First of all, the one-loop contribution to the temperature-dependent self-energy is depicted in Fig. \ref{one-loop self}
and is computed as
\begin{equation}
    \pi^{(1)} (0) = 3 \lambda T \sum_{n = - \infty}^{\infty} \int \frac{d^3 p}{(2\pi)^3} \, \frac{1}{\omega_n^2 + \vec{p}^{\, 2} + m^2(\phi_c)} = 3 \lambda I (m^2)  \, ,
\end{equation}
where the mass counterterm \(\delta m^2\) has canceled the zero-temperature contribution in the integral to lead to the last temperature-dependent expression \cite{Dolan:1973qd} and the function \(I (m^2)\) is given by \cite{Dolan:1973qd, Altherr:1989rk, Senaha:2020mop, Bellac:2011kqa}
\begin{equation}
    I (m^2) = \frac{T^2}{\pi^2}\frac{\partial J_B (y^2)}{\partial y^2} = \frac{T^2}{12} - \frac{T m}{4 \pi} + ...
\end{equation}
with \(y^2 = \beta^2 m^2 \). Now, we consider the two different two-loop diagrams in Fig. \ref{figure_8} and \ref{sunset}. The diagram in Fig. \ref{figure_8} consists of one 1-vertex bubble (VB) and one 2-VB. The 1-VB behaves like \(\sim \lambda T^2\) which is estimated from the one-loop self-energy and the 2-VB is computed as
\begin{equation}\label{one_loop_con}
    6 \lambda T \sum_{n = - \infty}^{\infty} \int \frac{d^3 p}{(2\pi)^3} \, \frac{1}{\left(\omega_n^2 + \vec{p}^{\, 2} + m^2 \right)^2} = - 6 \lambda \frac{\partial I (m^2)}{\partial m^2} \sim \frac{\lambda T}{m}
\end{equation}
In total, this two-loop diagram amounts to
\begin{equation}\label{fig_8_total}
    \sim \lambda T^2 \left(\frac{\lambda T}{m} \right)
\end{equation}
However, the so-called sunset diagram in Fig. \ref{sunset} is estimated in the high-temperature approximation as \cite{Arnold:1992rz, Espinosa:1992gq, Dolan:1973qd}
\begin{equation}\label{sunset_con}
    \sim \lambda^2 T^2 \ln \frac{m}{T} \, .
\end{equation}
In comparison, the sunset diagram is suppressed with respect to the two-loop diagram in Fig. \ref{figure_8} at the high-temperature limit. 

\begin{figure}[H]
    \centering
    \includegraphics[width=0.25\linewidth]{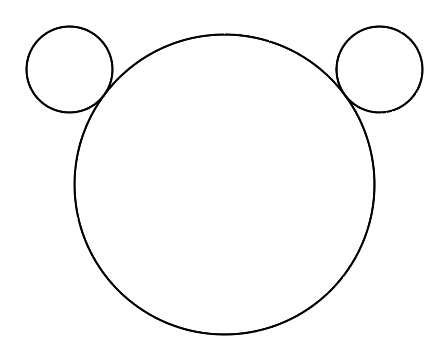}
    \caption{Mouse diagram.}
    \label{mouse}
\end{figure}

\begin{figure}[H]
    \centering
    \includegraphics[width=0.25\linewidth]{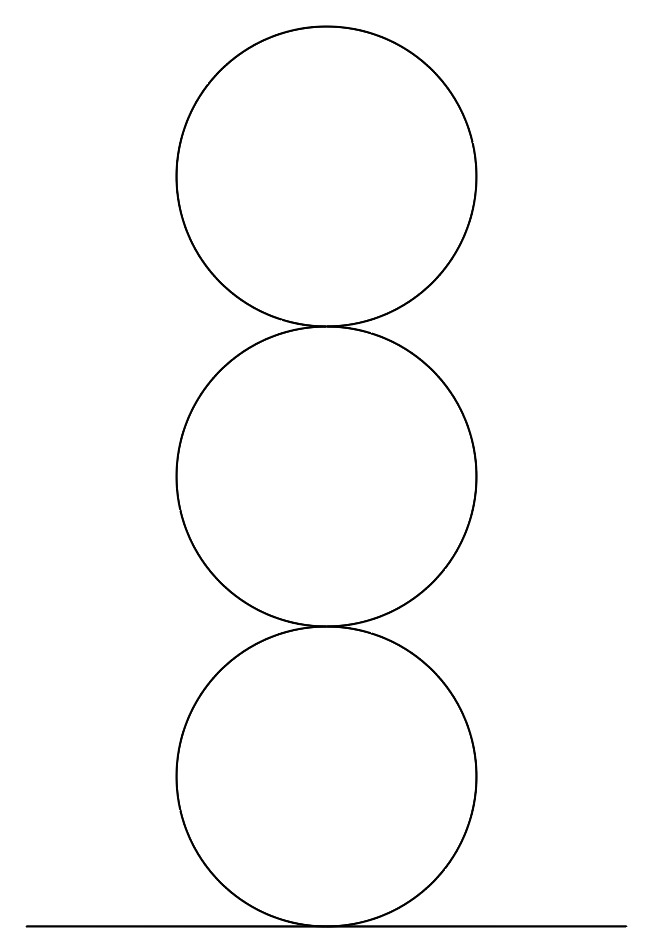}
    \caption{Cactus diagram.}
    \label{Cactus}
\end{figure}

Moving on, a higher-loop ring diagram is constructed by attaching one more 1-VB to the lower loop in the two-loop diagram Fig. \ref{figure_8}. The new contribution is computed as \cite{Dolan:1973qd, Arnold:1992rz}
\begin{equation}
     6 \lambda T \sum_{n = - \infty}^{\infty} \int \frac{d^3 p}{(2\pi)^3} \, \frac{1}{\left(\omega_n^2 + \vec{p}^{\, 2} + m^2 \right)^3} = 3 \lambda \frac{\partial^2 I(m^2)}{\partial (m^2)^2} \sim \frac{\lambda T}{m^3} 
\end{equation}
and in total the diagram in Fig. \ref{mouse}, which is called the mouse diagram, amounts for 
\begin{equation}\label{mouse_con}
    \sim (\lambda T^2)^2 \left( \frac{\lambda T}{m^3} \right) \, .
\end{equation}
On the other hand, one additional 1-VB could be also attached to the upper bubble of the two-loop diagram \ref{figure_8} which leads to the three-loop diagram, known as the cactus diagram \ref{Cactus}. In view of (\ref{one_loop_con}) and (\ref{fig_8_total}), the cactus diagram behaves like \cite{Dolan:1973qd}
\begin{equation}\label{cactus_con}
    \sim \lambda T^2 \left(\frac{\lambda T}{m} \right)^2 \, .
\end{equation}
Therefore, the mouse diagram has the dominant contribution at the high-temperature approximation compared to the cactus diagram since the ratio of the contributions (\ref{mouse_con}) and (\ref{cactus_con}) is \(T/m\). One could also draw a three-loop diagram by attaching one 1-VB to the sunset diagram, but such a diagram is less important than the mouse diagram in Fig. \ref{mouse}. Namely, a ring diagram could be constructed by attaching one 1-VB to the lower bubble in the two-loop diagram in Fig. \ref{figure_8}. As a result, the ring diagram in Fig. \ref{ring_2} which consists of (\(n-1\)) 1-VB and one n-VB can be 
estimated as
\begin{equation}
    \sim \frac{\lambda^2 T^3}{m} \left( \frac{\lambda T^2}{m^2} \right)^{n-2} \, .
\end{equation}
Namely, each additional 1-VB in a ring diagram contributes a factor of
\begin{equation}
    \alpha \equiv \frac{\lambda T^2}{m^2} \, .
\end{equation}
However, in the one-loop approximation, one observes in the previous section in (\ref{Veff_first}) and (\ref{Veff_first}) that the symmetry is restored at \(T_c \simeq m/\sqrt{\lambda}\) causing the breakdown of the perturbative expansion due to \(\alpha \simeq 1\). Hence, thermal resummation is necessary to resum all powers of \(\alpha\) and include all the relevant diagrams. 

Moreover, the thermal mass essentially incorporates the ring diagram contributions to all orders in the effective potential. Nevertheless, some important diagrams are not included in this thermal resummation such as the sunset and the cactus diagrams. Consequently, the reliability of the improved perturbative expansion is guaranteed by the condition
\begin{equation}
    \beta \equiv \frac{\lambda T}{m} \ll 1
\end{equation}
and the requirement for the ordinary one-loop approximation
\begin{equation}
    \lambda \ll 1 \, .
\end{equation}
Namely, as it was shown in (\ref{sunset_con}) and (\ref{cactus_con}), the non-ring diagram contributions to the self-energy are suppressed with respect to ring diagram contributions by \(\mathcal{O} (\beta)\), while these diagrams contribute to the effective potential corrections which are suppressed by \(\mathcal{O} (\beta^2)\) \cite{Espinosa:1992gq}.

Lastly, the previous results for scalar fields can be generalized similarly or directly to gauge fields. The main difference between the scalar and the gauge field theories in the framework of thermal resummation is based on the fact that thermal contributions to the transverse mode of the gauge bosons are suppressed by the gauge symmetry. In contrast, this resummation is not applied to the fermion fields since only the bosonic Matsubara frequencies can vanish and generate the infrared divergences, whereas the fermionic Matsubara frequencies, \(\omega_n = (2 n+ 1) \pi T\), do not vanish. The thermal resummation in gauge theories will be presented in a later chapter in the context of the Standard Model.

\subsection{Resummation Scheme}

In the previous paragraphs, thermal resummation mainly consisted of replacing the effective mass-squared by the thermal mass. This replacement in the full temperature-dependent effective potential is known as the Parwani scheme \cite{Parwani:1991gq} which was also adopted in the effective potential (\ref{VE}) and (\ref{T-dependent_V}). This scheme can be generally described by writing the effective potential as
\begin{equation}\label{Parwani}
    V_{\text{eff}} (\phi_c,T) = V_0 (\phi_c) + \sum_{i} \left[ V^i_1 \left(m^2_i(\phi_c) + \Pi_i(T)\right) + V^i_T \left(m^2_i(\phi_c)+ \Pi_i(T),T \right) \right]
\end{equation}
where \(i\) counts the fields of the theory. The first term represents the tree-level potential, while the second and the third terms correspond to the one-loop contributions in the effective potential at zero temperature and finite temperature, respectively. In the case of the scalar field theory, (\ref{usual_scalar}), the one-loop finite-temperature effective potential in the high-temperature approximation reads
\begin{equation*}
    V_{\text{eff}} (\phi_c, T)  =  - \frac{\mu^{2}}{2} \phi_c^2 + \frac{\lambda}{4} \phi_c^4 + \frac{m^2 (\phi_c)}{24}T^2 - \frac{T}{12 \pi} \left[M^2 (\phi_c,T) \right]^{3/2}
    + \frac{M^4(\phi_c, T)}{64 \pi^2}\left( \ln \frac{a_b T^2}{\mu^2_r} -\frac{3}{2} \right)
\end{equation*}
where \(M(\phi_c, T)\) is the thermal mass defined in (\ref{def_thermal_mass}) and the field-independent terms are neglected. One notices that the effective mass-squared in the logarithm in the high-temperature expansion is canceled by the same logarithm in the one-loop effective potential at zero temperature. Additionally, it is important to mention that the ultraviolet divergence in the one-loop zero-temperature effective potential requires temperature-dependent counterterms since this contribution becomes temperature-dependent due to the presence of the self-energy.

A similar method, developed by Arnold and Espinosa \cite{Arnold:1992rz}, resums only the zero Matsubara modes that cause the infrared divergence. In the Arnold-Espinosa scheme, the resummed full effective potential is written as follows,
\begin{equation}\label{Arnold-Espinosa}
     V_{\text{eff}} (\phi,T) = V_0 (h,\phi) + \sum_{i} \left[ V^i_1 \left(m^2_i(\phi)\right) + V^i_T \left(m^2_i(\phi),T \right) + V^i_{ring} \left(m^2_i(\phi),T\right) \right]
\end{equation}
The last term is added to incorporate the thermal resummation. In particular, this scheme is based on resumming the infrared divergent contributions only to the Matsubara zero-mode propagator. In this scheme, the higher-order corrections due to the ring diagrams are included only in the resummation of the zero Matsubara mode of the propagator. It is additionally argued that the Arnold-Espinosa scheme is more natural in the case of the gauge fields \cite{Arnold:1992rz}. For instance, in this scheme, the second term in the effective potential (\ref{VE}) is expressed differently as
\begin{equation}
\begin{split}
     V_{1 + ring} (\phi_c, T) = & \frac{T}{2} \sum_{n = - \infty}^{\infty} {}^{\prime} \int \frac{d^3 p}{(2 \pi)^3} \, \ln \left( \omega_n^2 + \vec{p}^{\, 2} + m^2 (\phi_c) \right) \\
     & + \frac{T}{2} \int \frac{d^3 p}{(2\pi)^3} \, \ln \left(\vec{p}^{\, 2} + M^2(\phi_c, T) \right) \, ,
\end{split} 
\end{equation}
where the prime in front of the frequency sum means that the value \(n = 0\) is excluded which implies the zero Matsubara frequency is not included in the summation. Then, this expression can be rewritten as
\begin{equation}
\begin{split}
      V_{1 + ring} (\phi_c, T) = & \frac{T}{2} \sum_{n = - \infty}^{\infty} \int \frac{d^3 p}{(2 \pi)^3} \, \ln \left( \omega_n^2 + \vec{p}^{\, 2} + m^2 (\phi_c) \right) \\
     & -  \frac{T}{2} \int \frac{d^3 p}{(2\pi)^3} \, \ln \left(\vec{p}^{\, 2} + m^2(\phi_c) \right) + \frac{T}{2} \int \frac{d^3 p}{(2\pi)^3} \, \ln \left(\vec{p}^{\, 2} + M^2(\phi_c, T) \right) ,
\end{split}
\end{equation}
where the first line shows the unresummed one-loop effective potential (\ref{initial_V1}) and the terms in the second line correspond to the ring corrections in the effective potential. Therefore, the integrals in the ring corrections are computed using the formula (\ref{useful_formula}) as
\begin{equation}
\begin{split}
    V_{ring} (\phi_c,T) & = \frac{T}{2} \int \frac{d^3 p}{(2\pi)^3} \, \ln \left[\frac{\vec{p}^{\, 2} + M^2(\phi_c, T)}{\vec{p}^{\, 2} + m^2(\phi_c)} \right] \\
    &= \frac{T}{2} \int \frac{d^3 p}{(2\pi)^3} \, \ln \left( 1 + \frac{\pi (0)}{\vec{p}^{\, 2} + m^2(\phi_c)} \right) \\
    & = \frac{T}{4 \pi^2} \int_{0}^{\infty} d x \, x^2 \ln \left( 1 + \frac{\pi (0)}{x^2 + m^2(\phi_c)}\right).
\end{split}
\end{equation}
After integration, the ring corrections in the effective potential read
\begin{equation}\label{rings}
    V_{ring} \left(\phi_c,T \right) = \frac{T}{12\pi} \left[m^3 (\phi_c) - \left(m^2(\phi_c) + \pi (0) \right)^{3/2} \right],
\end{equation}
where an infinite constant was omitted. Therefore, in the high-temperature limit, the full one-loop effective potential in the Arnold-Espinosa scheme can be written as
\begin{equation}\label{ring_correction_1}
    V_{\text{eff}} (\phi_c, T)  = V_0 (\phi_c)+ \frac{m^2 (\phi_c)}{24}T^2 - \frac{T}{12 \pi} \left[M^2 (\phi_c,T) \right]^{3/2}
    + \frac{m^4(\phi_c)}{64 \pi^2}\left( \ln \frac{a_b T^2}{\mu^2_r} -\frac{3}{2} \right),
\end{equation}
where the field-independent terms are neglected. The Arnold-Espinosa scheme differs from the Parwani scheme in the last term in the high-temperature limit. As a result, this scheme is equivalent to the Parwani scheme in the high-temperature or massless limit as the logarithmic term is very small at high temperatures or low masses compared to the other terms.

Moreover, one observes the cancellation of the cubic term \(m^3(\phi_c)\) in the unresummed temperature-dependent part with the \(m^3 (\phi_c)\) term in the ring correction (\ref{ring_correction_1}). This demonstrates precisely the importance of ring diagrams which have contributions of order \(\lambda^{3/2}\) and these contributions are important in the high-temperature approximation. These higher-loop ring diagrams due to the infrared divergences are generated by long-range fluctuations at finite temperatures \cite{Carrington:1991hz}. 

The ring diagram contributions are also crucial since the cubic term in the effective mass controls the strength of the first-order phase transition in the scalar field theory. This argument is also valid in the Standard Model, as it was clearly demonstrated in the previous section in which the presence of a cubic term in the ordinary perturbative expansion resulted in a first-order phase transition. This is also shown explicitly in Chapter \ref{Electroweak Phase Transition}.

\chapter{Cosmological Phase Transitions}\label{Cosmological Phase Transitions}

\section{Introduction}

 In the previous chapter, the finite-temperature effective potential showed that a spontaneously broken gauge symmetry, such as the gauge symmetry \(SU(2)_L \times U(1)_Y\), can be restored at high temperatures. In the early Universe, the vacuum state is described by a zero Higgs vacuum expectation value with an unbroken gauge symmetry in the Standard Model. However, a second local minimum appears in the effective potential, as the temperature decreases and below some critical temperature, the non-zero minimum is more energetically favored and stable. As a consequence, a phase transition occurs from the symmetric phase with \(\langle \hat{\phi}\rangle = 0\) to the broken phase with \(\langle \hat{\phi}\rangle \ne 0\). In particular, any phase transition is a process that involves a symmetry that is spontaneously broken before or after the phase transition \cite{Linde:1978px}. This relation between spontaneous symmetry breaking and phase transitions in quantum field theory is also apparent in condensed-matter physics which describes numerous phenomena such as ferromagnetism and superconductivity.

In general, a phase transition is associated with an order parameter that distinguishes the different phases of the system. More specifically, the order parameter may be identified as the vacuum expectation value of an order parameter operator which could vanish in one of the phases. In quantum field theory, the order parameter operator is usually an elementary or composite field operator. This operator is often defined as a scalar field operator \(\hat{\phi}(x)\) which is related to the mechanism of the phase transition. Moreover, phase transitions can be classified into two kinds: the first-order and the second-order phase transitions. In the first kind, the order parameter is a discontinuous function of temperature\footnote{The phase transition can be driven by various parameters such as pressure, temperature, magnetic field, etc, but we concentrate on temperature as the driving parameter.} between the two phases, whereas the order parameter is continuous during a second-order phase transition. For instance, in the context of the electroweak phase transition, the order parameter, denoted as \(\upsilon (T)\), is defined as the vacuum expectation value of the Higgs boson which corresponds to the global minimum of the finite-temperature effective potential. During a first-order electroweak phase transition, the order parameter, which is assumed to be non-negative without loss of generality, reads
\begin{equation}\label{VEV_DEF}
    \upsilon (T) = 
     \begin{cases}
       0, &\quad T>T_c,\\
       f(T), &\quad 0 \leq T \leq T_c,\\
     \end{cases}
\end{equation}
where \(f (T)\) is an arbitrary continuous function, considering \(f(0) = \upsilon\) and \(f(T_c) = \upsilon_c \neq 0\). This order parameter is clearly a discontinuous function at \(T = T_c\) as was also demonstrated by the Standard Model effective potential in the previous chapter.

\begin{figure}[h!]
    \centering
    \includegraphics[width=0.70\linewidth]{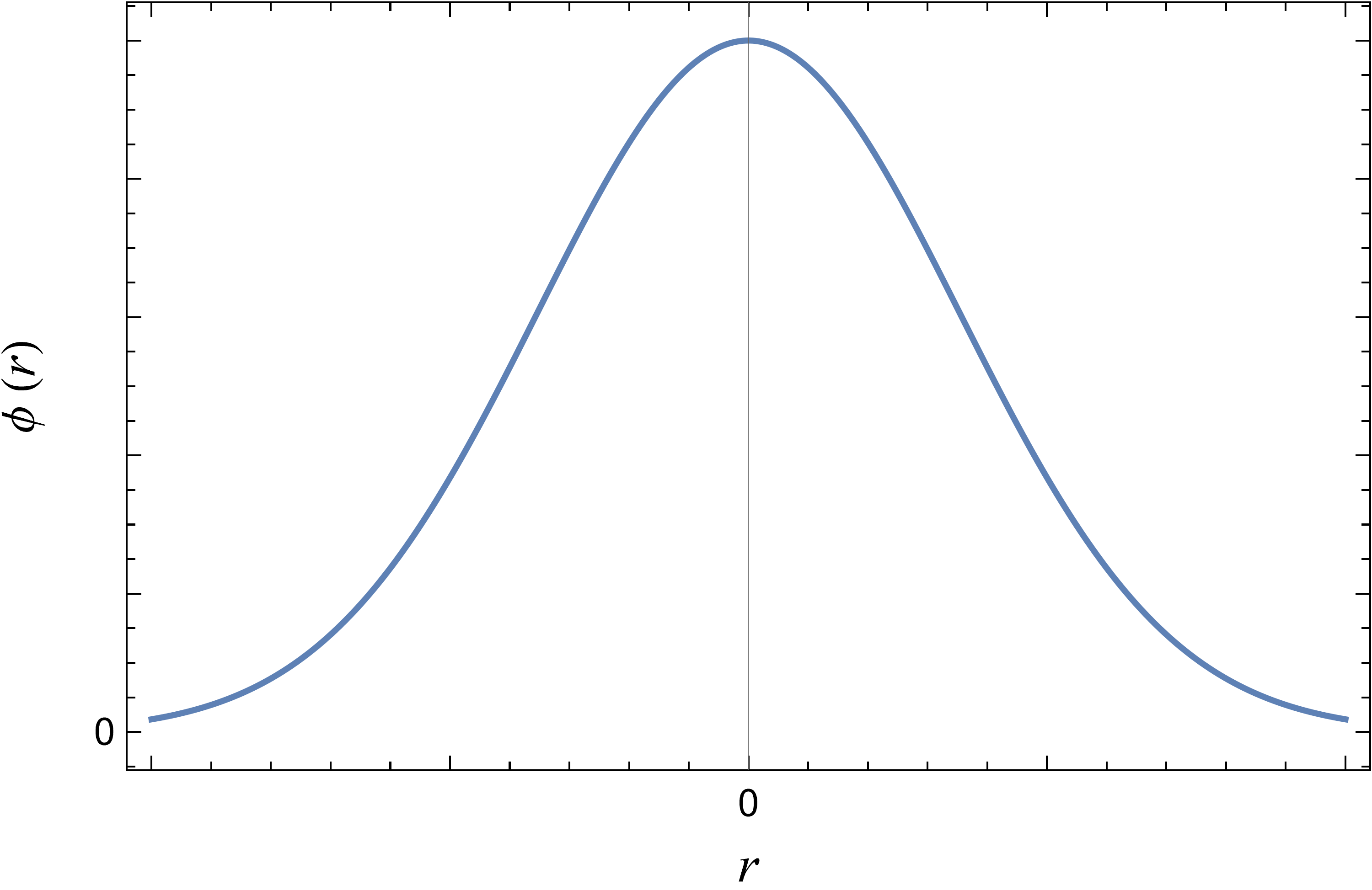}
    \caption{Bubble configuration.}
    \label{Bubble Conf}
\end{figure}

It was shown earlier that a potential barrier appears in a first-order phase transition. In quantum theory, a system could tunnel through any barrier between different vacuum states. At zero temperature, this tunneling is caused by quantum fluctuations, but thermal fluctuations are also present at finite temperatures. As a result, a phase transition from the false vacuum to the true vacuum proceeds by thermal tunneling which causes the formation of the so-called bubbles of the broken phase, analogous to the bubbles in boiling water. An arbitrary bubble configuration is illustrated in Fig. \ref{Bubble Conf}. In other words, phase transitions can occur anywhere in spacetime generating these bubbles. Then, the bubbles could spread throughout the Universe to convert the false vacuum into the true one. Nevertheless, the transition may not be completed since the expanding bubbles are impeded by the barriers between states, by the expansion of the Universe, and by interactions with the surrounding plasma of particles. In the next sections, we will study in detail the dynamics of phase transitions to explain further the concepts of transition rates and bubble nucleation in thermal quantum field theory and cosmology.

\section{Tunneling Rate}

In this section, οur main goal is to compute the transition probability per unit time per unit volume at finite temperatures \cite{Linde:1977mm, Linde:1980tt, Linde:1981zj}. This transition probability was first computed at zero temperature \cite{Coleman:1977py, Kobzarev:1974cp, Callan:1977pt}. In the context of quantum mechanics, it is well known that at zero temperature the tunneling amplitude for a one-dimensional system in the WKB approximation is written as \cite{Devoto:2022qen}
\begin{equation}
    A \exp \left(- \int_{q_1}^{q_2} dq\, \sqrt{2 \left(V(q) - E\right)} \right),
\end{equation}
where \(A\) is a constant prefactor,  \(V(q)\) is the potential and \(V(q_1) = V(q_2) = E\). The tunneling rate is the square of this amplitude. In general, the transition probability per unit time per unit volume can be cast into the form,
\begin{equation}
    p (t) \equiv \frac{\Gamma}{V} = |A| e^{-B}.
\end{equation}
At finite temperatures, the transition probability per unit time per unit volume between two different minima can be computed by applying the finite-temperature field theory \cite{Linde:1977mm, Linde:1980tt, Linde:1981zj}. Schematically, it can be defined as
\begin{equation}
    p (t;T) \equiv \frac{\Gamma}{V} = |A(T)| e^{-B(T)},
\end{equation}
where the factor \(A(T)\) depends on both the tunneling through the barrier and the fluctuations over the barrier at finite temperatures and the exponent can be written as \(B(T) = C(T)/T\). The exponent \(C(T)\) can be determined by the known bounce equation which is a non-linear ordinary equation of motion. A recent review on tunneling rates and cosmological phase transitions can be found in Ref. \cite{Athron:2023xlk}. A more pedagogical approach is adopted in Ref. \cite{Devoto:2022qen} which discusses in detail the tunneling phenomena in quantum mechanics and quantum field theory. A shorter analysis is presented in Refs. \cite{Sher:1988mj, Brandenberger:1984cz}, as well.

In the history of the Universe, a first-order phase transition could potentially occur only within a limited finite temperature interval, since the effective potential varies with temperature. During the progress of the phase transition, it is assumed that only two phases coexist across that entire temperature interval. 

In the previous chapter, during a first-order phase transition, the critical temperature \(T_c\) was defined as the temperature at which the local minima of the effective potential are degenerate. Nevertheless, if the barrier is high enough the tunneling phenomenon is suppressed and the phase transition effectively starts at a temperature lower than the temperature \(T_c\). Therefore, the phase transition which proceeds via thermal tunneling with the formation of bubbles of the broken phase starts at a critical temperature, denoted as \(T_n\), which is \(T_c > T_n > T_0\). In this chapter, it will be shown that in the electroweak phase transition the thermal tunneling proceeds sufficiently fast to fill the Universe with bubbles of the true vacuum only at the nucleation temperature \(T_n\). At this temperature, the corresponding Euclidean action \(S_E = S_3/T\), which suppresses the tunneling rate, becomes of the order \(\mathcal{O} (130 - 140)\) \cite{Dine:1991ck, McLerran:1990zh, Anderson:1991zb}. Initially, we introduce the three-dimensional Euclidean action, while the tunneling probability per unit time per unit volume at finite temperature is expressed as
\begin{equation}\label{bubble_rate}
    p (T) \simeq A(T) e^{-\frac{S_3}{T}},
\end{equation}
where the prefactor \(A(T)\) can be considered approximately as of order \(\mathcal{O}(T^4)\) and the three-dimensional Euclidean action \(S_3\) can be defined as
\begin{equation}\label{defS3}
    S_3 = \int d^3 x\, \left[ \frac{1}{2} \left(\nabla \phi \right)^2 + V_{\text{eff}} (\phi,T) \right].
\end{equation}
The finite-temperature effective potential appears in three-dimensional action (\ref{defS3}) to include the quantum corrections to the classical equations of motion \cite{Brandenberger:1984cz, Linde:1981zj}. In other words, the effective potential can be replaced by the classical potential to neglect the higher-order corrections\footnote{The corrections to the kinetic terms are omitted since they do not contribute significantly to the calculation of the tunneling probability per unit time per unit volume \cite{Linde:1981zj}.}. In fact, the solution of the bounce equation possesses a \(O(3)\) symmetry at very high temperatures \cite{Linde:1977mm, Linde:1980tt, Linde:1981zj} and the Euclidean action can be expressed as
\begin{equation}
    S_3 = 4 \pi \int_{0}^{\infty} r^2 dr\, \left[\frac{1}{2} \left(\frac{d \phi}{d r}\right)^2 + V_{\text{eff}}(\phi(r), T)  \right] \,,
\end{equation}
where \(r^2 = \vec{x}^{\, 2}\). As a result, the Euclidean equation of motion can be easily derived,
\begin{equation}
    \frac{d^2 \phi}{d r^2} + \frac{2}{r} \frac{d \phi}{d r} = \frac{\partial V_{\text{eff}}}{\partial \phi}
\end{equation}
with boundary conditions
\begin{equation}
    \lim_{r\to \infty} \phi(r) = 0
\end{equation}
and 
\begin{equation}
    \left.\frac{d \phi}{d r} \right|_{r = 0} = 0 \,.
\end{equation}

\begin{figure}[H]
    \centering
    \includegraphics[width=0.85\linewidth]{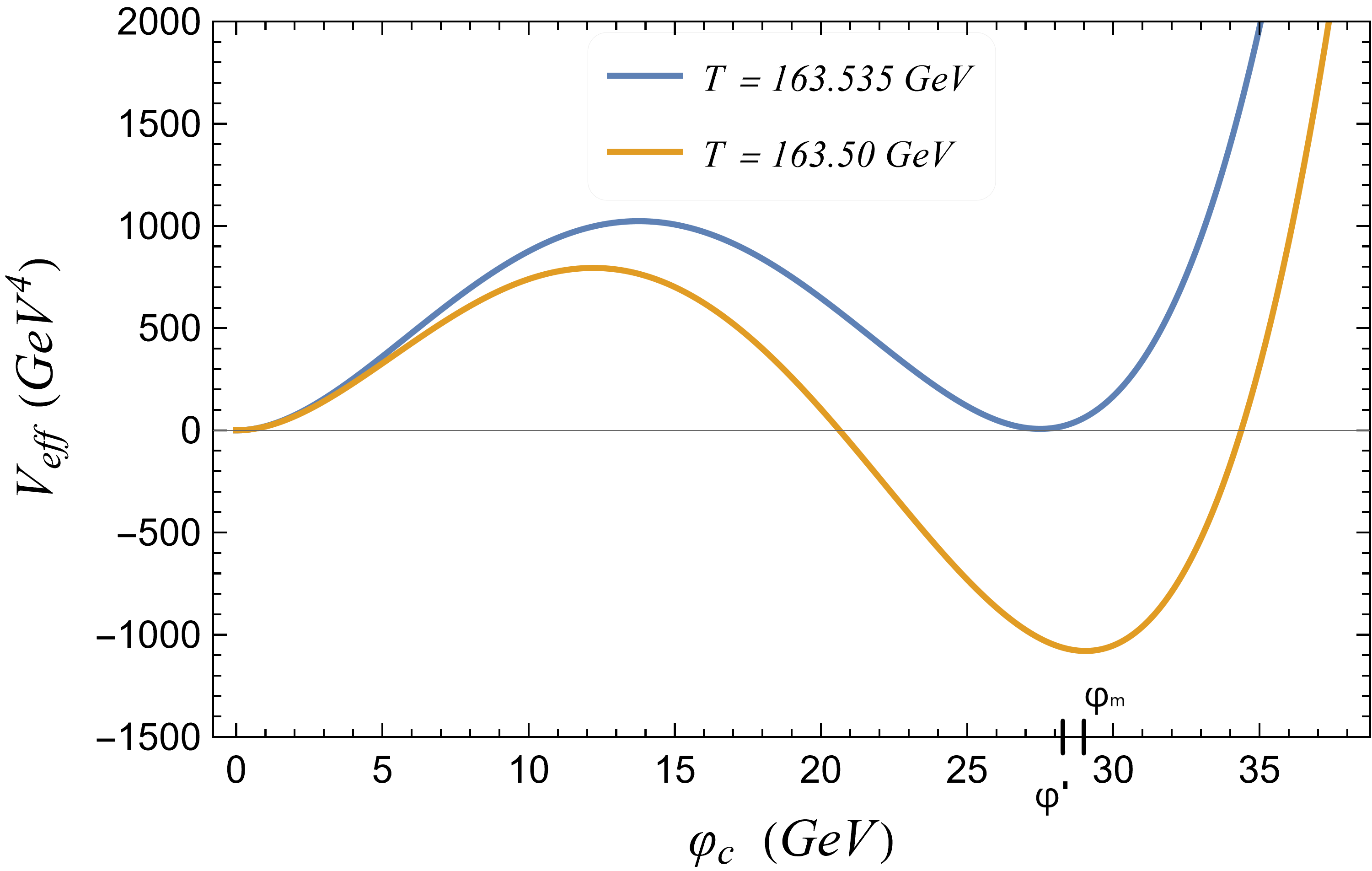}
    \caption{The finite-temperature one-loop effective potential (\ref{Veff_first}) around the critical temperature \(T_c\).}
    \label{Symmetry_Restoration_4}
\end{figure}

In order to demonstrate further the tunneling phenomenon in finite-temperature field theory we consider the effective potential (\ref{Veff_first}) which describes a first-order phase transition in the Standard Model. This transition occurs via the nucleation of the bubbles of the true vacuum state \(\phi_c = \phi_m (T)\) converting the Universe from the false vacuum into the true vacuum state. Accordingly, it is essential to determine the free-energy barrier that should be surmounted so that bubbles can expand \cite{Anderson:1991zb}. Consider a bubble of the broken phase in the sea of the false vacuum state \(\phi_c = 0\) and assume that the center of the bubble is described by \(\phi_c = \phi^{\prime}\) as shown in Fig. \ref{Symmetry_Restoration_4}. 
Now, the value of the effective potential is conventionally set to vanish at the origin as some field-independent terms have been omitted. Thus, the surplus free energy of a true vacuum bubble can be expressed as \cite{Anderson:1991zb}
\begin{equation}\label{surplus}
    S_3 = 4 \pi \int_{0}^{R} r^2 dr\, \left[\frac{1}{2} \left(\frac{d \phi}{d r}\right)^2 + V_{\text{eff}}(\phi(r), T)  \right] \,,
\end{equation}
where \(R\) is the radius of the bubble. One observes two contributions to this surplus free energy (\ref{surplus}): 1) a surface free energy \(F_S\) coming from the derivative term and 2) a volume term \(F_V\) which arises from the difference in the free-energy density inside and outside the bubble. As a result, these contributions scale as,
\begin{equation}
    S_3 \sim 2 \pi R^2 \left(\frac{\delta \phi}{\delta R}\right)^2 \delta R + \frac{4 \pi R^3 \langle V_{\text{eff}} \rangle}{3},
\end{equation}
where \(\delta R\) is the thickness of the bubble wall, \(\delta \phi = \phi^{\prime}\), and \(\langle V_{\text{eff}} \rangle\) is the expectation value of the effective potential in the bubble. 

At temperatures just below the critical temperature, the height of the barrier \(V_{\text{eff}} (\phi_M, T)\) is large compared to the depth of the effective potential at the minimum \(- V_{\text{eff}} (\phi^{\prime}, T)\)\footnote{For thin-wall bubbles \(\phi^{\prime}\) will lie at the absolute minimum, whereas in the case of thick-wall bubbles the possibility that the \(\phi^{\prime}\) is somewhat less than the minimum \(\phi_m\) is allowed.}. In that case, the solution of minimal action corresponds to minimizing the contribution to \(F_V\) which comes from the region around \(\phi_c = \phi_M\). In other words, when the height and the width of the barrier around \(\phi_M\) are large compared to the depth and width of the well at \(\phi_m\), for the minimal solution, \(\phi_c\) will change quickly between \(0\) and \(\phi^{\prime}\). Hence, this leads to a very small bubble wall \(\delta R \ll R\) which implies that the first bubbles that could be formed are thin-wall bubbles at temperatures lower than the critical temperature. 

As the temperature decreases towards the temperature \(T_0\) defined in Eq. (\ref{TT}), the height of the barrier \(V_{\text{eff}}(\phi_M, T)\) tends to zero compared to the depth of the effective potential at the minimum \(- V(\phi_m, T) \). Namely, the free-energy density between the states \(\phi_c = 0\) and \(\phi_c = \phi_m\) rises significantly. Then, the contribution to \(F_V\) from the region near \(\phi_c = \phi_M\) is negligible, and the minimal action results in the minimization of the surface term \(F_S\). This amounts to a configuration where \(\delta R\) is as large as possible which means that \(\delta R/R = \mathcal{O} (1)\) and we work in the thick-wall approximation. Therefore, whether the phase transition proceeds through the nucleation of thin or thick wall bubbles depends on how large the rate of bubble nucleation (\ref{bubble_rate}) becomes, or how small \(S_3\) is, before thick-wall bubbles are energetically favored.

Lastly, if the Standard Model is described by the effective potential (\ref{Veff_first}), the three-dimensional Euclidean action can be calculated without the assumption of the thin-wall approximation \cite{ Dine:1992wr, Dine:1992vs}. This analytic formula reads
\begin{equation}
    \frac{S_3}{T} = \frac{13.72}{E^2} \left[D \left(1 - \frac{T_0^2}{T^2} \right) \right]^{3/2} f \left(\frac{\lambda(T) D}{E^2 } \left(1 - \frac{T_0^2}{T^2} \right) \right),
\end{equation}
where the parameters above come from the effective potential (\ref{Veff_first}) defined in the previous chapter and the function, labeled as \(f(x)\), is given by \cite{Dine:1992vs}
\begin{equation}
    f(x) = 1 + \frac{x}{4} \left[ 1 + \frac{2.4}{1-x} + \frac{0.26}{(1-x)^2}\right] .
\end{equation}

\section{Bubble Nucleation}

We have determined the free energy and the critical radius of a bubble large enough to grow after formation. However, the evolution of a phase transition then depends on the ratio of the rate of production of bubbles of true vacuum, over the expansion rate of the Universe. For instance, if the latter is always larger than the former, the state will remain trapped in the false vacuum state. Otherwise, the phase transition starts at a temperature \(T_n\) by bubble nucleation. As the temperature drops from \(T_c\) to \(T_0\), a point is reached where thermal fluctuations are large enough to nucleate bubbles of true vacuum. These thermal fluctuations produce bubbles of true vacuum at a tunneling rate per unit volume \cite{Arnold:1992rz}
\begin{equation}\label{tunneling rate}
    p(T) = A(T) e^{-\frac{S_3}{T}},
\end{equation}
where the prefactor is given by \(A(T) = \omega T^4\), which is not very important as the exponential is dominant in the bubble nucleation rate and the effect of the parameter \(\omega \ne 1\) is very small.

Now we illustrate how to determine when the onset of the bubble nucleation occurs in the case of the electroweak phase transition. Initially, we describe the cosmology of the Universe at temperatures near the electroweak critical temperature since the progress of a phase transition depends on the expansion of the Universe. More specifically, a homogeneous, isotropic, and flat (\(k = 0\)) Universe is described by the Friedmann-Robertson-Walker (FRW) metric. The FRW metric in comoving coordinates is cast into the from
\begin{equation}
    ds^2 = dt^2 - a(t)^2 \left(dr^2 + r^2 d \Omega^2 \right) \, ,
\end{equation}
where the scale factor \(a(t)\) characterizes the relative size of spacelike hypersurfaces at different times. The Einstein Equations can determine the evolution of the scale factor of the FRW spacetime and lead to the Friedmann Equations, 
\begin{equation}\label{FRW_1}
    H^2 \equiv \left( \frac{\Dot{a}}{a}\right)^2 = \frac{8 \pi}{3 M^2_{Pl}} \rho\, 
\end{equation}
and
\begin{equation}\label{FRW_2}
    \dot{H} + H^2 = \frac{\Ddot{a}}{a} = -\frac{8 \pi}{6M^2_{Pl}} \left( \rho + 3 p \right)\, ,
\end{equation}
where \(H\) is the Hubble parameter, \(\rho\) is the energy density and \(p\) is the pressure\footnote{\(M_{Pl} = G^{-1/2}\) is the Planck mass, instead of the reduced Planck mass \(m_{Pl} = \left(8 \pi G\right)^{-1/2}\), which is used in the literature.}. The expansion of the Universe is described by the Hubble parameter. For instance, in the electroweak phase transition, the temperature is around \(100\) GeV and the Universe is radiation-dominated with the equation of state,
\begin{equation}
    \rho = 3 p \,.
\end{equation}
Then, the second FRW Equation (\ref{FRW_2}) yields to
\begin{equation}\label{rho_1}
    \dot{H} + H^2 = -\frac{8 \pi}{3 M^2_{Pl}} \rho
\end{equation}
and the Hubble parameter is given by
\begin{equation}
    H = \frac{1}{2t} \,.
\end{equation}
In addition, the energy density of the Universe is
\begin{equation}\label{rho_2}
    \rho = \frac{\pi^2}{30} g(T) T^4
\end{equation}
with
\begin{equation}\label{total_dof}
    g(T) = g_B (T) + \frac{7}{8} g_F(T) \,,
\end{equation}
where \(g_B\) (\(g_F\)) is the effective number of bosonic (fermionic) degrees of freedom with respect to the temperature \cite{Lyth:2009zz}. Namely, if we assume an adiabatic expansion of the Universe,
\begin{equation*}
    \frac{a(T_1)}{a(T_2)} = \frac{T_2}{T_1} \, ,
\end{equation*}
Eqs. (\ref{FRW_1}) and (\ref{rho_2}) result in the timescale \(t\),
\begin{equation}\label{time_temperature}
    \frac{M^2_{Pl}}{4 t^2} = \frac{8 \pi^3}{90} g(T) T^4 \quad \Rightarrow \quad t = \frac{1}{4 \pi} \sqrt{\frac{45}{g (T)\pi} } \frac{M_{Pl}}{T^2}.
\end{equation}
Thus, if the horizon size is \(d_H = 2t\), the size of the causal volume at a temperature \(T\) can be considered as \cite{Anderson:1991zb}
\begin{equation}\label{Horizon_volume}
     V_H(t) = 8 t^3 = \zeta^3 \frac{M^3_{Pl}}{T^6}
\end{equation}
with
\begin{equation}
    \zeta = \frac{1}{4 \pi} \sqrt{\frac{45}{g (T)\pi} }\,.
\end{equation}
In the Standard Model, the total number of the effective degrees of freedom (\ref{total_dof}) can be considered independent of temperature and equal to \(g = 106.75\) \cite{Husdal:2016haj} which leads to the numerical value \(\zeta \simeq 3 \times 10^{-2}\) \footnote{
One notices that the determinant of the FRW metric reads
\begin{equation}
    g = - a(t)^6 r^4 \sin^2 \theta,
\end{equation}
which leads to the factor in the volume form
\begin{equation}
    \sqrt{-g} = a (t)^3 r^2 |\sin \theta|.
\end{equation}
In the radiation-dominated era, this factor is written as
\begin{equation}
    \sqrt{-g} \sim r^2 t^{3/2} ,
\end{equation}
setting \(|\sin \theta| = 1\). In terms of the time-temperature relation (\ref{time_temperature}), the previous equation is expressed as
\begin{equation}
    \sqrt{-g} = \left(\zeta\frac{ M_{Pl}}{t_0}\right)^{3/2} \frac{r^2}{T},
\end{equation}
where \(t_0 = 4.3 \times 10^{17}\) s} is the age of the universe. 

The bubble nucleation is initiated at a nucleation temperature \(T_n\) such that the probability for a single bubble to be nucleated within one horizon volume (\ref{Horizon_volume}) is around one \cite{Anderson:1991zb, Moreno:1998bq},
\begin{equation}
     \int_{0}^{t_n} dt \, \Gamma V_H (t) = \int_{T_n}^{\infty} \frac{dT}{T} \left( \frac{2 \zeta M_{Pl}}{T}\right)^4 \exp \left(- \frac{S_3 (T)}{T} \right)= \mathcal{O} (1) \,.
\end{equation}
This expression can be computed numerically as
\begin{equation}
    \frac{S_3 (T_n)}{T_n} \simeq 137 + \ln \frac{10^2 E^2}{\lambda D} + 4 \ln \frac{100 \, \text{GeV}}{T_n},
\end{equation}
where the parameters are normalized as \(T_n \sim 100\) GeV and \(E^2/(\lambda D) \sim 10^{-2}\) \cite{Quiros:1999jp}. These values are typical values and are calculated in the context of the Standard Model as shown in the previous chapter.

\chapter{Electroweak Baryogenesis}\label{Electroweak Baryogenesis}
Nowadays, the baryon number of the Universe is connected with two significant unsolved puzzles in modern physics. Firstly, during the early stages of the Universe, matter and antimatter should have been produced in equal amounts. However, the observable Universe predominantly consists of matter, with almost no evidence of antimatter. For instance, antiprotons are produced as secondaries in collisions and there are antiprotons in the cosmic rays, but their proportions are very small compared to the abundance of matter. Secondly, the abundance of matter over antimatter in the current Universe is often described by the baryon-to-photon ratio,
\begin{equation}
    \eta \equiv \frac{n_B - n_{\bar{B}}}{n_{\gamma}}.
\end{equation}
As there is no important evidence of antimatter, \(n_B \gg n_{\Bar{B}}\), this parameter can be written as \cite{Lyth:2009zz}
\begin{equation}\label{b_t_e}
    \eta = \frac{n_B}{n_{\gamma}} \simeq (6.1 \pm 0.2) \times 10^{-10}.
\end{equation}
This ratio was initially computed from the Big Bang nucleosynthesis, which is the epoch where deuterium (D), helium (\(^3\)He and \(^4\)He), and lithium (\(^7\)Li) were produced\footnote{A more recent review on the value of the baryon-to-photon ratio can be found in Refs. \cite{Yeh:2022heq, 
ParticleDataGroup:2022pth}.}. It is possible that the baryon-to-photon ratio has not changed since nucleosynthesis \cite{Kolb:1990vq}. If there are no processes that would have produced entropy to alter the photon number, the baryon number is conserved at the energy scales around \(1\) MeV. Then, one computes the ratio between entropy density and photon density as
\begin{equation}
    \frac{s}{n_{\gamma}} = 3.91 \frac{\pi^4}{45 \zeta (3)} = 7.04,
\end{equation}
which results in the baryon-to-entropy ratio
\begin{equation}
    \frac{n_B}{s} = \frac{1}{7.04} \frac{n_B}{n_{\gamma}} = \frac{\eta}{7.04}.
\end{equation}
The value of the ratio (\ref{b_t_e}) cannot be explained in the framework of the standard cosmological model, if it is assumed that \(n_B = n_{\bar{B}}\) at the beginning of the Universe. Therefore, an initial asymmetry has to be imposed by hand as an initial condition or could have been produced during cosmological phase transitions. This chapter investigates in detail the connection between cosmological phase transitions and the baryon asymmetry.

\section{Criteria for Baryogenesis}
In the standard cosmological model, the size of the ratio (\ref{b_t_e}) is inconsistent with nucleosynthesis, and one could impose it as an initial condition. On the other hand, Sakharov suggested that a very small baryon asymmetry might have been generated in the early Universe resulting in (\ref{b_t_e}) after \(p \Bar{p}\) annihilations. Thus, this asymmetry can be established if the so-called Sakharov criteria are fulfilled \cite{Sakharov:1967dj}. In particular, these conditions for baryogenesis are i) baryon number violation, ii) \(C\) and \(CP\) violation, and iii) departure from thermal equilibrium.

First of all, the first condition is clearly understood because the Universe is initially baryon symmetric with \(\Delta B = 0\) and evolves to a state with non-zero \(\Delta B \). However, interactions that do not conserve the baryon number might mediate proton decay: \(p \to \pi^0 \, e^{+}\), but the proton lifetime is strongly constrained by the experimental data: \(\tau_p > 10^{33}\) years \cite{ParticleDataGroup:2022pth}.

Moving on, the combined action of the charge conjugation and parity matrix changes the sign of the baryon number as they interchange particles with antiparticles. This is clearly demonstrated if we consider the Dirac spinors,
\begin{equation*}
    \psi = \begin{pmatrix}
        \psi_L \\ \psi_R
    \end{pmatrix},
\end{equation*}
where \(\psi_L\) (\(\psi_R\)) is the left-handed (right-handed) spinor. The components of the fermion field \(\psi\) transform under the parity transformation as
\begin{equation}
    \psi_L \to \psi_R, \quad \psi_R \to \psi_L
\end{equation}
and the left-handed and right-handed fields transform under the charge conjugation as
\begin{equation}
    \psi_L \to \psi_L^{(c)} \equiv \sigma_2 \psi^{*}_R, \quad \psi_R \to \psi_R^{(c)} \equiv - \sigma_2 \psi_L^{*}.
\end{equation}
Therefore, the combined \(CP\) transformations result in
\begin{equation}
    \psi_L \to \psi^{(c)}_R, \quad \psi_R \to \psi_L^{(c)}.
\end{equation}
If there is no baryon asymmetry in the early Universe, and without a preferred direction of time as in the standard cosmological model, the Universe is then represented by a \(C\) and \(CP\) invariant state, \(\ket{\phi_0}\) with \(B = 0\). If \(C\) and \(CP\) were conserved,
\begin{equation}
    [C, H] = [CP, H] = 0,
\end{equation}
where \(H\) is the Hamiltonian. Then, the state of the Universe at a later time \(t\), 
\[\ket{\phi(t)} = e^{i H t} \ket{\phi_0},\]
would be \(C\) and \(CP\) invariant and, as a result, baryon number conserving. Hence, \(C\) and \(CP\) violating interactions are necessary to produce a net non-zero \(\Delta B\).

Finally, the departure from thermal equilibrium is the third criterion for baryogenesis. If every particle in the Universe remained in thermal equilibrium, then there would be no defined direction for time, and \(CPT\) invariance would prevent the emergence of any baryon excess, rendering \(CP\)-violating interactions irrelevant \cite{Kolb:1979qa, Kolb:1979ui}. In addition, in case all the interaction rates of a particle species are much faster than the expansion of the Universe, this particle species is in thermal equilibrium. Otherwise, a deviation from equilibrium is expected whenever a crucial rate for maintaining it is less than the expansion rate of the Universe. Moreover, departure from the thermal equilibrium could not happen in a homogeneous isotropic Universe consisting only of massless species as massive species are generally important for such deviations to occur. 

\section{Equilibrium Approximation}

An important aspect of a first-order phase transition is the departure from the thermal equilibrium, which is also the third condition to explain the baryon asymmetry in the Universe. During this phase transition, the Universe is out of equilibrium which is caused by the passage of the expanding bubble walls through the surrounding plasma in the Universe \cite{Morrissey:2012db}. However, our analysis in the finite-temperature field theory is based on the premise of the equilibrium state. Namely, the finite-temperature field theory can be applied using the thermal equilibrium approximation, examining its validity.

In the early Universe, the departure from the thermal equilibrium is described by the ratio of the interaction rate for different types of interactions and the rate of the Universe expansion \cite{Shaposhnikov:1996th}. Firstly, each interaction is either an interaction, which involves two or more initial particles, or a decay involving a single initial particle. Each process has a typical reaction time expressed as
\begin{equation}
    \tau_{r} \sim \langle \sigma n v \rangle^{-1},
\end{equation}
where \(\sigma\) is the corresponding cross section, \(n\) is the particle density, and \(v\) is the relative velocity of the interacting particles. Secondly, an essential time scale in cosmology is associated with the expansion rate of the Universe and is described by the inverse age of the Universe,
\begin{equation}
    t_{U}^{-1} = \frac{T^2}{M_0}
\end{equation}
with
\begin{equation}
    M_0 =  \frac{M_{Pl}}{1.66 \sqrt{N}},
\end{equation}
where \(N\) is the effective number of the massless degrees of freedom \cite{Lyth:2009zz}. Then, the age of the Universe can be written as
\begin{equation}
    t_U \simeq \frac{10^{18} \text{ GeV}}{T^2}.
\end{equation}
Hence, the rate of the Universe expansion can be considered as a unique non-equilibrium parameter of the system  \cite{Shaposhnikov:1996th} and a process is in thermal equilibrium if only if the interaction rate for a given interaction is much bigger than the expansion of the Universe \cite{Lyth:2009zz}, 
\begin{equation}
    d_i = \frac{t_U^{-1}}{\tau_i^{-1}} = \frac{\tau_i}{t_U} < 1,
\end{equation}
where \(i\) refers to the different types of interactions between the particles in the cosmological plasma.

In the electroweak phase transition, the deviation from the thermal equilibrium can be measured around the temperature \(T = 100\) GeV, calculating the reaction rate for the strong, weak, and electromagnetic interactions. In particular, the strong interactions have interaction rates given by
\begin{equation}
    \tau_s^{-1} \sim \alpha_s^2 T.
\end{equation}
Similarly, the typical weak interactions are described by the interaction rate
\begin{equation}
    \tau_w^{-1} \sim \alpha_W^2 T.
\end{equation}
On the other hand, the interactions that involve chirality flips for the lightest fermions, such as \(e_R H \to \nu W\), are the slowest interactions with interaction rate,
\begin{equation}
    \tau_e^{-1} \sim y_e^2 \alpha_W T,
\end{equation}
where \(y_e\) is the electron Yukawa coupling. Consequently, during the electroweak epoch, the ratio \(d_i\) approximately ranges from \(10^{-14}\) to \(10^{-2}\). One notices that the deviation from thermal equilibrium is maximal for the right-handed electron. In contrast, the particle distribution functions of quarks and gluons, vector bosons, Higgs boson, and left-handed charged leptons and neutrinos are identified with the equilibrium ones and the equilibrium approximation is valid with a high accuracy which is better than \(10^{-13}\). 

The thermal equilibrium approximation may be violated at higher temperatures than the critical temperature of the electroweak phase transition. It is fascinating that following the arguments above one may compute the temperature interval, where the process under study was in thermal equilibrium. This is clearly demonstrated in the electromagnetic interactions. The rate of these interactions is bigger than the expansion rate of the Universe at temperatures from few eV to \(\alpha^2 M_{Pl} \simeq 10^{15}\) GeV.

All in all, our approach to the finite-temperature field theory remains valid in a wide range of applications in cosmology, especially around the electroweak scale. Thus, electroweak baryogenesis can be sufficiently described by the methods in equilibrium dynamics which is a very precise approximation.

\section{Baryogenesis in the Standard Model}
 
The theory of electroweak baryogenesis describes the observed baryon asymmetry of the Universe. Electroweak baryogenesis is a physical mechanism in the early Universe that generates an asymmetry between baryons and anti-baryons in the electroweak phase transition, while baryon creation proceeds in the vicinity of the expanding bubble walls. In the early Universe, \(CP\)-violating interactions of the plasma with the bubble wall of the broken phase produce \(CP\) and \(C\) asymmetries in particle number densities ahead of the bubble wall \cite{Morrissey:2012db}. Then, these asymmetries diffuse into the symmetric phase in front of the bubble wall, where they are converted to baryons by electroweak sphalerons \cite{Cohen:1993nk}. In the broken phase, the rate of sphaleron transitions may be strongly suppressed to avoid washing out the generated baryons. Thus, it is necessary for a successful electroweak baryogenesis scenario, that the baryon asymmetry generated at the expanding bubble wall is not washed out by sphalerons within the broken phase. A detailed review of baryogenesis is provided in Refs. \cite{Trodden:1998ym, Riotto:1998bt, Riotto:1999yt, Cohen:1993nk, Morrissey:2012db} and the following sections are based on them as well.

It will be demonstrated later that electroweak baryogenesis requires a first-order electroweak phase transition, which satisfies
\begin{equation}\label{criterion_baryogenesis}
    \frac{\upsilon_c}{T_c} > 0.6 - 1.4 \,,
\end{equation}
which is related to a factor in the rate of sphaleron transitions in the broken phase. This criterion, which will be later called sphaleron rate criterion\footnote{This criterion is also called baryon-number preservation criterion in the literature.}, varies as certain uncertainties are introduced in the calculations \cite{Patel:2011th, Fuyuto:2014yia}.

In the Standard Model, the electroweak baryogenesis satisfies the three Sakharov conditions for baryogenesis \cite{Trodden:1998ym, Riotto:1998bt, Kuzmin:1985mm, Kuzmin:1987wn, Cohen:1993nk}. More specifically, the lepton and baryon numbers are anomalous global symmetries, which are violated by non-perturbative effects (instantons). In the 1970s, 't Hooft discovered the violation of the baryon number \cite{tHooft:1976rip, tHooft:1976snw}. Furthermore, \(CP\) violation can be induced by the CKM phases in the fermion mass matrix \cite{Kobayashi:1973}. However, \(CP\) violation is insufficient to produce large enough chiral asymmetries to explain the observed baryon-to-entropy ratio. Lastly, during the electroweak phase transition, the Universe is out of equilibrium only if the phase transition is a sufficiently strong first-order phase transition. Namely, if a strong first-order electroweak phase transition occurs, the \(B\)-violating interactions are out of equilibrium in the bubble walls and the baryon asymmetry can be generated during the phase transition \cite{Cohen:1993nk}.

\subsection{Baryon and Lepton Number Violation}
The baryon number is a global symmetry of the Standard Model Lagrangian, for which all quarks are described by \(B = 1/3\) and leptons by \(B = 0\). Another global symmetry is the lepton number, for which quarks have \(L = 0\) and leptons have \(L = 1\). Namely, the baryonic and leptonic currents are conserved at a classical level in the Standard Model. Nevertheless, in the quantum theory, that conservation is violated due to quantum corrections through the chiral anomaly which is related to the triangle fermionic loop in external gauge fields. In particular, the non-conserved currents lead to the following relations
\begin{equation}\label{non_con_cur}
    \partial_{\mu} j^{\mu}_B = \partial_{\mu} j^{\mu}_L = N_f \left( \frac{g^2}{32 \pi^2} W^{a \mu \nu} \tilde{W}_{\mu \nu}^{a} - \frac{g^{\prime 2}}{32 \pi^2} Y^{\mu \nu} \Tilde{Y}_{\mu \nu} \right),
\end{equation}
where \(N_f\) is the number of fermion generations (\(N_f = 3\) in the Standard Model), \(W^{a}_{\mu \nu}\) and \(Y_{\mu \nu}\) are the gauge field strength tensors\footnote{The gauge field tensor with the tilde in the expressions (\ref{non_con_cur}) denotes the dual tensor,
\begin{equation}
    \Tilde{W}^{\mu \nu} = \frac{1}{2}\epsilon^{\mu \nu \alpha \beta } W_{\alpha \beta}
\end{equation}.} for \(SU(2)_L\) and \(U(1)_Y\), respectively.

One observes that the difference \(B - L\) is strictly conserved and only the sum \(B+L\) is anomalous and can be violated. An additional significant feature of (\ref{non_con_cur}) is that fluctuations of the gauge field strengths can result in fluctuations of the corresponding current. The product of gauge field strengths on the right-hand side of Eq. (\ref{non_con_cur}) can be expressed in terms of four-divergences,
\begin{equation}
   W^{a \mu \nu} \tilde{W}_{\mu \nu}^{a} = \partial_{\mu} k^{\mu}_W, \quad Y^{\mu \nu} \Tilde{Y}_{\mu \nu} = \partial_{\mu} k_Y^{\mu},
\end{equation}
where the four-vectors are given by
\begin{equation}\label{fourvector}
\begin{split}
    k^{\mu}_Y &= \epsilon^{\mu \nu \alpha \beta} Y_{\nu \alpha} Y_{\beta} \\
    k^{\mu}_W &= \epsilon^{\mu \nu \alpha \beta} \left( W_{\nu \alpha}^{a} W_{\beta}^{a} - \frac{g}{3} \epsilon_{a b c} W^{a}_{\nu} W^{b}_{\alpha} W_{\beta}^{c} \right),
\end{split}
\end{equation}
where \(W_{\mu}\) and \(Y_{\mu}\) are the gauge fields of \(SU(2)_L\) and \(U(1)_Y\) gauge symmetries, respectively. The total derivatives are not generally observable since they can be integrated by part and be removed from the integrals. This also holds for the terms in the four-vectors (\ref{fourvector}) proportional to the field strengths of the gauge fields. This implies that regarding the Abelian subgroup \(U(1)_Y\), the current non-conservation induced by quantum corrections becomes non-observable, but this is not mandatory for gauge fields, for which the integral can be nonzero. Therefore, the current non-conservation induced by quantum corrections can become observable only for non-Abelian groups. More specifically, if we calculate the change in the total baryon and lepton number from time \(t = 0\) to the final time \(t = t_f\), their variations are written as
\begin{equation}\label{bef_NCS}
    \Delta B = \Delta L = N_f (\Delta N_{CS} - \Delta n_{CS}),
\end{equation}
where \(N_{CS}\) and \(n_{CS}\) are called the Chern-Simons numbers that describe the topology of the gauge field configuration \cite{Riotto:1998bt, Riotto:1999yt, Cohen:1993nk}. In view of Eqs. (\ref{fourvector}) and (\ref{bef_NCS}), the Chern-Simons numbers are given by
\begin{equation}\label{Chern-Simons}
    N_{CS} = \frac{g^2}{32 \pi^2} \int d^3 x \, \epsilon^{ijk}\left(W^{a}_{ij} W^{a}_{k} - \frac{g}{3} \epsilon_{abc} W^{a}_{i} W^{b}_{j} W^{c}_{k}\right) 
\end{equation}
and
\begin{equation}
    n_{CS} = \frac{g^{\prime 2}}{32 \pi^2} \int d^3 x \, \epsilon^{ijk} Y_{ij} Y_{k}.
\end{equation}
It is remarkable that the variation of this number, \(\Delta N_{CS} \), is gauge invariant, although the Chern-Simons number is not. 

At a classical level, the ground state should correspond to a field configuration that is time-independent with zero energy density. Thus, the field strength tensors \(W_{\mu \nu}^{a}\) vanish and one can calculate the \(\Delta B\) between an initial and a final configuration of gauge fields considering the temporal gauge, \(W_0 = 0\). However, the gauge potentials do not necessarily vanish and can be represented by purely gauge fields,
\begin{equation}\label{puregauge}
    W_{\mu} = \frac{i}{g} U(x) \partial_{\mu} U^{-1} (x).
\end{equation}
As a result, the Chern-Simons number \(n_{CS}\) vanishes, as it is proportional to the \(U_Y (1)\) field strength. There are two classes of gauge transformations respecting \(W_{\mu \nu} = 0\) \cite{Riotto:1998bt}. The first class refers to the continuous transformations of the potential leading to \(\Delta N_{CS} = 0\). Regarding the second class, if one attempts to produce \(\Delta N_{CS} \ne 0\) by a continuous variation of the potentials then one has to enter a region where \(W_{\mu \nu}\) is non-zero. Namely, the vacuum states with different topological charges are separated by potential barriers. 

In Euclidean spacetime, the trajectory in field space configuration that connects the vacua differing by a unit of topological charge is called the instanton. The tunneling probability can be computed in the semi-classical approximation as shown in Ref. \cite{Coleman:1977py}. This probability is computed by the Euclidean action evaluated at the aforementioned trajectory as
\begin{equation}
    \Gamma \simeq e^{- \frac{4 \pi}{\alpha_W}} \simeq 10^{-162}
\end{equation}
with \(\alpha_W = g^2/4\pi\). This number is so small that the prefactor does not change the result and the barrier to be penetrated has zero probability.

\subsection{Sphalerons}

At finite temperatures, a particle may classically go over the barrier due to thermal fluctuations with a probability obtained by the Boltzmann exponent. Let us consider the potential energy that depends on the gauge field configuration \(W_{\mu}\). This potential possesses an infinite number of degenerate minima, denoted as \(\Omega_n\), which are described by different values of the Chern-Simons number (\ref{Chern-Simons}). The configuration \(W_{\mu} = 0\) corresponds to the minimum \(\Omega_0\). By convention, the value of the potential is set to zero at this point. Other potential minima are described by gauge fields which are cast into the form (\ref{puregauge}). In the temporal gauge \(W_{0} = 0\), the gauge transformation \(U\) is time-independent owing to the gauge field configurations with \(W_{\mu \nu} = 0\). As a result, the gauge transformation is \(U = U(\vec{x})\), and defines maps
\begin{equation}
    U: S^3 \to SU(2).
\end{equation}
The potential energy vanishes at all the potential minima with \(W_{\mu \nu} = 0\), but those defined by a map \(U(\vec{x})\) with non-zero Chern-Simons number,
\begin{equation}
    n [U] = \frac{1}{24 \pi^2} \int d^3 x \, \epsilon^{ijk} Tr \left( U \partial_i U^{-1} U \partial_j U^{-1} U \partial_k U^{-1} \right),
\end{equation}
correspond to degenerate minima in the configuration space with non-zero baryon and lepton number\footnote{The gauge can be chosen such that \(W_i = N_{CS}=0\) at \(t = 0\) \cite{Cohen:1993nk}. This yields to the following change in the total baryon number in the time interval from \(t = 0\) to \(t = t_f\),
\begin{equation*}
    \Delta B = N_f N_{CS} (t_f) = \frac{N_f}{24 \pi^2} \int d^3 x \, \epsilon^{ijk} Tr \left( U \partial_i U^{-1} U \partial_j U^{-1} U \partial_k U^{-1} \right).
\end{equation*}
}.

During a first-order phase transition, the degenerate minima are separated by the effective potential barrier. The field configuration at the top of the barrier is known as the sphaleron, which is a static unstable solution to the classical equations of motion. In Ref. \cite{Kunz:1992uh}, the sphaleron solution is presented in detail in the case of an arbitrary value of \(\sin^2 \theta_W\). An ansatz for the sphaleron solution for the case of zero Weinberg angle neglecting terms of \(\mathcal{O}(g^{\prime})\) was investigated using the zero-temperature potential in Refs. \cite{Manton:1983nd, Klinkhamer:1984di} for the Standard Model with a single Higgs doublet, so that,
\begin{equation}\label{257}
    W_i^{a} \sigma^a dx^i = - \frac{2i}{g} f(\xi) dU \left(U^{-1}\right)
\end{equation}
for the gauge field, and
\begin{equation}\label{258}
    H = \frac{\upsilon}{\sqrt{2}} h(\xi) U \begin{pmatrix}
        0 \\ 1
    \end{pmatrix}
\end{equation}
for the Higgs field, where \(\xi = g \upsilon r\) is the dimensionless radial distance and the gauge transformation \(U\) is given by,
\begin{equation}\label{259}
    U = \frac{1}{r} \begin{pmatrix}
        z & x+ iy \\
        -x + iy & z
    \end{pmatrix}.
\end{equation}
Namely, if in the Standard Model, we consider the limit of vanishing mixing angle (\(\theta_W \to 0\)), the theory is a pure gauge \(SU(2)_L\) theory coupled to the Higgs field \(H\). The corresponding energy functional reads
\begin{equation}\label{enfunc}
    E = \int d^3 x \, \left( \frac{1}{4} W_{ij}^{a} W^{a}_{ij} + (D_i H)^{\dagger} (D_i H) + V (H^{\dagger} H) \right),
\end{equation}
where \(V (H^{\dagger} H)\) is the Higgs potential. The fermion fields and the time component of the gauge fields are set to vanish \cite{Klinkhamer:1984di}. Using the ansatz (\ref{257}), (\ref{258}) and (\ref{259}) the field equations yield to
\begin{equation}\label{260}
\begin{split}
    &\xi^2 \frac{d^2 f}{d \xi^2} = 2 f (1-f)(1-2f) - \frac{\xi^2}{4} h^2 (1-f)\\
    &\frac{d}{d \xi} \left( \xi^2 \frac{d h}{d \xi} \right) = 2 h (1-f)^2 + \frac{\lambda}{g^2} \xi^2 (h^2-1) h
\end{split}
\end{equation}
with the boundary conditions, \(f(0) = h(0) = 0\) and \(f(\infty) = h(\infty) = 1\). Then, the energy functional (\ref{enfunc}) is rewritten as
\begin{equation}\label{261}
    \begin{split}
        E =& \frac{4 \pi \upsilon }{g} \int_{0}^{\infty} d \xi \, \biggl\{4 \left( \frac{d f}{d \xi}\right)^2 + \frac{8}{\xi^2} f^2 (1-f)^2 + \frac{1}{2} \xi^2 \left(\frac{d h}{d \xi}\right)^2 \\
        &+ h^2 (1-f)^2 + \frac{1}{4} \left(\frac{\lambda}{g^2}\right) \xi^2 (h^2 - 1)^2\biggr\}.
    \end{split}
\end{equation}
The equations of motion (\ref{260}) can be solved numerically and their solutions depend on the Higgs self-coupling \(\lambda\) and gauge coupling \(g\). Then, the solutions are inserted in the energy functional (\ref{261}) and they result in the sphaleron energy \(E_\text{sph}\), which is the height of the barrier between different degenerate minima. The solution is commonly cast into the form,
\begin{equation}\label{262}
    E_{\text{sph}} = \frac{2 m_W}{\alpha_W} B \left(\frac{\lambda}{g^2}\right)
\end{equation}
where \(B\) is a constant that is evaluated numerically. In the Standard Model, this parameter is computed as\footnote{The formula (\ref{B(X)}) holds for Higgs masses in the interval \(25\, \text{GeV} \leq m_H \leq 250\, \text{GeV}\). It is noticeable that the experimental value of the Higgs mass is almost in the middle of this interval.}
\begin{equation}\label{B(X)}
    B \left(\frac{m_H}{m_W}\right) = 1.58 + 0.32 \frac{m_H}{m_W} - 0.05 \left(\frac{m_H}{m_W}\right)^2.
\end{equation}
The sphaleron energy (\ref{262}) was derived at zero temperature. In contrast, at a finite temperature, the sphaleron is discussed in Refs. \cite{Brihaye:1993ud, Moreno:1996zm} and the sphaleron energy obeys the approximate scaling law,
\begin{equation}\label{Esph(T)}
    E_{\text{sph}} (T) =  E_{\text{sph}} \frac{\langle \phi(T)\rangle}{\langle \phi(0) \rangle},
\end{equation}
where \(\langle \phi (T) \rangle\) is the vacuum expectation value of the Higgs boson in the broken phase, which essentially minimizes the finite-temperature effective potential. According to Eq. (\ref{262}), the energy (\ref{Esph(T)}) is expressed as
\begin{equation}\label{E_SPH}
    E_{\text{sph}} (T) = \frac{2 m_W (T)}{\alpha_W} B \left(\frac{\lambda}{g^2}\right)
\end{equation}
with the W-boson mass above defined as 
\begin{equation}
    m_W (T) = \frac{g}{2} \langle \phi(T) \rangle.
\end{equation}

The baryon violation rate is computed differently at temperatures below and above the critical temperature. In the symmetric phase, at \(\phi_c = 0\), the baryon violation rate cannot be computed using perturbation theory since the perturbative analysis breaks down due to the infrared divergences in this phase. In particular, at a high temperature \(T> T_c\), the Higgs boson decouples from the dynamics, and a pure \(SU(2)_L\) gauge theory is considered sufficiently. Non-perturbative magnetic field configurations, which are almost time-independent, are produced to generate a change in the Chern-Simons number and in the baryon number, \(\Delta B = N_f \Delta N_{CS}\). Thus, the infrared divergences are cut off by the non-perturbative generation of a magnetic mass,
\begin{equation}
    m_M \sim \alpha_W T,
\end{equation}
which implies a magnetic screening length,
\begin{equation}
    \xi \sim \frac{1}{\alpha_W T}.
\end{equation}
The baryon violation rate per unit time and unit volume does not have any exponential Boltzmann factor as it would disappear from (\ref{tunneling rate}) in the high-temperature limit (\(T\to \infty\)). Therefore, the pre-exponential is calculated from dimensional grounds \cite{Arnold:1987mh, Khlebnikov:1988sr} to write the baryon violation rate as
\begin{equation}\label{b_v_r}
    \Gamma = k (\alpha_W T)^4,
\end{equation}
where the coefficient \(k\) can be computed numerically as demonstrated in Refs. \cite{Ambjorn:1987qu, Ambjorn:1990wn, Zakharov:1990xt, Ambjorn:1992np, Ambjorn:1995xm} leading to the coeffcient\footnote{An additional factor in (\ref{b_v_r}) as \(k^{\prime} \alpha_W\) is provided by lattice calculations \cite{Arnold:1996dy, Arnold:1998cz,Arnold:1998cy, Moore:1998zk, Bodeker:1998hm, Moore:1998swa}.} \(0.1 \lesssim k \lesssim 1.0\).

On the other hand, below the critical temperature, \(T< T_c\), the baryon violation rate can be computed using the semi-classical approximation given by Eq. (\ref{tunneling rate}). More specifically, this rate per unit time and unit volume for fluctuations between neighboring minima includes a Boltzmann suppression factor,
\begin{equation}
    \exp{\left(-\frac{E_{\text{sph}}(T)}{T}\right)},
\end{equation}
and a prefactor that contains the determinant of all zero and non-zero modes. Therefore, following the theory of cosmological phase transitions presented in the last chapter, the prefactor is written as \cite{Carson:1990jm}
\begin{equation}\label{GAMMA_T}
    \Gamma \simeq 2.8 \times 10^5 \left(\frac{\alpha_W}{4 \pi} \right)^4 \frac{\kappa \, \zeta^7}{B^7 } T^4 \, e^{-\zeta (T)},
\end{equation}
where \(\kappa\) is the functional determinant related to the fluctuations about the sphaleron, \(B\) is given by Eq. (\ref{B(X)}) and \(\zeta (T)\) is defined as
\begin{equation}\label{zeta_def}
    \zeta (T) = \frac{E_{\text{sph}}(T)}{T}
\end{equation}
with \(E_{\text{sph}}(T)\) given by Eq. (\ref{E_SPH}). The functional determinant is computed to be in the interval \(10^{-4} \lesssim \kappa \lesssim 10^{-1}\) \cite{Dine:1991ck}. Furthermore, the dilution, denoted as \(S\), of the baryon asymmetry in the anomalous electroweak processes is governed by the following equation \cite{Shaposhnikov:1987tw, Shaposhnikov:1987pf, Bochkarev:1987wf},
\begin{equation}\label{EQ_S}
    \frac{\partial S}{\partial t} = - V_B (t) S,
\end{equation}
where \(V_B (t)\) is the rate of the baryon number non-conserving processes. The relation between the rate of baryon number non-conserving processes \(V_B(t)\) and the rate \(\Gamma\) per unit time and per unit volume is given by
\begin{equation}
    V_B  = \frac{13}{2} N_f \frac{\Gamma}{T^3},
\end{equation}
where it was assumed that the interactions with the Higgs boson are in thermal equilibrium implying that the concentrations of left- and right-handed fermions are equal to each other \cite{Bochkarev:1987wf}. This assumption is valid at temperatures below the critical temperature. Thus, after integration, the solution of Eq. (\ref{EQ_S}) yields to
\begin{equation}\label{S_e}
    S = e^{- V_B t}.
\end{equation}
On account of Eqs. (\ref{time_temperature}) and (\ref{GAMMA_T}), the exponent in (\ref{S_e}) can be expressed as
\begin{equation}
    V_B t \simeq  7.79 \times 10^{19} \frac{ \alpha_W^4}{B^7 T} \kappa \zeta^7 e^{-\zeta},
\end{equation}
The final baryon asymmetry should agree with the current observations. If the maximal baryon asymmetry in the current Universe is described by (\ref{b_t_e}) and is of the order \(10^{-10}\), then one requires the condition \cite{Bochkarev:1987wf, Bochkarev:1990gb, Ambjorn:1987qu, Shaposhnikov:1987pf, Shaposhnikov:1987tw, Ambjorn:1988gf},
\begin{equation}
    S \gtrsim 10^{-5} \quad \Rightarrow \quad V_B t \lesssim 10,
\end{equation}
which leads to the condition on (\ref{zeta_def}) and is then expressed as
\begin{equation}
    \zeta (T_c) \gtrsim 7 \ln \zeta (T_c) + 9 \ln 10 + \ln \kappa,
\end{equation}
where it was evaluated considering the following values\footnote{The values of the bounds may slightly differ from the values in the references as the experimental value of the Higgs mass is taken into account here, which was not measured before 2012.}: \(\alpha_W = 0.0339\), \(N_f = 3\), \(T_c \simeq 100\) GeV, \(m_H = 125\) GeV and \(m_W = 80.4\) GeV to compute the constant as \(B = 1.96\). If we also set \(\kappa = 0.1\) which is around its upper bound, the last condition leads to
\begin{equation}\label{upperbound}
    \zeta (T_c) \gtrsim 45
\end{equation}
and the lower bound, \(\kappa = 10^{-4}\), leads to
\begin{equation}\label{lowerbound}
    \zeta (T_c) \gtrsim 36.
\end{equation}

Finally, the lower and upper bounds (\ref{lowerbound}) and (\ref{upperbound}) can be expressed as bounds on the ratio \(\upsilon_c/T_c\) \cite{Bochkarev:1987wf, Bochkarev:1990gb, Shaposhnikov:1987pf, Shaposhnikov:1987tw, Ambjorn:1988gf}, where \(\upsilon_c \equiv \upsilon (T_c)\) is the Higgs vacuum expectation value (\ref{VEV_DEF}) at the critical temperature, which is degenerate with the minimum at the origin. Namely, the relation between the ratios is obtained using (\ref{E_SPH}),
\begin{equation}
    \frac{\upsilon_c}{T_c} = \frac{g}{4 \pi B} \frac{E_{\text{sph}} (T_c)}{T_c}.
\end{equation}
Therefore, the upper bound (\ref{upperbound}) is expressed as
\begin{equation}\label{upperbound_1}
     \frac{\upsilon_c}{T_c} \gtrsim 1.2,
\end{equation}
whereas the lower bound (\ref{lowerbound}) translates into,
\begin{equation}\label{lowerbound_1}
     \frac{\upsilon_c}{T_c} \gtrsim 1.0.
\end{equation} 
The lower bound (\ref{lowerbound_1}) is usually taken into account for calculations, but the upper bound (\ref{upperbound_1}) is also important to measure the uncertainty on the lower bound (\ref{lowerbound_1}). As a consequence, the electroweak phase transition is required to satisfy the conditions (\ref{upperbound_1}) and (\ref{lowerbound_1}) to describe the generation of the observed baryon asymmetry in the Universe. Namely, if these conditions are satisfied, the phase transition is sufficiently strong first-order. For instance, during a second-order phase transition, any generated baryon asymmetry is washed out since it is \(\upsilon_c = 0\), in this case.

A last comment on these results is that the bound (\ref{upperbound_1}) can be translated into a bound on the Higgs mass to require a strong first-order phase transition in the Standard Model. The effective potential of the Standard Model was briefly presented in Chapter \ref{Finite-Temperature Field Theory} and is given by (\ref{Veff_first}). At the critical temperature, the effective potential is minimized at the non-zero value,
\begin{equation}\label{before_mH}
    \upsilon_c = \frac{2 E T_c}{\lambda(T_c)},
\end{equation}
where \(\lambda(T_c)\) can be approximated by \(\lambda\). Thus, the expression (\ref{before_mH}) results in,
\begin{equation}
     \frac{\upsilon_c}{T_c} \simeq \frac{2 E T_c}{\lambda} \quad \Rightarrow \quad m_H \simeq \sqrt{\frac{4 E T_c}{\upsilon_c}}\upsilon
\end{equation}
and (\ref{upperbound_1}) yields to the bound on the Higgs mass,
\begin{equation}
    m_H \lesssim 44.04 \, \text{GeV}.
\end{equation}
Therefore, the electroweak phase transition described by the effective potential of the Standard Model (\ref{Veff_first}) is not strong enough to sustain the previously generated baryon asymmetry. The reader could assume that this problem comes from the loose approximations that were considered in the derivation of the effective potential (\ref{Veff_first}). However, this is not the problem and the Standard Model cannot describe a strong first-order transition in general as it is shown in the next chapter.

\chapter{Electroweak Phase Transition}\label{Electroweak Phase Transition}
 The Standard Model which describes the electroweak interactions is based on the spontaneous symmetry breaking of the gauge symmetry \(SU(2)_L \times U(1)_Y\). In the past decades, numerous studies showed that this symmetry between the weak and electromagnetic interactions could be restored at high temperatures. As a result, a phase transition is associated with the electroweak symmetry breaking, which is called electroweak phase transition. This phase transition is primarily studied by the effective potential at finite temperature in the context of the Standard Model. Hence, the effective potential in the Standard Model is presented below in terms of the Higgs field which determines the electroweak symmetry breaking.

The theory of electroweak baryogenesis is primarily based on Sakharov's third condition for baryogenesis, the out-of-equilibrium condition. In electroweak baryogenesis, this condition is satisfied by requiring a strong enough first-order electroweak phase transition. In the Standard Model, the electroweak phase transition is a first-order phase transition \cite{Anderson:1991zb, Quiros:1999jp, Arnold:1992rz, Carrington:1991hz, Morrissey:2012db, Dine:1992wr, Dolan:1973qd, Senaha:2020mop}, but it is insufficiently strong to generate the baryon asymmetry in the Universe. This is demonstrated in detail here and the problem of baryon asymmetry was discussed further in Chapter \ref{Electroweak Baryogenesis}.

\section{Effective Potential}
First of all, the effective potential in the Standard Model is expressed in terms of the Higgs field and is derived by the Lagrangian in the Standard Model which is invariant under the \(SU(3)_C \times SU(2)_L \times U(1)_Y\) gauge transformations. More specifically, the Higgs doublet is introduced as
\begin{equation}\label{Higgs_def_2}
    H = \frac{1}{\sqrt{2}}\begin{pmatrix}
        \chi_1 + i \chi_2 \\
        \phi + i \chi_3
        \end{pmatrix} \, ,
\end{equation}
where \(\phi\) is the Higgs boson, \(\chi_1, \chi_2\), and \(\chi_3\) are the three Goldstone bosons. In this work, we consider only the dominant contributions to the one-loop finite temperature effective potential coming from the gauge bosons, the top quark, and the Higgs and Goldstone bosons. Namely, the top quark as the heaviest fermion has the most significant contribution among all fermions. It is straightforward to prove that the zeroth contribution to the effective potential is the tree-level potential given by,
\begin{equation}\label{eq:1.3}
    V_0 (h) = - \frac{\mu^{2}_H}{2} h^2 + \frac{\lambda_H}{4} h^4 \, ,
\end{equation}
where \(h\) denotes the real constant background field, \(m_H = \sqrt{2}\mu_H \) is the Higgs mass, and \(\lambda_H > 0 \) is the Higgs self-coupling. The tree-level potential is minimized by the background field, which is the Higgs vacuum expectation value at tree level,
\begin{equation*}
    h = \upsilon = \frac{\mu_H }{\sqrt{\lambda_H}}\, .
\end{equation*}
By definition, the effective mass-squared of the Higgs field reads
\begin{equation}
    m_h^2 (h) = - \mu^2_H + 3\lambda_H h^2.
\end{equation}
The effective masses of the fields that are coupled to the Higgs field are derived from the Standard Model Lagrangian which was presented in section \ref{Standard Model} using the general definition of the effective mass-squared. For instance, the effective mass of the Goldstone boson \(\chi_i\) comes from the Higgs potential (\ref{Higgs_Goldstone}) using the Higgs doublet (\ref{Higgs_def_2}). As a result, the effective mass-squared of the Goldstone boson \(\chi_i\) is
\begin{equation}
    m_{\chi}^2 (h) = - \mu^2_H + \lambda_H h^2 \, .
\end{equation}
with \(i = 1, 2, 3\). In addition, the effective mass of the \(W\) bosons is derived using Eqs. (\ref{g_m_1}) and (\ref{m_W}), 
\begin{equation}\label{effectivemassW_1}
    m_W^2 (h) = \frac{g^2}{4} h^2,
\end{equation}
where \(g\) is the \(SU(2)_L\) gauge coupling. The \(Z\)-boson effective mass is similarly computed using  
(\ref{Z_A_MASS}) and (\ref{m_Z}),
\begin{equation}\label{effectivemassZ_1}
    m^2_Z (h) = \frac{g^2 + g^{\prime 2}}{4} h^2,
\end{equation}
where \(g^{\prime}\) is the \(U(1)_Y\) gauge coupling. As expected, the photon has zero effective mass-squared. Finally, the top-quark effective mass-squared is calculated using the relevant term in the Standard Model Lagrangian, which is similar to the term (\ref{quark_mass}) and leads to the up-quark mass (\ref{m_u}). Likewise, the effective mass-squared of the top quark reads
\begin{equation}\label{effectivemasstop_1}
    m^2_t (h) = \frac{y^2_t}{2} h^2,
\end{equation}
where \(y_t\) is the top quark Yukawa couplings. In the numerical results, the above masses at the vacuum state \(h = \upsilon\) are considered as \(m_H = 125\) GeV, \(m_W = 80.4\) GeV, \(m_Z = 91.2\) GeV and \(m_t = 173\) GeV \cite{ParticleDataGroup:2022pth}. 

\subsection{Zero-Temperature Corrections}
The CW potential in the Standard Model is simply computed by Eqs. (\ref{finalV1}), (\ref{fermi_eff}), and (\ref{gauge_eff}) in the Landau gauge\footnote{The one-loop effective potential in the Standard Model is comprehensively presented in the Fermi and background field \(R_\xi\) gauges in Refs. \cite{DiLuzio:2014bua, Devoto:2022qen, Patel:2011th, Kapusta:2006pm}. The background field \(R_\xi\) gauge is a useful gauge to compute the effective potential promoting the Higgs vacuum expectation value to the constant background field.}. However, these integrals have to be regularized since they are ultraviolet divergent as shown in Chapter \ref{Effective Action}. Namely, the divergent contributions are canceled by the counterterms
\begin{equation}\label{MS_Counter}
    V_1^{ct} (h) = \delta \Omega + \frac{\delta \mu^2}{2} h^2 + \frac{\delta \lambda}{4} h^4 \,.
\end{equation}
As a result, the renormalized effective potential is finite and dependent on the applied regularization and, correspondingly, on the renormalization conditions. This section simply continues the discussion on the \(\overline{\text{MS}}\) and OS renormalization scheme in Chapter \ref{Effective Action}.

\subsubsection{Modified Minimal Subtraction Renormalization}
We initially consider the Eqs. (\ref{bef_CUV}), (\ref{before_CUV_2}), and (\ref{before_CUV_3}) to demonstrate the \(\overline{\text{MS}}\) renormalization scheme in the Standard Model effective potential. This scheme was presented in Chapter \ref{Effective Action} and the following procedure is very similar. In particular, the terms, such as the term (\ref{Ccc}), are subtracted and canceled by the counterterms in Eq. (\ref{MS_Counter}). Therefore, the renormalized one-loop effective potential at zero temperature is written as
\begin{equation}
    V_{\text{eff}} (h, T = 0) = V_0 (h) + \sum_{i} V^i_1 (h)
\end{equation}
with
\begin{equation}
    V^i_1 (h) = (-1)^{F_i} n_i \frac{m^4_{i}(h)}{64 \pi^2} \left[ \ln{ \frac{m^2_{i}(h)}{\mu^2_R}} - C_i \right],
\end{equation}
where \(i = \{h, \chi, W, Z, t \}\) counts the fields that mainly contribute to the CW potential, \(F_i = 1\) \((0)\) for fermions (bosons), \(n_i\) is the number of degrees of freedom of each field \(i\), \(\mu_R\) denotes the renormalization scale, and \(C_i = 3/2\) \((5/6)\) for scalars and fermions (gauge bosons). The degrees of freedom of each particle \(i\) are,
\begin{equation}\label{eq:11}
n_h = 1, \quad n_{\chi} = 3, \quad n_{W} = 6, \quad n_{Z} = 3, \quad n_{t} = 12.
\end{equation}
The infinities are canceled by the counterterms which are given by \cite{Quiros:1999jp}
\begin{equation}
    \delta \Omega = \frac{\mu^4_H}{64 \pi^2} (n_h + n_{\chi}) \left( \frac{1}{2-\frac{n}{2}} - \gamma_E + \ln 4 \pi \right) \, ,
\end{equation}
\begin{equation}
    \delta m^2 = - \frac{3 \lambda_H \mu^2_H}{16 \pi^2} \left(n_h + \frac{1}{3} n_{\chi} \right) \left( \frac{1}{2-\frac{n}{2}} - \gamma_E + \ln 4 \pi \right) \, ,
\end{equation}
and
\begin{equation}
    \delta \lambda = \frac{3 }{16 \pi^2} \left[\frac{2g^4 + (g^2 + g^{\prime 2})^2}{16} - y^4_t +   \left(3n_h + \frac{1}{3} n_{\chi} \right) \lambda_H^2 \right] \left( \frac{1}{2-\frac{n}{2}} - \gamma_E + \ln 4 \pi \right) \, .
\end{equation}
It is interesting that the mass counterterms \(\delta \mu^2\) and \(\delta \Omega\) are generated by the Higgs sector.

\subsubsection{Cut-off Regularization}
First, the OS scheme involves regularizing the effective potential with a momentum-space cut-off \(p^2_E = \Lambda^2\). We additionally require that the Higgs minimum, \( h = \upsilon\), and the Higgs mass do not alter with respect to their tree level values \cite{Anderson:1991zb},
\begin{equation}\label{ren_con11}
    \left. \frac{d \left( V_1 + V_1^{ct}\right)}{d h} \right|_{\upsilon} = 0
\end{equation}
and
\begin{equation}\label{ren_con_22}
    \left. \frac{d^2 \left( V_1 + V_1^{ct}\right)}{d h^2} \right|_{\upsilon} = 0 \, .
\end{equation}
In view of (\ref{one_reg}), the one-loop correction to the classical potential at zero temperature is expressed as
\begin{equation}\label{infinity}
    V_1(h) = \frac{1}{32 \pi^2} \sum_i n_i \left[m_i^2 (h) \Lambda^2 + \frac{m_i^4 (h)}{2} \left( \ln \frac{m_i^2 (h)}{\Lambda^2} - \frac{1}{2}\right) \right]
\end{equation}
The infinities in (\ref{infinity}) cancel against those in the counterterms (\ref{MS_Counter}) by imposing the renormalization conditions (\ref{ren_con11}) and (\ref{ren_con_22}). Therefore, the one-loop effective potential at zero temperature in this scheme is
\begin{equation}
    V_{\text{eff}} (h, T = 0) = V_0 (h) + V_1 (h) \, ,
\end{equation}
where the CW potential is
\begin{equation}\label{OS}
     V_1 (h) = \sum_{i} \frac{(-1)^{F_i} n_i}{64 \pi^2} \left[  m^4_{i}(h)\left( \ln  \frac{m^2_{i}(h)}{m^2_{i}(\upsilon)} - \frac{3}{2} \right) + 2 m^2_{i}(h) m^2_{i}(\upsilon) \right],
\end{equation}
Now the counterterms in (\ref{MS_Counter}) are expressed as \cite{Quiros:1999jp}
\begin{equation}
    \delta \lambda = - \frac{1}{16 \pi^2} \sum_i n_i \left(\frac{m_i^2 (\upsilon) - b_i}{\upsilon^2}\right)^2 \left( \ln \frac{m_i^2 (\upsilon)}{\Lambda^2} + 1\right)\,,
\end{equation}
\begin{equation}\label{dm2}
    \delta \mu^2 = - \frac{1}{16 \pi^2} \sum_i n_i \left(\frac{m_i^2 (\upsilon) - b_i}{\upsilon^2}\right) \left[ \Lambda^2 - m_i^2 (\upsilon) + b_i\left( \ln \frac{m_i^2 (\upsilon)}{\Lambda^2} + 1\right)\right] \,,
\end{equation}
and
\begin{equation}\label{dO}
    \delta \Omega = \frac{\mu_H^2}{32 \pi^2} \sum_{k = h, \chi} n_k \left[ \Lambda^2 - m_k^2 (\upsilon) + \frac{b_k}{2}\left( \ln \frac{m_k^2 (\upsilon)}{\Lambda^2} + 1\right)\right] \, ,
\end{equation}
where \(b_W = b_Z = b_t = 0\) and \(b_h = b_{\chi} = - \mu_H^2\). One observes that the cosmological constant (\ref{dO}) does not appear if the contribution from the Higgs sector is neglected. In contrast with the \(\overline{\rm MS}\) scheme, the mass counterterm (\ref{dm2}) also comes from the contribution of the gauge boson and fermion loops in the effective potential.

Finally, the dependence of the effective potential on the renormalization scale in \(\overline{\rm MS}\) renormalization scheme is crucial as it introduces a theoretical uncertainty in our calculations. In particular, the choice of the renormalization scale leads to an uncertainty in the critical temperature \(T_c\) and other related quantities, especially in the extensions of the Standard Model \cite{Chiang:2018gsn,Athron:2022jyi, Croon:2020cgk, Gould:2021oba}. Different schemes, such as the OS scheme, can be followed to compute the CW potential and handle this uncertainty. In the literature, the OS-like scheme is commonly adopted, since it has a fixed prescription for the renormalization scales for each particle. The OS-like scheme differs from the OS scheme since the mass is not defined at the pole position of the Higgs boson propagator. The renormalization scale-dependence could be also eliminated by the Renormalization Group Equations (RGE) improvement for the CW potential \cite{Andreassen:2014eha, Andreassen:2014gha, Bando:1992np, Tamarit:2014dua, Quiros:1999jp}, and this approach is left to be implemented in future studies.

\subsection{Finite-Temperature Corrections}

The effective potential at finite temperature contains the effective potential at zero temperature which was presented in the previous section. The temperature-dependent term in the Standard Model effective potential is obtained by the contributions of the scalar, fermion, and gauge fields in Eqs. (\ref{V_scalar_1_T}), (\ref{eff-T-fermion}), and (\ref{eff-T-gauge}), respectively. In general, each field \(i\) contributes a temperature-dependent term as follows
\begin{equation}\label{T-potential1}
V_T^i(h,T) = (-1)^{F_i} \frac{n_iT^4}{2 \pi^2} \int_{0}^{\infty} dx \,    x^2 \ln \left[ 1 - (-1)^{F_i} \exp \left( - \sqrt{x^2 + \frac{m^2_{i}  (h)}{T^2} }\right) \right],
\end{equation}
where \(i = \{h, \chi, W, Z, t \}\). In the Standard Model, the high-temperature expansion of the thermal functions is commonly adopted to evaluate the full finite-temperature one-loop effective potential as it is valid and quite precise compared to the numerical evaluation.

Thermal resummation is essential to incorporate the terms that are of the same order in the coupling constant as the cubic term in the high-temperature expansion of the effective potential \cite{Carrington:1991hz}. In general, the thermal resummation introduces the thermal mass defined as,
\begin{equation}
    M^2_i (h,T) = m^2_i (h) + \Pi_i (h, T) \, ,
\end{equation}
where \(\Pi_i(T)\) is the temperature-dependent self-energy corresponding to the one-loop resummed diagrams to leading powers of the temperature. The thermal mass of each field in the Standard Model is obtained by following the same procedure as in Chapter \ref{Electroweak Phase Transition}. Our approach is based on Ref. \cite{Carrington:1991hz}. Firstly, the temperature-dependent self-energy for the scalar fields is decomposed into the contributions from the Higgs field, the top quark field, and the \(SU(2)_L\) and \(U(1)_Y\) gauge bosons,
\begin{equation}\label{T-higgs1}
    \Pi_h (T) = \Pi_{\chi} (T) =  \pi^{(A^{a}_{\mu})}_h (0) + \pi^{(B_{\mu})}_h (0) + \pi^{(h)}_h (0) + \pi^{(\psi)}_h (0),
\end{equation}
where the corresponding contributions are
\begin{equation}
    \pi^{(A^{a}_{\mu})}_h (0) = \frac{g^2}{8} T^2, \quad \pi^{(B_{\mu})}_h (0) = \frac{g^2 + g^{\prime 2}}{16} T^2,  \quad \pi^{(h)}_h (0) = \frac{\lambda_H}{2} T^2, \quad \pi^{(\psi)}_h (0) = \frac{y^2_t}{4} T^2 .
\end{equation}
Therefore, the temperature-dependent self-energy (\ref{T-higgs1}) reads
\begin{equation}\label{T-higgs11}
    \Pi_h (T) = \Pi_{\chi} (T) = \left(\frac{3g^2 }{16} + \frac{g^{\prime 2}}{16}  +\frac{y^2_t }{4} + \frac{\lambda_H}{2} \right) T^2 \, .
\end{equation}
The thermal corrections of the gauge boson masses require special treatment as the transverse gauge fields have zero thermal corrections \cite{Carrington:1991hz},
\begin{equation*}
    \Pi_{W_{T}} (h,T) = \Pi_{Z_{T}} (h,T) = \Pi_{\gamma_{T}} (h,T) = 0 \,.
\end{equation*}
In addition, the temperature-dependent self-energy of the longitudinal gauge bosons in the high-temperature limit can be computed as
\begin{equation}\label{T-W}
    \Pi_{W_L} (T) = \frac{11}{6}g^2 T^2
\end{equation}
\begin{equation}\label{T-Z}
    \Pi_{Z_L} (h, T) = \frac{11}{6} g^2T^2 + \frac{g^2}{4} h^2
\end{equation}
\begin{equation}\label{T-Photon}
    \Pi_{\gamma_L} (h,T) = \frac{11}{6}g^{\prime 2} T^2 - \frac{g^2}{4}h^2 \,.
\end{equation}
The gauge boson mass matrix in the basis \((A^{1}_{\mu}, A^{2}_{\mu}, A^{3}_{\mu}, B_{\mu})\) is non-diagonal and the inclusion of the thermal corrections \cite{Espinosa:1993bs} leads to
\begin{equation*}
    M^2_{GB} (h, T) = \begin{pmatrix}
                \frac{1}{4}g^2 h^2 + \frac{11}{6}g^2 T^2 & 0 & 0 & 0 \\
                0 & \frac{1}{4}g^2 h^2 + \frac{11}{6}g^2 T^2 & 0 & 0 \\
                0 & 0 & \frac{1}{4}g^2 h^2 + \frac{11}{6}g^2 T^2 & -\frac{1}{4}gg^{\prime}h^2 \\
                0 & 0 & -\frac{1}{4}gg^{\prime}h^2  & \frac{1}{4}g^{\prime 2} h^2 + \frac{11}{6}g^{\prime} T^2
                    \end{pmatrix}.
\end{equation*}
Nevertheless, the photon and the \(Z\) boson are not mass eigenstates at high temperatures, since there is an additional mixing term between the \(Z\) boson and the photon \cite{Croon:2020cgk}. Therefore, the gauge boson mass matrix has the following eigenvalues which correspond to the actual thermal masses for the longitudinal photon and \(Z\) boson: 
\begin{equation}\label{Z-thermalmass}
\begin{split}
    M^2_{Z_L} (h,T) = & \,  \frac{1}{2} \left[ \frac{1}{4} \left(g^2 + g^{\prime 2}\right) h^2 + \frac{11}{6} \left(g^2 + g^{\prime 2} \right) T^2\right] \\
    & +  \frac{1}{2}\sqrt{\left(g^2 - g^{\prime 2} \right)^2 \left( \frac{1}{4} h^2 + \frac{11}{6}  T^2 \right)^2  + \frac{g^2 g^{\prime 2}}{4} h^4 }
\end{split}  
\end{equation}
and
\begin{equation}\label{Photon-thermalmass}
\begin{split}
     M^2_{\gamma_L} (h, T) = & \, \frac{1}{2} \left[ \frac{1}{4} \left(g^2 + g^{\prime 2}\right) h^2 + \frac{11}{6} \left(g^2 + g^{\prime 2} \right) T^2\right] \\
     & -  \frac{1}{2}\sqrt{\left(g^2 - g^{\prime 2} \right)^2 \left( \frac{1}{4} h^2 + \frac{11}{6}  T^2 \right)^2  + \frac{g^2 g^{\prime 2}}{4} h^4 } \,.
\end{split}
\end{equation}
In the limit \(m_W^2 (h)/T^2 \ll 1 \), the actual resummed mass eigenvalues (\ref{Z-thermalmass}) and (\ref{Photon-thermalmass}) can be effectively approximated by those in Eqs. (\ref{T-Z}) and (\ref{T-Photon}). The numerical difference between these expressions is small, which indicates that one could treat the photon and \(Z\) boson as mass eigenstates \cite{Croon:2020cgk}.

It has been previously mentioned that thermal resummation can be implemented using either the Parwani \cite{Parwani:1991gq} or the Arnold-Espinosa scheme \cite{Arnold:1992rz}. A comparison between the two schemes is presented in Refs. \cite{Athron:2022jyi, Croon:2020cgk}, where it is evident that the numerical differences between the two schemes in the critical temperature and the ratio \(\upsilon_c/T_c\) are small. In the Arnold-Espinosa scheme, the resummed full effective potential generalizes the result in (\ref{ring_correction_1}),
\begin{equation}\label{Arnold-Espinosa_1}
     V_{\text{eff}} (h,T) = V_0 (h) + \sum_{i} \left[ V^i_1 \left(m^2_i(h)\right) + V^i_T \left(m^2_i(h),T \right) + V^i_{ring} \left(m^2_i(h),T\right) \right],
\end{equation}
where the last term includes the ring diagram contributions,
\begin{equation}\label{rings_1}
    V^i_{ring} \left(m^2_i(h),T \right) = \frac{\overline{n}_i T}{12\pi} \left[m^3_i(h) - \left(m^2_i(h) + \Pi_i(T) \right)^{3/2} \right],
\end{equation}
where \(i = \{h, \chi, W, Z, \gamma\}\) and \(\overline{n}_i = \{1, 3, 2, 1, 1\}\) is the modified number of degrees of freedom which takes into account that only the longitudinal polarizations of the gauge bosons contribute to the temperature-dependent self-energy \cite{Arnold:1992rz, Carrington:1991hz}.

Finally, it is significant to mention that the finite-temperature one-loop effective potential can have imaginary contributions if the squared effective masses become negative. This occurs especially for the scalar fields due to the logarithmic and cubic terms in the thermal functions (\ref{bosonthermalfunction_1}) and (\ref{fermionthermalfunction_1}), since the gauge bosons and the top quark have always a positive squared effective mass. In general, the effective mass in the logarithmic term is canceled by its counterpart in the one-loop zero-temperature correction. Moreover, the thermal resummation could potentially cure the imaginary part originating from the cubic term which signals the breakdown of the perturbative expansion\footnote{In Ref.\cite{Weinberg:1987vp}, the authors argue that the imaginary part of the effective potential may be interpreted as a decay rate of a state of the scalar fields.}. Nevertheless, the \(m^2_i(h) + \Pi_i(T)\) in the ring correction can be negative for certain temperatures and field values. As a result, the effective potential is still complex and we consider only the real part of the full effective potential and ensure the field's stability during the phase transition, provided its imaginary counterpart remains sufficiently insignificant.

\section{Electroweak Phase Transition: Full Analysis}
The dynamics of the electroweak phase transition are described by the finite-temperature effective potential in terms of the Higgs background field including the dominant contributions of the gauge bosons, the top quark, and the Goldstone bosons. In the following, the \(\overline{\rm MS}\) renormalization scheme is adopted to avoid the issue with the infrared divergence that originates from the Goldstone bosons in the OS schemes. Then, in the Arnold-Espinosa scheme, the full one-loop finite-temperature effective potential is generally written as
\begin{equation}\label{eff_SM}
\begin{split}
    V^{SM}_{eff} (h, T)  =& - \frac{\mu^{2}_H}{2} h^2 + \frac{\lambda_H}{4} h^4 + \sum_{i} (-1)^{F_i} n_i \frac{m^4_{i}(h)}{64 \pi^2}\left[ \ln \left( \frac{m^2_{i}(h)}{\mu^2_R}\right) - C_i \right] \\
    & + \sum_{i} (-1)^{F_i} \frac{n_i T^4}{2 \pi^2} J_{B/F} \left(\frac{m^2_i (h)}{T^2}\right) \\
    & + \sum_{k} \frac{\overline{n}_k T}{12\pi} \left[m^3_k(h) - \left(M^2_k (h,T) \right)^{3/2} \right],
\end{split}
\end{equation}
where \(k = \{h, \chi, W, Z, \gamma\}\) corresponds to the bosons in the Standard Model.

The high-temperature expansion of the thermal functions (\ref{bosonthermalfunction_1}) and (\ref{fermionthermalfunction_1}) for the temperature-dependent one-loop effective potential is valid with very high accuracy in the case of the electroweak phase transition in the Standard Model since the ratio \(m_i/T\) was computed for each particle and satisfied the high-temperature limit given by the numerical analysis, which was discussed in Chapter \ref{Finite-Temperature Field Theory}. Hence, the high-temperature expansion is implemented at the full effective potential which reads
\begin{equation}\label{eff_SM_HT}
\begin{split}
    V^{SM}_{eff} (h, T)  =& - \frac{\mu^{2}_H}{2} h^2 + \frac{\lambda_H}{4} h^4 \\
    & + \frac{m^2_h (h)}{24}T^2 - \frac{T}{12 \pi} \left[M^2_h (h, T)\right]^{3/2} + \frac{m^4_h(h)}{64 \pi^2} \left[\ln \left(\frac{a_b T^2}{\mu^2_R}\right) -\frac{3}{2} \right] \\
    & + \frac{3 m^2_{\chi} (h)}{24}T^2 - \frac{3T}{12 \pi} \left[ M^2_{\chi} (h, T) \right]^{3/2} + \frac{ 3 m^4_{\chi} (h)}{64 \pi^2} \left[\ln \left(\frac{a_b T^2}{\mu^2_R}\right) -\frac{3}{2} \right]  \\
    & + \frac{6 m^2_{W} (h)}{24}T^2 - \frac{4T}{12 \pi} m^3_{W} (h) - \frac{2T}{12 \pi}\left[ M^2_{W_L} (h,T)\right]^{3/2} + \frac{ 6 m^4_{W} (h)}{64 \pi^2} \left[\ln \left(\frac{a_b T^2}{\mu^2_R}\right) -\frac{5}{6} \right] \\
    & + \frac{3 m^2_{Z} (h)}{24}T^2 - \frac{2T}{12 \pi} m^3_{Z} (h) - \frac{T}{12 \pi} \left[  M^2_{Z_{L}} (h, T) \right]^{3/2} + \frac{ 3 m^4_{Z} (h)}{64 \pi^2} \left[\ln \left(\frac{a_b T^2}{\mu^2_R}\right) -\frac{5}{6} \right] \\
    & + \frac{12 m^2_{t} (h)}{48}T^2 - \frac{ 12 m^4_{t} (h)}{64 \pi^2} \left[\ln \left(\frac{a_f T^2}{\mu^2_R}\right) -\frac{3}{2} \right] - \frac{T}{12 \pi}\left[ M^2_{\gamma_{L}} (h, T) \right]^{3/2},
\end{split}
\end{equation}
where each term is defined above. It is interesting to notice that only the cubic terms of the gauge bosons do not cancel each other in the high-temperature expansion of the effective potential and the ring corrections because the rest of the fields have the same degrees of freedom in the ring correction. 

\begin{figure}[h!]
\centering
\includegraphics[width=34pc]{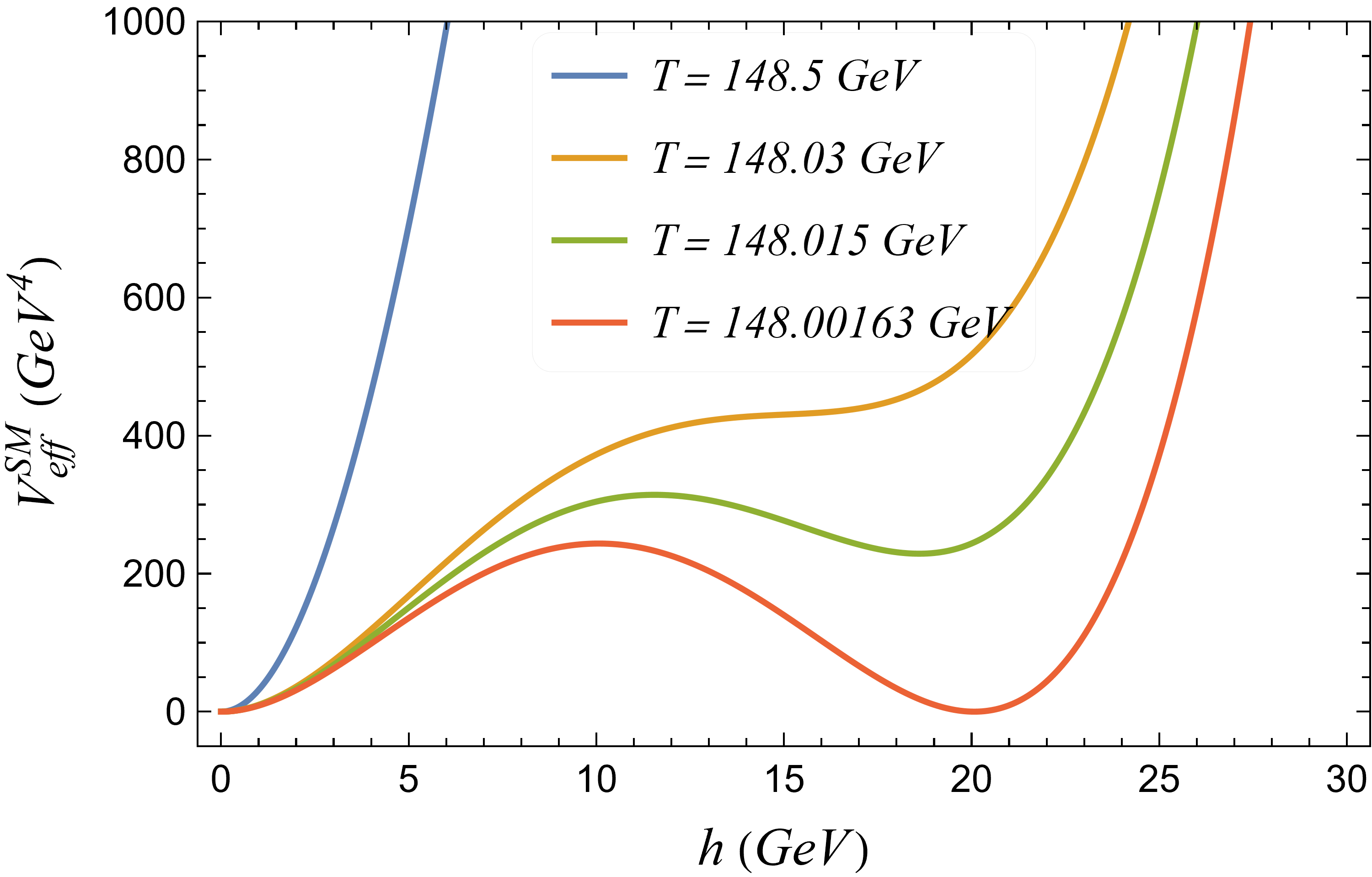}
\caption{The finite-temperature effective potential of the Standard Model computed for temperatures higher than the critical temperature using \(\mu_R = \upsilon\).}\label{electroweak phase transition_in_SM1}
\end{figure}

\begin{figure}[h!]
\centering
\includegraphics[width=34pc]{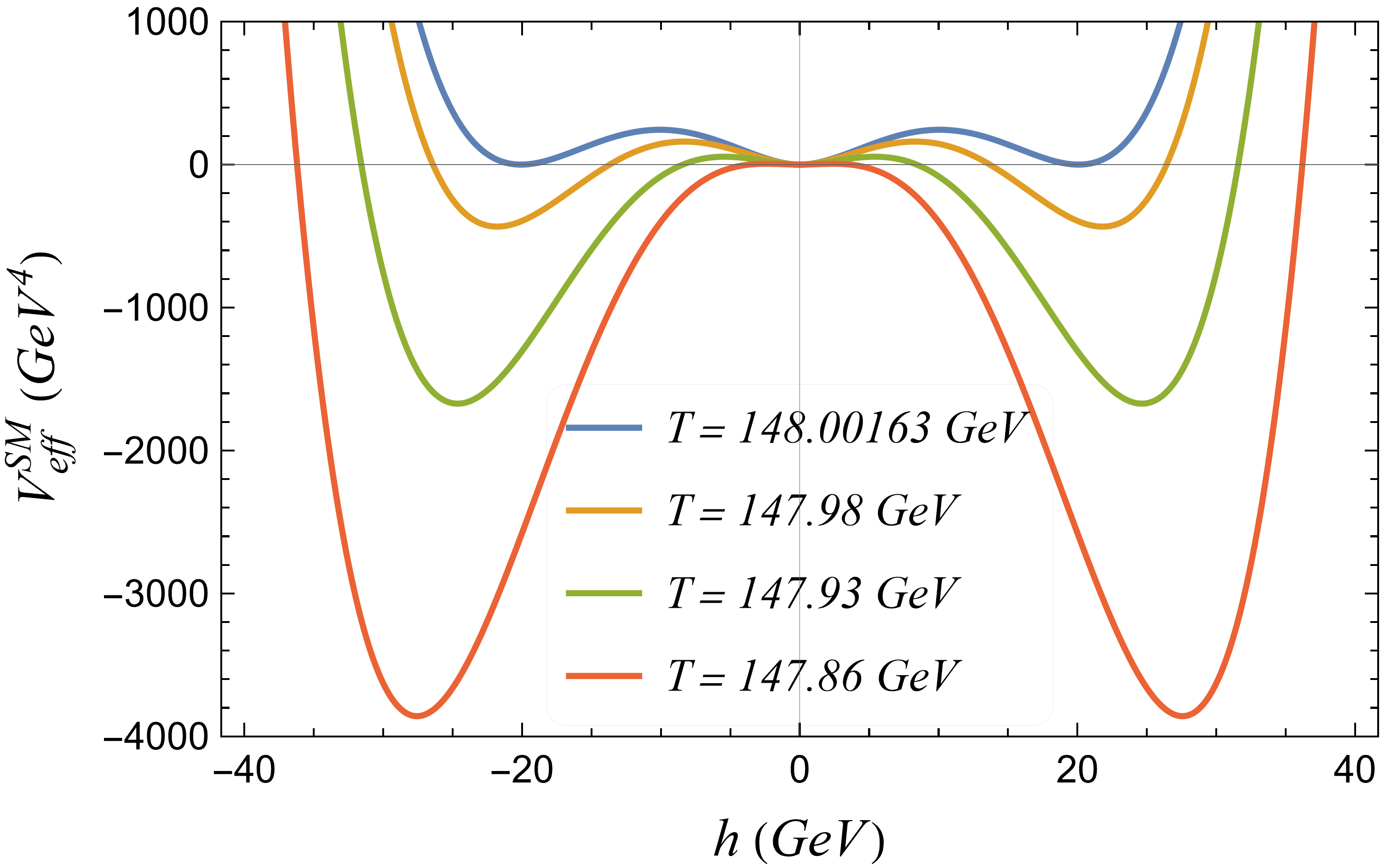}
\caption{The finite-temperature effective potential of the Standard Model computed for temperatures around the critical temperature using \(\mu_R = \upsilon\).}\label{electroweak phase transition_in_SM2}
\end{figure}

\begin{figure}[h!]
\centering
\includegraphics[width=33.4pc]{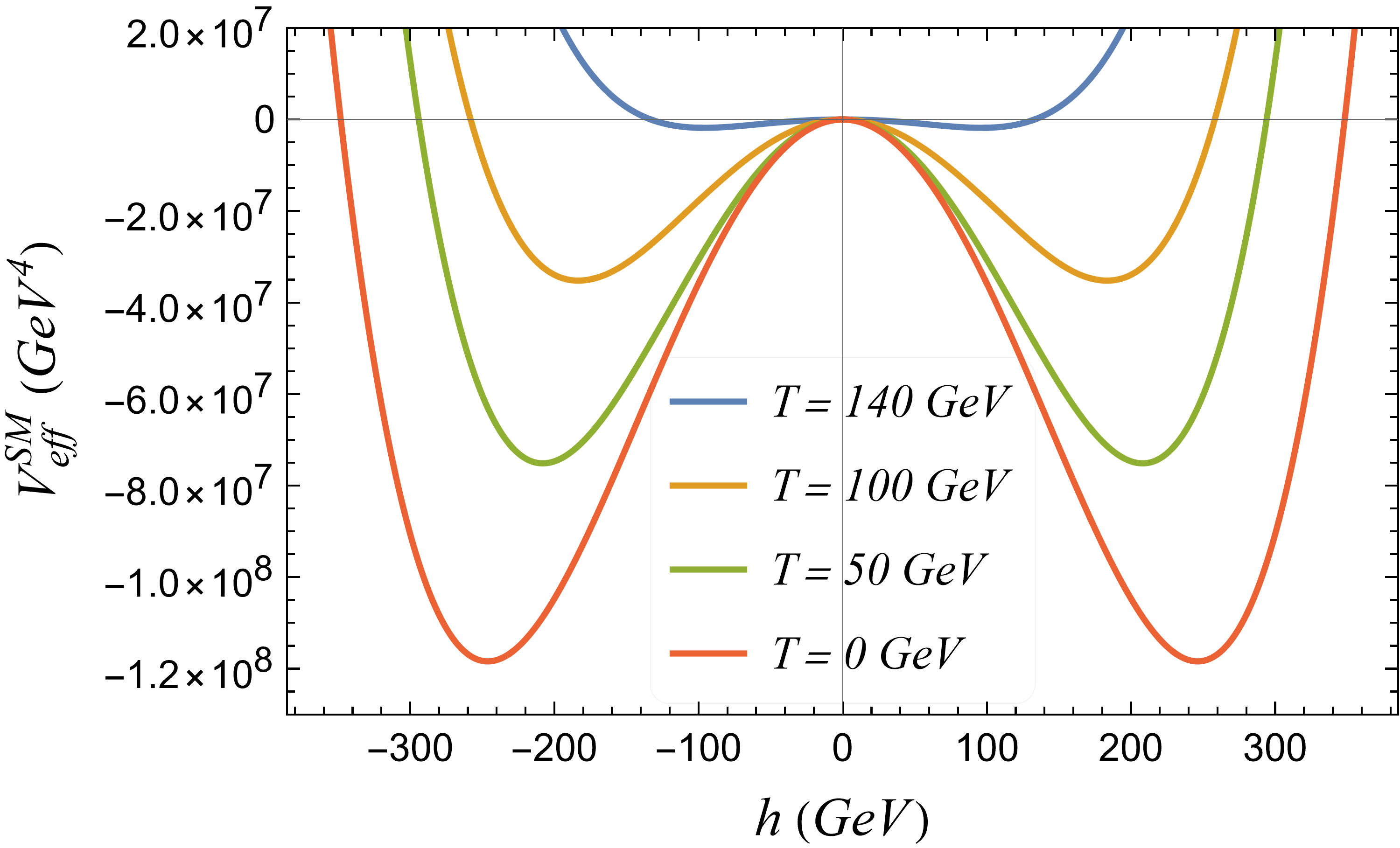}
\caption{The finite-temperature effective potential of the Standard Model computed for temperatures lower than the critical temperature using \(\mu_R = \upsilon\).}\label{electroweak phase transition_in_SM3}
\end{figure}

\begin{figure}[h!]
\centering
\includegraphics[width=34pc]{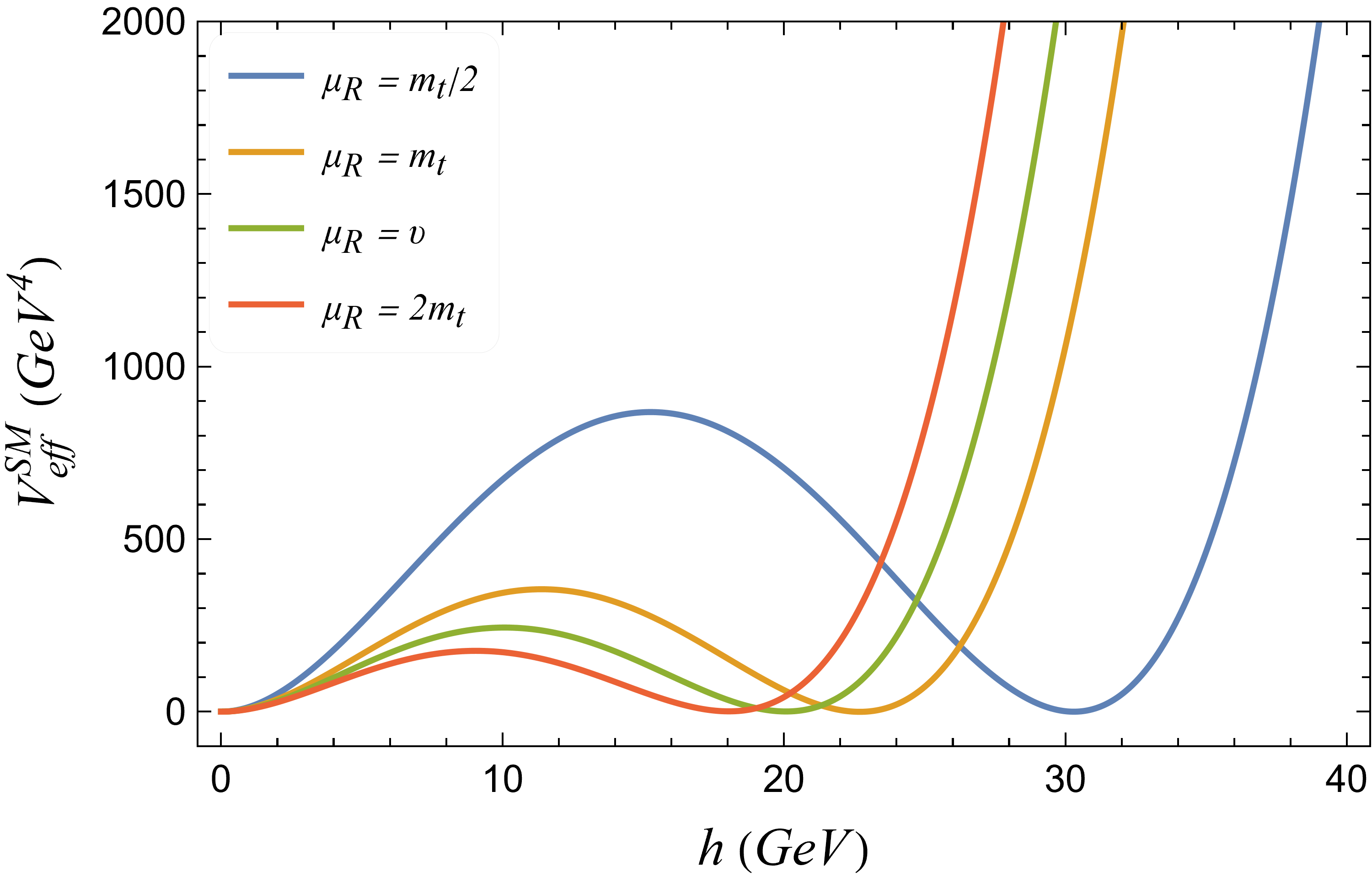}
\caption{The finite-temperature effective potential of the Standard Model at the critical temperature for different renormalization scales \(\mu_R = m_t/2\) (blue), \(m_t\) (orange), \(\upsilon\) (green) and \(2m_t\) (red).}\label{electroweak phase transition_in_SM}
\end{figure}

Now, our main goal is to analyze in detail the phase transition that occurs at a finite temperature. The electroweak phase transition could be studied by varying the renormalization scale for different values, but the renormalization scale \(\mu_R = \upsilon\) is commonly used in the Standard Model as a natural choice. In Figs. \ref{electroweak phase transition_in_SM1}, \ref{electroweak phase transition_in_SM2}, and \ref{electroweak phase transition_in_SM3}, the one-loop finite-temperature effective potential is presented using the renormalization scale \(\mu_R = \upsilon\) to show its general evolution as the temperature decreases gradually.
Furthermore, the finite-temperature effective potential (\ref{eff_SM_HT}) at the critical temperature is illustrated in Fig. \ref{electroweak phase transition_in_SM} for \(\mu_R = m_t/2\), \(m_t\), \(\upsilon\) and \(2m_t\) and Table \ref{table:1} presents the dependence of the critical temperature and the critical Higgs vacuum expectation value on the renormalization scale.

\begin{table}[H]
\centering
\begin{tabularx}{0.5\textwidth} {
  | >{\centering\arraybackslash}X
  | >{\centering\arraybackslash}X
  | >{\centering\arraybackslash}X| }
 \hline
 \(\mu_R\) (GeV) & \(T_c\) (GeV) & \(\upsilon_c\) (GeV)\\
 \hline
 86.5 & 148.987 &  30\\
 \hline
 173.0 & 148.326 & 23\\
 \hline
 246.2 & 148.002 & 20\\
 \hline
 346.0 & 147.693 & 18\\
 \hline
\end{tabularx}
\caption{The critical temperature of the electroweak phase transition (\(T_c\)) and the Higgs vacuum expectation value (\(\upsilon_c\)) for different values of the renormalization scale.} \label{table:1}
\end{table}
In view of Table \ref{table:1}, the electroweak phase transition is not a strong enough first-order phase transition since the criterion (\ref{criterion_baryogenesis}) is not satisfied. The maximum ratio between the Higgs vacuum expectation value and the critical temperature is evaluated for \(\mu_R = m_t/2\) with \(\upsilon_c/T_c \simeq 0.2 < 0.6\). Similarly, the authors of Ref. \cite{Anderson:1991zb} argued that the sphaleron rate condition is not satisfied for \(m_H = 125\) GeV in the Standard Model. In particular, they considered the high-temperature expansion to calculate the ratio \(\upsilon_c/ T_c \simeq 0.2\) for \(m_H = 120\) GeV and \(m_t = 170\) GeV, without taking into account the thermal resummation. Moreover, similar calculations can be found in Refs. \cite{Anderson:1991zb, Quiros:1999jp,Carrington:1991hz,Morrissey:2012db,Senaha:2020mop}. In Table \ref{table:1}, one also observes that while at the critical temperature, the height of the barrier in the effective potential and the Higgs vacuum expectation value change for various renormalization scales, the critical temperature is slightly dropped by less than \(1 \%\) for higher values of the renormalization scale.

In Chapter \ref{Electroweak Baryogenesis}, we demonstrated that the criterion for electroweak baryogenesis (\ref{criterion_baryogenesis}) can be easily translated into an upper bound on the Higgs in order to require a strong first-order phase transition. On the other hand, in lattice calculations, this bound can be computed as \(m_h \lesssim 72 - 80\) GeV in lattice calculations \cite{Gurtler:1997hr,Laine:1998jb,Csikor:1998eu,Aoki:1999fi}. Therefore, the lack of a strong electroweak phase transition in the Standard Model can be understood by these results as well.

\begin{figure}[h!]
\centering
\includegraphics[width=35pc]{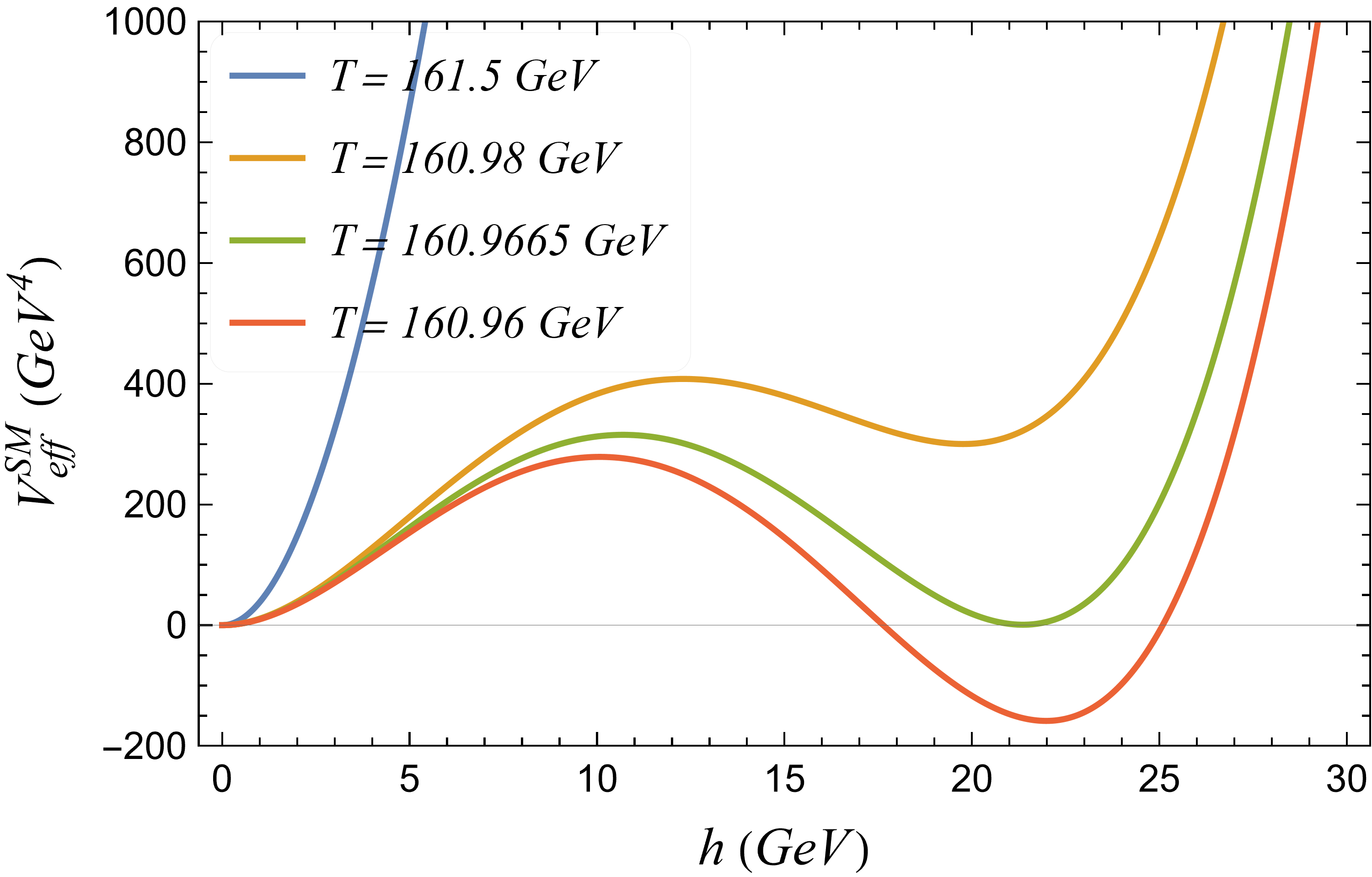}
\caption{The finite-temperature effective potential of the Standard Model in the OS renormalization scheme computed for different temperatures.}\label{electroweak phase transition_in_SM_OS}
\end{figure}

For the sake of completeness, the one-loop finite-temperature effective potential in the OS renormalization scheme is briefly presented to showcase the differences between the renormalization schemes. More specifically, the one-loop finite-temperature effective potential can be then written as
\begin{equation}\label{eff_SM_2}
\begin{split}
    V^{SM}_{eff} (h, T)  = & - \frac{\mu^{2}_H}{2} h^2 + \frac{\lambda_H}{4} h^4 \\
    & + \sum_{i} \frac{(-1)^{F_i} n_i}{64 \pi^2} \left[  m^4_{i}(h)\left( \ln  \frac{m^2_{i}(h)}{m^2_{i}(\upsilon)} - \frac{3}{2} \right) + 2 m^2_{i}(h) m^2_{i}(\upsilon) \right]\\
    & + \sum_{i} (-1)^{F_i} \frac{n_i T^4}{2 \pi^2} J_{B/F} \left(\frac{m^2_i (h)}{T^2}\right) \\
    & + \sum_{k} \frac{\overline{n}_k T}{12\pi} \left[m^3_k (h) - \left(M^2_k(h,T) \right)^{3/2} \right],
\end{split}
\end{equation}
where \(k = \{h, \chi, W, Z, \gamma\}\) and the one-loop contribution to the effective potential at zero temperature in the \(\overline{\rm MS}\) scheme was simply replaced by the CW potential given by (\ref{OS}) in the OS scheme. However, the effective mass of the Goldstone bosons vanishes at the Higgs vacuum expectation value \(h = \upsilon\). In practice, the infrared divergences can be canceled by the addition of the following term as shown in Ref. \cite{Delaunay:2007wb},
\begin{equation}
    \delta V_1 (h) = \frac{3 m^4_{\chi} (h)}{64 \pi^2} \ln m^2_{\chi},
\end{equation}
which makes the final result finite. In this scheme, the effective potential is presented in Fig. \ref{electroweak phase transition_in_SM_OS} for various temperatures in the early Universe and we computed the critical temperature \(T_c = 160.9665\) GeV and the ratio \(\upsilon_c/T_c = 0.13\) numerically. It is remarkable that the numerical results for the critical temperature do not differ significantly from the ones in the \(\overline{\rm MS}\) scheme. Regarding the Higgs vacuum expectation value at the critical temperature and the height of the barrier, the results of the OS scheme approach the results of the \(\overline{\rm MS}\) scheme for \(\mu_R = m_t\).

Lastly, I would like to comment that the Higgs tree-level potential could be extended by higher-order operators of \(H^{\dagger} H\) as discussed in Refs. \cite{Delaunay:2007wb, Chala:2018ari, Bodeker:2004ws, Noble:2007kk, Huang:2016odd, Hashino:2022ghd}. However, these higher-order terms would have coefficients with dimensions of inverse masses to some power and lead to non-renormalizable theories.

\chapter{Singlet Extensions of the Standard Model}\label{Singlet Extensions to the Standard Model}

\section{Introduction}
High energy physics is now increasingly relying on gravitational wave experiments and astrophysical observations to tackle fundamental challenges such as electroweak baryogenesis and dark matter. This shift comes in response to the lack of new particle detections at the LHC since the discovery of the Higgs boson. For instance, the stochastic gravitational wave background produced by a first-order phase transition could provide insights into electroweak baryogenesis. Therefore, gravitational wave experiments could play a crucial role in discovering new physics \cite{Apreda:2001us,Schabinger:2005ei,Kusenko:2006rh,McDonald:1993ex,Chala:2018ari,Davoudiasl:2004be,Baldes:2016rqn,Noble:2007kk,Zhou:2020ojf,Weir:2017wfa,Hindmarsh:2020hop,Han:2020ekm,Child:2012qg,Fairbairn:2013uta,LISACosmologyWorkingGroup:2022jok,Caprini:2015zlo,Huber:2015znp,Chung:2012vg,Barenboim:2012nh,Grojean:2006bp,Katz:2014bha,Alves:2018jsw,Athron:2023xlk}. 

A wide range of fundamental problems in particle physics and cosmology have motivated physicists to look for new physics beyond the Standard Model. Extensions to the Standard Model could potentially explain dark energy, dark matter, the strong \(CP\) problem, and the matter-antimatter asymmetry. In previous chapters, it was shown that the Standard Model cannot explain the observed baryon asymmetry in the Universe, as the electroweak phase transition is too smooth and the \(CP\) violation is insufficient. However, various extensions to the Standard Model may address these issues, changing the Higgs potential to predict a strong electroweak phase transition and introducing new sources of \(CP\) violation. This has been shown through approaches like the singlet scalar extensions \cite{Profumo:2007wc,Damgaard:2013kva,Ashoorioon:2009nf,OConnell:2006rsp,Gonderinger:2012rd,Profumo:2010kp,Gonderinger:2009jp,Barger:2008jx, Cheung:2012nb,Barger:2007im,Cline:2013gha,Burgess:2000yq,Kakizaki:2015wua,Enqvist:2014zqa,Katz:2014bha,Espinosa:1993bs,Alanne:2014bra,Cline:2012hg,Beniwal:2017eik,Curtin:2014jma,Dev:2019njv,Ghorbani:2018yfr,Ghorbani:2020xqv,Espinosa:2011ax,Espinosa:2007qk, Kurup:2017dzf,Alves:2018jsw, Zhang:2023nrs, Jain:2017sqm, Vaskonen:2016yiu, Huang:2018aja, Jiang:2015cwa, Grzadkowski:2018nbc}, the two-Higgs doublet models \cite{Cline:1995dg, Dorsch:2016nrg, Wang:2019pet, Zhou:2020irf, Goncalves:2021egx, Biekotter:2022kgf, Morais:2019fnm}, the composite Higgs models \cite{Espinosa:2011eu, Bruggisser:2018mrt, Chala:2016ykx, Xie:2020bkl, Bian:2019kmg, Aziz:2013fga}, and the higher-order operators in the Standard Model \cite{Delaunay:2007wb, Chala:2018ari, Bodeker:2004ws, Noble:2007kk, Huang:2016odd, Hashino:2022ghd}.

The simplest extension to the Standard Model is the real singlet extension, which adds a real singlet scalar field \(S\) coupled with the Higgs doublet \(H\) \cite{Profumo:2007wc,Damgaard:2013kva,Ashoorioon:2009nf,OConnell:2006rsp,Gonderinger:2012rd,Profumo:2010kp,Gonderinger:2009jp, Cheung:2012nb,Barger:2007im,Cline:2013gha,Burgess:2000yq,Kakizaki:2015wua,Enqvist:2014zqa,Katz:2014bha,Espinosa:1993bs,Alanne:2014bra,Cline:2012hg,Beniwal:2017eik,Curtin:2014jma,Dev:2019njv,Ghorbani:2018yfr,Ghorbani:2020xqv,Espinosa:2011ax,Espinosa:2007qk, Kurup:2017dzf,Alves:2018jsw, Zhang:2023nrs, Jain:2017sqm, Vaskonen:2016yiu, Huang:2018aja}. More specifically, in this extension, the Lagrangian density can be invariant under a \(\mathbb{Z}_2\) symmetry transformation,
\begin{equation}
    S \to - S,
\end{equation}
if it does not include cubic interaction terms which may appear in the most general renormalizable tree-level potential \cite{Profumo:2007wc, Gonderinger:2009jp, Zhang:2023nrs, Alanne:2014bra, Noble:2007kk}. This \(\mathbb{Z}_2\) symmetry allows us to study the consequences of a cosmological phase transition in the singlet sector for the electroweak phase transition \cite{Profumo:2007wc,Damgaard:2013kva,Ashoorioon:2009nf,OConnell:2006rsp,Gonderinger:2012rd,Profumo:2010kp,Gonderinger:2009jp, Cheung:2012nb,Barger:2007im,Cline:2013gha,Burgess:2000yq,Kakizaki:2015wua,Enqvist:2014zqa,Katz:2014bha,Espinosa:1993bs,Alanne:2014bra,Cline:2012hg,Beniwal:2017eik,Curtin:2014jma,Dev:2019njv,Ghorbani:2018yfr,Ghorbani:2020xqv,Espinosa:2011ax,Espinosa:2007qk, Kurup:2017dzf,Alves:2018jsw, Zhang:2023nrs, Jain:2017sqm, Vaskonen:2016yiu, Huang:2018aja}. Then, the tree-level potential, which respects this \(\mathbb{Z}_2\) symmetry, reads
\begin{equation}\label{treepot1}
    V (H,S) = - \mu_H^{2} |H|^2 + \lambda_H |H|^4 - \frac{\mu^2_S}{2} S^2 + \frac{\lambda_S}{4} S^4 + \lambda_{HS} |H|^2 S^2  .
\end{equation}
This singlet extension provides a viable explanation for electroweak baryogenesis and the theoretical and phenomenological implications of this extension have been extensively studied over the past decades. It is remarkable and appealing that this extension offers a solution for electroweak baryogenesis, following the principle of Occam's razor.

In this study, we could also consider that a dimension-six operator of the singlet scalar field is weakly coupled to the Higgs doublet. This higher-order operator comes from an effective theory, which is active at a scale \(M\) beyond \(15\) TeV. This is also supported by the absence of new particle observations at the LHC. Therefore, the dimension-six operator is included in the tree-level potential (\ref{treepot1}),
\begin{equation}\label{eq:1.1}
    V (H, S) = - \mu_H^{2} |H|^2 + \lambda_H |H|^4 - \frac{\mu^2_S}{2}  S^2 + \frac{\lambda_S}{4} S^4 + \lambda_{HS} |H|^2 S^2 + \frac{\lambda}{M^2}|H|^2 S^4,
\end{equation}
where the dimensionless coupling \(\lambda\) is the Wilson coefficient of the effective theory. This singlet extension can be presented at the same time as the usual singlet extension, which is described by the tree-level potential (\ref{treepot1}), by setting the Wilson coefficient to zero.


This chapter is based on research conducted by the author and his supervisor Prof. V. K. Oikonomou, published in Physical Review D and can be found in Ref. \cite{Oikonomou:2024jms}. It focuses on the real singlet extension to the Standard Model with higher-order interactions between the Higgs doublet and a real singlet scalar field to examine their impact on electroweak baryogenesis. This study shows that in this extended Standard Model, the Universe undergoes two phase transitions: an initial phase transition in the singlet sector at high temperatures, followed by the electroweak phase transition. The nature of the singlet phase transition—either second order or first order—depends on the mass of the singlet scalar and its couplings with the Higgs boson. Finally, we conclude that the presence of the dimension-six operator strengthens the electroweak phase transition in certain regions of the parameter space, which were previously excluded in singlet extensions without these higher-order operators.

\section{Effective Potential}
In this model, the effective potential can be derived following the methods demonstrated earlier and is similar to the one in the Standard Model, but it is important to note that the real singlet scalar field is also associated with a constant background field as the Higgs scalar field. Thus, the zero-temperature contribution and finite-temperature contribution to the one-loop effective potential will be shortly shown in this section.

\begin{figure}[h!]
\centering
\includegraphics[width=30pc]{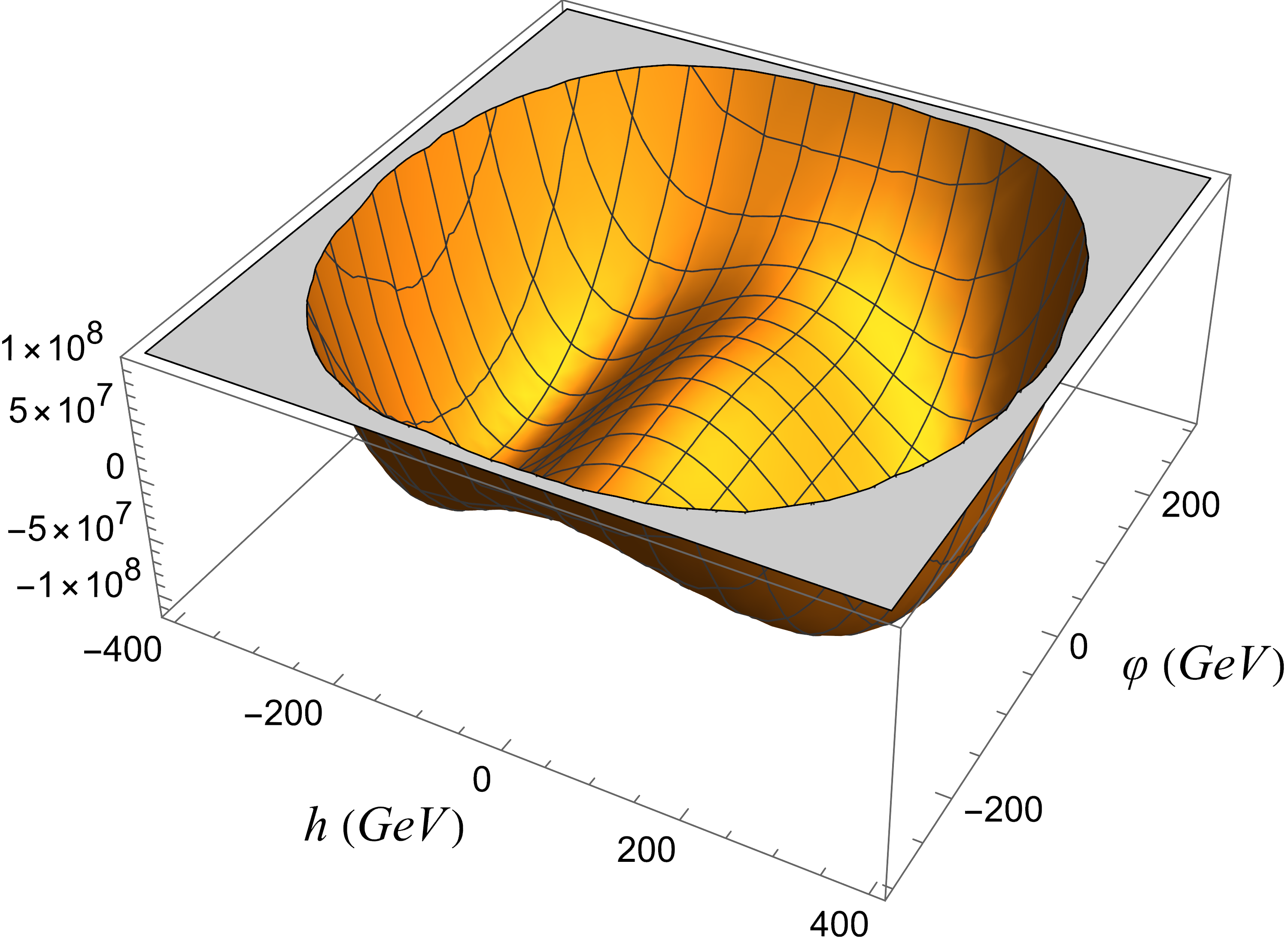}
\includegraphics[width=30pc]{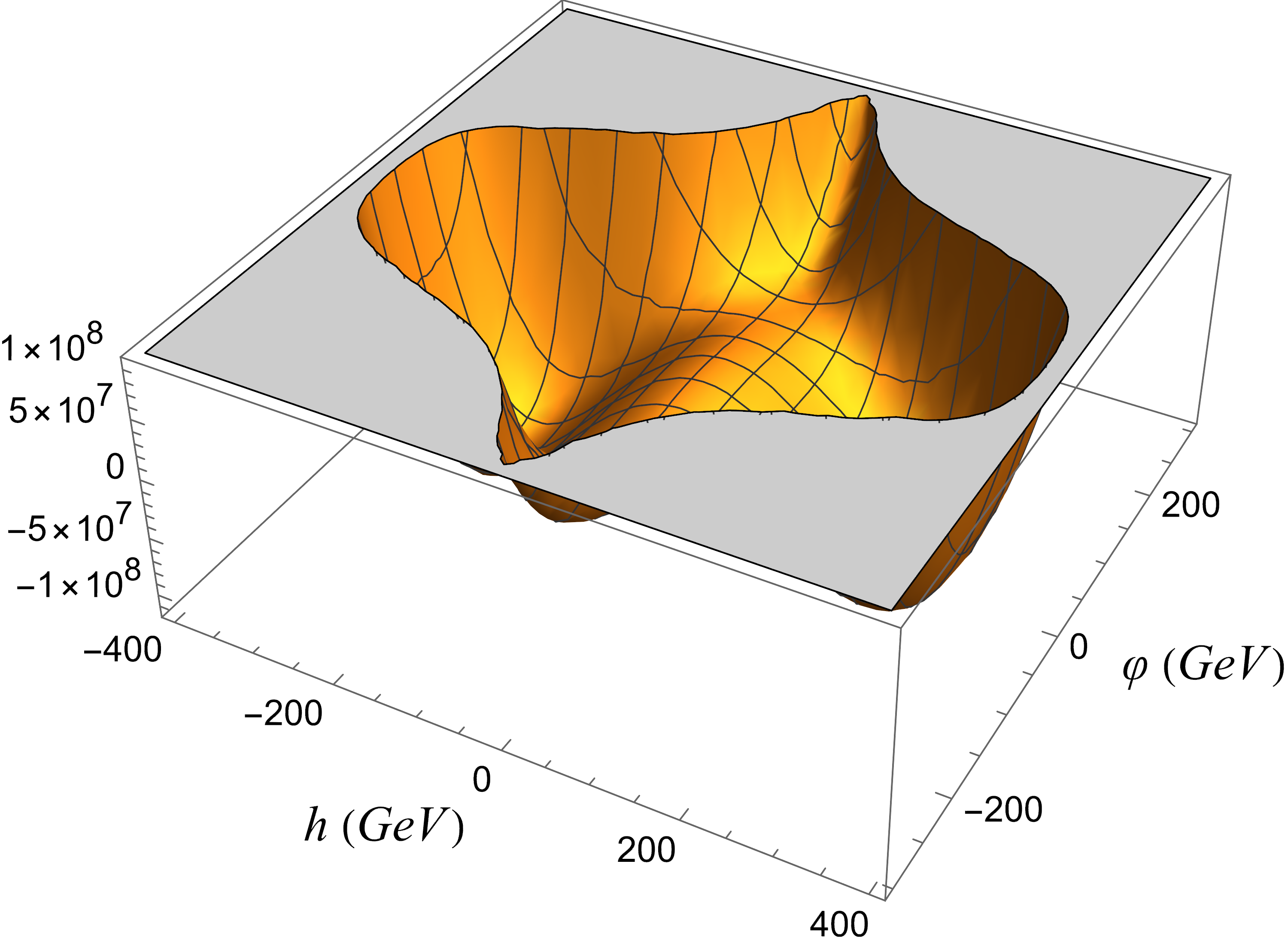}
\caption{The tree-level potential for \(\lambda
= 0\) (upper) and \(\lambda/M^2 = 2 \times 10^{-5}\) GeV\(^{-2}\)
(lower), considering \(m_S = 62.5\) GeV, \(\lambda_{HS} = 0.1\) and \(a =
0.1\).}\label{treelevelpotential}
\end{figure}

To begin with, the zero-loop correction to the classical potential (\ref{eq:1.1}) is written in terms of the background fields as,
\begin{equation}\label{tree-level_singlet}
    V_0 (h,\phi) = - \frac{\mu^{2}_H}{2} h^2 + \frac{\lambda_H}{4} h^4 - \frac{\mu^2_S}{2}  \phi^2 + \frac{\lambda_S}{4} \phi^4 + \frac{\lambda_{HS}}{2} h^2 \phi^2 +  \frac{\lambda}{2M^2}h^2 \phi^4,
\end{equation}
where the Higgs doublet was written as in Eq. (\ref{Higgs_def_2}) and \(h\) and \(\phi\) are the real constant background fields associated with the Higgs boson and the real singlet scalar field, respectively. The tree-level potential (\ref{tree-level_singlet}) is plotted in Fig. \ref{treelevelpotential} for \(\lambda = 0\) and \(\lambda/M^2 = 2 \times 10^{-5}\) GeV\(^{-2}\).
In this work, only the dominant contributions to the one-loop finite temperature effective potential are considered, including the gauge bosons, the top quark, and the Higgs and Goldstone bosons. The effective masses of the Higgs boson, the singlet scalar field, and those fields which are coupled to the background fields \(h\) and \(\phi\) are given by,
\begin{equation}\label{effectivemasshiggs}
    m^2_h (h,\phi) = - \mu^2_H + 3\lambda_H h^2 + \lambda_{HS} \phi^2 + \frac{\lambda}{M^2}\phi^4
\end{equation}
\begin{equation}\label{effectivemasschi}
    m^2_{\chi} (h,\phi) = - \mu^2_H + \lambda_H h^2 + \lambda_{HS} \phi^2 +\frac{\lambda}{M^2}\phi^4
\end{equation}
\begin{equation}\label{effectivemasssinglet}
    m^2_S (h,\phi) = - \mu^2_S + 3\lambda_S \phi^2 + \lambda_{HS} h^2 + \frac{6\lambda}{M^2}h^2 \phi^2
\end{equation}
\begin{equation}\label{effectivemassW}
    m^2_W (h) = \frac{g^2}{4} h^2
\end{equation}
\begin{equation}\label{effectivemassZ}
    m^2_Z (h) = \frac{g^2 + g^{\prime 2}}{4} h^2
\end{equation}
\begin{equation}\label{effectivemasstop}
    m^2_t (h) = \frac{y^2_t}{2} h^2.
\end{equation}

\subsection{Zero-Temperature Corrections}
In the \(\overline{\rm MS}\) renormalization scheme, the zero-temperature one-loop contribution to the effective potential reads,
\begin{equation}\label{zero-temperature_singlet}
    V^i_1 (h,\phi) = (-1)^{F_i} n_i \frac{m^4_{i}(h,\phi)}{64 \pi^2} \left[ \ln{\left( \frac{m^2_{i}(h,\phi)}{\mu^2_R}\right)} - C_i \right],
\end{equation}
where \(i = \{ h, \chi, S, W, Z, t \}\) counts the particles
that contribute to the effective potential and their degrees of freedom are,
\begin{equation}
n_h = 1, \quad n_{\chi} = 3, \quad n_{S} = 1, \quad n_{W} = 6, \quad n_{Z} = 3, \quad n_{t} = 12.
\end{equation}
The choice of the renormalization scale \(\mu_R\) causes an uncertainty in the critical temperature \(T_c\) and the Higgs vacuum expectation value \cite{Chiang:2018gsn,Athron:2022jyi, Croon:2020cgk, Gould:2021oba}. In Ref. \cite{Chiang:2018gsn}, it was concluded that varying \(\mu_R\) from \(m_t/2\) to \(2 m_t\) the OS-like and \(\overline{\rm MS}\) schemes agreed with each other within the scale uncertainties which were \((3.8 - 6.2) \% \) in the critical temperature \(T_c\) and \((10 - 23)\% \) in the ratio \(\upsilon_c/T_c\). When the renormalization scale was equal to \(\mu_R = m_t/2\), the OS-like and \(\overline{\rm MS}\) results were almost identical. In our numerical study, the renormalization scale  \(\mu_R = m_t/2\) is chosen. The authors in Ref. \cite{Athron:2022jyi} also analyzed the implications of different renormalization schemes.

\subsection{Finite-Temperature Corrections}
The temperature-dependent component of the one-loop finite-temperature effective potential for each particle \(i\) is given by
\begin{equation}\label{T-potential}
V_T^i(h,\phi,T) = (-1)^{F_i} \frac{n_iT^4}{2 \pi^2} \int_{0}^{\infty} dx \,    x^2 \ln \left[ 1 - (-1)^{F_i} \exp \left( - \sqrt{x^2 + \frac{m^2_{i}  (h,\phi)}{T^2} }\right) \right],
\end{equation}
where the thermal functions are given by (\ref{bosonthermalfunction_1}) and (\ref{fermionthermalfunction_1}). 

The thermal resummation in this model is very similar to the Standard Model, which was presented earlier. In the Arnold-Espinosa scheme, the effective potential is cast into the form (\ref{Arnold-Espinosa_1}) as in the Standard Model effective potential. Only the temperature-dependent self-energies of the scalar fields are modified in the singlet-extended Standard Model and are expressed as,
\begin{equation}\label{T-higgs}
    \Pi_h (T) = \Pi_{\chi} (T) = \left(\frac{3g^2 }{16} + \frac{g^{\prime 2}}{16}  +\frac{y^2_t }{4} + \frac{\lambda_H}{2} + \frac{\lambda_{HS}}{12}\right) T^2,
\end{equation}
and
\begin{equation}\label{T-singlet}
    \Pi_S (T) = \left( \frac{\lambda_S}{4} + \frac{\lambda_{HS}}{3} +  \frac{\lambda \upsilon^2}{2M^2} \right) T^2.
\end{equation}
The thermal masses of the gauge bosons remain the same as in the Standard Model. Finally, the effective potential is commonly preferred to be computed using the Arnold-Espinosa scheme in Beyond the Standard Model physics, as a result, the Arnold-Espinosa scheme is adopted throughout this chapter.

In conclusion, in view of Eqs. (\ref{tree-level_singlet}), (\ref{zero-temperature_singlet}), and (\ref{T-potential}), the one-loop finite-temperature effective potential can be written as,
\begin{equation}\label{generalfullpotential}
\begin{split}
    V_{\text{eff}} (h, \phi, T)  =& - \frac{\mu^{2}_H}{2} h^2 + \frac{\lambda_H}{4} h^4 - \frac{\mu^2_S}{2}  \phi^2 + \frac{\lambda_S}{4} \phi^4 + \frac{\lambda_{HS}}{2} h^2 \phi^2 + \lambda \frac{h^2 \phi^4}{2M^2} \\
    & + \sum_{i} (-1)^{F_i} n_i \frac{m^4_{i}(h,\phi)}{64 \pi^2}\left[ \ln \left( \frac{m^2_{i}(h,\phi)}{\mu^2_R}\right) - C_i \right] \\
    & + \sum_{k} \frac{n_k T^4}{2 \pi^2} J_{B} \left(\frac{m^2_k (h,\phi)}{T^2}\right) - \frac{n_t T^4}{2 \pi^2} J_{F} \left(\frac{m^2_t (h)}{T^2}\right) \\
    & + \sum_{k} \frac{\overline{n}_k T}{12\pi} \left[m^3_k(h,\phi) - \left(M^2_k(h,\phi, T) \right)^{3/2} \right],
\end{split}
\end{equation}
where \(i = \{h, \chi, S, W, Z, t, \gamma\}\) and \(k = \{h, \chi, S, W, Z, \gamma \}\). Hence, the effective potential (\ref{generalfullpotential}), using the high-temperature expansion of the thermal functions, is cast into the form,
\begin{equation}\label{HT-fullpotential}
\begin{split}
    V^{\text{HT}}_{\text{eff}} (h, \phi, T)  = & - \frac{\mu^{2}_H}{2} h^2 + \frac{\lambda_H}{4} h^4 - \frac{\mu^2_S}{2}  \phi^2 + \frac{\lambda_S}{4} \phi^4 + \frac{\lambda_{HS}}{2} h^2 \phi^2 +  \frac{\lambda}{2M^2}h^2 \phi^4 \\
    &+ \frac{m^2_h (h,\phi)}{24}T^2 - \frac{T}{12 \pi} \left[M^2_h (h,\phi, T) \right]^{3/2} + \frac{m^4_h(h,\phi)}{64 \pi^2} \left[\ln \left(\frac{a_b T^2}{\mu^2_R}\right) -\frac{3}{2} \right] + \\
    & + \frac{3 m^2_{\chi} (h,\phi)}{24}T^2 - \frac{3T}{12 \pi} \left[ M^2_{\chi} (h,\phi, T) \right]^{3/2} + \frac{ 3 m^4_{\chi} (h,\phi)}{64 \pi^2} \left[\ln \left(\frac{a_b T^2}{\mu^2_R}\right) -\frac{3}{2} \right] \\
    &+  \frac{m^2_{\phi} (h,\phi)}{24}T^2 - \frac{T}{12 \pi} \left[ M^2_{\phi} (h,\phi, T) )\right]^{3/2} + \frac{ m^4_{\phi} (h,\phi)}{64 \pi^2} \left[\ln \left(\frac{a_b T^2}{\mu^2_R}\right) -\frac{3}{2} \right]  + \\
    &+  \frac{6 m^2_{W} (h)}{24}T^2 - \frac{4T}{12 \pi} m^3_{W} (h) - \frac{2T}{12 \pi}\left[ M^2_{W_{L}} (h, T)  \right]^{3/2} + \frac{ 6 m^4_{W} (h)}{64 \pi^2} \left[\ln \left(\frac{a_b T^2}{\mu^2_R}\right) -\frac{5}{6} \right] \\
    &+ \frac{3 m^2_{Z} (h)}{24}T^2 - \frac{2T}{12 \pi} m^3_{Z} (h) - \frac{T}{12 \pi} \left[  M^2_{Z_{L}} (h, T) \right]^{3/2} + \frac{ 3 m^4_{Z} (h)}{64 \pi^2} \left[\ln \left(\frac{a_b T^2}{\mu^2_R}\right) -\frac{5}{6} \right] \\
    &+ \frac{12 m^2_{t} (h)}{48}T^2 - \frac{ 12 m^4_{t} (h)}{64 \pi^2} \left[\ln \left(\frac{a_f T^2}{\mu^2_R}\right) -\frac{3}{2} \right] - \frac{T}{12 \pi}\left[ M^2_{\gamma_{L}} (h, T) \right]^{3/2}.
\end{split}
\end{equation}

Now the dynamics of the phase transitions in the singlet extensions of the Standard Model can be analyzed using the finite-temperature effective potential (\ref{HT-fullpotential}). However, this model should comply with all the experimental and theoretical constraints in cosmology and particle physics. As a result, the parameter space is reduced significantly as will be presented in the next sections.

\section{Electroweak Baryogenesis in the Singlet Extensions}

The theory of electroweak baryogenesis was discussed in detail in Chapter \ref{Electroweak Baryogenesis}. In this section, we briefly comment on the implications of the singlet extensions to the electroweak baryogenesis. First and foremost, one of the main ingredients for electroweak baryogenesis requires a strong first-order electroweak phase transition which is characterized by the sphaleron rate criterion (\ref{criterion_baryogenesis}). However, in the Standard Model, this criterion is not fulfilled and additional \(CP\)-violating sources are required to realize electroweak baryogenesis, a fact that encouraged physicists to formulate theories beyond the Standard Model. We will show that the singlet extensions of the Standard Model can provide us with numerous insights to explain the observed baryon asymmetry of the Universe.

In the real singlet extension, a \(CP\)-violating source has been proposed in Ref. \cite{Cline:2012hg}. In particular, a dimension-six operator is introduced and originates from an effective field theory at a new physics scale. This operator couples the singlet scalar field with the top quark and the top-quark Yukawa interaction term is written as
\begin{equation}\label{CP_source}
     y_t \overline{Q}_L H \left( 1 + \frac{\eta}{\Lambda^2} S^2 \right) t_R +  \text{h.c.}\,,
\end{equation}
where \(\eta\) is a complex phase and \(\Lambda\) is the new physics scale. Subsequently, this higher-order operator changes the effective mass of the top quark which now reads
\begin{equation}
    m_t^2 (h, \phi) = \frac{y_t^2}{2} h^2 + \frac{y^2_t}{2} \frac{\eta^{*}}{\Lambda^2} h^2 \phi^2 + \frac{y^2_t}{2} \frac{\eta}{\Lambda^2} h^2 \phi^2 + \frac{y_t^2}{2} \frac{\eta^{*} \eta}{\Lambda^4} h^2 \phi^4
\end{equation}
Therefore, the top-quark mass acquires a complex phase that varies spatially along the profile of the bubble wall, providing the \(CP\)-violating source required to explain the matter-antimatter asymmetry in the Universe \cite{Cline:2012hg}.

It is essential to mention that the effective potential (\ref{HT-fullpotential}) changes due to this dimension-six operator in  (\ref{CP_source}), and its contribution to the effective potential at leading order in the high-temperature expansion is given by\footnote{This is the real part of the contribution of the dimension-six operator in (\ref{CP_source}).}
\begin{equation}
    \delta V = \frac{T^2}{4} \frac{y_t^2}{2} h^2 \left(\frac{\phi}{\Lambda}  \right)^4 
\end{equation}
However, the dimension-six operator in the effective potential does not impact the critical temperature or the vacuum expectation values at the critical temperature, as it vanishes in both the \(h\) and \(\phi\) directions. This operator primarily increases the height of the barrier at the critical temperature \(T_c\), resulting in a thinner bubble wall. Furthermore, we assume that \(\phi/\Lambda\) remains very small, leading to a negligible contribution to the effective potential. Consequently, it does not significantly influence the phase transition dynamics and is not taken into account in the finite-temperature one-loop effective potential (\ref{generalfullpotential}). This is further discussed in Refs. \cite{Cline:2012hg, Vaskonen:2016yiu}.

\section{Physical Parameter Space}
In this singlet extension to the Standard Model, the parameter space is defined by \(\mu_S\), \(\lambda_S\), \(\lambda_{HS}\), and the coefficient \(\lambda\). For the numerical results, it is assumed that the effective theory is active at a scale of \(M = 15\) TeV. Moreover, the model's parameter space is mainly constrained by the vacuum structure, the validity of perturbation theory, and experimental measurements of the invisible Higgs decay width. If the model predicts a strong electroweak phase transition, the parameter space is further restricted by the sphaleron rate criterion (\ref{criterion_baryogenesis}).

The vacuum state of the current Universe is described by the Higgs vacuum expectation value at zero temperature, where the \(SU(2)_L \times U(1)_Y\) symmetry is broken, and the zero vacuum expectation value of the singlet scalar field, where the \(\mathbb{Z}_2\) symmetry remains unbroken. Specifically, at zero temperature, the singlet mass-squared is written as,
\begin{equation}\label{singlet_mass}
   m^2_{S}  \equiv m^2_{S} (\upsilon, 0) = - \mu^2_S + \lambda_{HS} \upsilon^2 > 0. 
\end{equation} 
Consequently, the Higgs minimum in the effective potential at zero temperature must be the global minimum, expressed as
\begin{equation}\label{global_min}
    V_0 (\upsilon, 0) < V_0 (0, \upsilon_s) \Rightarrow \lambda_S > \lambda_H \frac{\mu^4_S}{\mu^4_H},
\end{equation}
where \(\upsilon_s\) is the singlet vacuum expectation value at zero temperature in the \(\phi\) direction. During the two-step electroweak phase transition, the parameter \(\mu^2_S\) is assumed to be positive. Additionally, it is required that \(\lambda_S > 0\) and \(\lambda > 0\) to ensure the tree-level potential is bounded from below.

Based on Eqs. (\ref{global_min}) and (\ref{singlet_mass}), the minimum value of the singlet self-coupling \(\lambda_S\) can be determined as
\begin{equation}\label{lambda_min}
    \lambda^{min}_S = \frac{\lambda_H}{\mu^4_H} \left(m^2_S - \lambda_{HS} \upsilon^2 \right)^2.
\end{equation}
The singlet self-coupling can be then expressed as
\begin{equation}\label{lambda_definition}
    \lambda_S = \lambda^{min}_S + a,
\end{equation}
where \(a\) is a positive parameter, typically set to \(a = 0.1\) for singlet masses \(m_S \geq m_H/2\) \cite{Curtin:2014jma, Chiang:2018gsn, Kurup:2017dzf, Jain:2017sqm, Senaha:2020mop}. Therefore, the parameter space can be equivalently described in terms of the singlet mass, the Higgs-singlet interaction coupling, and the Wilson coefficient for a given parameter \(a\).

Furthermore, the one-loop perturbative analysis is not valid for large coupling constants. Therefore, to ensure the perturbativity of the couplings at high energy scales, the RGEs are solved at one loop for the couplings \(\lambda_{HS}\), \(\lambda_H\), and \(\lambda_S\), as detailed in Appendix \ref{Appendix B}. The RGE evolution of the gauge couplings and the top quark Yukawa coupling remains consistent with that of the Standard Model. The contribution of the higher-order operator to the RGEs is neglected, given that the effective field theory is weakly coupled with \(\lambda/M^2 < 10^{-4}\) GeV\(^{-2}\). According to Ref. \cite{Curtin:2014jma}, the RGE evolution indicates that the model remains perturbative up to scales of \(10 - 100\) TeV, depending on the coupling constant \(\lambda_{HS}\). For a reliable perturbative analysis, the constraint \(\lambda^{min}_S < 8\) is imposed, which corresponds to \(\lambda_{HS} < 5\) with the singlet mass ranging from \(0 - 550\) GeV.

Finally, the tree-level potential possesses a \(\mathbb{Z}_2\) symmetry, and the singlet scalar may act as a dark matter candidate \cite{Alanne:2014bra, Cline:2013gha, Ghorbani:2018yfr, Ghorbani:2020xqv, Cline:2012hg, GAMBIT:2017gge,Athron:2018ipf, Feng:2014vea, Beniwal:2017eik}. Thus, the unbroken \(\mathbb{Z}_2\) symmetry at zero temperature ensures the stability of the dark matter particle and forbids the mixing between the Higgs boson and the singlet scalar field.

\section{Invisible Higgs Decay}

The existence of a singlet particle could have a number of consequences in particle physics. These consequences are highly determined by the value of the singlet mass, which plays an important role in the progress of the phase transitions in the extended Standard Model. If the singlet mass is \(m_S < m_H/2\), the decay \(H \to S S\) will be kinematically allowed, contributing to the invisible decay width of the Higgs boson. According to the latest collider experiments, the branching ratio of the Higgs to the invisible sector is set to \(BR_{inv} < 0.11 - 0.19\) at \(95 \% \) CL, where the upper and lower limits of this range correspond to the results from the ATLAS \cite{ATLAS:2020kdi, ATLAS:2023tkt} and the CMS collaboration \cite{CMS:2018yfx}, respectively.

\begin{figure}[H]
\centering
\includegraphics[width=30.9pc]{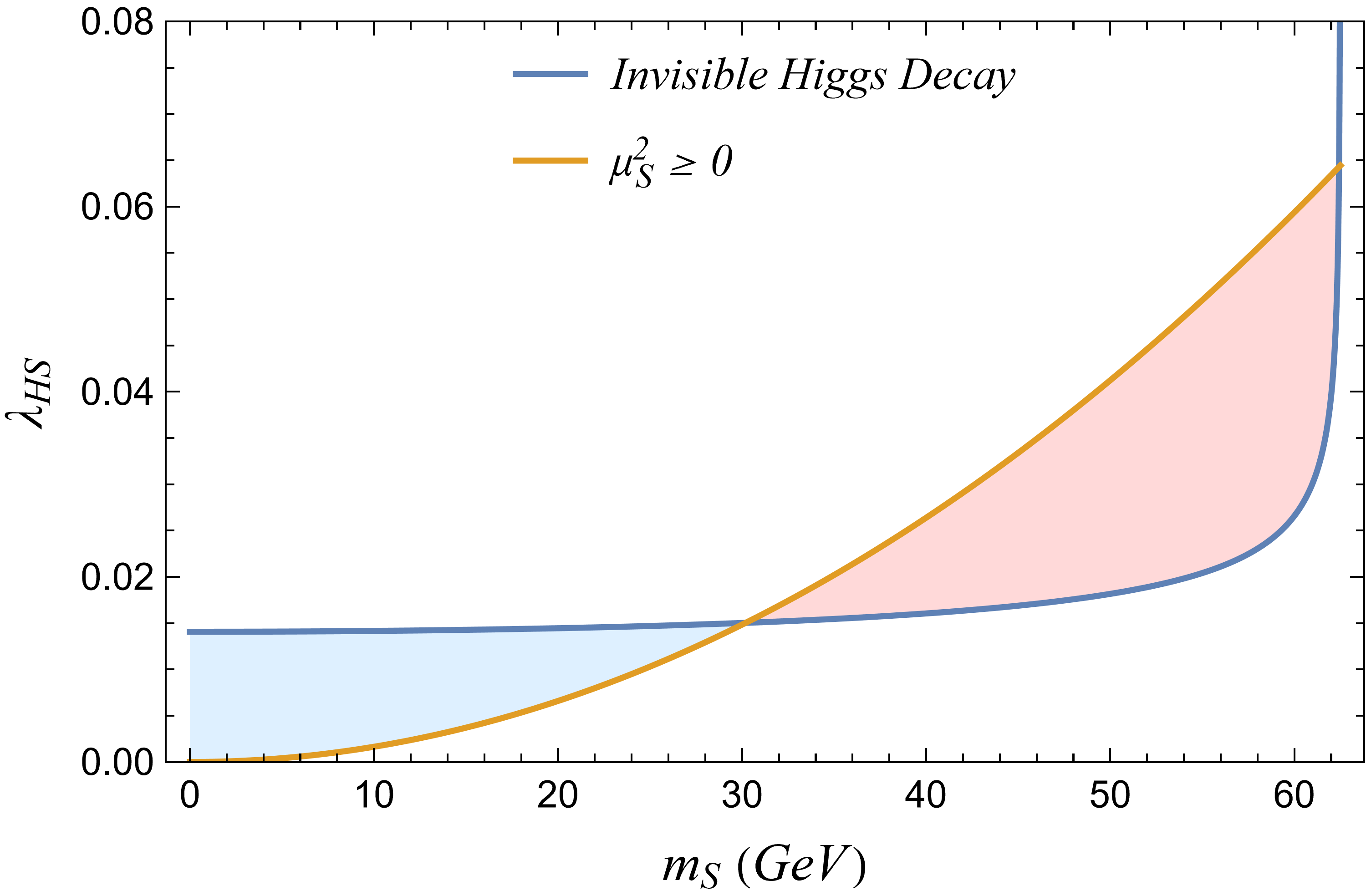}
\includegraphics[width=30.9pc]{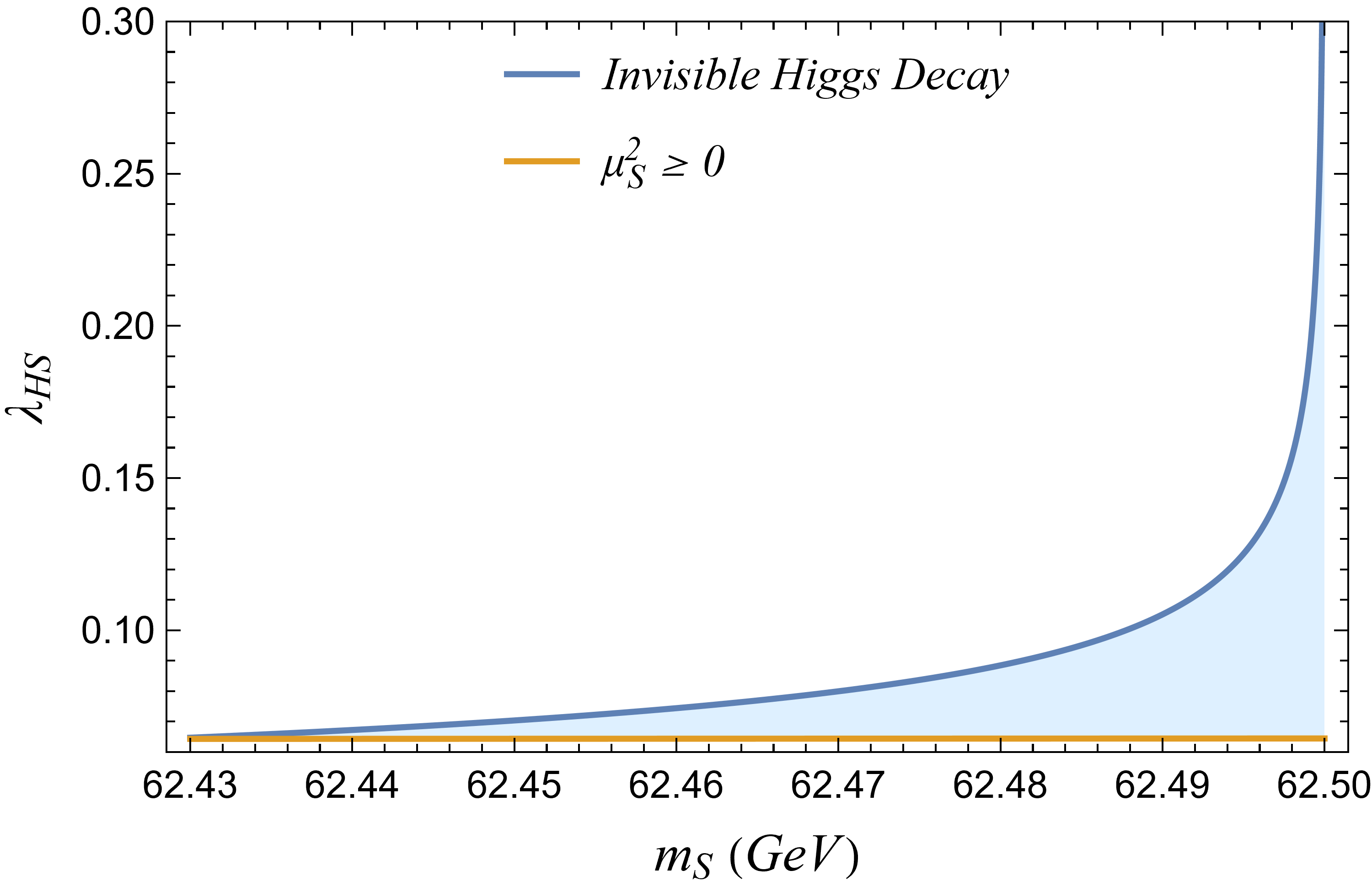}
\caption{The parameter space (blue shaded) for
\(m_S < m_H/2\) (upper) and the higher mass region (lower) which satisfy the constraint (\ref{decay_width_0.19}) and \(\mu_S^2 \geq 0\) setting \(BR_{inv} <
0.19\).} \label{BR_0.19}
\end{figure}

First of all, the branching ratio of the Higgs to the invisible particles reads
\begin{equation}
    BR_{inv} = \frac{\Gamma_{inv}}{\Gamma_{inv} + \Gamma_{vis}}.
\end{equation}
Now if the branching ratio of the Higgs to the invisible particles is set at \(BR_{inv} < 0.19\) and the Higgs decay width to visible channels is \(\Gamma_{vis} = 4.07\) MeV, the upper bound on the invisible decay width of the Higgs boson is,
\begin{equation}\label{decay_width_0.19}
    \Gamma_{inv} < 0.955 \text{ MeV},
\end{equation}
where this decay width is computed as
\begin{equation}\label{decay_width}
    \Gamma_{inv} = \frac{\lambda^2_{HS} \upsilon^2}{32 \pi m_H} \sqrt{1 - \frac{4 m^2_S}{m^2_H}}.
\end{equation}
Subsequently, the Higgs-singlet coupling can be constrained using the inequality (\ref{decay_width_0.19}),
\begin{equation}\label{special_condition_coupling}
    \lambda_{HS} < \sqrt{ \frac{32 \pi m_H}{\upsilon^2}\left({1 - \frac{4 m^2_S}{m^2_H}} \right)^{-1/2}\Gamma_{m} (H \to SS) },
\end{equation}
where \(\Gamma_m (H \to SS) \) is the upper bound on the invisible decay width of the Higgs boson. Owing to \(\mu^2_S \geq 0\), Eq. (\ref{special_condition_coupling}) leads to
\begin{equation}\label{condition_coupling}
    \frac{m_S^2}{\upsilon^2} < \lambda_{HS} < \sqrt{\frac{32 \pi m_H}{\upsilon^2}\left({1 - \frac{4 m^2_S}{m^2_H}} \right)^{-1/2} \Gamma_{m} (H \to SS) }.
\end{equation}
Namely, the singlet masses which are allowed by the constraints (\ref{condition_coupling}) are separated into the following regions, as shown in Fig. \ref{BR_0.19},
\[m_S \leq 30.19 \text{ GeV} \quad \text{ and } \quad m_S \geq 62.43 \text{ GeV}.\]
In general, the lower mass region is restricted by the upper bound of \(\lambda_{HS} = 0.014\), while the higher mass region is characterized by a wider range of values for \(\lambda_{HS}\), but it has almost a fixed singlet mass. 

\par The invisible Higgs boson branching ratio can be alternatively considered \(BR_{inv} < 0.11\) at \(95 \% \) CL with invisible decay width,
\begin{equation}\label{decaywidth0.11}
    \Gamma_{inv} < 0.503 \text{ MeV}
\end{equation}
As a result, the allowed singlet masses are
\[m_S \leq 25.45 \text{ GeV} \quad \text{ and } \quad m_S \geq 62.48 \text{ GeV},\]
which satisfy the condition (\ref{condition_coupling}).
\begin{figure}[H]
\centering
\includegraphics[width=30.9pc]{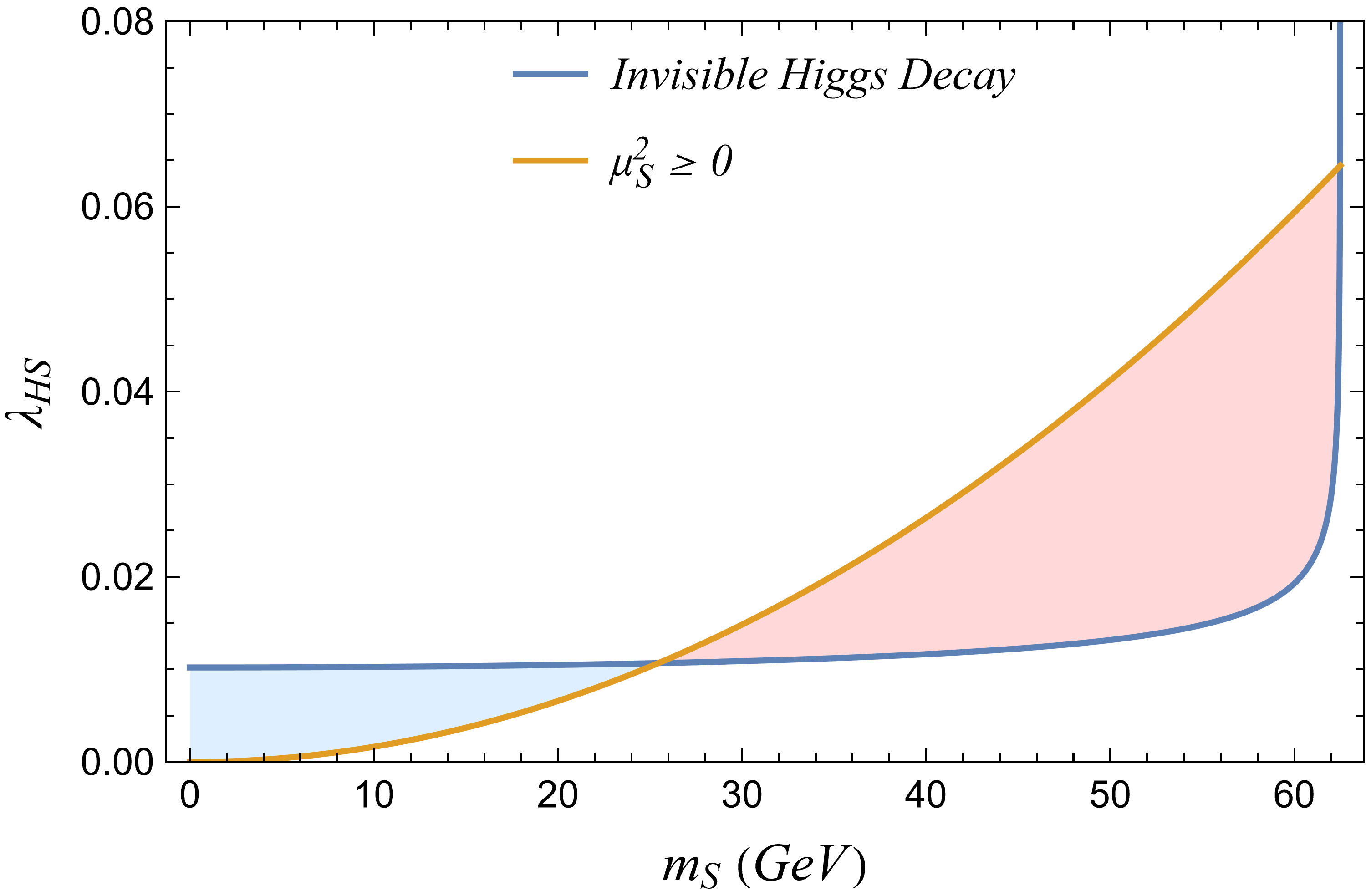}
\includegraphics[width=30.9pc]{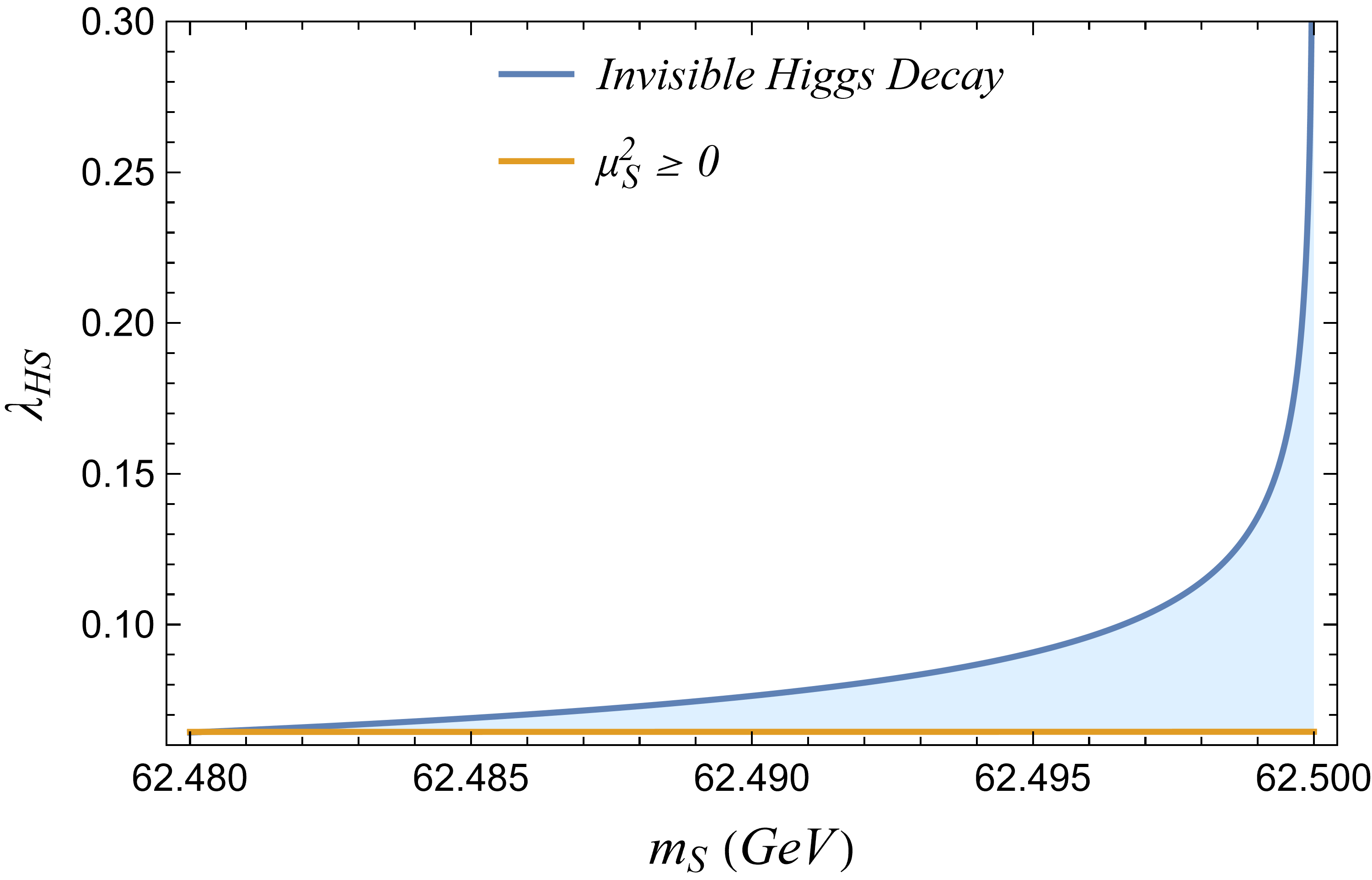}
\caption{The parameter space (blue shaded) for
\(m_S < m_H/2\) (upper) and the higher mass region (lower) which satisfy the constraint (\ref{decaywidth0.11}) and \(\mu_S^2 \geq 0\) setting \(BR_{inv} <
0.11\).} \label{BR_0.11}
\end{figure}

\section{The Electroweak Phase Transition in the Singlet Extensions}
In this section, the behavior of the one-loop finite-temperature effective potential (\ref{HT-fullpotential}) is analyzed in the two-dimensional configuration space \((h, \phi)\) spanned by the background fields associated with the Higgs and the singlet scalar field.

\begin{figure}[h!]
\centering
\includegraphics[width=19pc]{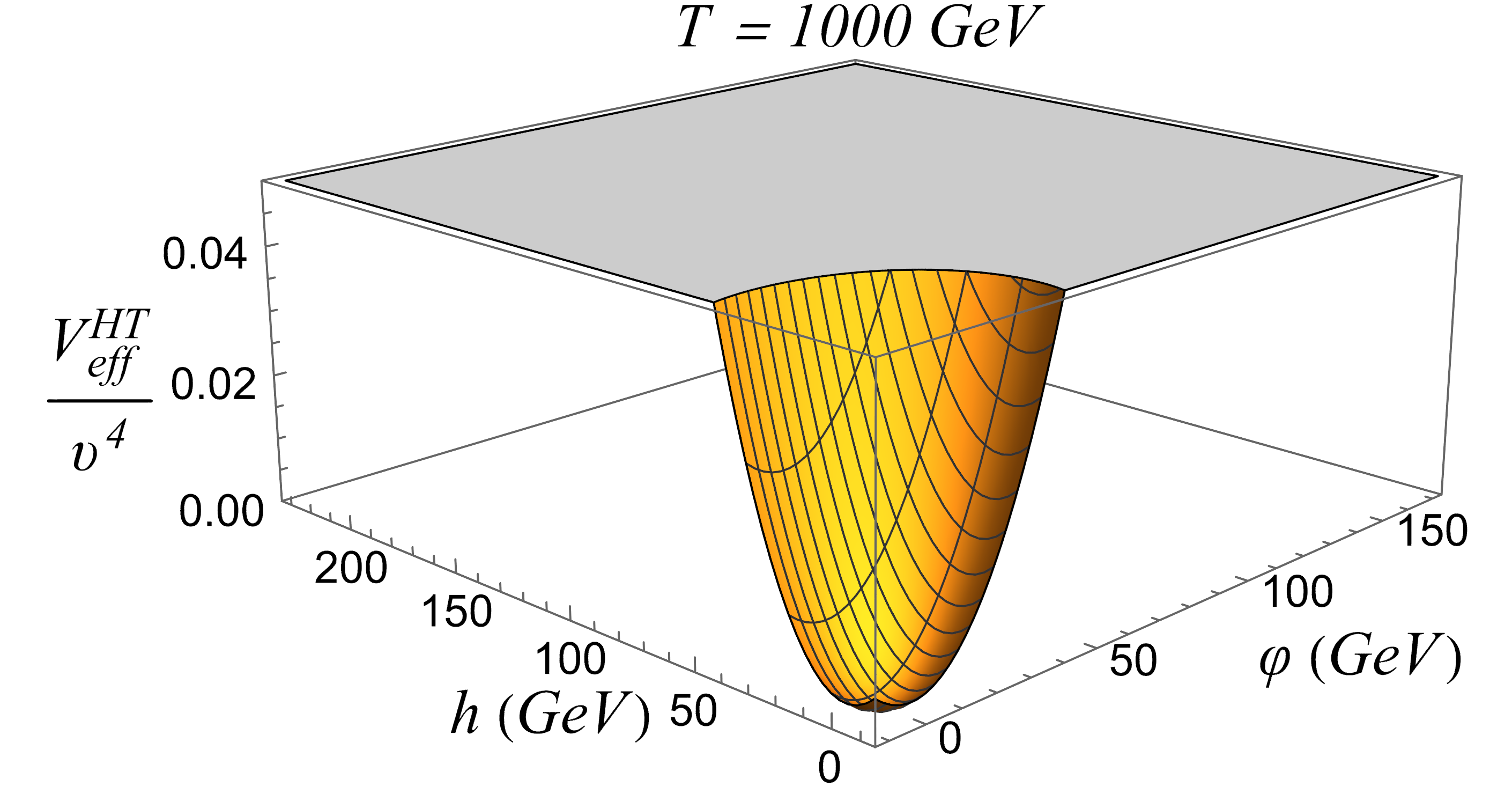}
\includegraphics[width=19pc]{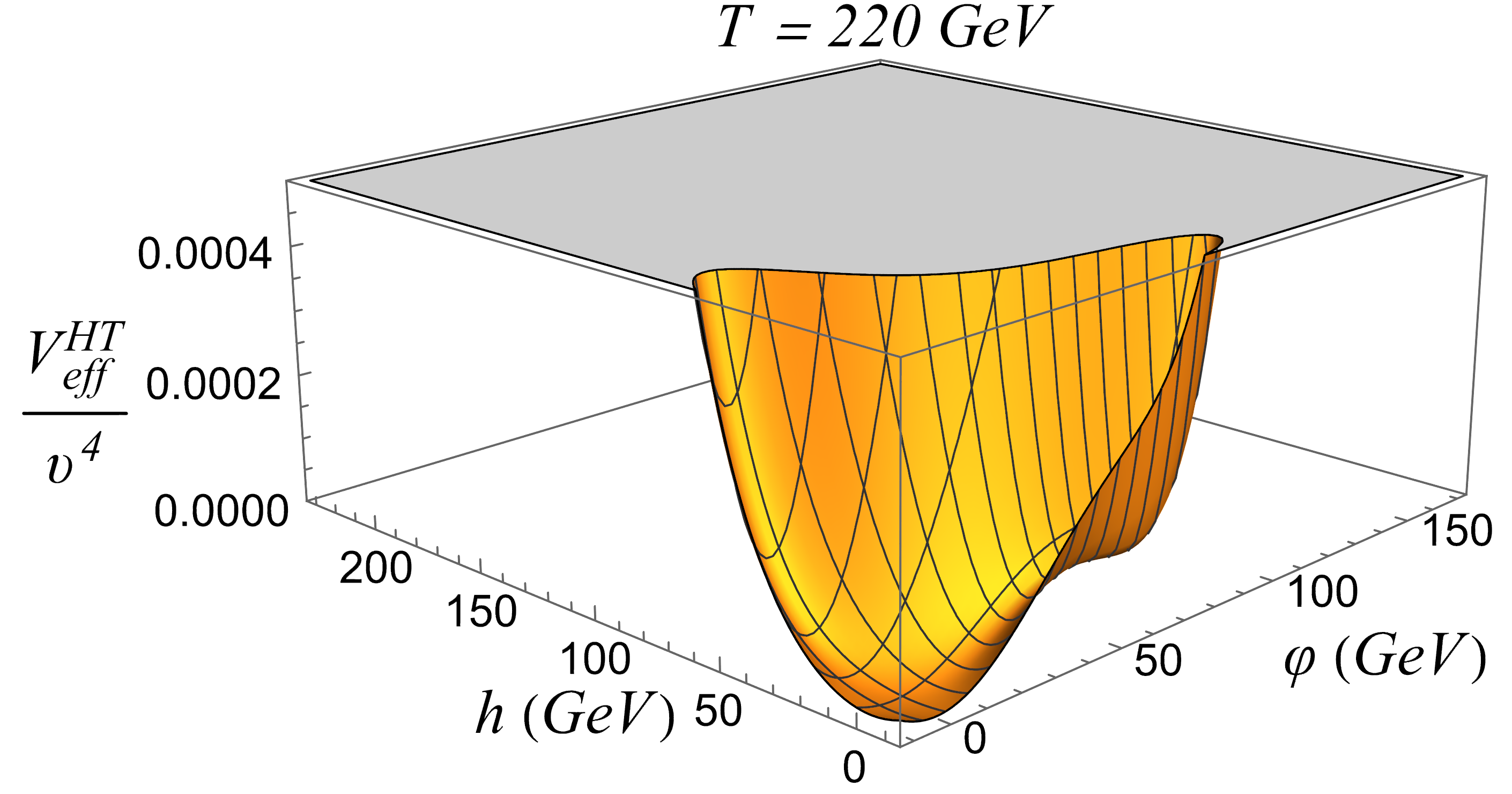}
\includegraphics[width=19pc]{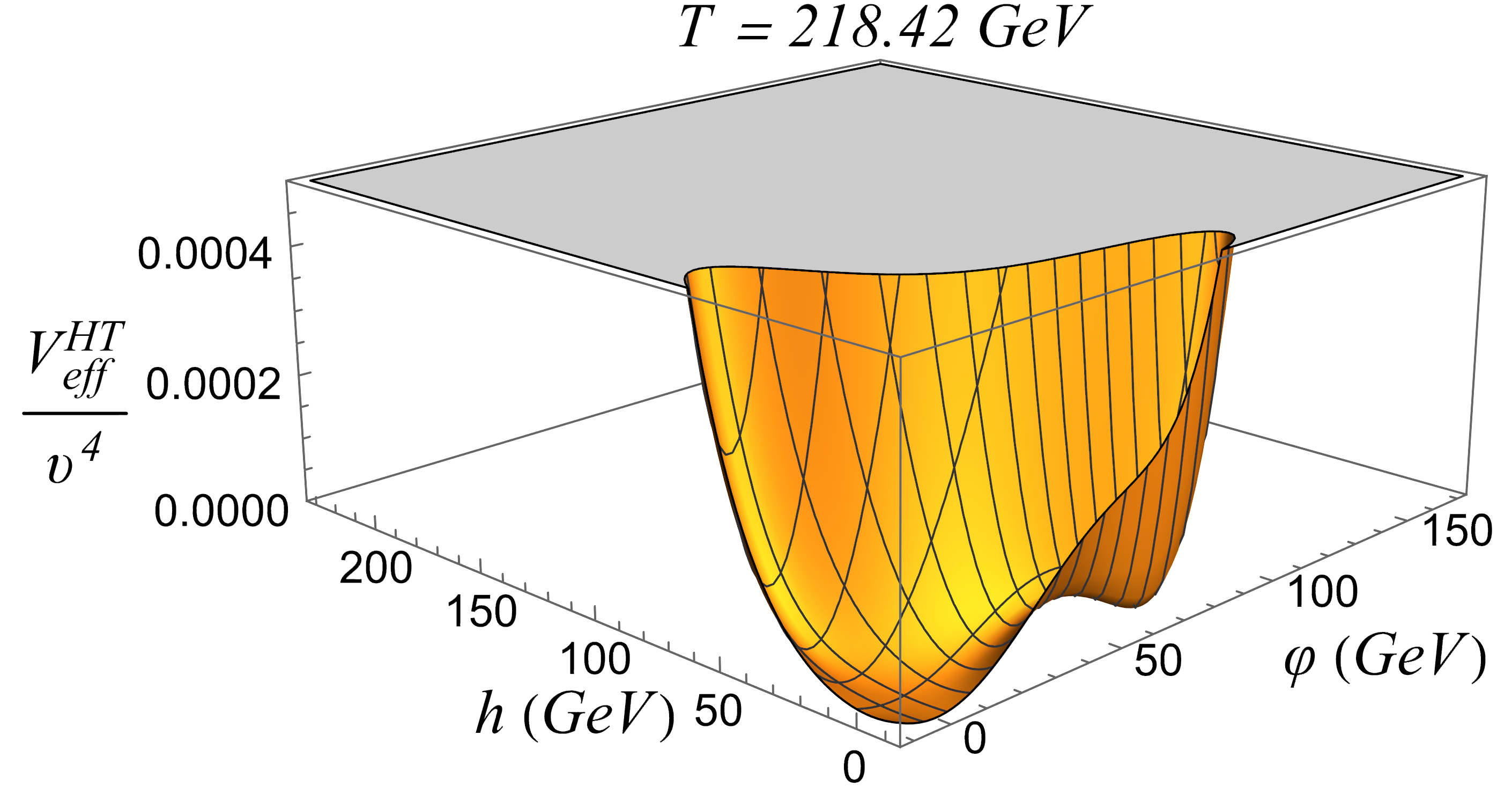}
\includegraphics[width=19pc]{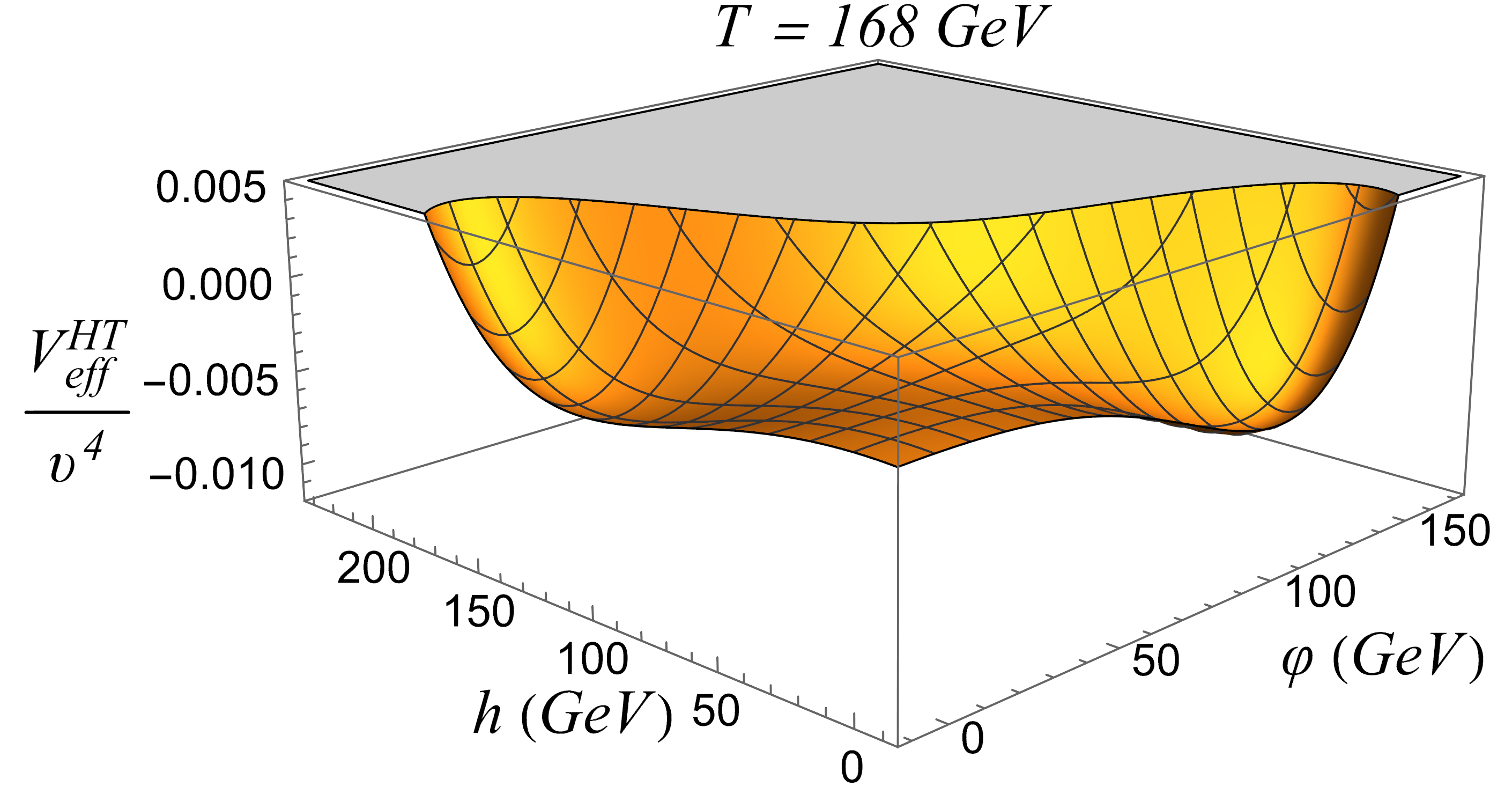}
\includegraphics[width=19pc]{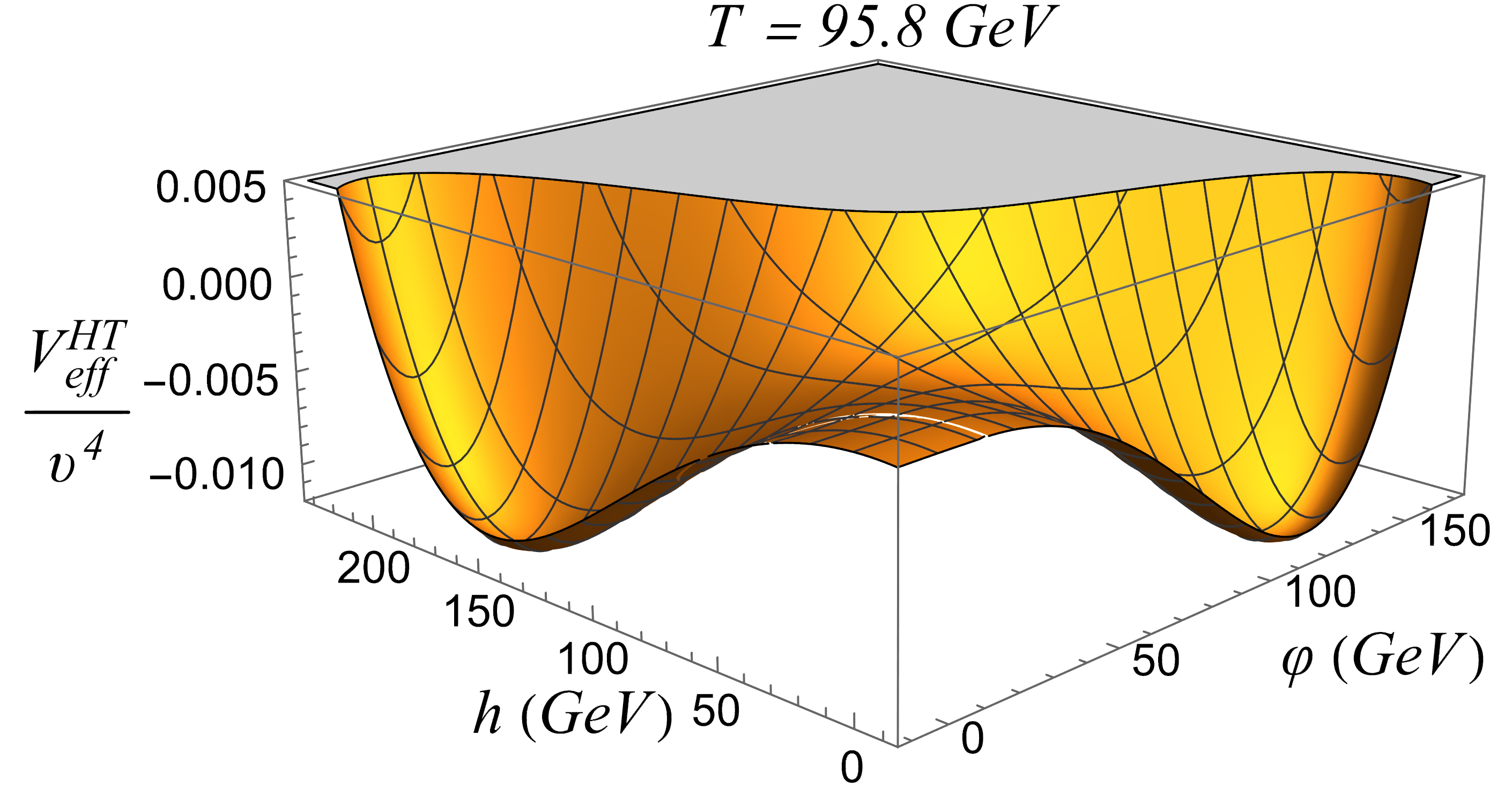}
\includegraphics[width=19pc]{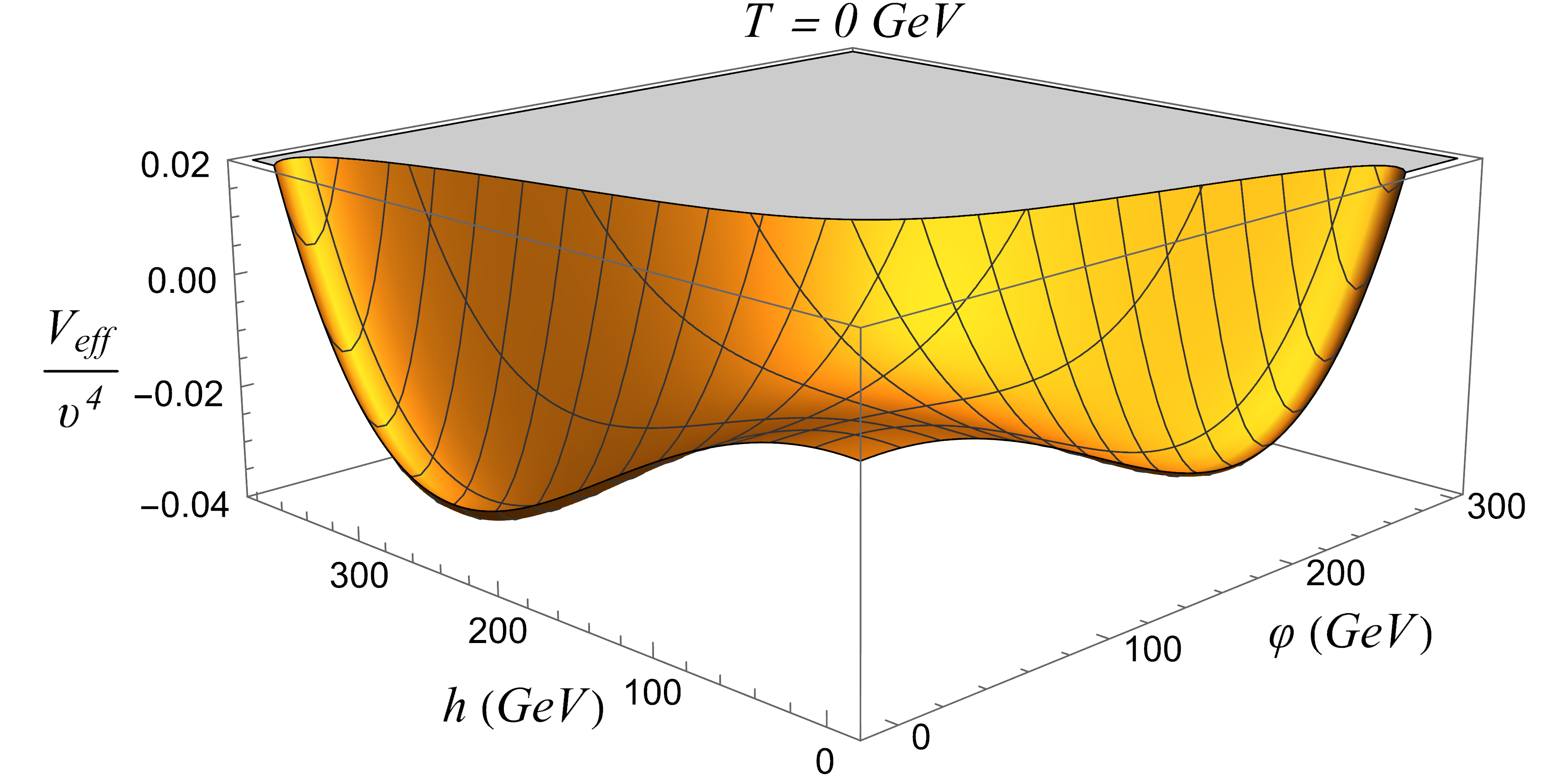}
\caption{The full effective potential during the strong
electroweak phase transition as the temperature decreases. In this
example, the singlet's phase transition is first-order using a
point of the parameter space with \(m_S = 500\) GeV,
\(\lambda_{HS} = 4.3\), \(\lambda/M^2 \simeq 2 \times 10^{-5}\)
GeV\(^{-2}\) and \(a = 0.1\).} \label{T-potential_2}
\end{figure}

\begin{figure}[h!]
\centering
\includegraphics[width=19pc]{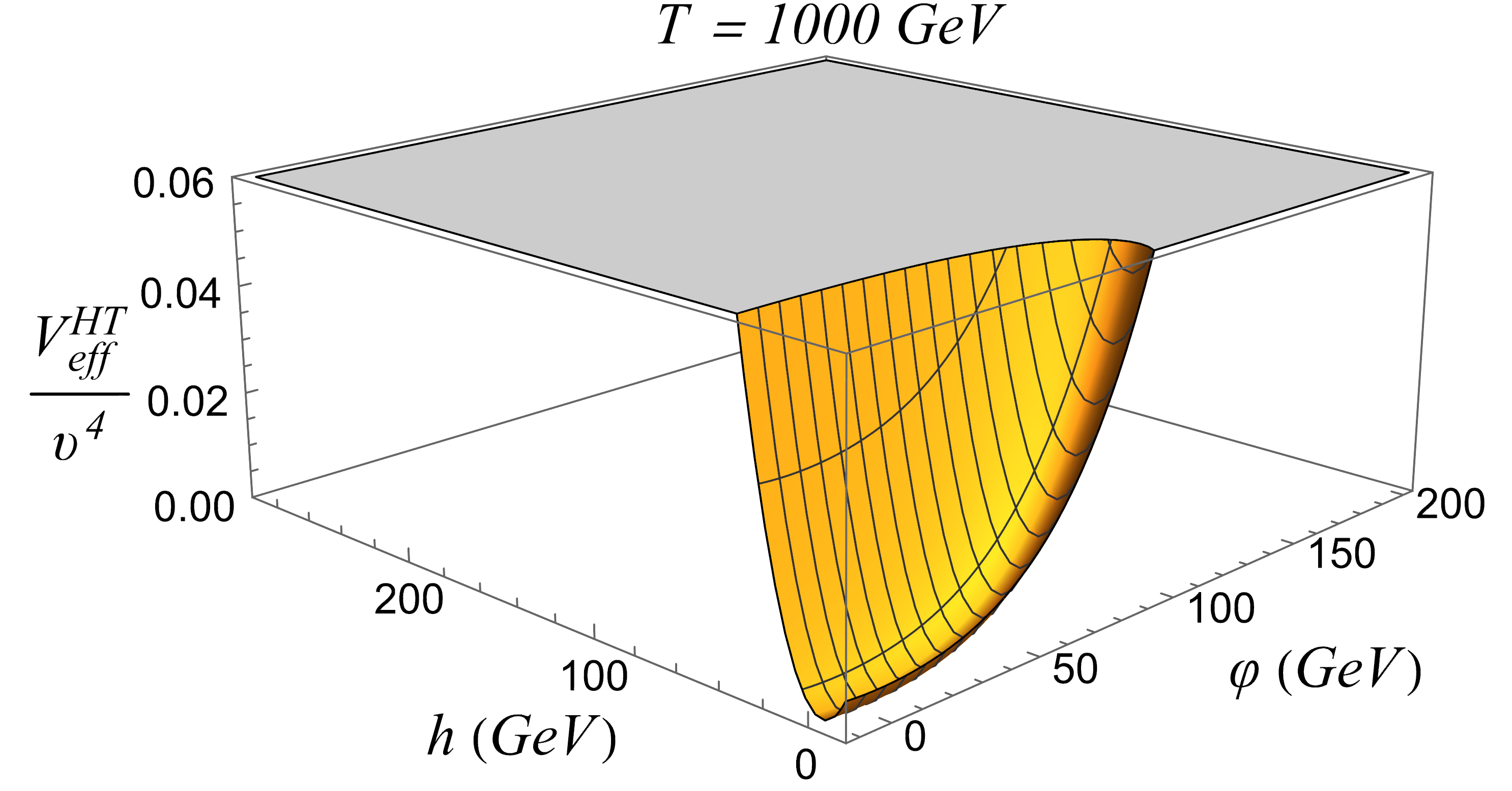}
\includegraphics[width=19pc]{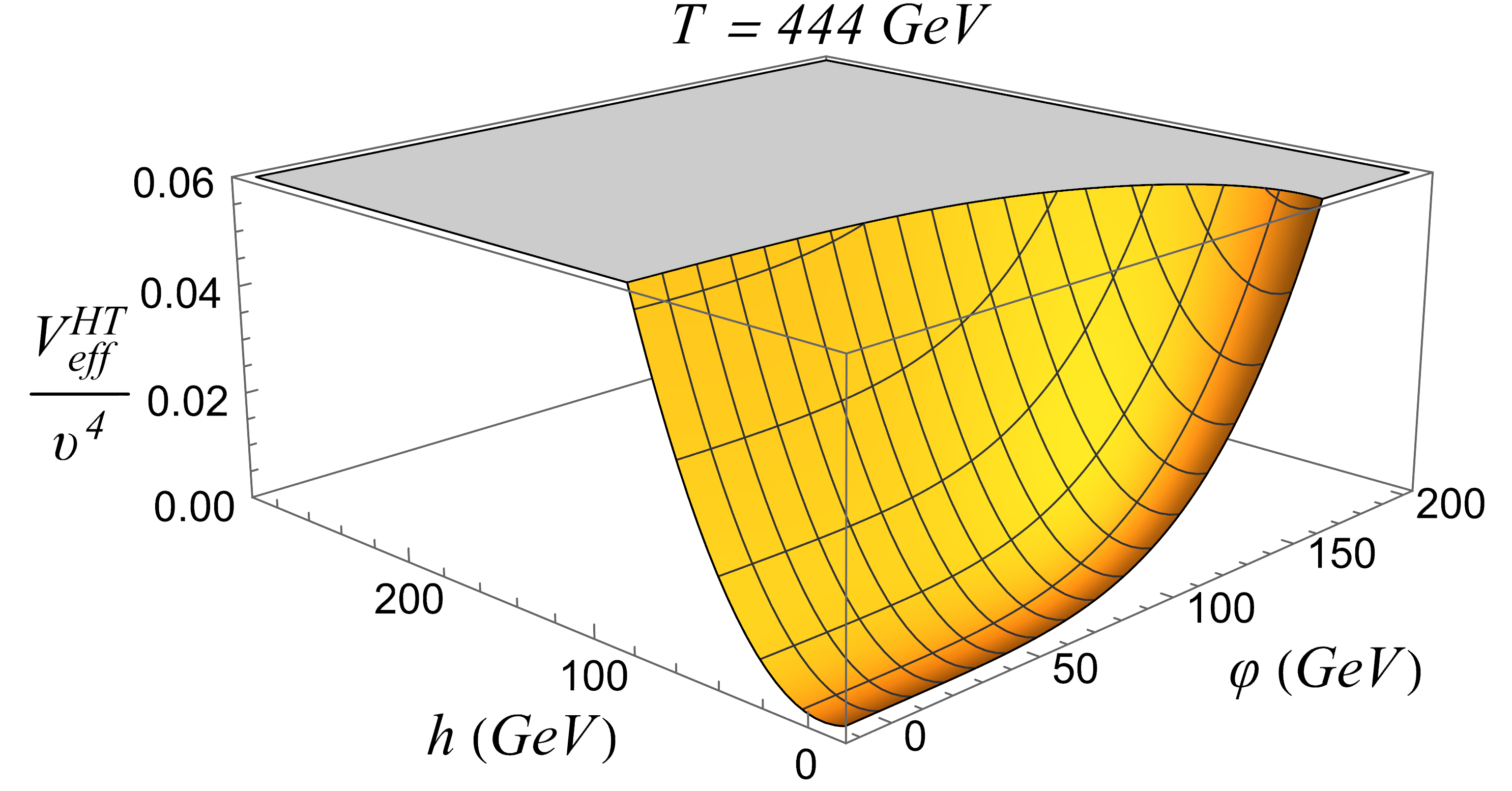}
\includegraphics[width=19pc]{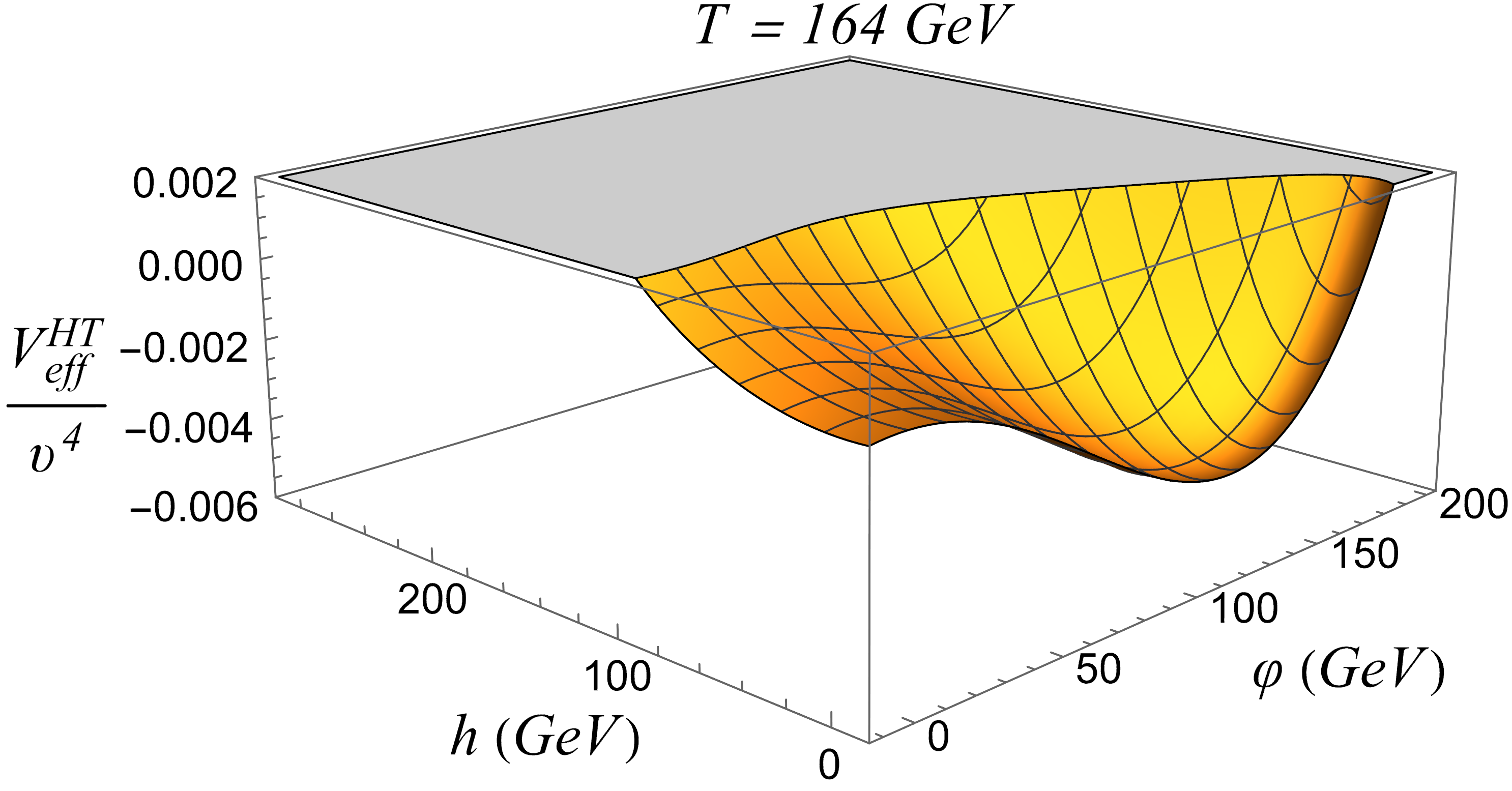}
\includegraphics[width=19pc]{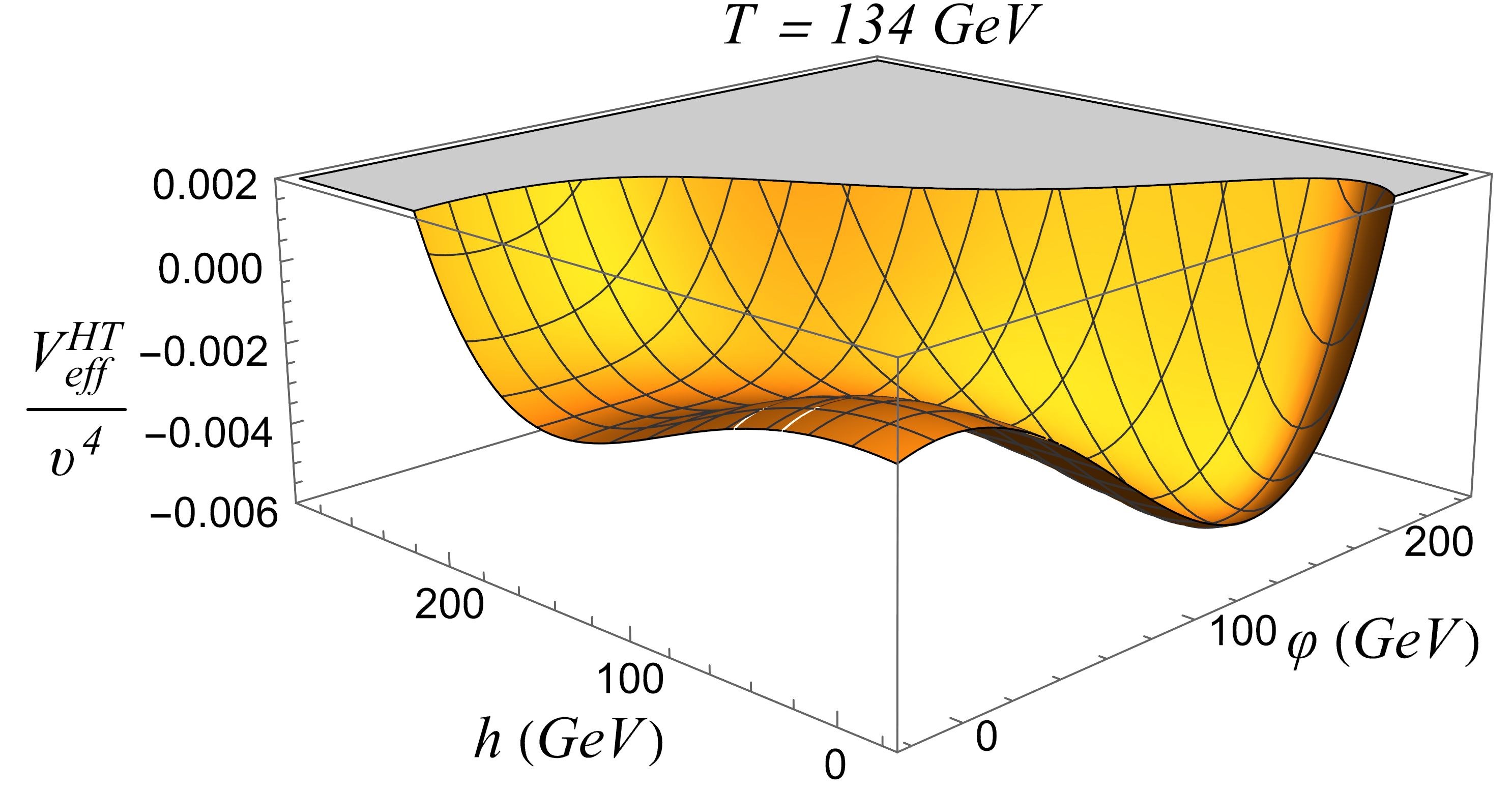}
\includegraphics[width=19pc]{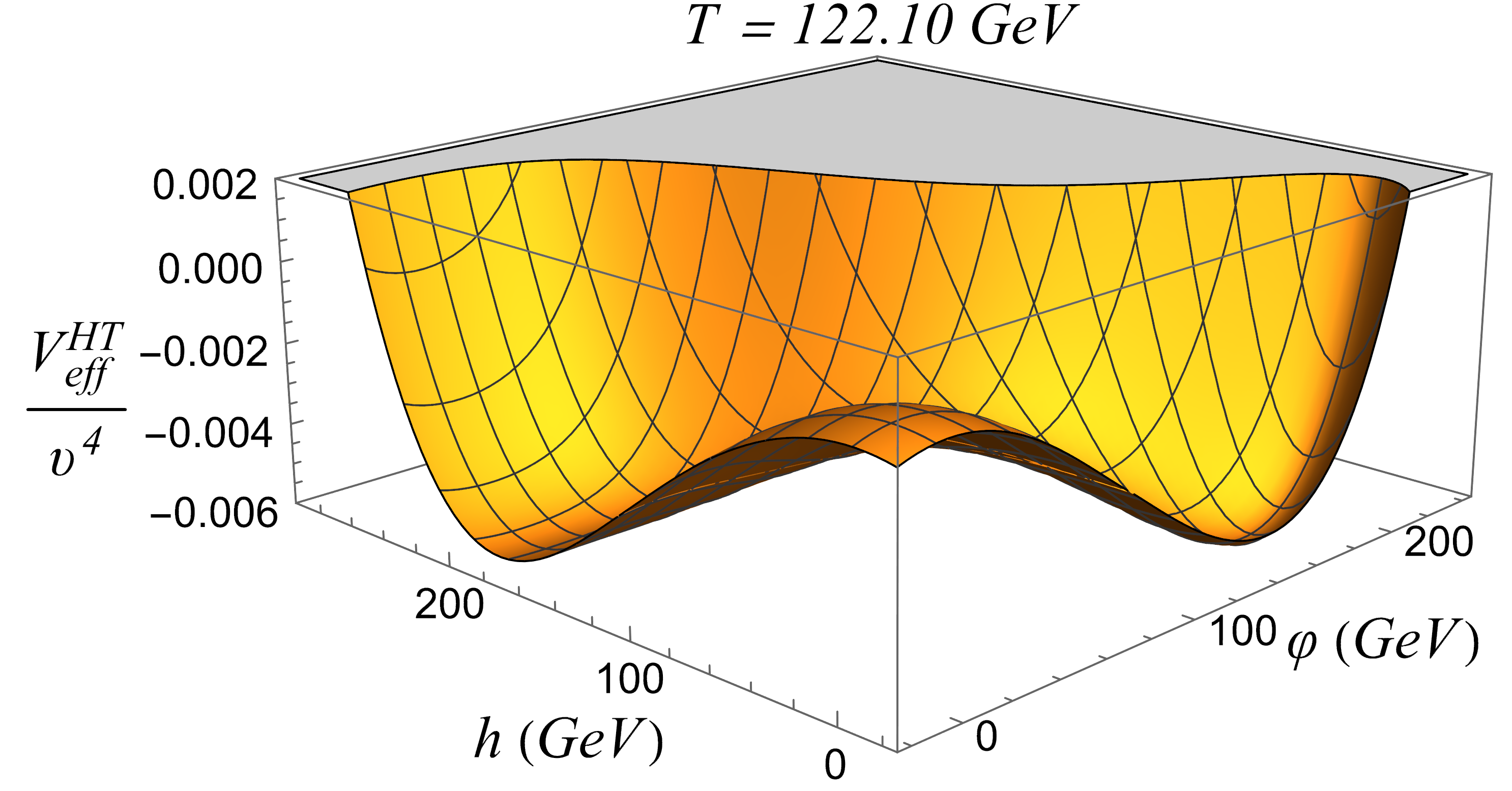}
\includegraphics[width=19pc]{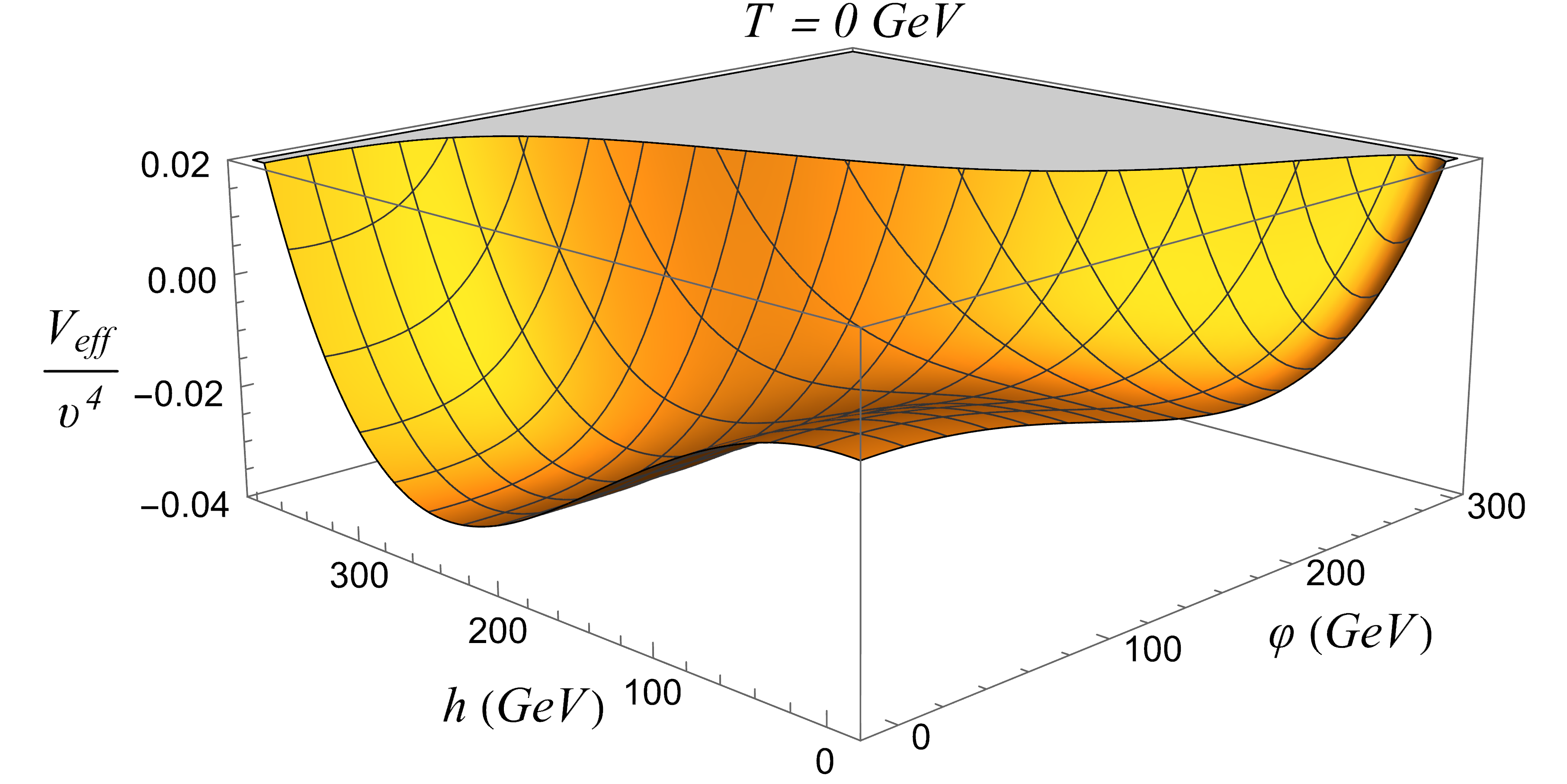}
\caption{The full effective potential during the strong
electroweak phase transition as the temperature decreases. In this
example, the singlet's phase transition is second-order using a
point of the parameter space with \(m_S = 62.5\) GeV,
\(\lambda_{HS} = 0.15\), \(\lambda/M^2 \simeq 2 \times 10^{-5}\)
GeV\(^{-2}\) and \(a = 0.1\).} \label{T-potential_1}
\end{figure}

\par
First of all, the electroweak phase transition consists of an initial first-order or second-order phase transition in the \(\phi\) direction from the origin \((h, \phi) = (0, 0)\) to a non-zero singlet vacuum expectation value and a subsequent first-order phase transition from \((0, \upsilon^{\prime}_{s})\) to the Higgs vacuum \((\upsilon_c, 0)\). At first, the vacuum of the Universe resides at the origin \((h, \phi) = (0, 0)\). At a high temperature, denoted as \(T_s\), a phase transition proceeds in the \(\phi\) direction from the origin to a non-zero singlet vacuum expectation value\footnote{It is important to mention that the \(\mathbb{Z}_2\) symmetry is never restored for certain large values of the couplings \(\lambda\), \(\lambda_{HS}\) and \(\lambda_{S}\). The singlet phase transition may occur at extremely high temperatures, although it is uncertain if such a temperature can be reached in the electroweak epoch. This is also a feature in the singlet extensions without the dimension-six operator and does not affect our discussion since the electroweak phase transition is realized with \(T_c \ll T_s\).}. As the temperature falls, in the finite-temperature effective potential, a barrier is formed, separating the global minimum and a second local minimum as shown in Figs. \ref{T-potential_2} and \ref{T-potential_1}. When the initial global minimum in the \(\phi\) direction and the second local minimum in the \(h\) direction are degenerate at a critical temperature, a first-order phase transition occurs from the non-zero singlet vacuum expectation value to the non-zero Higgs vacuum expectation value. Consequently, the phase transition proceeds via thermal tunneling, and the bubbles of the broken phase nucleate within the surrounding plasma of the false vacuum. 

Furthermore, the spontaneous symmetry breaking of the discrete \(\mathbb{Z}_2\) symmetry causes the emergence of topological defects, while the scalar field \(S\) acquires a non-zero vacuum expectation value at a high temperature \cite{Zeldovich:1974uw}. In particular, the domain walls, which are produced after the \(\mathbb{Z}_2\) symmetry breaking, can importantly influence the electroweak phase transition. In Ref. \cite{Angelescu:2021pcd}, it is proposed that this issue can be addressed by the inclusion of a dimension-six operator in the ordinary real singlet extensions, considering a scenario in which the \(\mathbb{Z}_2\) symmetry was never a symmetry of the vacuum state. This scenario is not presented in this study, and the implications of topological defects can be studied in future works.

In the next paragraphs, the study of the two-step electroweak phase transition is completed by dividing the parameter space into three mass regions: the low-mass region, the Higgs resonance region, and the high-mass region.
\subsection{High-mass Region}

\begin{figure}[h!]
\centering
\includegraphics[width=35pc]{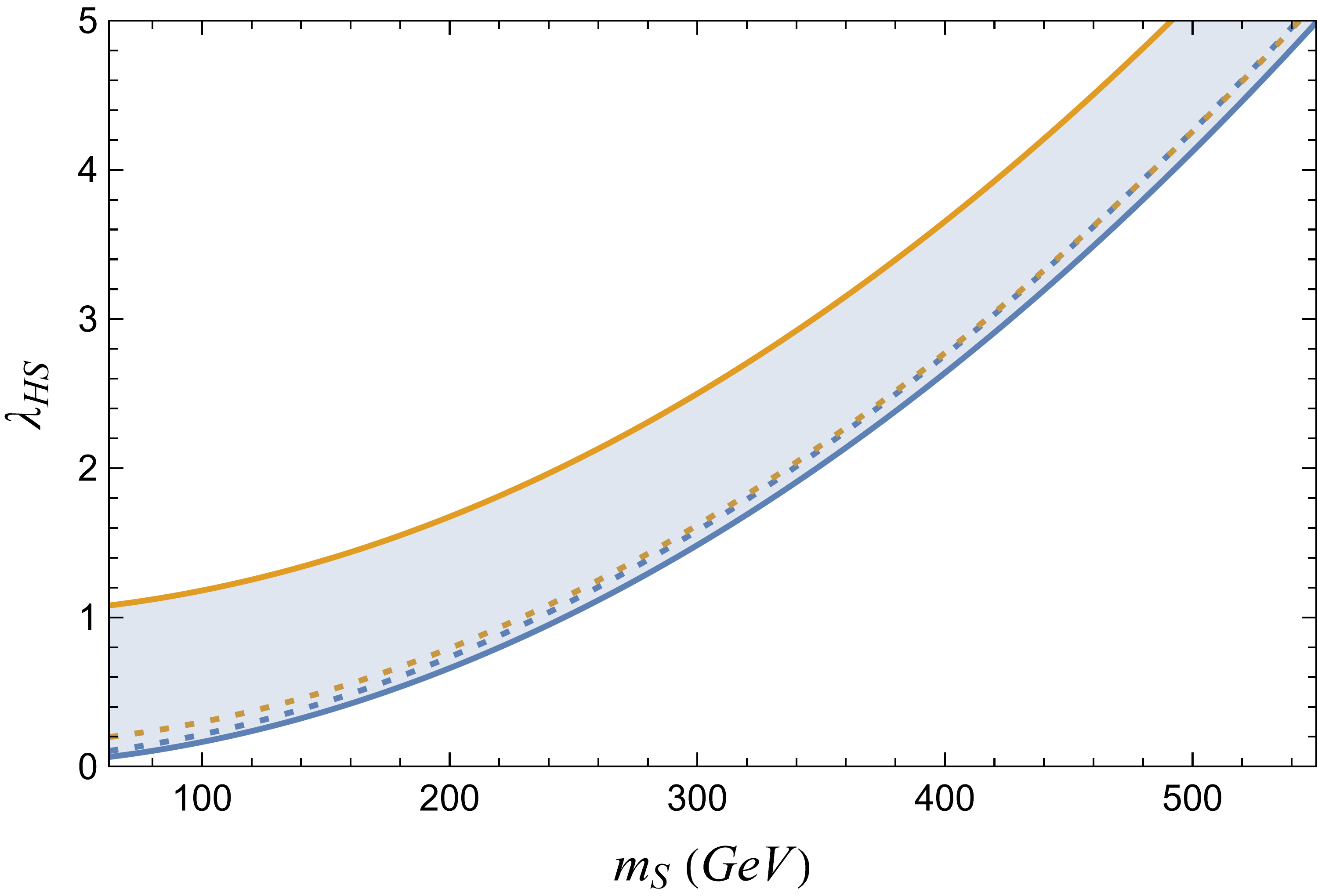}
\caption{The blue region is the parameter space of the singlet
model for a two-step electroweak phase transition with \(m_S >
m_H/2\), \(a = 0.1\) and \(\lambda = 0\). The orange dotted line
describes the constant \(\mu_S = 90\) GeV. The parameter space of
the strong two-step electroweak phase transition with
\(\upsilon_c/T_c > 1\) corresponds to the blue region which is
bounded from below by the blue dotted line. The critical
temperature in this parameter space varies from \(T_c \simeq 30 -
140\) GeV.}\label{ParameterSpace1}
\end{figure}

Initially, the parameter space for \(m_H < 2 m_S\) in Fig.
\ref{ParameterSpace1} with zero Wilson coefficient refers to the two-step electroweak phase transition, adopting the common parametrization with \(a = 0.1\). This parameter space includes numerous small regions with a one-step electroweak phase transition. More specifically, it was computed that the parameter space with \(\mu_S \lesssim 90\) GeV corresponds to a region with both scenarios. In contrast, after we impose the criterion for a strong electroweak phase transition, the parameter space with \(\lambda = 0\) is eliminated further. In particular, this occurs primarily when \(m_S > 200\) GeV as illustrated by the parameter space and the blue dotted line in Fig. \ref{ParameterSpace1}. These findings are consistent with other \cite{Curtin:2014jma, Beniwal:2017eik, Kurup:2017dzf, Jain:2017sqm}. However, the perturbative analysis has certain uncertainties related to the renormalization scale dependence and other theoretical aspects, as discussed in references \cite{Chiang:2018gsn, Athron:2022jyi, Croon:2020cgk, Gould:2021oba}. These uncertainties could lead to small differences between our results and those in the literature.

\begin{figure}[h!]
\centering
\includegraphics[width=35pc]{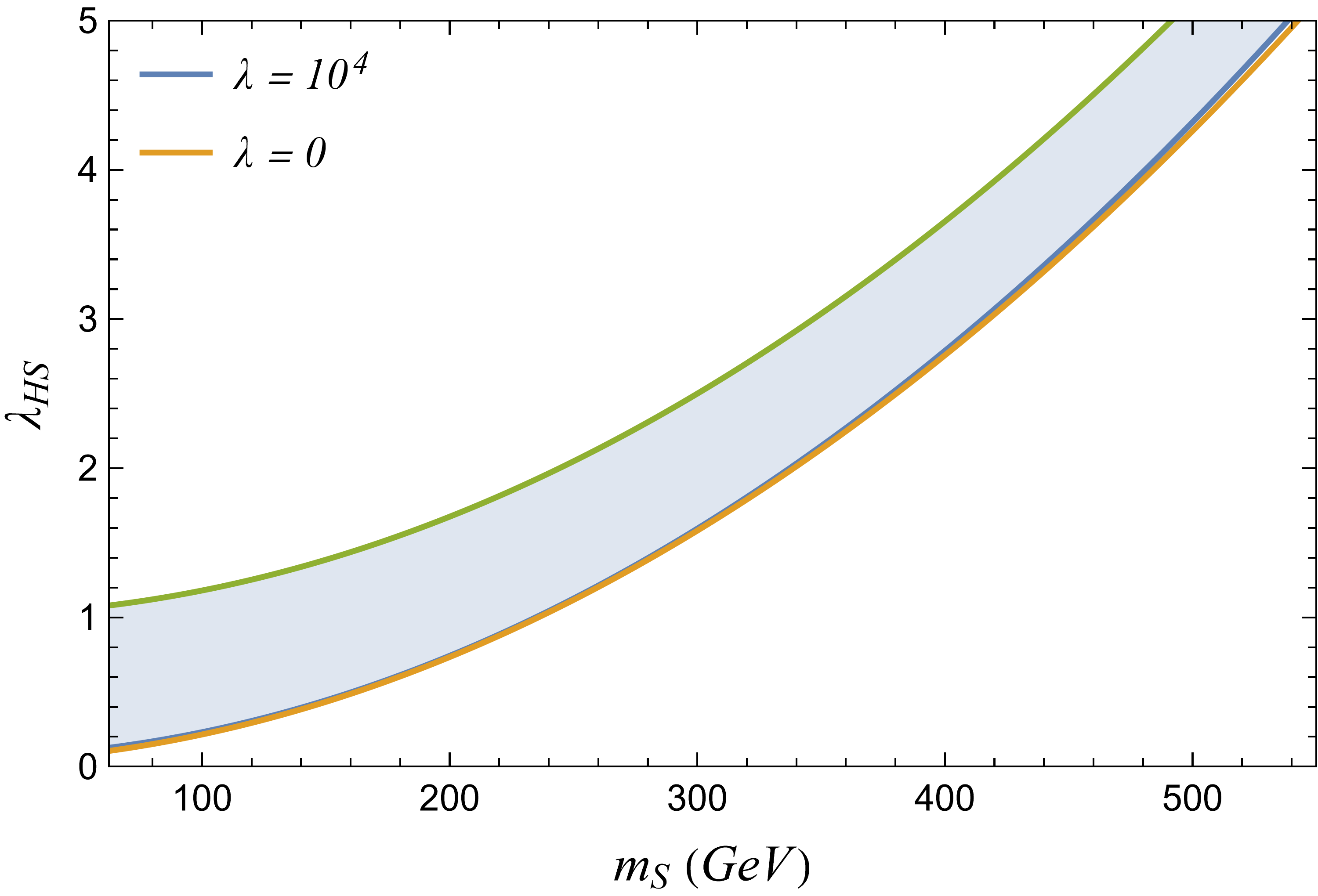}
\caption{The parameter space (blue region) of the singlet
extension with a dimension-six operator (\(\lambda = 10^4\)) in
order to realize a strong two-step electroweak phase transition
(\(\upsilon_c/T_c > 1\)). The orange line shows the lower bound of
the parameter space of the singlet extension which is represented
by a blue dotted line in Fig. \ref{ParameterSpace1}. In both
cases, we take \(a = 0.1\).}\label{ParameterSpace2}
\end{figure}

The parameter space for the strong electroweak phase transition in Fig. \ref{ParameterSpace1} changes slightly if the Wilson coefficient is \(\lambda < 10^{2}\). Lower singlet masses weaken the impact of the non-zero Wilson coefficient when the Wilson coefficient increases. A comparison between the parameter spaces of the singlet extension with \(\lambda = 0\) and \(\lambda = 10^4\) is shown in Fig. \ref{ParameterSpace2}. Notably, for \(a = 0.1\), the influence of the dimension-six operator is observed to be insignificant. However, this behavior changes significantly for higher values of the parameter \(a\) and is further investigated for lower singlet masses.

Fig. \ref{T-potential_2} illustrates the behavior of the finite-temperature effective potential during the thermal history of the Universe. In this specific case, the phase transition of the singlet is first order, characterized by parameters \(m_S = 500\) GeV, \(\lambda_{HS} = 4.3\), \(\lambda/M^2 \simeq 2 \times 10^{-5}\) GeV\(^{-2}\), and \(a = 0.1\). However, for lower singlet masses and Higgs-singlet couplings, it can also be a second-order phase transition. As depicted, at high temperatures, the effective potential is symmetric. As the Universe cools down, the effective potential develops a second minimum in the \(\phi\) direction, which becomes the global minimum at temperatures below \(218\) GeV. Then, a minimum appears in the Higgs direction, and a barrier is formed between this minimum and the global minimum. At the critical temperature, \(T_c = 95.8\) GeV, the two minima are degenerate and below this temperature the Higgs vacuum becomes deeper than the singlet vacuum, leading to the electroweak phase transition.

\subsection{Higgs Resonance Region}

The Higgs resonance region is of particular interest in dark matter physics because the singlet can be considered a viable dark matter candidate with a mass around \(m_S = 62.5\) GeV \cite{GAMBIT:2017gge, Athron:2018ipf, Feng:2014vea}. This region, however, is highly constrained by direct dark matter searches conducted by experiments such as LUX, XENON1T, and XENONnT \cite{Cline:2013gha, Cline:2012hg, GAMBIT:2017gge, Athron:2018ipf, Feng:2014vea, Beniwal:2017eik}.

\begin{figure}[h!]
\centering
\includegraphics[width=35pc]{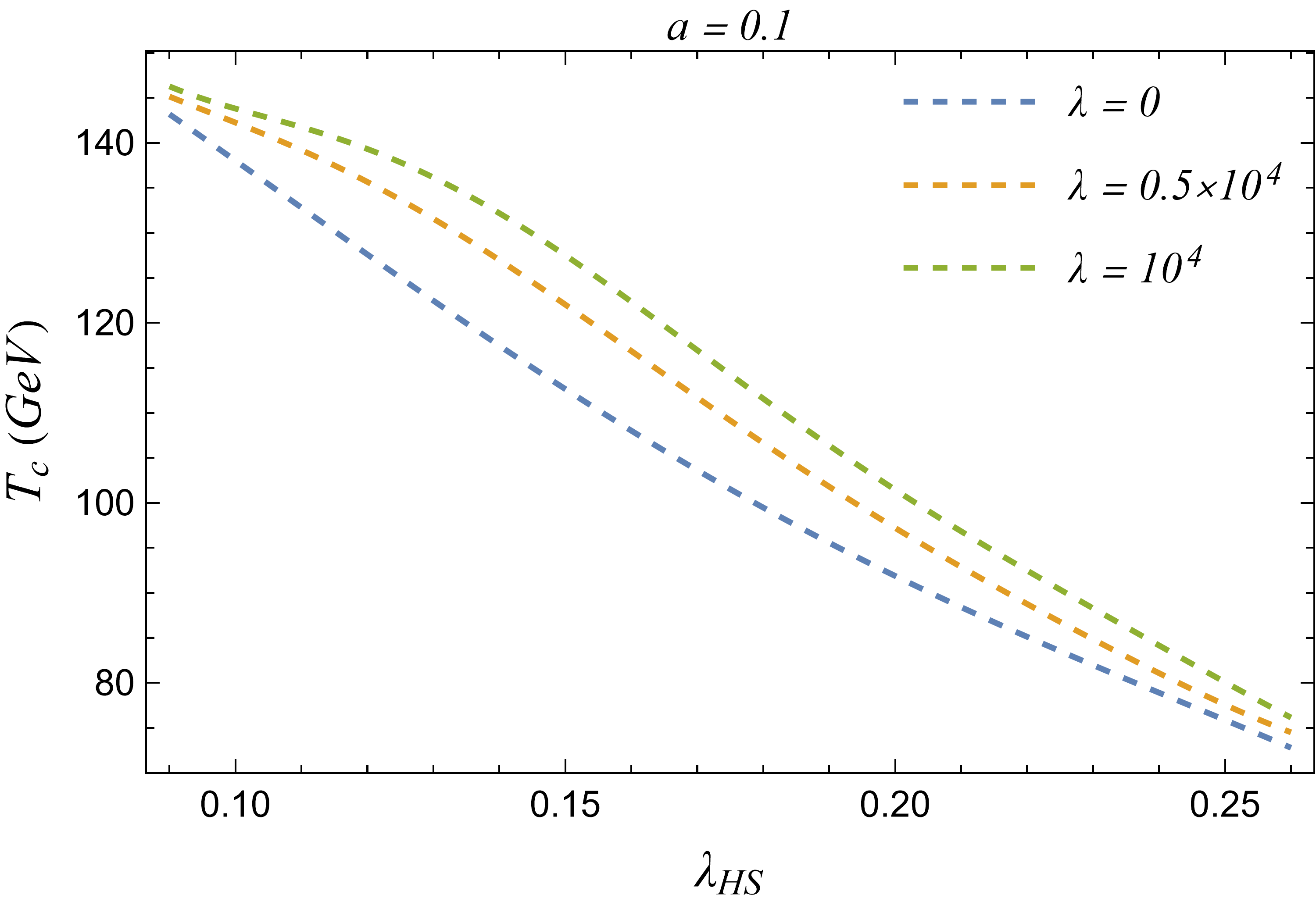}
\caption{The critical temperature (\(T_c\)) as a function of the
Higgs-singlet coupling (\(\lambda_{HS}\)) for $m_S = 62.5 $ GeV
and \(a = 0.1\).}\label{Tc_ms_625_a_01}
\end{figure}

\begin{figure}[h!]
\centering
\includegraphics[width=30.5pc]{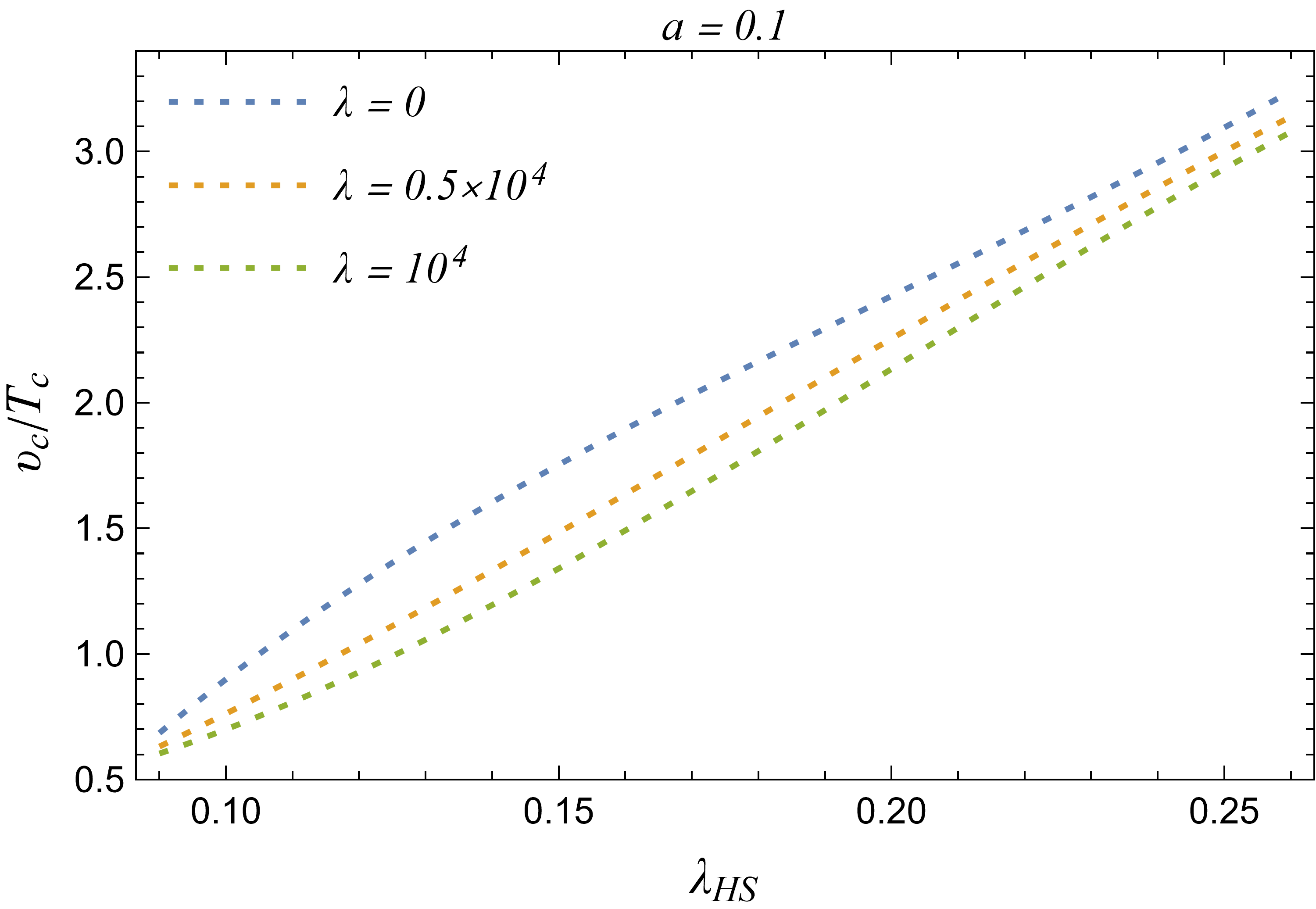}
\includegraphics[width=30.5pc]{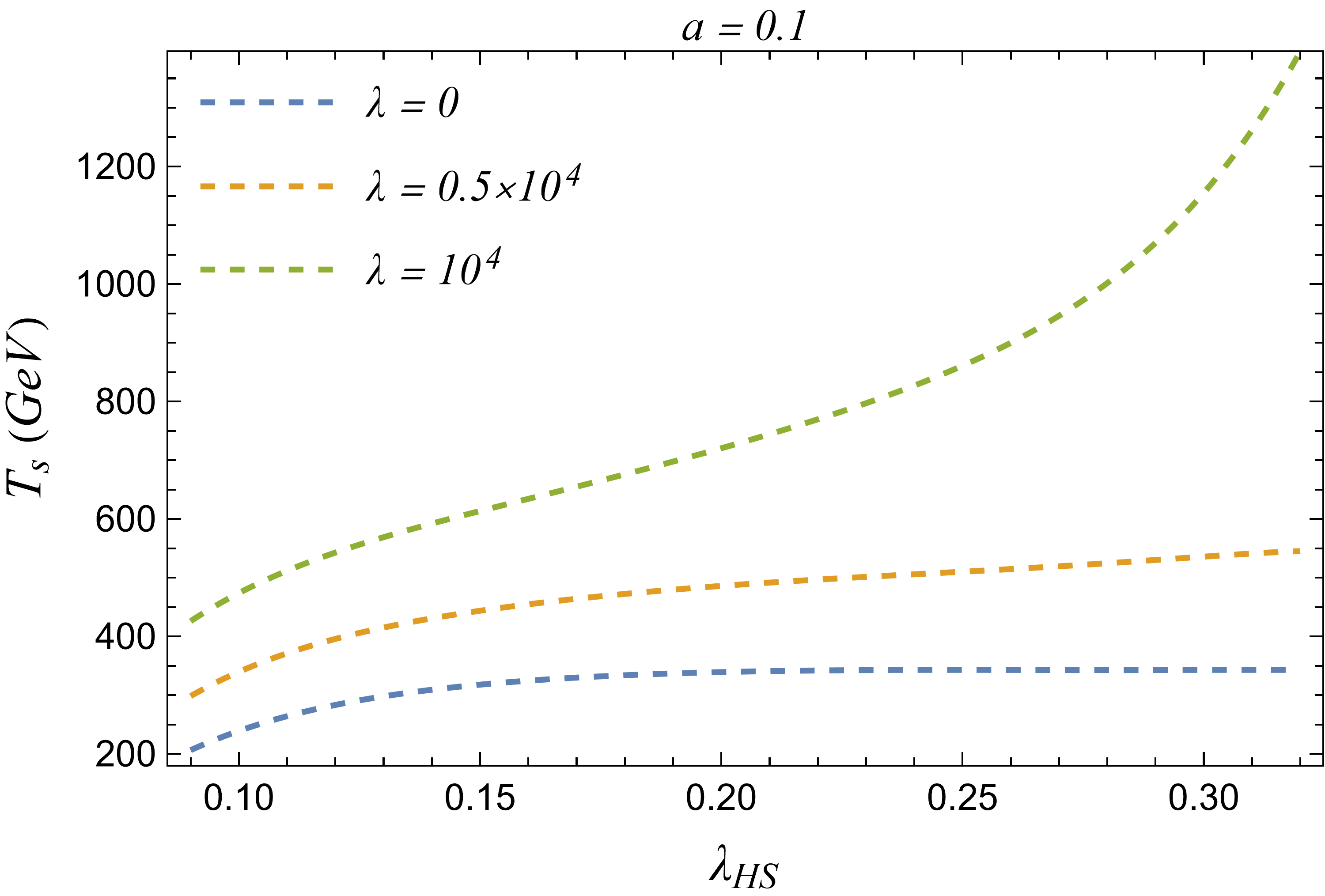}
\caption{\textbf{Upper}: The sphaleron rate criterion as a function
of the coupling \(\lambda_{HS}\) for $m_S = 62.5 $ GeV and \(a =
0.1\) in the case of zero and non-zero Wilson coefficient.
\textbf{Lower}: The critical temperature of the singlet's
second-order phase transition \(T_s\) as a function of the
coupling \(\lambda_{HS}\) for $m_S = 62.5 $ GeV and \(a =
0.1\).}\label{sr_Ts_ms_625_a_01}
\end{figure}

\begin{figure}[h!]
\centering
\includegraphics[width=30.5pc]{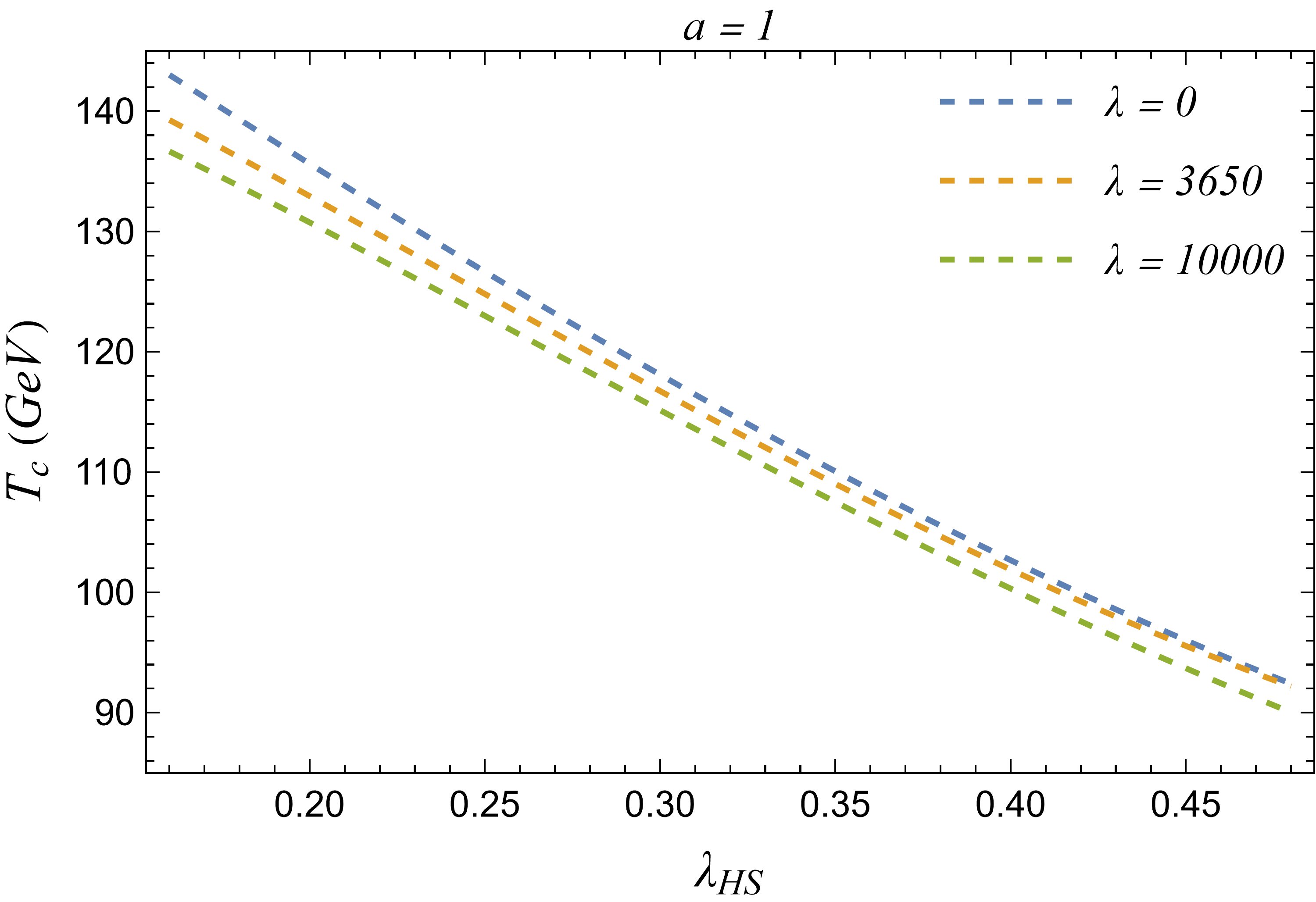}
\includegraphics[width=30.5pc]{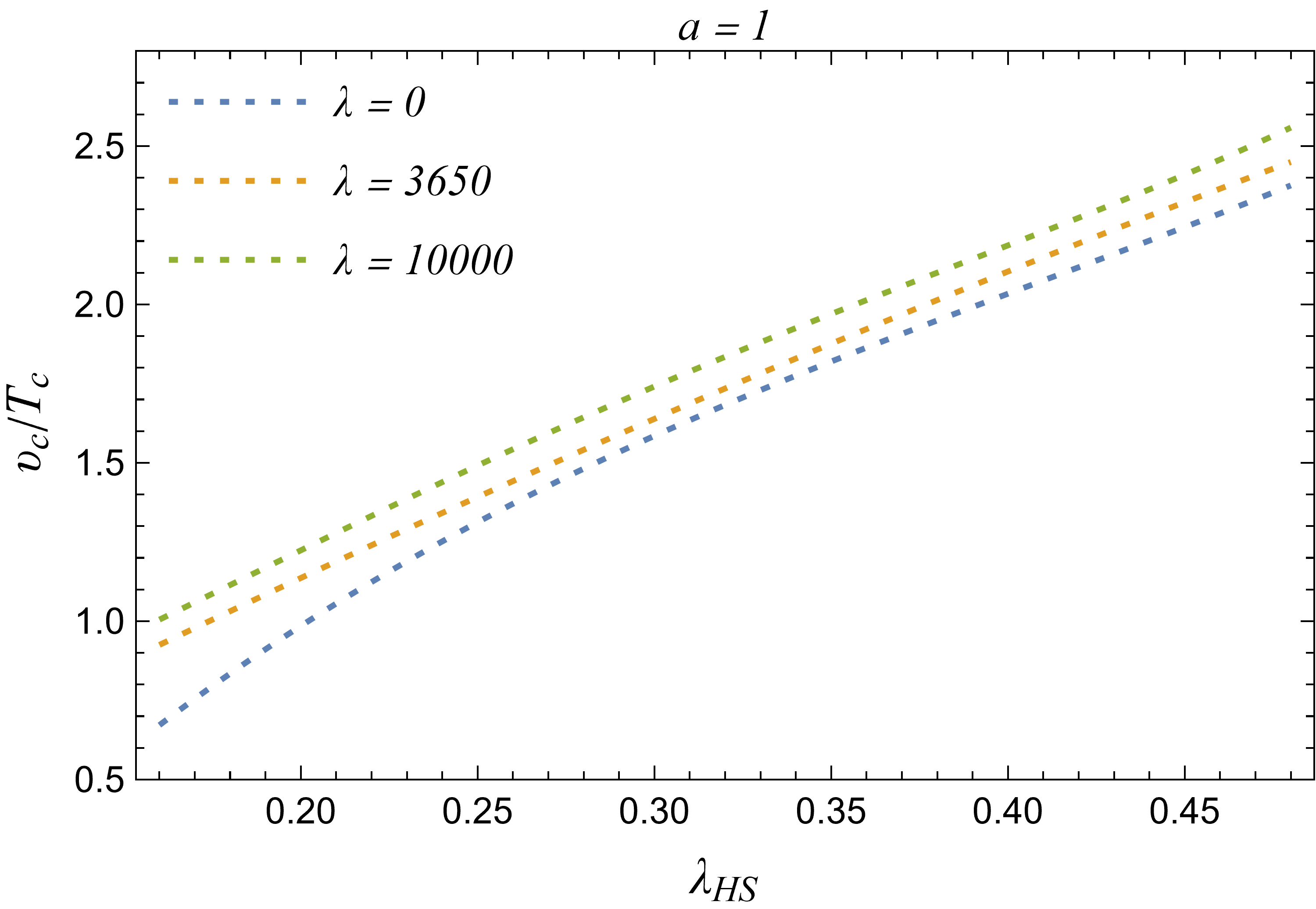}
\caption{The electroweak phase transition with $m_S = 62.5 $ GeV
and \(a = 1\): \textbf{Upper}: The critical temperature (\(T_c\))
as a function of the Higgs-singlet coupling (\(\lambda_{HS}\)).
\textbf{Lower}: The sphaleron rate criterion as a function of the
coupling \(\lambda_{HS}\) in the case of zero and non-zero Wilson
coefficient.}\label{Tc_sr_ms_625_a_1}
\end{figure}

\begin{figure}[h!]
\centering
\includegraphics[width=35pc]{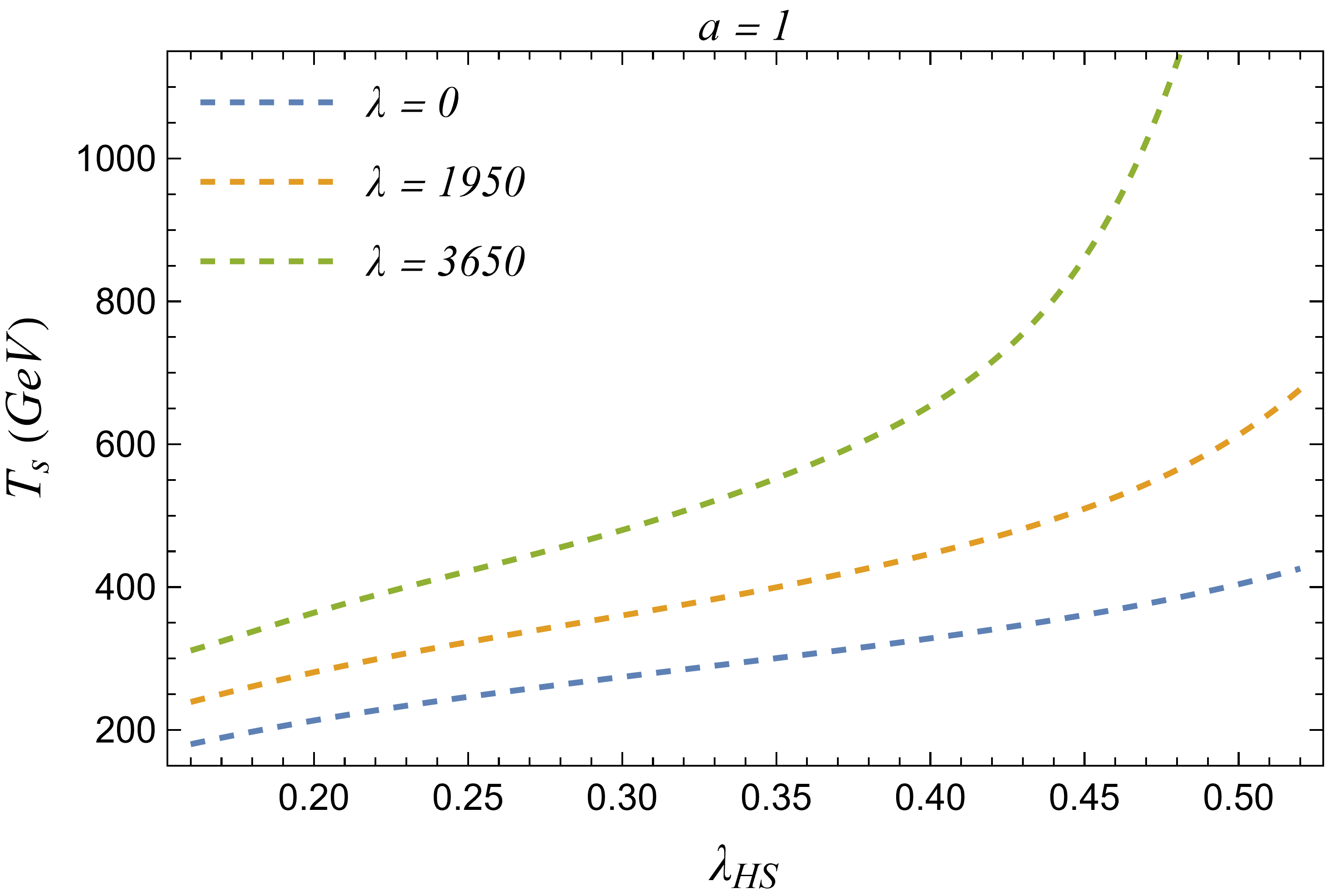}
\caption{The critical temperature of the singlet's second-order
phase transition \(T_s\) as a function of the coupling
\(\lambda_{HS}\) for $m_S = 62.5 $ GeV and \(a = 1\) in the
case of zero and non-zero Wilson
coefficient.}\label{Ts_ms_625_a_1}
\end{figure}

To begin with, we adopt \(a = 0.1\) to calculate the critical temperatures \(T_c\) and \(T_s\), along with evaluating the sphaleron rate criterion for various Wilson coefficient values and Higgs-singlet couplings. These results are shown in Figs. \ref{Tc_ms_625_a_01}, \ref{sr_Ts_ms_625_a_01}, \ref{Tc_sr_ms_625_a_1}, and \ref{Ts_ms_625_a_1}. The presence of the higher-order operator significantly influences the electroweak phase transition when \(\lambda > 10^{3}\), leading to a reduced parameter space where the ratio \(\upsilon_c/T_c\) decreases due to the non-zero Wilson coefficient and a higher Higgs-singlet \(\lambda_{HS}\) is required to realize a strong electroweak phase transition.


\par On the other hand, the parameter space for a strong electroweak phase transition expands when \(a > 0.4\). As demonstrated for \(a = 1\) in Fig. \ref{Tc_sr_ms_625_a_1}, the criterion \(\upsilon_c/T_c > 1\) can be met with much lower \(\lambda_{HS}\) than in the case with \(\lambda = 0\). Consequently, if \(\lambda/M^2 \gtrsim 10^{-4}\) GeV\(^{-2}\), a strong electroweak phase transition can occur for every low Higgs-singlet couplings (satisfying the condition \(\mu^2_S > 0\)). Therefore, the higher-order operator could assist a strong electroweak phase transition in parameter space regions that were previously excluded in earlier singlet extensions of the Standard Model.

\par In Figs. \ref{Tc_sr_ms_625_a_1} and \ref{Tc_ms_625_a_01}, the critical temperature with \(\lambda \ne 0\) converges with the critical temperature for \(\lambda = 0\) as \(\lambda_{HS}\) increases. This indicates that the higher-order operator has a more dominant effect at lower \(\lambda_{HS}\) values. This trend is also evident in the ratio \(\upsilon_c/T_c\) shown in Figs. \ref{sr_Ts_ms_625_a_01} and \ref{Tc_sr_ms_625_a_1}. Nevertheless, for \(a = 0.1\), the influence of the higher-order operator peaks at an intermediate low value of the Higgs-singlet coupling, rather than at the lowest values.

\subsection{Low-mass Region}
In the low mass region, the effective potential exhibits behavior similar to that presented in the Higgs resonance region in Fig. \ref{T-potential_1}. Meanwhile, the singlet's phase transition is predominantly second-order within the parameter space which complies with the experimental and theoretical constraints.

\begin{figure}[h!]
\centering
\includegraphics[width=35pc]{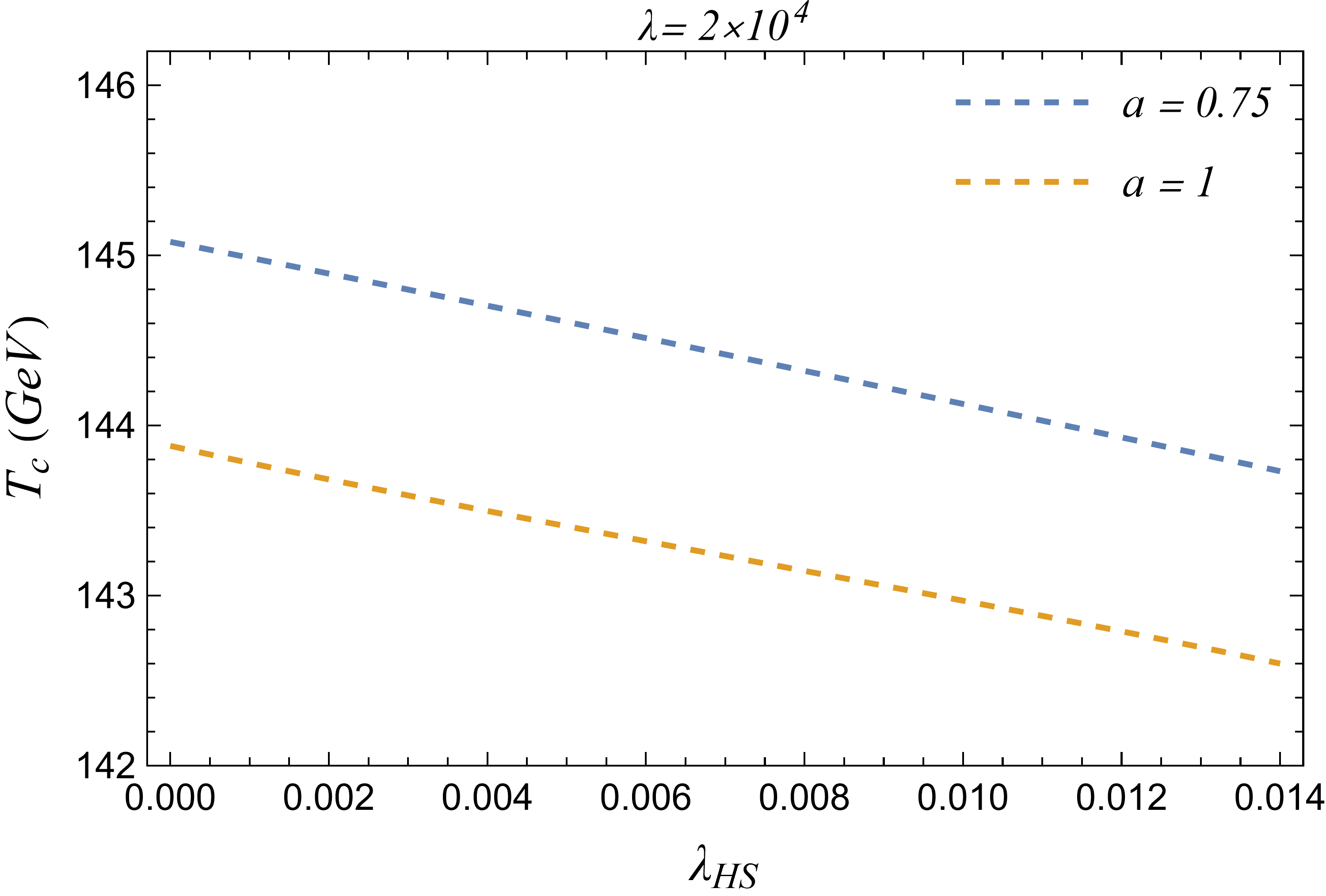}
\caption{The critical temperature as a function of the Higgs-singlet coupling for \(a = 0.75\) and \(1\) and \(m_S \leq 0.1\) GeV.}\label{T_c_a_1_0.75}
\end{figure}

In previous sections, we discussed that the parameter space for \(m_S < m_H/2\) is significantly constrained by the invisible decay width of the Higgs boson and the condition \(\mu^2_S > 0\). First, it is assumed that the branching ratio of the Higgs boson to invisible particles is \(BR_{inv} < 0.19\). The lower mass region \(m_S \lesssim 30\) GeV in Fig. \ref{BR_0.19} is eliminated by the sphaleron rate criterion (\ref{criterion_baryogenesis}) for \(\lambda = 0\) and \(a \gtrsim 0.05\). In contrast, large values of \(\lambda > 10^3\) and \(a \gtrsim 0.05\) can support a strong phase transition for \(m_S \leq 1\) GeV. This is evident when \(\lambda = 2 \times 10^4\) and \(a = 0.75\) are considered, because a strong electroweak phase transition occurs for any coupling \(\lambda_{HS}\) and \(m_S < 10\) GeV, as shown in Fig. \ref{T_c_a_1_0.75}. In particular, the lower bound of the Higgs-singlet coupling \(\lambda_{HS}\) with a singlet mass \(m_S = 0.1\) GeV and Wilson coefficient \(\lambda = 2 \times 10^4\) can induce a strong electroweak phase transition and is at least \(10^6\) times smaller than the corresponding value in the singlet extension with \(\lambda = 0\). Thus, the non-zero Wilson coefficient results in a strong electroweak phase transition for very low allowed Higgs-singlet couplings. This holds even if the branching ratio is \(BR_{inv} < 0.10\).

\par
\begin{figure}[h!]
\centering
\includegraphics[width=35pc]{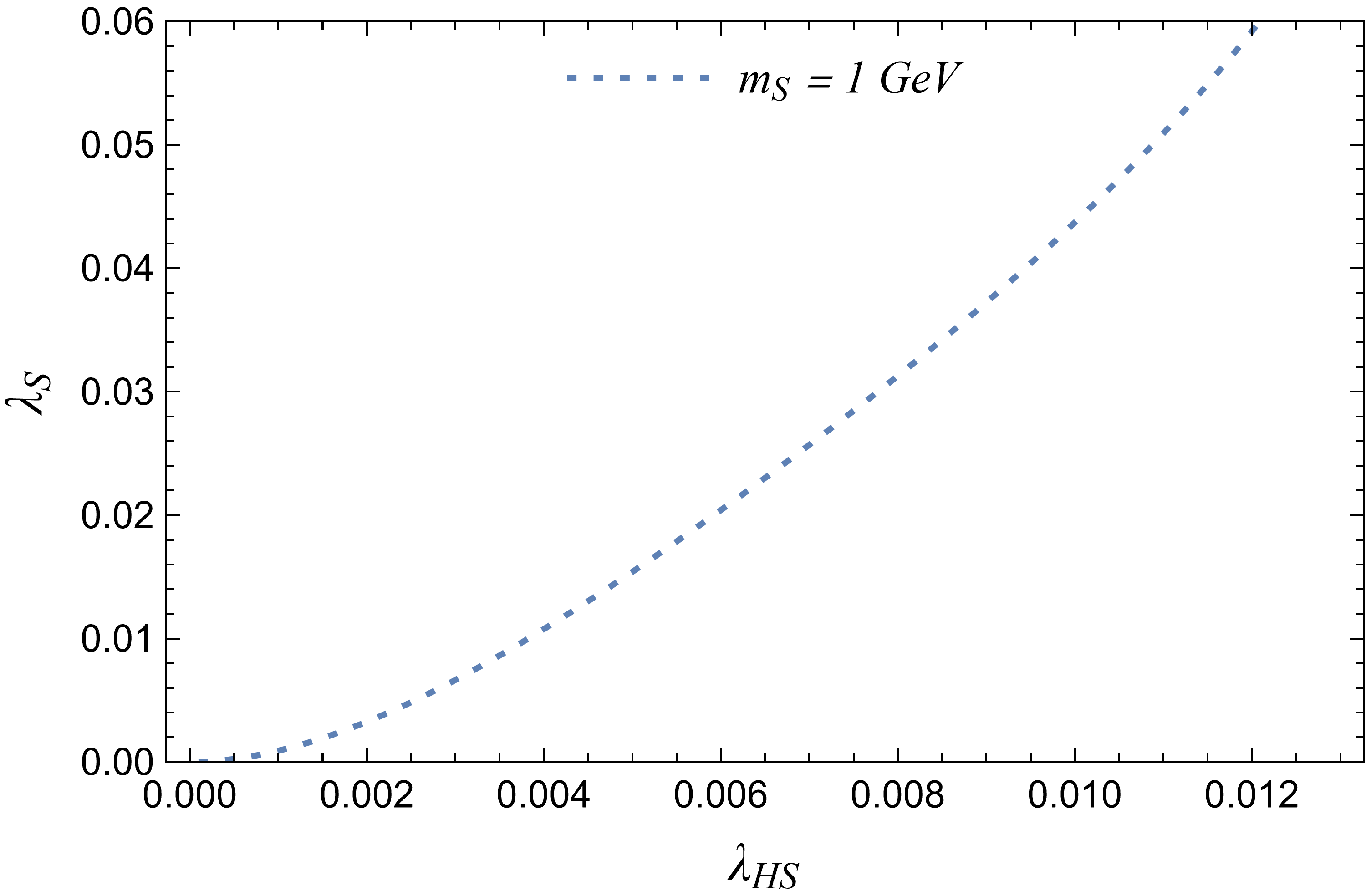}
\caption{The dependence of the lowest \(\lambda_{HS}\), which is
allowed by all the previous constraints imposing the criterion
\(\upsilon_c/T_c > 0.6\), on the parameter\(a\) for $m_S = 1 $GeV
and \(\lambda = 0\).}\label{lambda_s_k_lowest_ms_1_GeV}
\end{figure}

 On the other hand, the previous trend is reversed for \(a \lesssim 0.05\). In this range, \(\lambda = 0\) results in a strong first-order phase transition, while \(\lambda \ne 0\) weakens the phase transition. With \(\lambda = 0\), the electroweak phase transition primarily occurs for \(m_S < 10\) GeV to achieve a two-step phase transition. However, this upper mass limit increases for very low values of \(a\) and \(\lambda_{HS}\). This behavior is demonstrated in Fig. \ref{lambda_s_k_lowest_ms_1_GeV}, which shows that the lowest allowed value of \(\lambda_{HS}\) decreases as \(a\) decreases for a given singlet mass.

\par
\begin{figure}[h!]
\centering
\includegraphics[width=35pc]{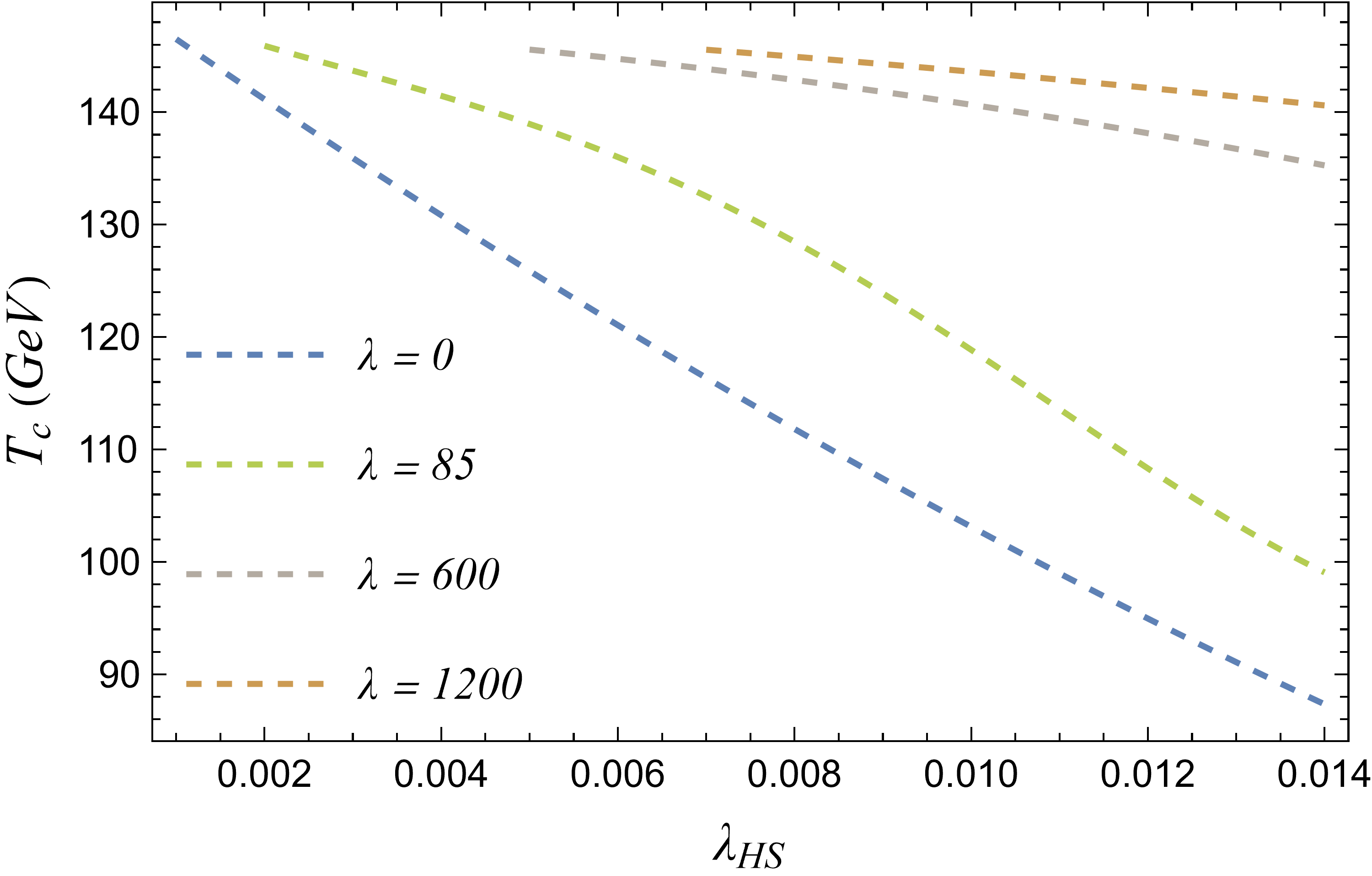}
\caption{The critical temperature of a strong electroweak phase transition as a function of the coupling
\(\lambda_{HS}\) for $m_S \leq 1 $ GeV and \(a = 0.001\) in the
case of zero and non-zero Wilson coefficient \(\lambda = 0\)
(blue), \(85\) (green), \(600\) (gray), \(1200\)
(brown).}\label{critical_temperature_ms_0.1_a_0.001}
\end{figure}

\begin{figure}[h!]
\centering
\includegraphics[width=30.5pc]{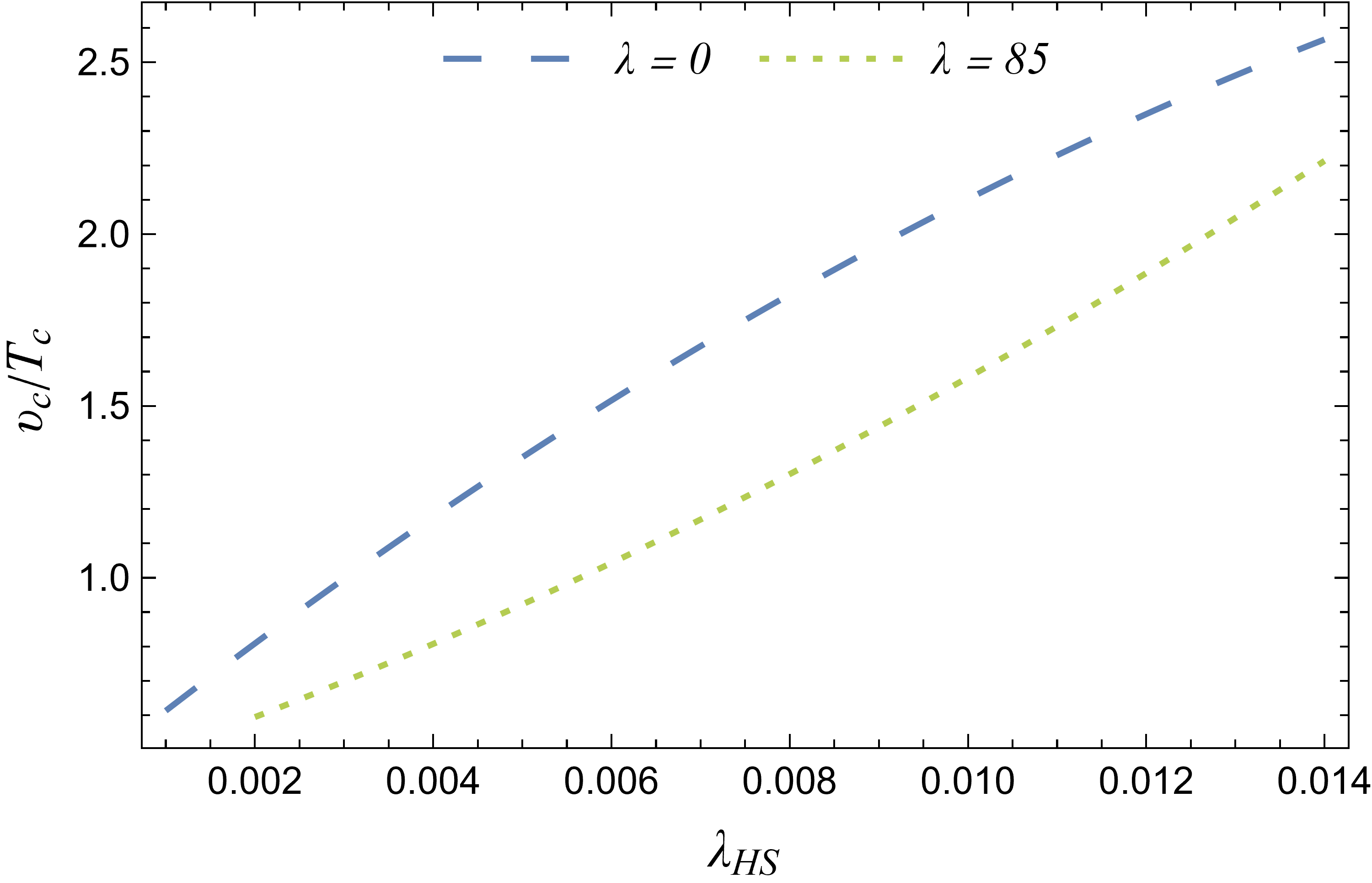}
\includegraphics[width=30.5pc]{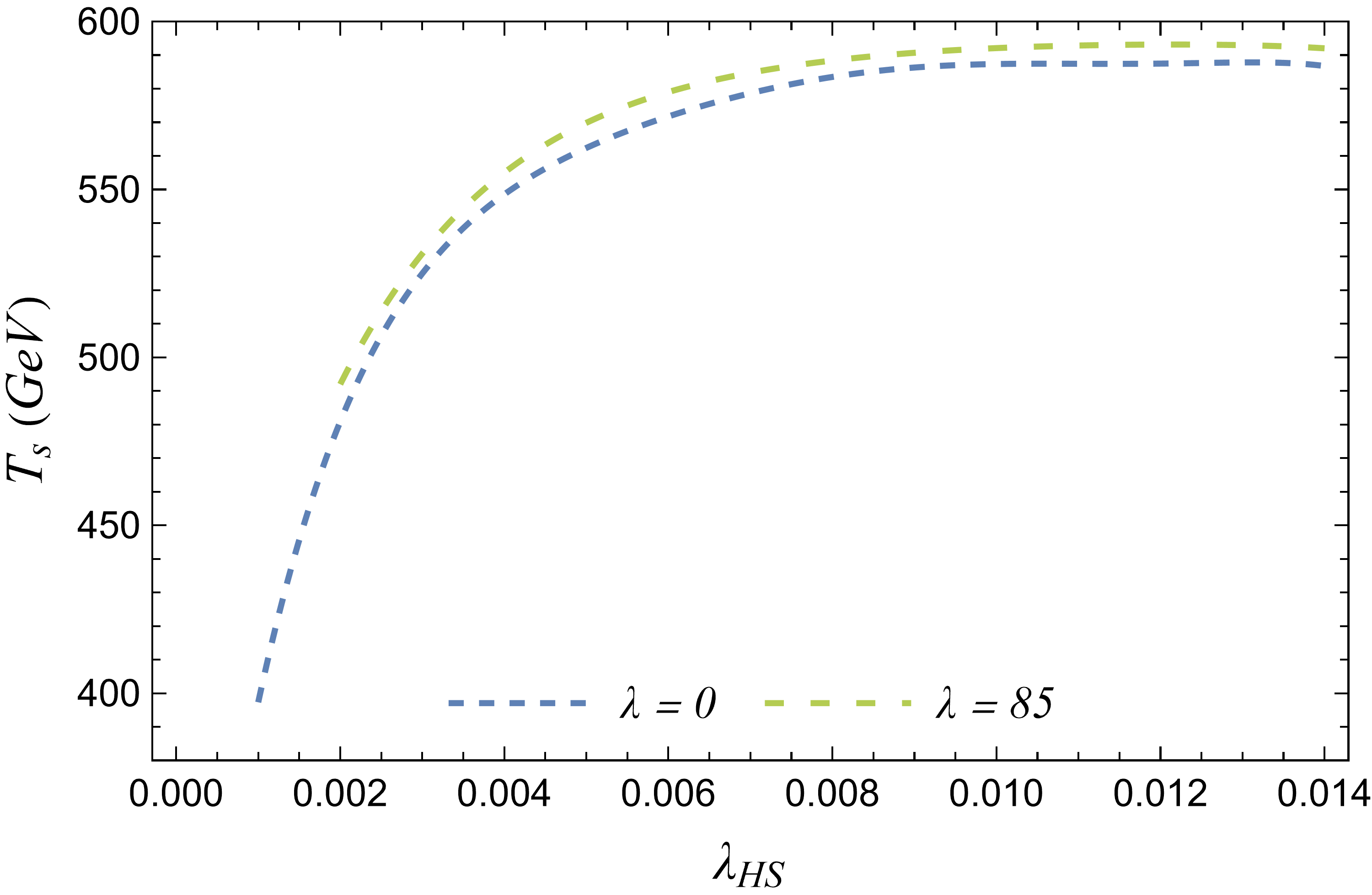}
\caption{\textbf{Upper}: The sphaleron rate criterion as a function
of the coupling \(\lambda_{HS}\) for $m_S \leq 1 $GeV and \(a =
0.001\) in the case of zero and non-zero Wilson coefficient.
\textbf{Lower}: The critical temperature of the singlet's
second-order phase transition \(T_s\) as a function of the
coupling \(\lambda_{HS}\) for $m_S \leq 0.1 $GeV and \(a = 0.001\)
in the case of zero and non-zero Wilson
coefficient.}\label{sr_Ts_ms_0.1_a_0.001}
\end{figure}

 In Figs. \ref{critical_temperature_ms_0.1_a_0.001} and \ref{sr_Ts_ms_0.1_a_0.001}, one observes the behavior of the critical temperatures and the ratio of the sphaleron rate condition for \(m_S \leq 1\) GeV and \(a = 0.001\). Initially, it is obvious that the lower bound of \(\lambda_{HS}\), which induces a strong electroweak phase transition, decreases considerably for \(\lambda/M^2 \simeq 10^{-7} - 10^{-5}\) GeV\(^{-2}\). In Fig. \ref{critical_temperature_ms_0.1_a_0.001}, the pattern of the critical temperature \(T_c\) is similar to that seen in the previous mass region, while in Fig. \ref{sr_Ts_ms_0.1_a_0.001}, the temperature \(T_s\) stabilizes at high values of \(\lambda_{HS}\), which contrasts with the behavior discussed earlier.


\par In the scenario where the branching ratio is \(BR_{inv} < 0.11\), the earlier conclusions for the lower mass region are unaffected. This constraint only drops the maximum Higgs-singlet coupling to \(\lambda^{max}_{HS} = 0.01\), which in turn further reduces the parameter space to describe a strong electroweak phase transition.

Regarding the second region of the parameter space for \(m_S < m_H/2\), the singlet mass is highly constrained to nearly half the mass of the Higgs boson due to the condition (\ref{condition_coupling}) (see Fig. \ref{BR_0.19} and Fig. \ref{BR_0.11}). Consequently, there is no significant difference between this scenario and the Higgs resonance scenario. More specifically, the results for the critical temperature \(T_c\) and the Higgs VEV \(\upsilon_c\) remain consistent with those from the previous section. Additionally, if we assume that \(BR_{inv} < 0.19\), the parameter space for a strong electroweak phase transition almost vanishes for zero Wilson coefficient due to the invisible Higgs decay width constraint (\ref{condition_coupling}). However, the presence of a higher-order operator can reduce the lower bound on \(\lambda_{HS}\) due to the sphaleron rate criterion discussed earlier. It is important to comment that in the higher mass region, an allowed value of \(\lambda_{HS} > 0.065\) can induce a viable electroweak phase transition for \(\lambda/M^2 \simeq 9 \times 10^{-5}\) GeV\(^{-2}\) and \(a = 0.75\). Thus, the dimension-six operator can lead to a strong first-order phase transition in the higher mass region of the parameter space for \(m_S < m_H/2\). This discussion also applies to \(BR_{inv} < 0.11\), as the lower bound of the Higgs-singlet coupling remains unchanged, but the parameter space for a strong two-step electroweak phase transition significantly reduces for \(m_S < 62.499\) GeV.

\section{Conclusions}

In this section, we present the key findings of our research on the real singlet extensions of the Standard Model to describe a strong electroweak phase transition and the observed baryon asymmetry. In this research, we particularly investigated the real singlet extensions, including dimension-six operators that couple the real singlet scalar field with the Higgs doublet. This study showed that the electroweak phase transition proceeds as a two-step transition: an initial phase transition involving the singlet scalar field at high temperatures, followed by a strong first-order phase transition in the Higgs sector. On the other hand, the parameters of our model should respect all the experimental and theoretical constraints in cosmology and particle physics, such as experimental constraints on the invisible Higgs decay width and the theoretical condition for a strong first-order phase transition. As a result, we examined the two-step electroweak phase transition that complies with these constraints for \(m_S = 0 - 550\) GeV. In this model, the critical temperature of the electroweak phase transition ranges from around \(30 - 200\) GeV and depends on the parameters of the model. Initially, we demonstrated that the parameter space of our model for \(m_S > m_H/2\) and \(a = 0.1\) remains approximately the same as in the singlet extension without a higher-order operator. Then, it was demonstrated that in the Higgs resonance region, the parameter space is expanded by taking a parameter \(a > 0.4\) and a non-zero Wilson coefficient. This is clearly illustrated in the case of \(a = 1\) in Fig. \ref{Tc_sr_ms_625_a_1}, where the criterion \(\upsilon_c/T_c > 1\) can be satisfied by much lower \(\lambda_{HS}\) compared to the case with zero Wilson coefficient. Namely, it is evident that the higher-order operator could assist a strong electroweak phase transition in regions of the parameter space that were previously excluded in the literature. Moreover, in the low-mass region, the parameter space is primarily excluded by the invisible Higgs decay width and the criterion for a strong first-order phase transition. If we assume that the branching ratio of the Higgs boson to invisible particles is set to \(BR_{inv} < 0.19\), the lower mass region \(m_S < 30\) GeV, is completely excluded by the sphaleron rate criterion considering zero Wilson coefficient and \(a > 0.05 \). In contrast, large values of the Wilson coefficient with the same value for \(a\) can assist the strong phase transition for singlet masses \(m_S \leq 1 - 10\) GeV. Consequently, a strong electroweak phase transition can be generated for \(m_S < m_H/2\) with Wilson coefficients \(\lambda/M^2 > 5 \times 10^{-6}\) GeV\(^{-2}\) and parameters \(a \gtrsim 0.05\).

\chapter{Conclusions and Discussion}
This present thesis explores the cosmological phase transitions in the early Universe, with an emphasis on the electroweak phase transition in the Standard Model and its extensions. Our study comprehensively investigates the role of the effective potential in determining the true vacuum of the theory, describing phase transitions, and providing numerous insights into electroweak baryogenesis.

After a brief introduction to spontaneous symmetry breaking, we proceeded to a discussion about the Higgs mechanism and the Standard Model which relies on the electroweak symmetry breaking to describe the massive gauge bosons and fermions in nature. We extended the discussion to the spontaneous symmetry breaking induced by quantum corrections to the classical potential. These radiative corrections are taken into account in the effective potential which is minimized at the vacuum expectation value of the quantum field operator in the true vacuum of the theory. Thus, the one-loop effective potential in scalar field theories and non-Abelian gauge theories was derived by evaluating the 1PI Feynman diagrams with a single loop and zero external momenta. We also proceeded to the regularization and renormalization in the aforementioned theories to deal with the ultraviolet divergences in the one-loop effective potential. 

Additionally, this thesis demonstrates how a spontaneously broken symmetry can be restored at high temperatures due to finite-temperature effects. This symmetry restoration is thoroughly examined using the finite-temperature effective potential. The finite-temperature field theory was essential in formulating the Feynman rules at finite temperatures and computing the effective potential for various quantum field theories. This approach immediately shows that the electroweak symmetry in the Standard Model is restored at high temperatures due to thermal effects and is associated with a cosmological phase transition in the early hot Universe. However, the previous perturbative analysis to compute the effective potential breaks down near the critical temperature and requires thermal resummation to incorporate the dominant contributions from the ring diagrams.

In the next part, we investigated the theory of cosmological phase transitions to describe the thermal tunneling and the bubble nucleation during a phase transition. In particular, this theory was developed to understand further the tunneling rate between two different vacua in finite-temperature field theory and determine the evolution of the phase transition in the expanding Universe. After the formation of the bubbles, we showcased that the bubble nucleation is initiated at a temperature such that the probability for a single bubble to be nucleated within one horizon volume is around unity.

Moving on, the study examined the physical mechanism to describe the observed matter-antimatter asymmetry of the Universe. More specifically, the three conditions for baryogenesis were presented in detail and electroweak baryogenesis was proved a viable description of the baryon asymmetry of the Universe. Then we emphasized the importance of a strong first-order phase transition during electroweak baryogenesis and formulated a criterion to be satisfied by this phase transition at the critical temperature. Subsequently, the finite-temperature one-loop effective potential of the Standard Model was computed, including the ring diagram corrections to study in detail the electroweak baryogenesis. In this perturbative analysis, it was shown that the electroweak phase transition is first order and the critical temperature is around \(T_c = 148\) GeV varying the renormalization scale from \(86.5\) GeV to \(346\) GeV in the \(\overline{\rm MS}\) scheme and \(T_c = 161\) GeV in the OS scheme. However, it was concluded that the electroweak phase transition in the Standard Model is not sufficiently strong to account for the observed baryon asymmetry in the Universe. In order to address the shortcomings of the Standard Model, the thesis discussed the potential of Beyond Standard Model physics in explaining fundamental problems in cosmology and particle physics.

In the last chapter, we aimed to provide a viable solution to the problem of electroweak baryogenesis. We particularly explored the electroweak phase transition within the framework of real singlet extensions of the Standard Model, including dimension-six operators that couple the real singlet scalar field with the Higgs doublet. This study showed that the electroweak phase transition proceeds as a two-step transition: an initial phase transition involving the singlet scalar field at high temperatures, followed by a strong first-order phase transition in the Higgs sector. Additionally, we considered the role of a CP-violating source, such as a dimension-six operator coupling the singlet to the top-quark mass. As a result, electroweak baryogenesis can be realized to account for the baryon asymmetry observed in the current Universe. This scenario is also compatible with the singlet particle being a dark matter candidate with a mass close to half of the Higgs mass. Therefore, we examined the two-step electroweak phase transition for singlet masses \(m_S = 0 - 550\) GeV that complies with various experimental and theoretical constraints. In this model, it was computed that the critical temperature of the electroweak phase transition ranges from around \(30 - 200\) GeV, which depends on the parameters of the model. Last but not least, it was concluded that the presence of the higher-order operator could assist a strong electroweak phase transition in regions of the parameter space that were excluded in the singlet extensions without the dimension-six operator, especially for singlet masses lower than half of the Higgs mass.

Finally, an important aspect we did not cover in our discussion is the stochastic gravitational wave background associated with the first-order phase transitions in the Standard Model and its extensions. During such phase transitions, the collisions between expanding bubbles of vacua are expected to produce a stochastic gravitational wave background detectable by current and future gravitational wave experiments \cite{Kosowsky:1991ua, Kamionkowski:1993fg, Huber:2008hg, Jinno:2016vai, Caprini:2007xq, Kosowsky:1992vn, Kosowsky:1992rz}. In fact, the sound waves generated after the phase transition and the magneto-hydrodynamical turbulence in the plasma could also source gravitational waves associated with a first-order phase transition in the early Universe \cite{Hindmarsh:2013xza, Hindmarsh:2015qta, Giblin:2013kea, Giblin:2014qia, Kahniashvili:2008pe, Kahniashvili:2009mf, Caprini:2009yp}. In an upcoming study, we plan to investigate this stochastic gravitational wave background in the singlet extension of the Standard Model with dimension-six operators. In this work, we also mentioned certain theoretical uncertainties in the perturbative analysis that can propagate to the calculations for the gravitational-wave background from first-order phase transitions \cite{Croon:2020cgk}. As a result, these uncertainties can be eliminated by following various methods, such as a gauge-invariant framework for computing the bubble nucleation rate \cite{Hirvonen:2021zej, Lofgren:2021ogg}, which are left to be implemented in future studies.

In conclusion, while the Standard Model does not provide a viable description for the observed baryon asymmetry of the Universe, certain extensions, particularly those involving additional singlet scalar fields and higher-dimensional operators, present a possible path forward. Further exploration of these extensions and their implications for particle physics and cosmology remains a promising area of research.

\appendix
\chapter{Chemical Potential}\label{Appendix A}

In Chapter \ref{Finite-Temperature Field Theory}, it was mentioned that if there is a set of conserved charges in a system, the chemical potential for each conserved charge should be introduced in the partition function. In this work, it was mainly considered that the chemical potentials related to conserved charges vanish or can be omitted compared to the temperature (\(\mu_i \ll T\)). However, in certain conditions in particle physics, such as in heavy ion collision experiments, the chemical potentials play an essential role and are included in the effective potential.

To begin with, we show the inclusion of the chemical potential in the effective potential for a complex scalar field theory. The classical Lagrangian is written as
\begin{equation}
    \mathcal{L} = \partial^{\mu} \phi^{*} \partial_{\mu} \phi - m^2 \phi^{*} \phi - \lambda (\phi^{*} \phi)^2
\end{equation}
and the Lagrangian is invariant under the \(U(1)\) global transformations
\begin{equation}
    \phi (x) \to e^{- i a} \phi (x) \, ,
\end{equation}
where \(a \in \mathbb{R} \). According to Noether's theorem, the global \(U(1)\) symmetry implies the conserved current,
\begin{equation}
    j_{\mu} = \frac{\partial \mathcal{L}}{\partial (\partial^{\mu} \phi ) } \frac{\delta \phi}{\delta a} + \frac{\partial \mathcal{L}}{\partial (\partial^{\mu} \phi^{*} ) } \frac{\delta \phi^{*}}{\delta a} = i \left(\phi^{*} \partial_{\mu} \phi - \phi\,  \partial_{\mu} \phi^{*} \right).
\end{equation}
The conserved charge is then defined as
\begin{equation}\label{Q_com}
     Q = \int d^3 x\, j_0 (x) = i \int d^3 x \, \left( \phi^{*} \partial_t \phi - \phi \, \partial_t \phi^{*} \right) \, ,
\end{equation}
The partition function is easily computed by (\ref{gen_Z}) if we express the complex scalar field in terms of two real scalar fields,
\begin{equation}
    \phi (x) = \frac{\phi_1 (x) + i \phi_2 (x)}{\sqrt{2}}
\end{equation}
with the Hamiltonian density
\begin{equation}\label{Ham_com}
    \mathcal{H} = \frac{1}{2} \left( \pi_1^2 + \pi_2^2 + \left( \partial_i \phi_1 \right)^2 + \left( \partial_i \phi_2 \right)^2 + m^2 \phi_1^2 + m^2 \phi_2^2 \right) + \frac{\lambda}{4} \left( \phi_1^2 + \phi_2^2 \right)^2 \, ,
\end{equation}
where the conjugate momentum is \(\pi_i = \partial_t \phi_i = i \partial_{\tau} \phi_i\) and the conserved charge is
\begin{equation}
    Q = i \int d^3 x \, \left( \pi_1 \phi_2 - \pi_2 \phi_1\right) \, .
\end{equation}
Therefore, the partition function in the complex scalar field theory is identical to the partition function (\ref{gen_Z})  in the real scalar field theory by replacing the Hamiltonian density with (\ref{Ham_com}) and the conserved charge with (\ref{Q_com}).  Likewise, the path integral can be written carrying out the integrals over the conjugate momenta \(\pi_1\) and \(\pi_2\) according to the partition function (\ref{part_SE}). As a result, the Lagrangian path integral in the imaginary time formalism can be expressed as \cite{Kapusta:2006pm, Bellac:2011kqa, Laine:2016hma}.
\begin{equation*}
    Z = \int_P \mathcal{D}\phi^{*} \mathcal{D}\phi  \, \exp\biggl\{ - \int_{0}^{\beta} d \tau \int d^3 x \left[ \left(\partial_{\tau} + \mu\right) \phi^{*} \left(\partial_{\tau} - \mu \right) \phi + \partial_i \phi^{*} \partial_i \phi + m^2 \phi^{*} \phi + \lambda (\phi^{*} \phi)^2 \right] \biggl\}
\end{equation*}
where the usual normalization factor is omitted and \(\mu\) is the chemical potential associated with the conserved charge.  The partition function can also expressed as
\begin{equation}
\begin{split}
    Z = & \int_P \mathcal{D}\phi^{*} \mathcal{D}\phi  \, \exp\biggl\{ \int_{0}^{\beta} d \tau \int d^3 x \left[ \left(\partial_{\nu} + i \mu \delta_{\nu 0}\right) \phi^{*} \left(\partial^{\nu} - i \mu \delta^{\nu 0} \right) \phi \right] \\
    & + \int_P \mathcal{D}\phi^{*} \mathcal{D}\phi  \, \exp\biggl\{ \int_{0}^{\beta} d \tau \int d^3 x \left[ - m^2 \phi^{*} \phi - \lambda (\phi^{*} \phi)^2 \right] \biggl\}
\end{split}
\end{equation}
in the Minkowski space-time. It is very interesting that the chemical potential appears in the action as a gauge field \(A_0\). Moreover, the imaginary time formalism that was developed in the previous chapters could be easily extended by introducing the zeroth-component of the four-momentum as \(p_0 = i \omega_n + \mu\).  As a result, one proves that the one-loop contribution in the effective potential of the free theory is given by \cite{Kapusta:2006pm, Bellac:2011kqa, Laine:2016hma}
\begin{equation}
    \frac{1}{\beta V} \ln Z = \int \frac{d^3 p}{(2 \pi)^3} \left[ \omega + \frac{1}{\beta} \ln \left(1 - e^{- \beta (\omega- \mu)} \right) + \frac{1}{\beta} \ln \left( 1 - e^{- \beta (\omega + \mu)} \right) \right],
\end{equation}
where \(\omega = \sqrt{\vec{p}^{\, 2}+ m^2}\) and the momentum integral converges only if  \(|\mu| \leq m\). This expression coincides with (\ref{last_V1}) substituting \(\omega \to \omega - \mu\) and \(\omega \to \omega + \mu\) since it corresponds to two real scalar fields. This result can be obtained by expanding the complex scalar field into a Fourier transform in terms of \(\zeta\) which carries the full infrared behavior of the field. The Matsubara mode with  \(\omega_{n} = 0\) and \(\vec{p} = 0\) is called the condensate and is described by \(\phi = \zeta\) \cite{Laine:2016hma, Kapusta:1981aa}. As a result, the finite-temperature one-loop effective potential can be identified as the free-energy density which equals to \cite{Laine:2016hma, Kapusta:1981aa, Haber:1981ts, Benson:1991nj}
\begin{equation}
\begin{split}
    V_{\text{eff}} (\zeta, \mu,  T)  =& \left(m^2 - \mu^2  \right) \zeta^2 - \int \frac{d^3 p}{(2 \pi)^3} \omega \\
    & - \int \frac{d^3 p}{(2 \pi)^3} \left[ \frac{1}{\beta} \ln \left(1 - e^{- \beta (\omega- \mu)} \right) + \frac{1}{\beta} \ln \left( 1 - e^{- \beta (\omega + \mu)} \right) \right],
\end{split}
\end{equation}
where \(\zeta\) is determined by the minimization of the effective potential. Therefore, 
\begin{equation}
    \frac{\partial V_{\text{eff}}}{\partial \zeta} = 2 \left(m^2 - \mu^2 \right) \zeta = 0,
\end{equation}
which leads to \(\zeta = 0\) unless \(|\mu| = m\).  The physical properties of a system are often determined by an intensive variable conjugate to the chemical potential which is defined as the number density of the conserved charge. If \(|\mu| = \zeta \), the number density is
\begin{equation}
    \rho = \frac{1}{\beta V} \left( \frac{\partial \ln Z}{\partial \mu} \right)_{T, V} = -2 m \zeta ^2 + \rho^{*} ( \beta, \mu = m),
\end{equation}
where 
\begin{equation}
    \rho^{*} = \int \frac{d^3 p}{(2 \pi)^3} \left( \frac{1}{ e^{\beta (\omega- m)} -1 }  + \frac{1}{ e^{\beta (\omega + m)} -1}  \right)
\end{equation}
where the case \(\mu = - m\) is similar. If the temperature decreases with fixed number density, the chemical potential will decrease until the point in which \(|\mu| = m\) holds \cite{ Kapusta:1981aa}. If the temperature declines further, then the number density is greater than \(\rho^{*}\) which implies that
\begin{equation}
    \zeta^2 = \frac{ \rho^{*} (\beta, \mu = m) - \rho}{2 m}
\end{equation}
when the temperature is \(T < T_c\), where \(T_c\) is the critical temperature obtained by the equation
\begin{equation}
    \rho = \rho^{*} ( \beta_c, \mu = m).
\end{equation}
In the non-relativistic limit, the critical temperature is 
\begin{equation}
    T_c = \frac{2\pi}{m} \left(\frac{\rho}{\zeta(3/2)}\right)^{2/3} , \quad \rho \ll m^3.
\end{equation}
In the ultra-relativistic limit, one obtains \cite{Kapusta:1981aa}
\begin{equation}
    T_c = \left(\frac{3 \rho}{m} \right)^{1/2}  , \quad \rho \gg m^3.
\end{equation}
When the bosons are massless, all the charge resides in the condensate at all temperatures. Finally, the behavior of chemical potential as a function of the number density and the temperature indicates that a second-order phase transition takes place at the critical temperature \cite{Laine:2016hma, Kapusta:1981aa, Haber:1981ts}. Namely, the order parameter \(\zeta\) declines continuously to zero as the temperature rises from zero to the critical temperature, while \(\zeta\) vanishes for higher temperatures. At zero temperature, all the conserved charge resides in the zero-momentum mode as a Bose-Einstein condensate. However, as the temperature increases, a part of the charge is excited out of the condensate. At some point, the temperature is high enough to thermally disorder the condensate.

Could we associate a chemical potential with a gauge symmetry? We shall consider the Lagrangian density (\ref{lag_Hi}) in the Higgs model to showcase the answer to this important question. The Lagrangian density (\ref{lag_Hi}) is invariant under the \(U(1)\) gauge transformations and a conserved current exists
\begin{equation}
    j_{\mu} = i\phi \left( \partial_{\mu}  - i e A_{\mu} \right) \phi^{*} - i \phi ^{*} \left( \partial_{\mu} + i e A_{\mu} \right) \phi
\end{equation}
As a result, the chemical potential is associated with the charge,
\begin{equation}
    Q = i \int d^3 x \, \left[ \phi \left( \partial_{t}  - i e A_{0} \right) \phi^{*} -  \phi ^{*} \left( \partial_{t} + e A_{0} \right) \phi \right]
\end{equation}
Analogously, the partition function in the Higgs model can be cast into the form \cite{Kapusta:1981aa}
\begin{equation*}
   Z = \int_P \mathcal{D} A_{\mu} \, \mathcal{D}  \phi \, \mathcal{D} \phi^{*}  \, \det \left( \frac{\partial F}{\partial \omega} \right)\delta (F) \, \exp\biggl\{\int_{0}^{\beta} d \tau \int d^3 x \, \mathcal{L}_{c} \biggl\} \, .
\end{equation*}
where the usual normalization factor is omitted, the determinant relates the integration measure between different gauges and the Lagrangian density is
\begin{equation}\label{lag_L_c}
     \mathcal{L}_{c} = - \frac{1}{4} F_{\mu \nu} F^{\mu \nu} +\left(\partial^{\nu} - i eA^{\nu} - i \mu \delta^{\nu 0}\right) \phi^{*} \left(\partial_{\nu} +i e A_{\nu}+ i \mu \delta_{\nu 0} \right) \phi - m^2 \phi^{*} \phi - \lambda (\phi^{*} \phi)^2 \, .
\end{equation}
which differs from the Lagrangian density (\ref{lag_Hi}) due to the inclusion of the chemical potential. The partition function does not depend on the chemical potential if the gauge field is shifted as follows
\begin{equation}
    A_{\nu} \to A_{\nu} - \frac{\mu}{e} \delta_{\nu 0} \, .
\end{equation}
However, an additional term should be introduced in the Lagrangian density (\ref{lag_L_c}) to compensate for the charge density of the scalar field and lead to an electrically neutral system \cite{Kapusta:1981aa}. The form of this term should be \(e A^{\nu} J_{\nu}\), where \(J_{\nu} = J_{0} \delta_{\nu 0}\) is a constant background charge density. Namely, this term makes the system neutral without considering vanishing chemical potential and thermodynamic equilibrium can be established. The zeroth component of the background charge density is obtained by imposing 
\begin{equation}
    \langle \hat{A}_{\nu} \rangle = 0 \, .
\end{equation}
The complex scalar field is parameterized as
\begin{equation}\label{d_p}
    \phi = \zeta + \frac{\phi_1 + i \phi_2}{\sqrt{2}}
\end{equation}
with expectation value
\begin{equation}
   \langle \hat{ \phi} \rangle = \langle \hat{\phi^{*}} \rangle = \zeta \, .
\end{equation}
Now the equations of motion can be directly derived by the Lagrangian density (\ref{lag_L_c}) \cite{Kapusta:1981aa}:
\begin{equation}\label{eq11}
    \left[\partial^2 + 2 i \mu \partial_0 - \mu^2 + m^2 + i e (\partial^{\nu} A_{\nu}) + 2 i e A_{\nu} \partial^{\nu} - e^2 A^{\nu} A_{\nu} - 2 e \mu A_0  \right] \phi = 2 \lambda (\phi^{*} \phi)\phi 
\end{equation}
and 
\begin{equation}\label{eq22}
    - \partial^2 A^{\nu} = i e \left( \phi \, \partial^{\nu} \phi^{*} - \phi^{*} \partial^{\nu} \phi \right) + 2 e \phi \left(e A^{\nu} + \mu \delta^{\nu 0} \right) \phi^{*} + e J^{\nu} \, .
\end{equation}
Thus, we substitute (\ref{d_p}) into the equations of motion (\ref{eq11}) and (\ref{eq22}) and then compute the grand canonical average of the equations assuming no mixing to obtain
\begin{equation}
    J^{\nu} = - \mu \delta^{\nu0} \left( 2 \zeta^2 + \langle \phi_1^2 \rangle + \langle \phi_2^2 \rangle \right)
\end{equation}
and
\begin{equation}
    \zeta^2 = 0
\end{equation}
or
\begin{equation}
    \zeta^2 = \frac{1}{2 \lambda} \left( - m^2 + \mu^2 + e^2 \langle A^2 \rangle -3 \lambda \langle \phi_1^2 \rangle - 3 \lambda  \langle \phi_2^2 \rangle  \right)
\end{equation}
A similar approach was followed by Linde and Kirzhnits \cite{Kirzhnits:1976ts}, but they also added the QED Lagrangian as a fermion sector in the Higgs model. Subsequently, a non-vanishing fermion chemical potential was introduced on account of the QED conserved current, \(\overline{\psi} \gamma_{\mu} \psi\). On the other hand, the chemical potential of bosons was considered to vanish. More specifically, a bosonic current is conserved in this theory and a chemical potential can be associated with this current which is inserted in the partition function \cite{Kirzhnits:1976ts}. In this case, this additional term implies the addition of a background current in the Lagrangian density, similar to the one in the Higgs model above. However, no charged particles are present in the Lagrangian density to create this additional background source. Hence, in this theory, the chemical potential of bosons is required to vanish as a necessary condition \cite{Kirzhnits:1976ts}. In summary, in the Higgs model \cite{Kapusta:1981aa}, a chemical potential is introduced on account of a conserved current and it is included in the partition function. As a consequence, the zeroth component of the gauge field is shifted by a term that cancels the chemical potential from the partition function which then appears by the addition of a constant background current.

In the framework of the Standard Model, four currents are conserved above the electroweak critical temperature as the gauge group \(SU(2)_L \times U(1)_Y\) has four independent generators which were introduced in the first chapter. On the other hand, after spontaneous symmetry breaking the electromagnetic current is the only conventionally conserved current. In Ref. \cite{Kapusta:1981aa} the chemical potential associated with the electromagnetic current is considered. The procedure is similar to the Higgs model but by far more complex. It was shown that a second-phase transition occurs in this context and the critical temperature increases with the electric charge density neglecting the radiative corrections.

At very high temperatures, if a charge is violated at a rate lower than the Hubble expansion rate, it can be considered conserved \cite{Bodeker:2015zda}. In the Standard Model with zero neutrino masses, the baryon number \(B\) and the \(n_f\) flavor lepton numbers \(L_i\) are global conserved charges below the critical temperature of the electroweak phase transition. Above this critical temperature, the difference
\begin{equation}
   N_i = \frac{B}{n_f} - L_i
\end{equation}
is only conserved, where \(i = 1, ..., 3\) counts the fermionic generations. One could also assign a chemical potential to the charge formed by the addition of the baryon and lepton number, but electroweak sphalerons rapidly violate this charge. If the early Universe experienced a period of inflation, it can be assumed that the initial values of the conserved charges are practically zero. However, a number of mechanisms, such as leptogenesis and baryogenesis should have generated at least the non-zero charges that one observes in the current Universe, such as the baryon number. In addition, the values of the charges and equivalently the chemical potentials are small, so that the free energy\footnote{The dependence of the free energy on the chemical potential is determined by its second derivatives at vanishing chemical potential, which are known as susceptibilities.} is obtained at the lowest non-trivial order \(\mathcal{O} (\mu^2)\) \cite{Bodeker:2015zda, Khlebnikov:1996vj}. For instance, the chemical potential associated with the baryon number is estimated to have a very small cosmological value \cite{Weinberg:1974hy}, of the order \(10^{-9}\). As a result, the absence of the chemical potential does not significantly affect our calculations. 

Above the critical temperature, the hypercharge and the weak isospin are conserved gauge charges that are related to the generators of the gauge group \(SU(2)_L \times U(1)_Y\). The gauge group has four generators, which have been presented in the first chapter. However, one selects two mutually commuting conserved charges out of the four gauge charges in order to be associated with chemical potentials. The corresponding chemical potentials cannot be considered arbitrarily since in thermodynamic equilibrium the system is neutral with respect to the gauge charges. This implies that the values of these chemical potentials are fixed and functions of the temperature and the chemical potentials associated with global charges \cite{Gynther:2003za}.

Taking into consideration the above discussion, the absence of chemical potentials in the Standard Model effective potential shown in the previous chapter can be a reliable approach, especially in cosmology. This has been further motivated by the predictions of Big Bang Nucleosynthesis which rely on this assumption, while the observed baryon asymmetry of the Universe corresponds to a chemical potential \(\mu \simeq 10^{-10} T\) \cite{Laine:2016hma}.


\chapter{One-loop Beta Functions}\label{Appendix B}

In the \(\overline{\rm MS}\) renormalization scheme, the RGEs for the parameters of the real singlet extension are given by
\begin{equation}\label{beta_lamdbaH}
    16 \pi^2 \beta_{\lambda_H} = 24 \lambda^2_H - 3 \left(3 g^2 + g^{\prime 2} - 4 y^2_t \right)\lambda_H + \frac{1}{2} \lambda^2_{HS} + \frac{3}{8} \left(3 g^4 + 2 g^2 g^{\prime 2} + g^{\prime 2} \right) - 6 y^4_t
\end{equation}
\begin{equation}\label{beta_lamdbaHS}
    16 \pi^2 \beta_{\lambda_{HS}} = 4 \lambda^2_{HS} + \left(12 \lambda_H + 6 \lambda_S + 6 y^2_t - \frac{9}{2} g^2 - \frac{3}{2} g^{\prime 2}  \right) \lambda_{HS}
\end{equation}
\begin{equation}\label{beta_lamdbaS}
    16 \pi^2 \beta_{\lambda_S} = 18 \lambda^2_S + 2 \lambda^2_{HS}
\end{equation}
\begin{equation}\label{beta_g}
    16 \pi^2 \beta_{g} = -\frac{19}{6} g^3
\end{equation}
\begin{equation}\label{beta_g1}
    16 \pi^2 \beta_{g^{\prime}} = \frac{41}{6} g^{\prime 3}
\end{equation}
\begin{equation}
    16 \pi^2 \beta_{g_s} = - 7 g^3_s
\end{equation}
\begin{equation}\label{beta_yt}
    16 \pi^2 \beta_{y_t} = \frac{9}{2} y^3_t - \left( \frac{9}{4} g^2 + \frac{17}{12} g^{\prime 2} + 8 g^2_s \right) y_t
\end{equation}
where the one-loop beta function can be written as
\begin{equation}\label{beta}
    \beta_g = \mu \frac{dg}{d \mu}.
\end{equation}

\end{document}